\documentclass{jfm}
\usepackage[T1]{fontenc}
\usepackage[utf8]{inputenc}
\usepackage{graphicx}
\usepackage{epstopdf, epsfig}
\usepackage{subfigure}
\usepackage{multirow}
\usepackage{amssymb}
\usepackage{amsmath}
\usepackage{mathrsfs}
\usepackage{ esint }
\usepackage{color, soul, framed}

\shorttitle{On mixing enhancement by SBV in SBI}
\shortauthor{H. Liu, B. Yu, B. Zhang and Y. Xiang}

\title{On mixing enhancement by secondary baroclinic vorticity in shock-bubble interaction}

\author{
Hong Liu\aff{1,\corresp{\email{hongliu@sjtu.edu.cn}} },
 Bin Yu\aff{1,\corresp{\email{kianyu@sjtu.edu.cn}} },
 Bin Zhang\aff{1}
\and
Yang Xiang\aff{1}}

\affiliation{
\aff{1}School of Aeronautics and Astronautics, Shanghai Jiao Tong University, Shanghai 200240,\\
 P.R. China}

\begin{document}

\maketitle

\begin{abstract}
    To investigate the intrinsic mechanism for mixing enhancement by variable density behaviour, a canonical variable density (VD) mixing extracted from a supersonic streamwise vortex protocol, shock bubble interaction (SBI), is numerically studied and compared with a counterpart of passive scalar (PS) mixing.
    It is meaningful to observe that the maximum concentration decays much faster in VD SBI than in PS SBI regardless of the shock Mach number ($Ma=1.22\sim 4$).
    The quasi-Lamb-Oseen type velocity distribution in the PS SBI is found by analyzing the azimuthal velocity that stretches the bubble. Meanwhile, for the VD SBI, an additional stretching enhanced by the secondary baroclinic vorticity (SBV) production contributes to the faster-mixing decay.
    The underlying mechanism of the SBV-enhanced stretching is further revealed through the density and velocity difference between the shocked light bubble and the heavy ambient air.
    By combining the SBV-accelerated stretching model and the initial shock compression, a novel mixing time estimation for VD SBI is theoretically proposed by solving the advection-diffusion equation under a deformation field of an axisymmetric vortex with the additional SBV induced azimuthal velocity.
    Based on the mixing time model, a mixing enhancement number defined by the ratio of VD and PS mixing time further reveals the contribution from the variable-density effect, which implies a better control of density distribution for mixing enhancement in a supersonic streamwise vortex. \\
\end{abstract}

\begin{keywords}
Authors should not enter keywords on the manuscript, as these must be chosen by the author during the online submission process and will then be added during the typesetting process (see http://journals.cambridge.org/data/\linebreak[3]relatedlink/jfm-\linebreak[3]keywords.pdf for the full list)
\end{keywords}

\section{Introduction}
\label{sec: intro}
%%% Marble introduces the shock induced streamwise vortex and its evolution
Of particular importance to mixing enhancement design in any combustion devices is the satisfactory prediction of the time that takes fuel to a well-mixed extent.
Among all combustion propulsion-based strategies, mixing enhancement in supersonic flows of a scramjet is a notoriously challenging problem due to the extremely short residence time for mixing.
A prototype of shock-enhanced mixing is first proposed by \citet{Marble1989Progress}, and it offers a new way of utilizing shock-induced streamwise vortex formed from the baroclinic vorticity production through the oblique shock/jet interaction to enhance mixing in supersonic flow~\citep{waitz1993investigation}. The streamwise vortex not only avoids the disadvantage of reduced mixing for parallel fuel injection from Kelvin-Helmholtz instability at high convective Mach number~\citep{curran1996fluid}, but also has the potential to shorten the length of the scramjet combustor~\citep{Chan2010Numerically} due to the rapid mixing rate. Thus, understanding the effect of a supersonic streamwise vortex on mixing is vital for mixing enhancement in the scramjet.

%%%%% difficultes--> passive scalar mixing in a point vortex
Studying the mixing enhancement mechanisms under supersonic streamwise vortex suffers from complexity of three-dimensional flow structures such as shock wave structures, density gradient between fuel and ambient air \citep{urzay2018supersonic}. Therefore, passive mixing and pseudo-combustion behaviour in flow field of an ideal Lamb-Oseen type vortex was pioneeringly studied by \citet{marble1985growth}.  Unsteady mixing process as well as diffusion flame growth is combined effect of advection and diffusion controlling based on the defined P\'eclet number $\Pen=\Gamma/\mathscr{D}$, where $\Gamma$ is the circulation of a streamwise vortex and $\mathscr{D}$ is the scalar diffusivity.
\citet{cetegen1993experiments} conducted a series of experimental studies in water to produce a single two-dimensional vortex with scalar concentration. The mixing indicator `mixedness' is proposed to study the mixing enhancement extent, which is defined as $f=4c(1-c)$ where $c$ is the scalar concentration. It is found that the mixedness grows linearly with the total circulation of the vortex, and the scalar dissipation follows this trend as well, showing a derivative relationship between mixedness and scalar dissipation.
\citet{basu2007computational} further studied the scalar mixing in the gaseous laminar line vortex. Empirical correlations considering the variable temperature ratio, circulation strength and time of interaction were built.
Systematic research of mixing time in a point vortex is further extended by \cite{meunier2003vortices} who theoretically pointed out the critical dependence of mixing time to decay a passive scalar on 1/3 scaling of $\Pen$ number.
\citet{sau2007passive} conducted the first DNS simulation of passive scalar mixing in a three-dimensional vortex ring. The optimal mixing from a vortex ring is also confirmed when the vortex grows to pinch-off status \citep{gharib1998universal}, which was also validated in the formation of a supersonic vortex ring recently \citep{qin2020formation,lin2020passive}.
Such simple passive scalar mixing protocol offers ample phenomena that are resolvable to a near-exact solution, which attracts researchers and gains more in-depth mechanisms of stretching enhanced diffusion characteristic of mixing \citep{villermaux2019mixing}.

%%% Streamwise to SBI; transition to The mixing study of SBI and RMI
The passive scalar mixing behaviour in a vortical flow has been applied in the design of a lobed mixer \citep{waitz1997enhanced} or a strut mixer \citep{vergine2016turbulent}, which generates the interacting streamwise vortices to enhance mixing~\citep{wang2021kinematic}.
It is noteworthy that the flow inside scramjet combustors invariably involves inhomogenous compressible mixing.
Therefore, it is desirable to understand the role of density difference \citep{schetz2010molecular} and shock compression \citep{tew2004impact} play in mixing enhancement.
Furthermore, whether the understanding of variable density effect will lead to the better control of shock enhanced mixing strategy is still an open question.
In order to simplify the streamwise vortex production from a shocked variable density jet, \citet{yang1994model} proposed that the steady three-dimensional jet shock interaction can be an analogy to a simple two-dimensional unsteady shock bubble interaction (SBI) under the slender body approximation.
If the supersonic inflow is high enough ($Ma>3$) that the spanwise velocity is much smaller than the streamwise velocity, the dynamics of spanwise flow can be related to a two-dimensional SBI \citep{yu2020two,zhang2021numerical}.
More importantly, SBI is a canonical prototype of Richtmyer-Meshkov instability (RMI), which is a fascinating research field to investigate the fundamental variable-density mixing problem in shock-accelerated flows~\citep{brouillette2002richtmyer,ranjan2011shock}.
Hence, SBI is a suitable research target to study the vortical mixing behaviour under the condition of the compressible and variable density environment.

%%% general introduction of SBI
After shock passage across a bubble,  baroclinic vorticity is formed by the misalignment of density gradient and pressure gradient, which enhances mixing between the density inhomogeneity~\citep{zabusky1999vortex}.
Delicate vortex formation and mixing evolution appear in the SBI~\citep{ranjan2008shock}.
Despite such simple initial conditions, a wealthy of physical phenomena still occur and have been studied numerically~\citep{quirk1996dynamics} and experimentally~\citep{Layes2003Distortion}.
The secondary vortex ring in the shock-light bubble interaction~\citep{ranjan2007experimental} and the late time turbulence behaviour in the shock heavy bubble interaction~\citep{ranjan2005experimental, niederhaus2008computational} are found continuously.
Moreover, due to the characteristic of vortex structure, lots of concentration is put on the circulation model building in the study of SBI~\citep{picone1988vorticity,yang1994model,samtaney1994circulation,
niederhaus2008computational,liu2020contribution} and the related shock dynamics of SBI~\citep{zhai2011evolution,luo2015interaction,si2015experimental,Ding2017On,igra2020shock}.
\begin{table}
  \begin{center}
        \begin{tabular}{llll}
        Reference                                            & Cases & Dimensionless mixing time & Mixing status \\
        \citet{marble1990shock}            & Air-He($Ma$=1.1)      &
        $c_0t/D(Ma^2-1)<$2.86      & well-mixed    \\
        \citet{jacobs1992shock}                         & Air-He($Ma$=1.2)      & $\Gamma^{\frac{2}{3}}\mathscr{D}^{\frac{1}{3}}t/D^2<$0.25   & well-mixed    \\
        \citet{vorobieff1998power}                        &  Air-SF$_6$($Ma$=1.2)      & $u'_1t/\lambda\approx6.6$    & well-mixed   \\
        \citet{kumar2005stretching}                             & Air-SF$_6$($Ma$=1.2)     & $c_0t/R=$50 &  well-mixed     \\
        \citet{niederhaus2008computational}  & Air-He($Ma$=1.22-3)    & $u'_1t/R=$10  & well-mixed  \\
                                                                      & N$_2$-Ar($Ma$=1.33-3.38)   & $u'_1t/R=$15 & partially-mixed  \\
                                                                      & Air-Kr($Ma$=1.2-3)    & $u'_1t/R=$12  & well-mixed \\
        \citet{tomkins2008experimental}                                         &   Air-SF$_6$($Ma$=1.2)    & $\Gamma^{\frac{2}{3}}\mathscr{D}^{\frac{1}{3}}t/D^2<$0.36   &  partially-mixed \\
   % Thornber 2011  & ${At}^+$=0.49($Ma$=1.85)  &  $\Delta ut/W_0>400$   & well-mixed   \\
    \citet{oggian2015computing}  & ${At}$=0.5($Ma=$1.84)  &  ${At}^+\Delta ut/\lambda_{\min}\approx250$   & well-mixed
        \end{tabular}
        \caption{The mixing behaviour of shock accelerated flows from different literature under different dimensionless time.}
  \label{tab: comp-1}
  \end{center}
\end{table}

%%% mixing study of SBI
As for the mixing studies of SBI, two crucial characteristics of mixing behaviours are studied.
The first is to measure the extent of stirring mixing due to the vortex stretching, such as mass fraction contour area and material stretching rate.
Pioneering experimental work conducted by \citet{marble1990shock} showed that the mixing of the distorted bubble became stable at approximately 1 ms under the weak shock condition ($Ma=1.1$). They proposed a characteristic scaling time based on dimensional analysis as $D^2/\Gamma\sim D/[c_0(Ma^2-1)]$ ($c_0$ is the sound speed, $D$ is the bubble diameter, and $\Gamma$ is the vortex circulation), demonstrating that small increase of shock strength will lead to a large reduction of the mixing time.
Further, \citet{jacobs1992shock} experimentally measured the mass fraction contour area by the PLIF technique for the first time. The mixing is controlled by the dimensionless time, $\Gamma^{\frac{2}{3}}\mathscr{D}^{\frac{2}{3}}t/D^2$, proposed from passive scalar mixing by \cite{marble1985growth}. It is found that the mixing due to vortex will happen at a faster rate before a dimensionless time of 0.25 is reached.
The stretching rate of the material line contour \citep{yang1993applications} are pervasive methods to render the spatial mixing measurement and presents the exponent growth of material line at the incipient stage after shock. However, the material line contour is valid, and the material stretching rate is appropriate only when the boundary of the mass fraction contour is clear to recognise~\citep{kumar2005stretching}, i.e. dimensionless time $c_0t/R<50$ ($R$ is the radius of the bubble).
The second is to measure the extent of the molecular diffusion mixing in the presence of the baroclinic vortex formed by shock impact.
The scalar dissipation caused by concentration gradient is examined in SBI mixing by detailed PLIF measurement~\citep{tomkins2008experimental}. The stable region such as bridge structures tends to be the high mixing rate contributor. Interestingly, the unstable region such as vortex formed after the shock impact and secondary instabilities offers less mixing than the stable region due to the high strain in bridge structure. Mixing time was also estimated by the theory proposed by \citet{marble1985growth} and was concluded that mixing continues inside the theoretical prediction.

It is noteworthy that significant process has been achieved in the past 15 years in RMI theories (see \citet{zhou2021rayleigh,thornber2010influence,hahn2011richtmyer,thornber2011growth} and references therein), which is beneficial for understanding SBI mixing mechanism. \citet{vorobieff1998power} found that mixing transition happens in shock-gas curtain interaction when a dimensionless time $u_1't/\lambda\approx6.6$ is achieved ($u_1'$ is the post-shock gas speed and $\lambda$ is perturbation wavelength of gas curtain). Following \citet{vorobieff1998power}, \citet{niederhaus2008computational} investigated typical SBI cases under dimensionless time $u_1't/R$, which scales the mixing process under different shock Mach numbers in one gas pair. Recently, based on impulsive model from \citet{richtmyer1960taylor}, \citet{oggian2015computing} found that self-similar growth is obtained at ${At}^+\Delta ut/\lambda_{\min}\approx250$ in multi-mode RMI (${At}^+$ is post-shock Atwood number, $\Delta u$ is velocity impulse imparted by shock and $\lambda_{\min}$ is minimal wavelength of perturbed interface), in accordance to the simulations from~\citet{lombardini2012transition}.

%%% table1
Table \ref{tab: comp-1} summarizes the mixing behaviour of different cases from the literature investigating the shock accelerated flows. It can be found that the definition of mixing time that is widely accepted and well-predictable is absent in shock accelerated variable-density mixing.
%Dimensionless timescale, without the consideration of diffusivity, is only applied to remark the process of flow structures, which lacks the fundamental expression of mixing behaviour.
Dimensionless times, related to post-shock velocity, are able to scale the temporal mixing width or bubble morphology evolution, while they lack the inherent expression of mixing behaviour, i.e. diffusivity $\mathscr{D}$.
The mixing time proposed by \citet{marble1985growth} and \citet{meunier2003vortices}, who considered the passive scalar mixing characteristic, can hardly predict the mixing time of the concentration decay in the variable density SBI, as analysed in the present paper.
Although the time scaling dependence on Mach number proposed by \citet{marble1990shock} is beneficial in preliminary mixing enhancement design, its theoretical basis is still the passive scalar mixing, which omits the variable density contribution on the mixing enhancement.
We will show that coupling mechanism from the variable density characteristic and the shock compression \citep{giordano2006richtmyer} can offer a well-posed mixing time prediction for SBI.

%%% Motivation and scopes
%In summary, the study of coupling between flow structures and mixing process, as the core characteristic of variable density mixing, is still lacking in SBI.
In this work, we are devoted to understanding and revealing the effect of secondary baroclinic vorticity (SBV), which is ubiquitous in variable density flows, on mixing enhancement.
By focusing on shock helium bubble interaction in a wide range of shock Mach numbers from 1.22 to 4, a variable density (VD) SBI and a passive scalar (PS) counterpart are compared for the first time to show the importance of SBV accelerated stretching on mixing behaviour.
Based on solving the advection-diffusion equation under a deformation field of an axisymmetric vortex with the additional SBV increased azimuthal velocity, a theoretical model for mixing time estimation is proposed.
This model reveals the role that density difference plays in shock-accelerated inhomogeneity mixing.
Our results provide insights into the nature and mechanisms of variable density mixing in general and yield the direction for predicting a well-mixed state in supersonic flows.

The organization of the present paper is as follows: The governing equations and numerical setup are introduced in $\S$\ref{sec: num and IC}. The vortex formation after the shock impact on the bubble and the mixing process of both VD SBI and PS SBI are examined in $\S$\ref{sec: faster mixing}. Section~\ref{sec: SBV enhanced stretch} introduces the mechanism of the baroclinic accelerated stretching. Based on the analysis on the baroclinicity, the characteristic time for mixing enhancement is derived for both PS SBI and VD SBI in $\S$\ref{sec: theo}.
Further, a mixing enhancement number is proposed to illustrate the contribution from variable-density effect in~$\S$\ref{sec: modulation}.
Important conclusions are summarized in~$\S$\ref{sec: conclusions}.

\section{Numerical approach and setup}
\label{sec: num and IC}
\subsection{Governing equations and numerical method}
\label{subsec: gover eq}
The multi-component flows concerned in present study are two-dimensional shock-cylindrical bubble interaction.
The two-dimensional compressible Navier-Stokes equations with multi-components are as follows:
\begin{equation}\label{eq: NS}
    \frac{\partial \boldsymbol{U}}{\partial t}+\frac{\partial \boldsymbol{F}}{\partial x}
    +\frac{\partial \boldsymbol{G}}{\partial y}
    =\frac{\partial \boldsymbol{F}_{v}}{\partial x}+\frac{\partial \boldsymbol{G}_{v}}{\partial y},
\end{equation}
where vector $\boldsymbol{U}$ is the conserved term, and vectors $\boldsymbol{F}$, $\boldsymbol{G}$ and $\boldsymbol{F}_{v}$, $\boldsymbol{G}_{v}$ represent the convection and diffusion terms repetitively. The vectors are given by:
\begin{equation}
\left\{
        \begin{array}{l}
        \boldsymbol{U}=[\rho,\rho u,\rho v,\rho e_0,\rho Y_1,\cdots\rho Y_{NS-1}]^T,\\
        \boldsymbol{F}=[\rho,\rho u^2+p,\rho uv,(\rho e_0+p)u,\rho_1 u,\cdots\rho_{NS-1}u]^T,\\
        \boldsymbol{G}=[\rho,\rho uv,\rho v^2+p,(\rho e_0+p)v,\rho_1 v,\cdots\rho_{NS-1}v]^T,\\
        \boldsymbol{F}_v=[0,\tau_{xx},\tau_{xy},u\tau_{xx}+v\tau_{xy}-q_x,
            -\rho \mathscr{D}_1\frac{\partial Y_1}{\partial x},\cdots
            -\rho \mathscr{D}_{NS-1}\frac{\partial Y_{NS-1}}{\partial x}]^T,\\
        \boldsymbol{G}_v=[0,\tau_{xy},\tau_{yy},u\tau_{xy}+v\tau_{yy}-q_y,
            -\rho \mathscr{D}_1\frac{\partial Y_1}{\partial y},\cdots
            -\rho \mathscr{D}_{NS-1}\frac{\partial Y_{NS-1}}{\partial y}]^T,
        \end{array}
\right.
\end{equation}
where $\rho,p,e_0$ represent the mixture's density, pressure, and the total energy per unit mass respectively, $u,v$ are the speed of mixture, $\rho_i$ and $Y_i$ denote the density and mass fraction of the species $i$.
$NS$ denotes the total number of species.
The viscous tensor $\tau$ and heat flux $q$ are defined as follows:
\begin{equation}
\left\{
        \begin{array}{l}
        \tau_{xx}=\frac{2}{3}\mu\left(
                2\frac{\partial u}{\partial x}
                -\frac{\partial v}{\partial y}\right),\\
        \tau_{yy}=\frac{2}{3}\mu\left(
                2\frac{\partial v}{\partial y}
                -\frac{\partial u}{\partial x}\right),\\
        \tau_{xy}=\mu\left(
                \frac{\partial u}{\partial y}
                +\frac{\partial v}{\partial x}\right),
        \end{array}
        \quad\textrm{and }
        \begin{array}{l}
        q_x=\lambda\frac{\partial T}{\partial x}
            +\rho\sum_{i=1}^{NS}\mathscr{D}_i h_i\frac{\partial Y_i}{\partial x},\\
        q_y=\lambda\frac{\partial T}{\partial y}
            +\rho\sum_{i=1}^{NS}\mathscr{D}_i h_i\frac{\partial Y_i}{\partial y},
        \end{array}
\right.
\end{equation}
in which $T$ is the temperature, and $h_i$ is the specific enthalpy of species $i$. $\mu,\lambda$ are the dynamic viscosity and thermal conductivity of the mixed gas, given by the Wike's semi-empirical formula~\citep{Gupta1989A}. For high-speed flow numerical simulations, mass diffusion can be simplified by ignoring pressure and temperature diffusion, and it is assumed to be constant for different components:
\begin{equation}
    \mathscr{D}_i=\mathscr{D}=\frac{\mu}{\rho Sc}.
\end{equation}
Moreover, the Schmidt number is assumed as constant $Sc=0.5$ \citep{Gupta1989A}.
For the thermal conductivity, $\lambda$ is calculated from the Prandtl number as $Pr=0.72$ \citep{houim2011low}.
In order to close the system of equations, the equation of ideal gas state is needed:
\begin{equation}
    p=\rho R_u T\sum_{i=1}^{NS}\frac{Y_i}{M_i},
\end{equation}
where $R_u$ is the ideal gas constant and $M_i$ is the molar mass of component $i$.
The total energy $e_0$, that contains the specific internal energy $e$ and kinetic energy, is given by:
%\begin{equation}
%    E=H-\frac{p}{\rho}=\sum_{i=1}^{NS}Y_i\left(h_{fi}^0+\int_{T_0}^{T}\frac{C_{pi}}{M_i}\mathrm{d}T\right)
%    -\frac{p}{\rho}+\frac{1}{2}\left(u^2+v^2\right),
%\end{equation}
\begin{equation}
    e_0=e+\frac{1}{2}\left(u^2+v^2\right)=h-\frac{p}{\rho}+\frac{1}{2}\left(u^2+v^2\right).
\end{equation}
Here, $h$ is the specific enthalpy of the gas mixture:
\begin{equation}
  h=\sum_{i=1}^{NS}Y_i h_i=\sum_{i=1}^{NS}Y_i\left(h_{fi}^0+\int_{T_0}^{T}\frac{C_{pi}}{M_i}\mathrm{d}T\right),
\end{equation}
in which $h_{fi}^0$ is heat generated by the component $i$ at the reference temperature $T_0$. The constant-pressure specific heat value $C_{pi}$ is fitted by the following temperature-dependent polynomial function:
\begin{equation}
    C_{pi}=R_u\left(a_{1i}+a_{2i}T+a_{3i}T^2+a_{4i}T^3+a_{5i}T^4\right),
\end{equation}
where the coefficients $a_{1i},\cdots,a_{5i}$ can be obtained from the NASA thermochemical polynomial fit coefficient data~\citep{kee1996chemkin}.
After the mathematical model is non-dimensionalized, the finite volume method is used for discretization.
Time is marched by the third-order TVD Runge-Kutta method~\citep{gottlieb1998total}; convection terms are discretized using the fifth-order WENO scheme~\citep{liu1994weighted,jiang1996efficient}, while viscous terms are discretized by using the central difference method.
The general platform of high-resolution calculation method in present study is implemented as in-house code \emph{ParNS}, which has been validated in our previous work and is suitable for this study~\citep{wang2018scaling,li2019gaussian,liu2020optimal}.
Moreover, the numerical uncertainty originated from differen WENO schemes is essential for capturing the physical values, which is discussed in appendix~\ref{App: WENO}.

\subsection{Initial conditions for VD SBI}
\label{subsec: IC}
\begin{figure}
  \centering
  \includegraphics[clip=true,trim=0 180 200 0,width=0.8\textwidth]{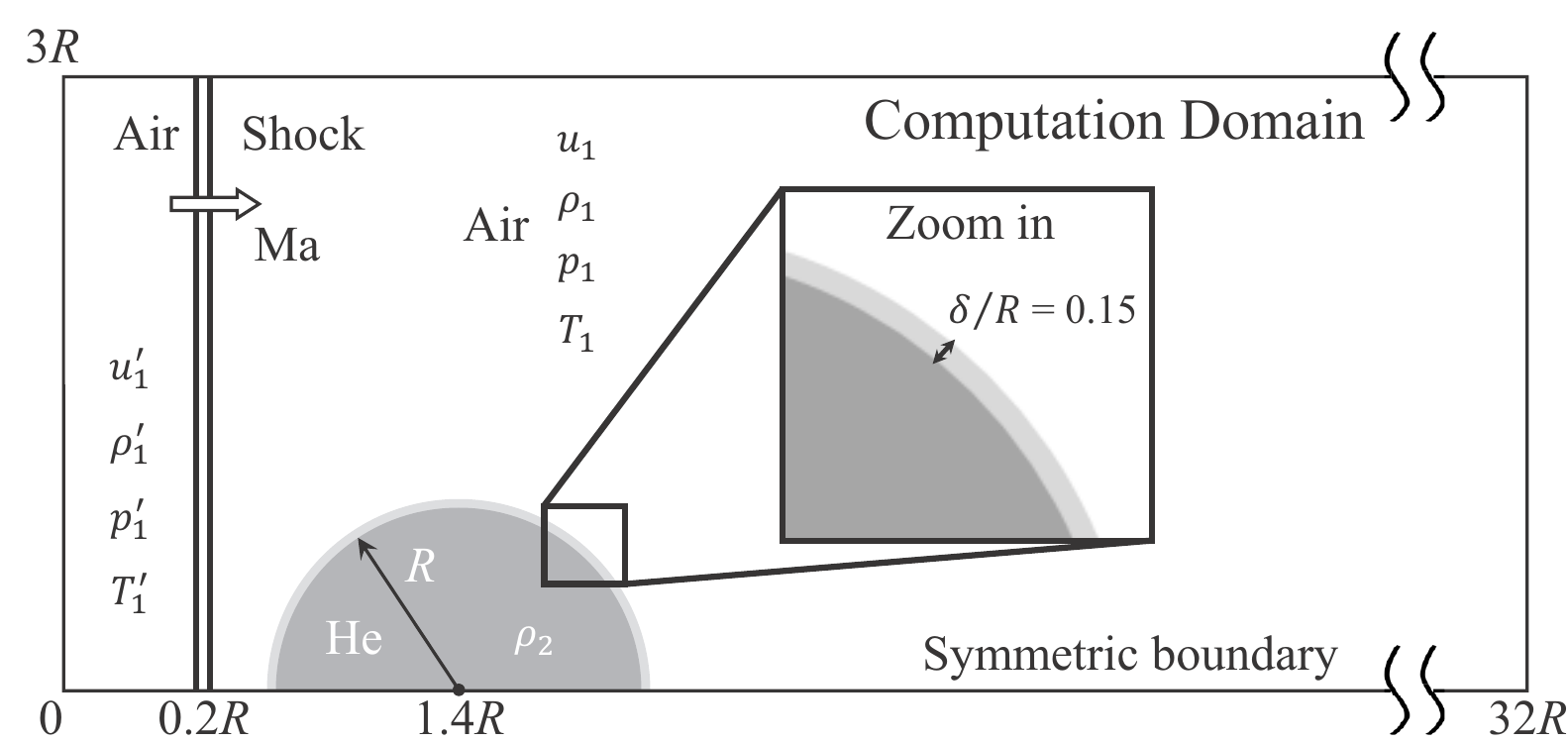}\\
  \caption{Schematic of the initial conditions for SBI.}\label{fig:initial-conditions}
\end{figure}
The schematic of the initial conditions of VD SBI is depicted in figure \ref{fig:initial-conditions}.
The cylindrical bubble is filled with pure helium.
A wide range of shock $Ma$ number cases ($Ma$ = 1.22 to 4) are simulated.
Same pre-shock conditions for all cases are quiescent air at $p_1=101325$~Pa and $T_1=293$~K.
The detailed initial conditions for post-shock air, obtained from one-dimensional shock dynamics, are listed in table~\ref{tab:initial-condi}.
To avoid spurious vorticity generated by the mesh discretization \citep{niederhaus2008computational}, we set the initial interface of the cylindrical bubble smoothed by a diffuse interfacial transition layer defined as:
\begin{equation}\label{eq:Gaussian}
    {Y_{\textrm{He}}}(r)=\left\{ \begin{array}{ll}
    Y_{\max} & {r\leq{R}}, \\
    Y_{\max}\mathrm{e}^{-\alpha[(r-R)/\delta]^2} & {R<r\leq{R+\delta}}, \\
    0 & r>R+\delta,
    \end{array}\right.
\end{equation}
where $\delta/R=0.15$, $\alpha=5$, and $R=2.6$ mm.
Since the reflecting shock from the top wall will influence the cylindrical bubble formation and mixing process, we mainly focus on the mixing mechanism from the once-shocked interface in a shock-free environment. Thus, the interpolation boundary conditions are chosen at the top and two sides of the computational domain. The centerline is chosen as the symmetry boundary condition.
The calculation domain in the streamwise direction $L=32R$ ($R$ is the radius of the cylindrical bubble) is longer than most other studies to study the long time evolution of bubble deformation and mixing growth. The calculation domain in the spanwise direction $H=3R$ is sufficient to avoid the reflected waves from the upper side. The mesh independence study can be found in appendix~\ref{App: mesh}. Here, using the mesh resolution $\Delta=2.5\times10^{-5}$ m leads to grid independence for the problem studied in this paper.
\begin{table}
  \begin{center}
\def~{\hphantom{0}}
  \begin{tabular}{lcccccc}
      $Ma$  & ${p}_1'$(Pa)   &   ${T}_1'$(K) & ${u}_1'$(m/s)  & $W_t$(m/s) & ${At}^+$ (-) \\[3pt]
       1.22   & 159036.9   & 334.06   &  114.71 & 1121.89  &  $-0.789$ \\
       1.8     & 366015.0   & 448.28   &  356.58 & 1400.00  &  $-0.831$ \\
       2.4     & 663791.3   & 596.87   &  568.31 & 1678.08  &  $-0.845$ \\
       3        & 1046646.5   & 783.41   &  764.11 & 1958.78 &  $-0.852$  \\
       4        & 1873802.9   & 1182.91   &  1074.53 & 2439.15  &  $-0.852$ \\
  \end{tabular}
  \caption{Parameters for different shock Mach number cases, including post-shock pressure $p_1'$, post-shock temperature $T_1'$, and the transmitted shock wave speed $W_t$, calculated from one-dimensional gas dynamics. Post-shock Atwood number ${At}^+=(\rho_2'-\rho_1')/(\rho_2'+\rho_1')$ is obtained from post-shock air density $\rho_1'$ and post-shock helium density $\rho_2'$. }
  \label{tab:initial-condi}
  \end{center}
\end{table}

\subsection{Initial conditions for PS SBI}
\label{subsec: IC for PS}
\begin{figure}
  \centering
  \includegraphics[clip=true,trim=0 250 0 0,width=.98\textwidth]{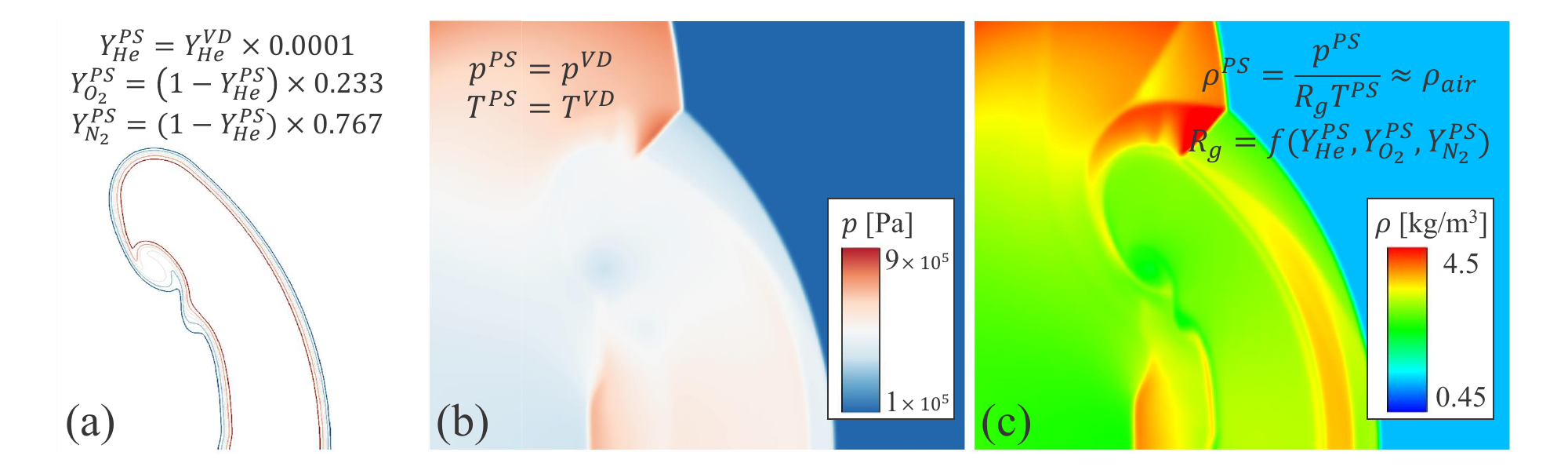}\\
  \caption{Initial conditions for PS SBI shown by (a) isolines of mass fraction, (b) pressure contour, and (c) density contour at $t=7.2$~$\umu$s of $Ma$=2.4 case.}\label{fig: PS_init}
\end{figure}

%%%  目的；方法；如何验证
To study the variable density effect on mixing, we set a counterpart of the PS SBI without density gradient to be compared with the original VD SBI.
One needs to maintain the same circulation, compression, and diffusivity as the SBI cases except that the density is desired to be the same as the shocked ambient air to illustrate the passive scalar mixing under a vortical flow. In that scenario, the density effect can be nearly ignored when the vortex grows, and the mixing process will not alter the flow dynamics as the passive scalar mixing~\citep{dimotakis2005turbulent}.
However, if a PS bubble is set before the shock impact, no vorticity will be deposited along with the bubble interface for the absence of density gradient between the cylindrical bubble and ambient air, which is essential for the baroclinic vorticity production.
Thus, a natural selection is to artificially make the cylindrical bubble's mass component equal to the air immediately after the shock impact, so the bubble density will increase to the value of the shocked air. Moreover, a small amount of helium is maintained to reflect the scalar mixing:
\begin{equation}\label{eq: initial condi for PS}
    \left\{ \begin{array}{l}
    Y_{\textrm{He}}^{ps}=Y_{\textrm{He}}^{vd}\times 0.0001 , \\
    Y_{\textrm{O}_2}^{ps}=(1-Y_{\textrm{He}}^{ps})\times 0.233,\\
    Y_{\textrm{N}_2}^{ps}=(1-Y_{\textrm{He}}^{ps})\times 0.767.
    \end{array}\right.
\end{equation}
If pressure and temperature are set the same as the VD SBI case, one obtains:
\begin{equation}\label{eq: IC for PS}
    \left\{ \begin{array}{l}
    p^{ps}=p^{vd},\\
    T^{ps}=T^{vd}.
    \end{array}\right.
\end{equation}
Then, the bubble density will rise to a value similar to the shocked air, as shown in figure~\ref{fig: PS_init}:
\begin{equation}\label{eq: IC for PS-2}
  \rho^{ps}=\frac{p^{ps}}{R_gT^{ps}}\approx \rho_{air} \textrm{   and   } R_g=R_u \sum_{i=1}^{NS}\frac{Y_i^{ps}}{M_i}.
\end{equation}
This set of values of $p,\rho,T$ and $Y_i$ compose the initial conditions for the PS SBI.
The calculation of the PS bubble diffusivity is introduced separately in appendix~\ref{App: diffusivity} to meet the requirement of the same level of diffusivity with a variable density bubble.
The discretized version of NS equations (\ref{eq: NS}) are then solved numerically by the same way as introduced in $\S$\ref{subsec: gover eq}. The flow will evolve without an evident density gradient due to the low concentration of helium.

Since the passive scalar mixing obeys advection-diffusion equation \citep{villermaux2019mixing}, that is different from species transport in the NS equations (\ref{eq: NS}), we here prove that if the density gradient can be eliminated, the multi-components transport equations can degenerate to the advection-diffusion equation, which offers the basis for studying the passive scalar mixing of SBI.
Referring back to (\ref{eq: NS}), the multi-components transport equations for an arbitrary component follow the Fickian's law of diffusion, which can be expressed as:
\begin{equation}\label{eq: species trans}
  \frac{\partial(\rho Y_s)}{\partial t}+\frac{\partial(\rho u_j Y_s)}{\partial x_j}=\frac{\partial}{\partial x_j}
  \left(\rho\mathscr{D} \frac{\partial Y_s}{\partial x_j} \right). \\
\end{equation}
When assuming the constant diffusion, it can be further derived as:
\begin{equation}\label{eq: species trans2}
  Y_s\left(\frac{\partial\rho}{\partial t}+\frac{\partial (\rho u_j) }{\partial x_j}\right)+
  \rho\left(\frac{\partial Y_s}{\partial t}+u_j\frac{\partial Y_s}{\partial x_j}-\mathscr{D}\frac{\partial^2 Y_s}{\partial x_j^2}\right)=\mathscr{D}\frac{\partial Y_s}{\partial x_j}\frac{\partial\rho}{\partial x_j}.
\end{equation}
Due to the conservation of mass, the term of the first parenthesis is zero, then we obtain:
\begin{equation}\label{eq: species trans3}
  \frac{\partial Y_s}{\partial t}+u_j\frac{\partial Y_s}{\partial x_j}-\mathscr{D}\frac{\partial^2 Y_s}{\partial x_j^2}=
  \frac{\mathscr{D}}{\rho} \left(\frac{\partial Y_s}{\partial x_j}\frac{\partial\rho}{\partial x_j}\right).
\end{equation}
Thus, when density gradient $\nabla\rho$ is negligible, the species equation can degenerate to the standard advection-diffusion equation.
%Also it can be noted that when light gas is considered, density gradient and mass fraction gradient is reversed to each other, which makes the right term in (\ref{eq: species trans3}) strictly negative. In that case, the mass fraction will be dissipated in a faster pattern by the additional right term in variable density flows than in passive scalar mixing, which is defined as density gradient accelerated mixing by \cite{yu2020scaling}. This is another evidence for the faster decay of maximum mass fraction observed in $\S$3.

\begin{figure}
    \centering
    \subfigure[]{
    \label{fig: cir-PS-VD}
    \includegraphics[clip=true,trim=15 15 40 50, width=.43\textwidth]{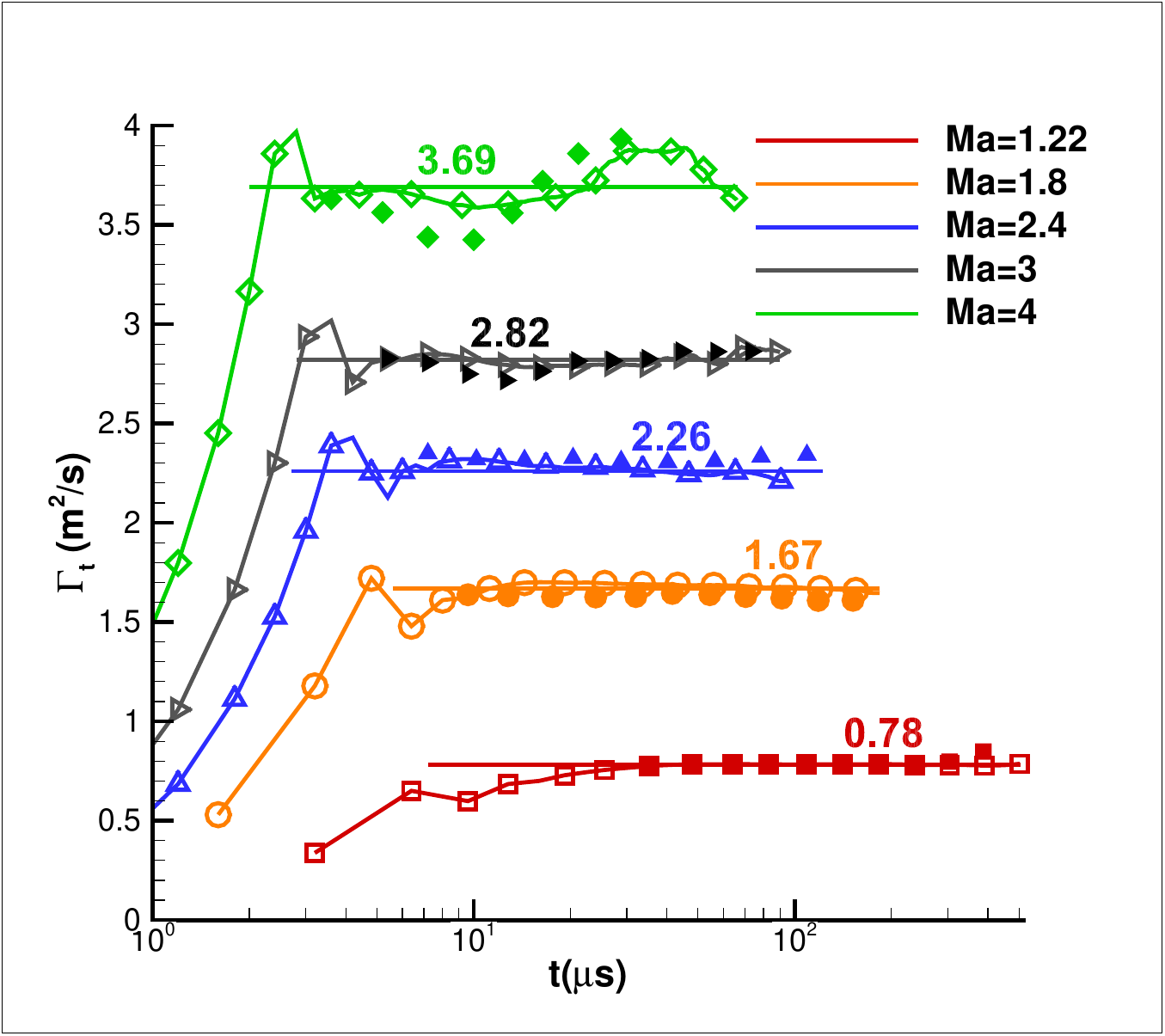}}
    \subfigure[]{
    \label{fig: com-PS-VD}
    \includegraphics[clip=true,trim=15 15 40 50,width=.43\textwidth]{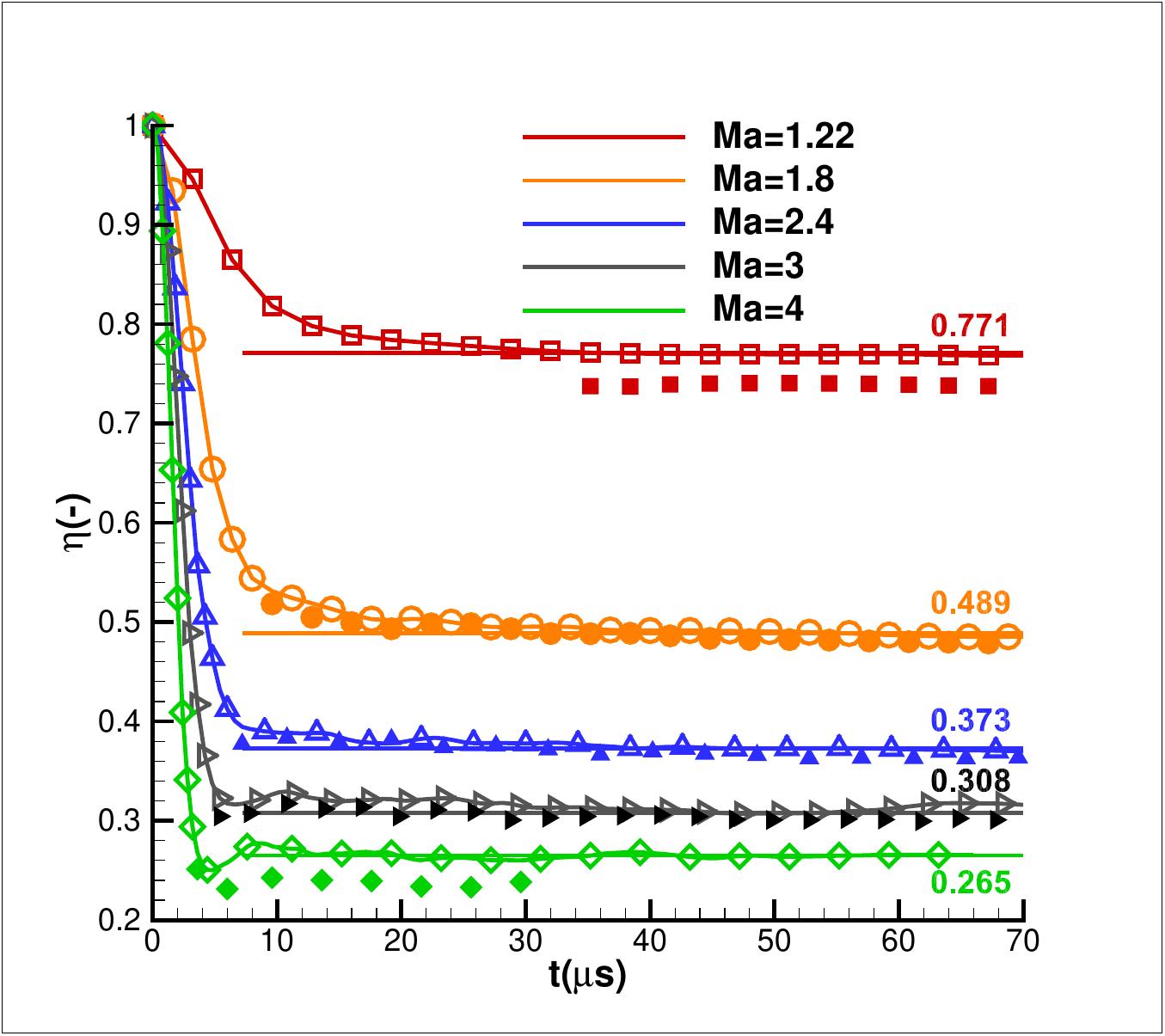}}
    \caption{Comparison of circulation (a) and compression rate (b) for PS (solid points) and VD SBI (hollow points) for different shock Mach number cases.  \label{fig: cir-com} }
\end{figure}

To check whether the circulation and compression are the same as in the variable density cases, we compare these two variables between PS SBI and VD SBI, as shown in figure~\ref{fig: cir-com}.
The total circulation is calculated as:
\begin{equation}\label{eq: circu}
    \Gamma_t=\int_{Y_{\textrm{He}}<1\%Y_{\max}} \omega \textrm{d}V,
\end{equation}
where $\omega$ is the magnitude of vorticity vector $\boldsymbol{\omega}=\nabla\times\boldsymbol{u}$.
The compression rate is calculated as \citep{giordano2006richtmyer}:
\begin{equation}\label{eq: com}
    \eta=\frac{\int \mathcal{V}_\textrm{He} \textrm{d}V}{\mathcal{V}_0}\equiv\frac{\mathcal{V}_b}{\mathcal{V}_0},
\end{equation}
where $\mathcal{V}_\textrm{He}$ is the volume fraction of the helium cylindrical bubble and $\mathcal{V}_b$ is the volume of compressed cylindrical bubble. For the initial volume of the PS bubble $\mathcal{V}_0$, it is calculated in the same way by decreasing the helium mass fraction multiplied by 0.0001.
Figure \ref{fig: cir-com} shows that a higher shock Mach number leads to higher circulation and lower compression rate.
The circulation and compression for both PS SBI and VD SBI are nearly the same, facilitating the study of variable-density effect on mixing behaviour.
The fluctuation of circulation magnitude at the late time, especially for high Mach number, is because the boundary chosen based on the mass fraction threshold contains negative vorticity forming from the slipstream of the triple point of the Mach stem. However, this negative vorticity has little effect on mixing.

\section{Faster mixing decay in VD SBI}
\label{sec: faster mixing}
\subsection{Temporal evolution of cylindrical bubble morphology}
\label{subsec: bubble morph}
\begin{figure}
  \centering
  \includegraphics[clip=true,trim=0 65 0 0,width=1.0\textwidth]{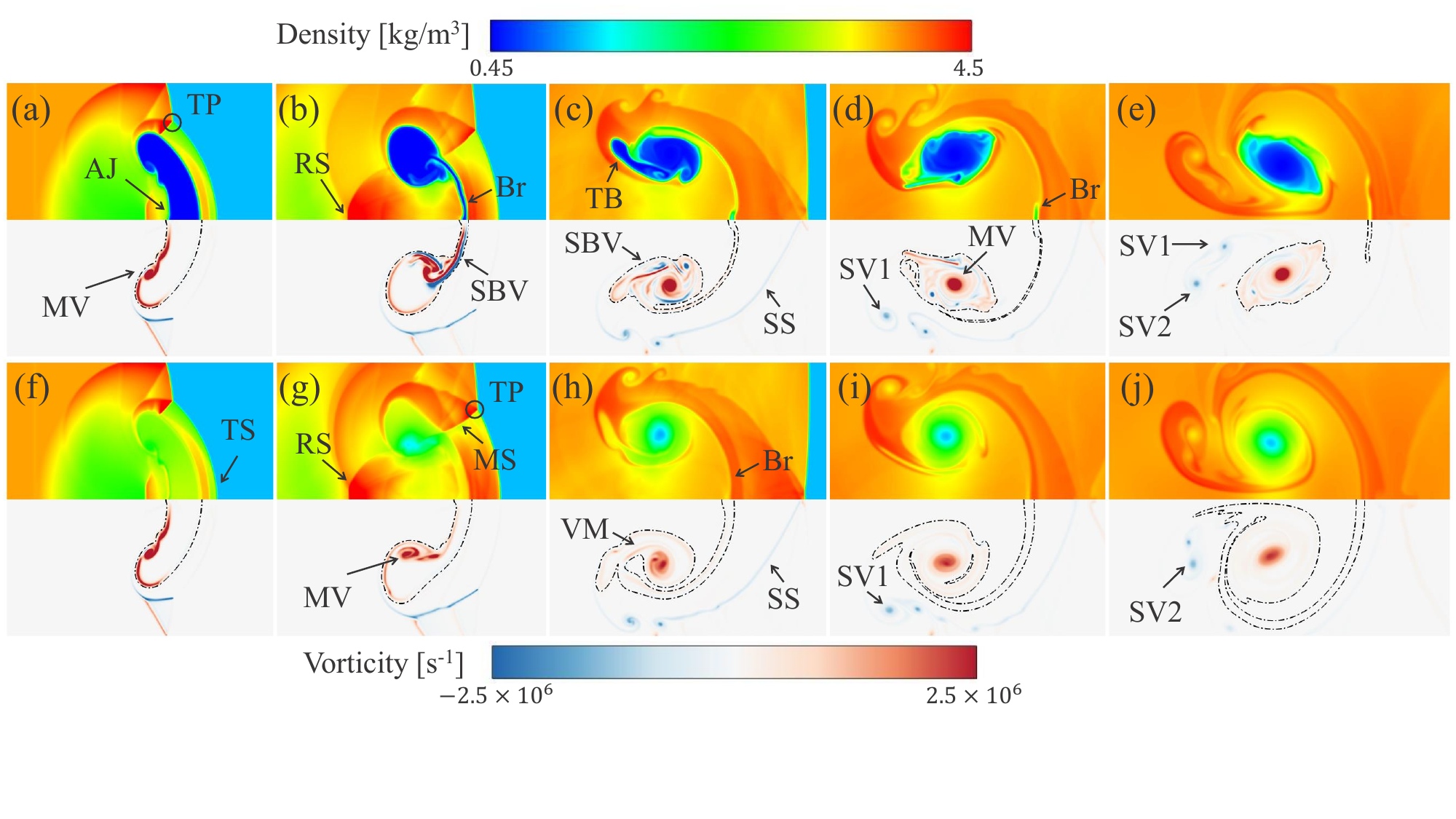}\\
  \caption{Density contour (top) and vorticity contour (bottom) of $Ma$=2.4 VD SBI case at (a) $t=7.2$~$\umu$s; (b) $t=13.2$~$\umu$s; (c) $t=27.6$~$\umu$s; (d) $t=42$~$\umu$s; (e) $t=69.6$~$\umu$s.
  $Ma$=2.4 PS SBI case is presented at the same moments as that of VD SBI in the bottom line from (f) to (j).
  Dashed dot line is the isoline of $Y_{\textrm{He}}=1\%Y_{\max}^0$ ($Y_{\max}^0=1$ for VD SBI and $Y_{\max}^0=0.0001$ for PS SBI). TP, triple point; AJ, air jet; TS, transmitted shock; MV, main vortex; RS, relfected shock; MS, mach stem; Br, bridge; SBV, secondary baroclinic vorticity; VM, vorticity merging; TB, trailing bubble; SS, slip stream; SV1 and SV2, secondary vortex 1 and secondary vortex 2.}\label{fig: den-vor-VD}
\end{figure}

%%VD：激波结构-->气泡形态变化（与低密度区域紧密相关，补充浓度梯度与密度梯度紧密相关公式）--> 涡结构特征--> 环量守恒性；
%PS：激波结构类似--> 气泡形态变化也类似（但此时密度变化与浓度变化不明显了，密度变化与涡结构紧密相关，涡心位置密度较小，符合一般涡结构特征）
The temporal evolution of the shocked cylindrical bubble morphology is generally similar for different shock strengths~\citep{Bagabir2001Mach}. Here, the impact of a shock with $Ma=2.4$ is particularly examined.
For shock-related structures in VD SBI and PS SBI, canonical shock wave structures such as Mach stem (MS), reflected shock (RS), and triple points (TP) are observed in figure~\ref{fig: den-vor-VD}.
As for the bubble deformation, the air jet (AJ) connects with the bubble's downstream edge, forming the bridge structures (Br), as reported in the shock-heavy bubble interaction \citep{tomkins2008experimental}. A trailing bubble (TB) is attached after the main vortex and gradually stirred, as illustrated by figures~\ref{fig: den-vor-VD}(c-d). Finally, the bubble turns into a vortex pair, as indicated by figures \ref{fig: den-vor-VD}(d) and \ref{fig: den-vor-VD}(e). The flow structures of PS and VD SBI are quite similar in general.

It is noteworthy that the low-density region in VD SBI is closely related to the helium mass fraction $Y$. After the shock impact, the flow can be assumed as incompressible variable-density flow. Therefore, the mass fraction is only a function of density $Y=Y(\rho)$ \citep{weber2012turbulent}:
\begin{equation}\label{eq: den-MF}
  \frac{1}{\rho}=\frac{Y}{\rho_2}+\frac{1-Y}{\rho_1}\Rightarrow \nabla Y=\frac{\rho_1\rho_2}{\rho_1-\rho_2}\frac{\nabla \rho}{\rho^2}.
\end{equation}
This equation shows that when mixing is happening, the bubble's density increases due to the stirring from the shocked ambient air with high density. The relation between mass fraction and density is the main characteristic of VD flows: the mixing process changes the flow structures, i.e. density distribution.
However, the density distribution in PS SBI can hardly be related to concentration and mixing process but is related to vortex formation since it follows the typical compressible vortex formation as studied by \citet{moore1987compressible}: low density at the vortex centre, as illustrated in figure~\ref{fig: den-vor-VD}(h-j).

%%  vorticity structure
%%PS：涡结构特征（无SBV存在）-->环量守恒性；
In VD SBI, the main baroclinic vorticity is produced from the misalignment of the density gradient of bubble and pressure gradient from the shock, as shown in the vorticity contour in the lower half of figure~\ref{fig: den-vor-VD}.
During the vortex growth of VD SBI, the vortex-bilayer structure, which exhibits the dominant-negative vorticity intensifying and rolling up the positive vorticity, occurs~\citep{gupta2003shock}.
\citet{peng2003vortex} named it the secondary baroclinic vorticity (SBV), which plays a dominant role in the flows structures in VD flows~\citep{peng2021mechanism,peng_yang_xiao_2021}.
However, different from the VD case, in the PS case only the initial positive vorticity without SBV merges (as indicated by ``VM'') into the main vortex at the late time.
Therefore, it can be observed that the vortex structure grows more steadily in PS SBI than in VD SBI from $t=42$~$\umu$s to $t=69.6$~$\umu$s: single vortex in PS SBI comparing to main vortex with negative SBV around.
This difference in the negative SBV behaviour in PS SBI and VD SBI will reveal the intricate mechanism of VD mixing, as discussed in $\S$\ref{sec: SBV enhanced stretch}.

\subsection{Mixing characteristic of PS and VD SBI}
\label{subsec: mixing chara of PS and VD}
%%  混合指标定义；VD混合特征（结合下线图）；PS 混合特征（结合线图）；总结
%% 混合指标定义
\begin{figure}
  \centering
  \includegraphics[clip=true,trim=0 65 0 0,width=1.0\textwidth]{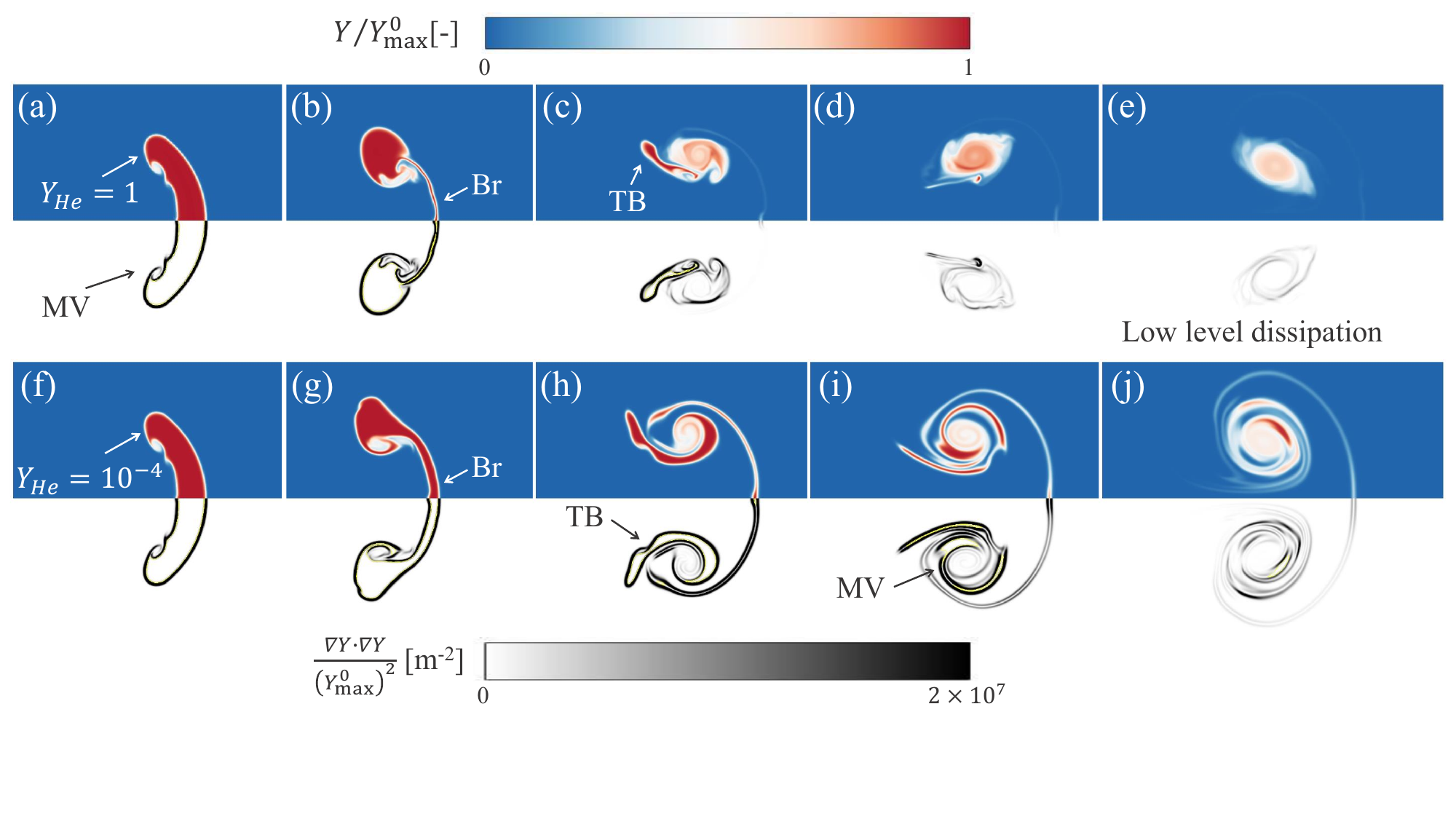}\\
  \caption{Normalised mass fraction (top) and scalar dissipation rate (bottom) of $Ma$=2.4 VD SBI case from (a) to (e).
  $Ma$=2.4 PS SBI case is presented in the bottom line from (f) to (j). The moments captured are the same ones as that of figure~\ref{fig: den-vor-VD}. }\label{fig: SDRMF-VD}
\end{figure}
As for mixing characteristics, it is appropriate to check the canonical advection-diffusion equation to obtain an objective mixing indicator:
\begin{equation}\label{eq: ADE}
  \left[\frac{\partial}{\partial t}+\boldsymbol{u}\cdot \nabla -\mathscr{D}\nabla^2\right]Y(x,t)=0.
\end{equation}
The scalar energy $\frac{1}{2}Y^2(x,t)$ behaviour can be obtained accordingly:
\begin{equation}\label{eq: scalar energy}
  \left[\frac{\partial}{\partial t}+\boldsymbol{u}\cdot \nabla -\mathscr{D}\nabla^2\right]\frac{1}{2}Y^2(x,t)=
              -\mathscr{D}\nabla Y(x,t)\cdot\nabla Y(x,t).
\end{equation}
The right term of this equation is defined as the scalar dissipation, which generally indicates the mixing rate, introduced by \citet{buch1996experimental}. Also, scalar dissipation was studied by \citet{tomkins2008experimental} to investigate the mixing behaviour of shock-heavy bubble interaction.
Here, the time history of scalar dissipation is studied by defining its area integral normalised by the maximum helium mass fraction at $t=0$ \citep{shankar2011two}:
\begin{equation}\label{eq:TMR}
  \chi=\frac{1}{{\left(Y^0_{\textrm{max}}\right)}^2}\int\nabla{Y}\cdot\nabla{Y} \mathrm{d}{V}.
\end{equation}
Another mixing indicator investigated in this study is the normalised maximum concentration:
\begin{equation}\label{eq: NMF}
  \overline{Y}=\frac{Y_{\max}}{Y_{\max}^0}.
\end{equation}
\citet{meunier2003vortices} found that the mixing time is reached when $\overline{Y}<1$, meaning that mixing turns from stretching enhancement to quasi-equilibrium diffusion stage. Moreover, the maximum concentration can illustrate the mixing performance in a scramjet combustor \citep{waitz1993investigation,lee1997computational}. Here, we use these two indicators to investigate the mixing in VD and PS SBI.

Figure~\ref{fig: SDRMF-VD} compares the temporal evolution of the helium mass fraction and scalar dissipation rate between VD and PS SBI. The general trend of scalar dissipation is increasing at the beginning and decreasing after the Br and TB structures are dissipated by vortex stretching into a steady-state of `well-mixed' stage.
From the normalised maximum helium concentration, $\overline{Y}$ maintains a value of 1 during the stage of $\chi$ with high value, meaning that the high scalar dissipation rate region is closely related to the maximum concentration. After maximum concentration is stretched, the vortical mixing enters the diffusion controlling regime. Therefore, the most evident difference between VD and PS cases from the mixing indicators evolution is that the maximum concentration attains a steady state in the VD case much faster than in the PS case. Thus, the SBI mixing time can be objectively defined when maximum helium concentration decreases from one, and low-level dissipation is attained.

\subsection{Faster decay of VD mixing for different shock Mach numbers}
\label{subsec: faster decay}
\begin{figure}
    \centering
    \subfigure[]{
    \label{fig: cir-PS-VD-2.4}
    \includegraphics[clip=true,trim=15 15 15 50, width=.5\textwidth]{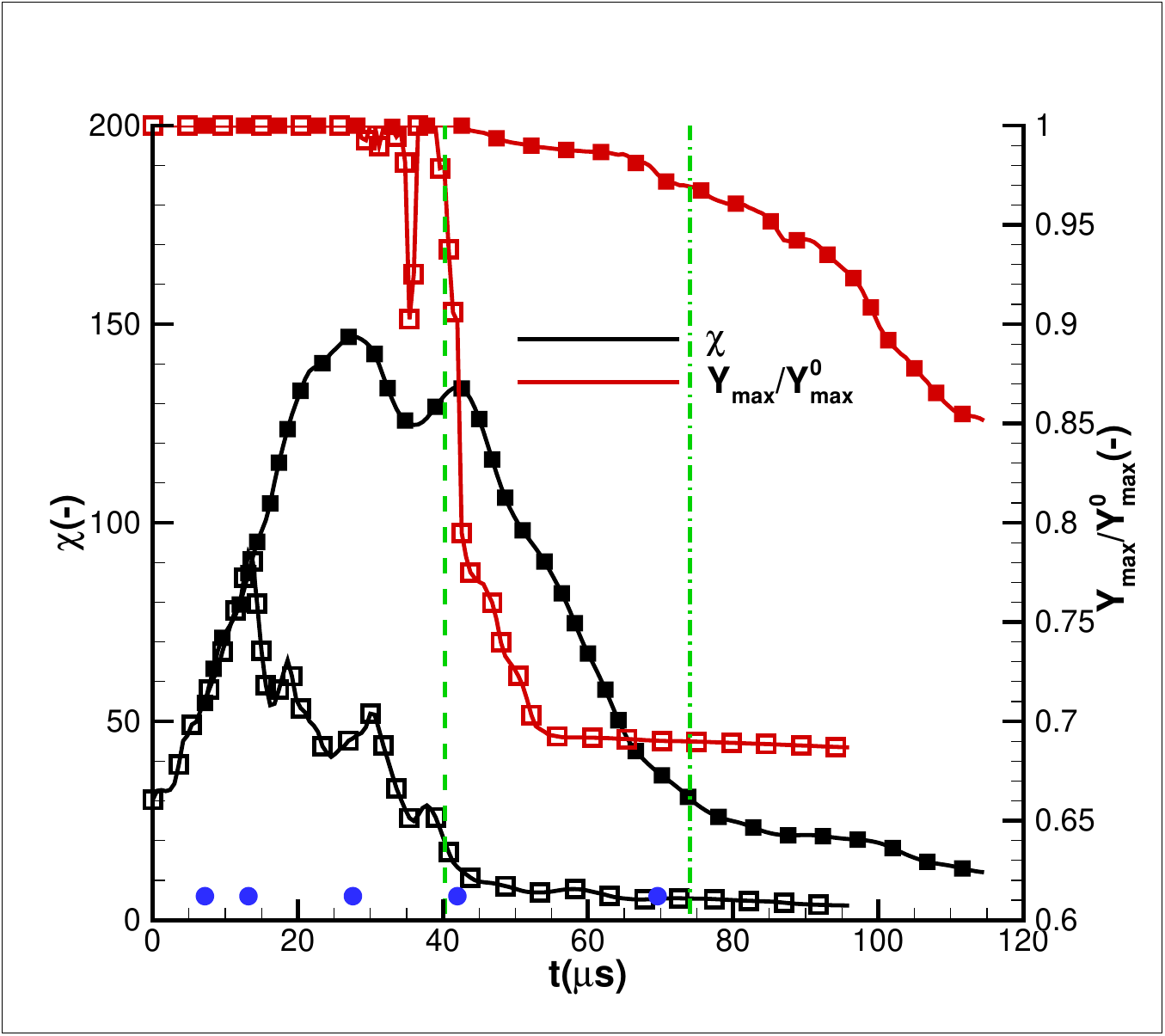}} \\
    \subfigure[]{
    \label{fig: cir-PS-VD-1.22}
    \includegraphics[clip=true,trim=15 15 15 50, width=.35\textwidth]{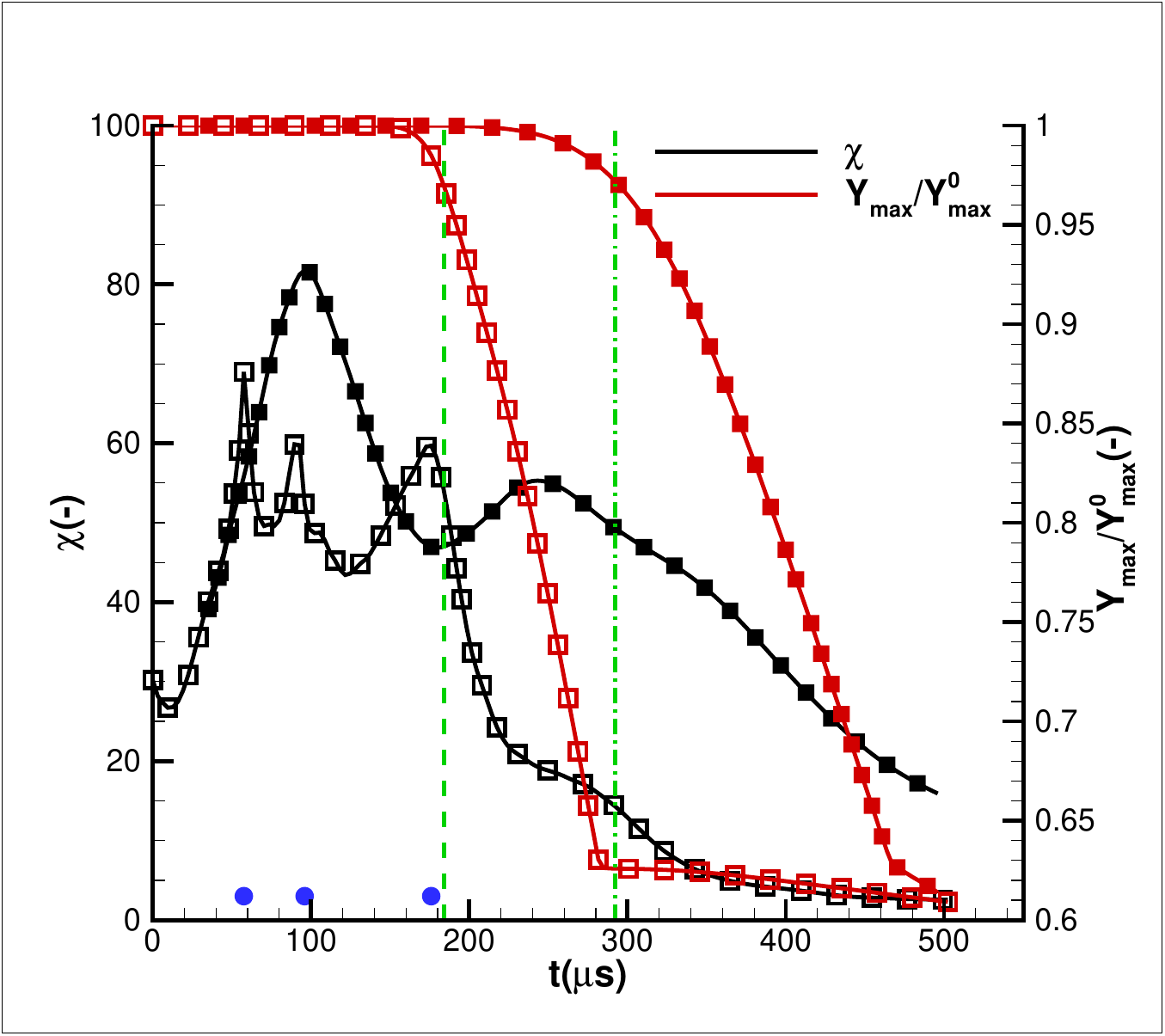}}
    \subfigure[]{
    \label{fig: cir-PS-VD-1.8}
    \includegraphics[clip=true,trim=15 15 15 50, width=.35\textwidth]{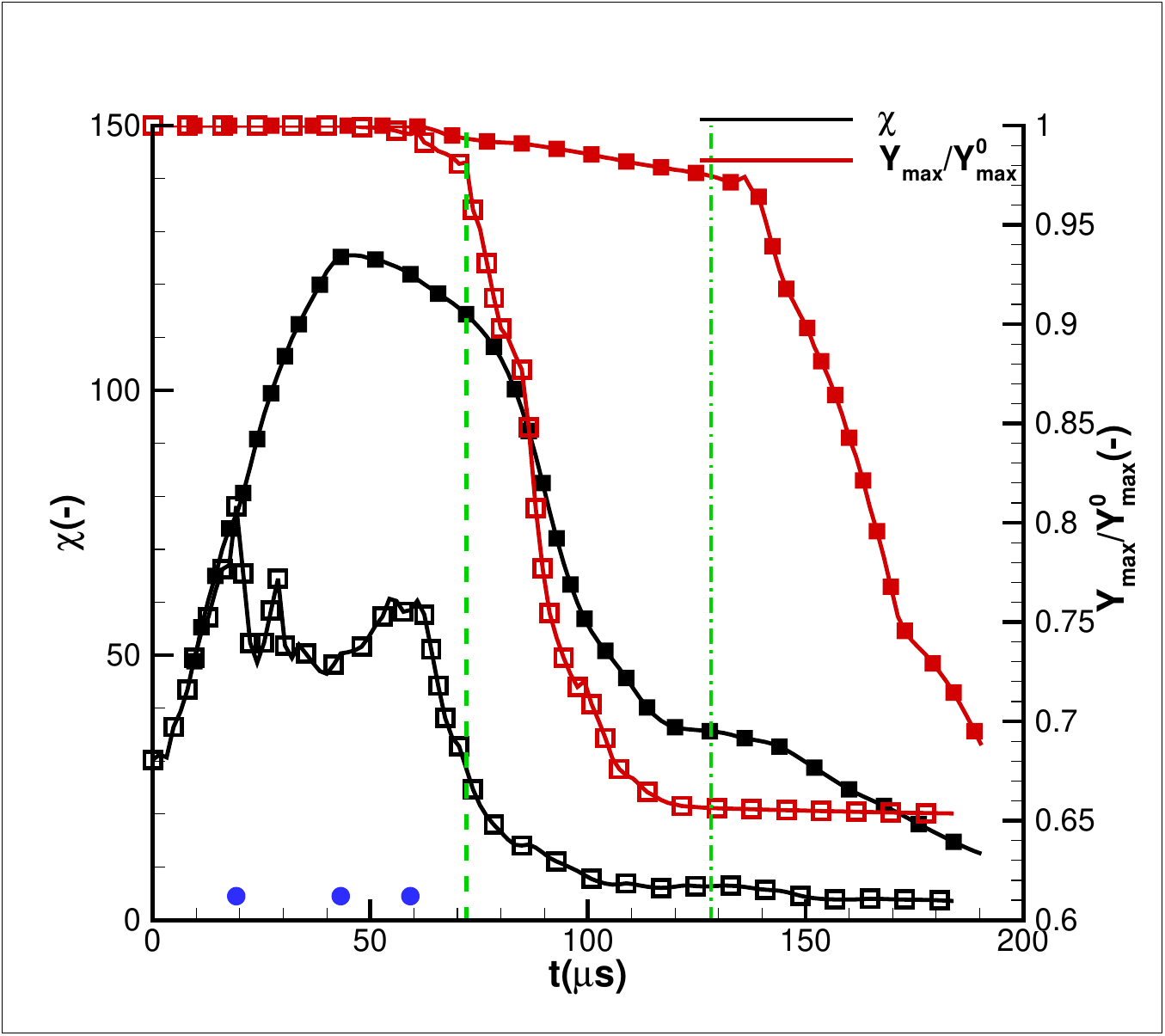}} \\
    \subfigure[]{
    \label{fig: cir-PS-VD-3}
    \includegraphics[clip=true,trim=15 15 15 50, width=.35\textwidth]{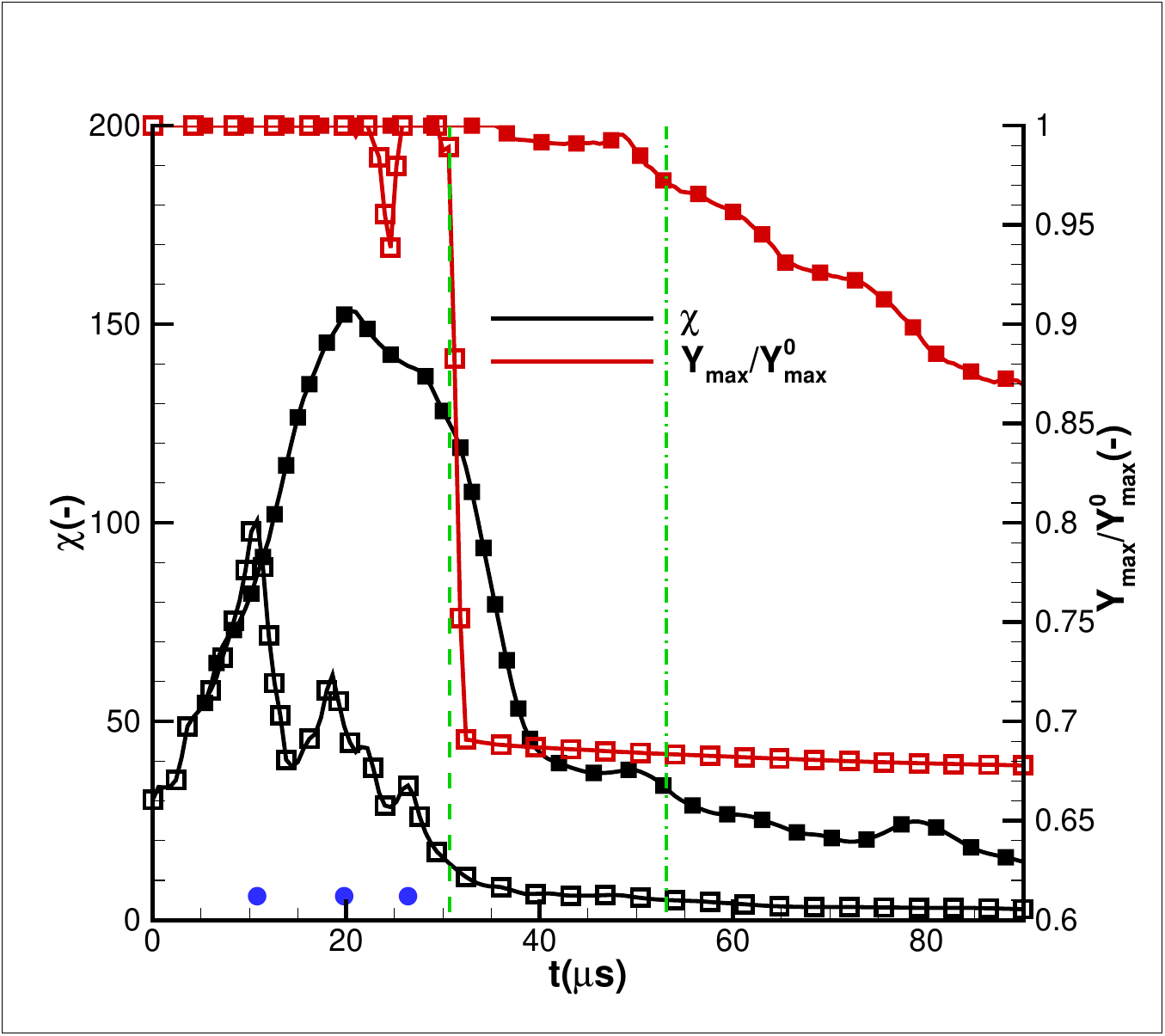}}
    \subfigure[]{
    \label{fig: cir-PS-VD-4}
    \includegraphics[clip=true,trim=15 15 15 50, width=.35\textwidth]{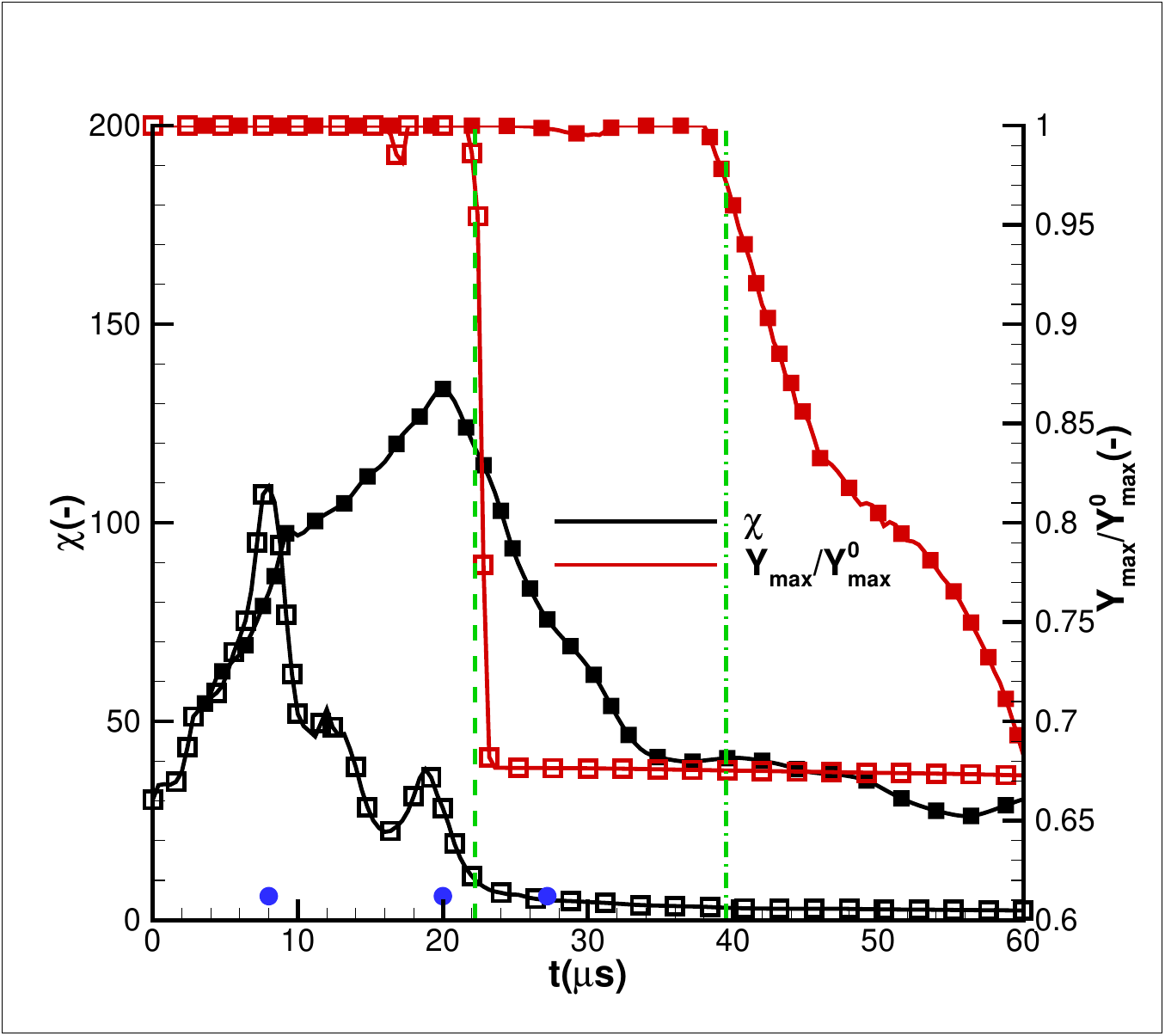}}
    \caption{Time history of scalar dissipation rate $\chi$ (\ref{eq:TMR}), and maximum normalised mass fraction $\overline{Y}$ (\ref{eq: NMF}), for (a) $Ma=2.4$; (b) $Ma=1.22$; (c) $Ma=1.8$; (d) $Ma=3$; (e) $Ma=4$. For all figures, hollow points represent VD case and solid points represent PS case. Green dashed lines show the mixing time when $\overline{Y}=0.97$ is reached. Five blue circles in (a) indicate the five different moments in figures~\ref{fig: den-vor-VD} and \ref{fig: SDRMF-VD}.
    Three blue circles presented separately in (b) to (e) indicate the three different moments for each Mach number in figure~\ref{fig: SDR evolu diff Ma} of appendix~\ref{App: SDR evolu}.
     \label{fig: mixing time} }
\end{figure}
\begin{table}
  \begin{center}
\def~{\hphantom{0}}
  \begin{tabular}{lccccc}
      $t_\mathfrak{m}$($\umu$s)        & $Ma$=1.22  & $Ma$=1.8  & $Ma$=2.4  & $Ma$=3  & $Ma$=4\\[3pt]
       VD case    & 184.1       & 72.1        &  40.3      &  30.7    &  22.2      \\
       PS case     & 292.2      & 128.3       &  74.1       &   53.1    &  39.5       \\
  \end{tabular}
  \caption{Mixing time $t_\mathfrak{m}$ of variable density and passive scalar cases when $\overline{Y}=0.97$ is reached. Data are measured from figure \ref{fig: mixing time}.}
  \label{tab: t_m}
  \end{center}
\end{table}
%%1. 三阶段；2. 差异从第二阶段开始；3. 最大浓度下降SDR稳定
Figure~\ref{fig: cir-PS-VD-2.4} shows three stages of the mixing process separated from the profile of $\chi$.
Five blue circles in figure~\ref{fig: cir-PS-VD-2.4} indicate the five different moments in figures~\ref{fig: den-vor-VD} and \ref{fig: SDRMF-VD}.
For VD SBI, the first stage is the scalar dissipation growth due to the vortex stretching before strong SBV occurs.
After the Br is dissipated, the scalar dissipation rate enters the second stage, decreasing with time.
Local peaks of scalar dissipation denote the stretching of TB structure by SBV during the decrease of dissipation, as shown in figure~\ref{fig: den-vor-VD}(c).
Mixing enters the third stage at a late time, the steady diffusion stage, with a low value of $\chi$. This steady mixing state can be regarded as the final `well-mixed' state, in which mixing no longer happens at a fast rate~\citep{weber2014experimental}.

For PS SBI, one can find a trend of the scalar dissipation $\chi$ similar to that for the VD case from 7.2~$\umu$s to 13.2~$\umu$s, meaning that the variable density effect on mixing has not been illustrated. After the SBV at the Br structure is formed in the VD case at $13.2$~$\umu$s, the mixing behavior of PS and VD SBI begins to diverge, which makes PS SBI mixing maintain a much longer time than VD SBI. Two $\chi$ peaks happen during the increase of scalar dissipation in PS case, which is different from VD cases where $\chi$ peaks are related to SBV. $\chi$ peaks in PS SBI is driven by strain from vorticity merging (VM) as indicated in figure~\ref{fig: den-vor-VD}(h). After the first $\chi$ peak, bridge structure dissipates and VM from TB structure contributes to the second $\chi$ peak, as shown in figure~\ref{fig: cir-PS-VD-2.4}.

Figures \ref{fig: mixing time}(b-e) validate the similar growth trend in scalar dissipation $\chi$ and normalised maximum concentration $\overline{Y}$ for different shock Mach numbers.
For higher shock Mach number, mixing continues to require a shorter time for both VD SBI than for PS SBI due to the higher compression and larger circulation.
Interestingly, the differences between PS and VD SBI for all Mach numbers start from the second stage, the dissipation decrease in VD SBI, which leads to the shorter mixing time in VD SBI than the one in PS SBI.
The differences of $\chi$ peaks behaviour between PS and VD SBI of other Mach numbers are further explained in appendix~\ref{App: SDR evolu}, which shows similar mechanism with that of $Ma=2.4$ case.

The maximum concentration $\overline{Y}$ decrease faithfully tracks the beginning of the steady diffusion stage in both VD SBI and PS SBI.
It is noteworthy that some valley phenomena occur in high Mach number VD cases. The appearance of valley is a bit earlier than the decay of maximum concentration. We conjuncture that valley results from the competition between density gradient accelerated dissipation and density gradient redistributed diffusion in variable density flows~\citep{yu2020scaling}. However, when valley happens, we observe that scalar dissipation is still at a relatively high value. Therefore, we denote the determinate decrease of maximum concentration as mixing time and the valley phenomena will be the focus of future study.
Figure~\ref{fig: mixing time} shows the mixing time by dashed lines (VD SBI) and dash-dot lines (PS SBI) when $\overline{Y}=0.97$, as tabulated in table~\ref{tab: t_m}.
Since the faster mixing is closely related to the scalar dissipation decrease in VD SBI when the SBV of Br structure forms, the SBV production mechanism and its effect on mixing are revealed in the next section.

\section{Secondary baroclinic vorticity enhanced stretching mechanism}
\label{sec: SBV enhanced stretch}
\subsection{PS mixing mechanism: Vortical stretching mixing}
\label{subsec: PS mixing}
%% PS混合由流动主导，对涡结构分析，在涡结构坐标系上看速度场特征
%% 要说道从ADE方程出发的公式，SDR 就是去拉伸最大浓度的，所以后面关注最大浓度如何变化的建模
\begin{figure}
  \centering
  \subfigure[]{
    \label{fig: Vthe-PS-contour}
    \includegraphics[clip=true,trim=0 300 0 0,width=.99\textwidth]{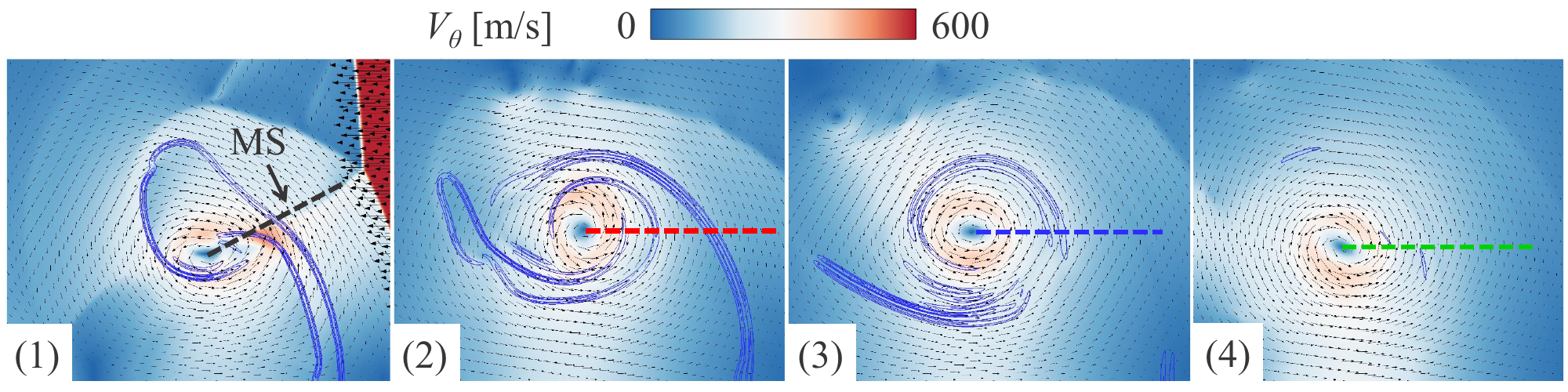}}\\
    \subfigure[]{
    \label{fig: Vthe-PS-line1}
    \includegraphics[clip=true,trim=10 10 15 15,width=.45\textwidth]{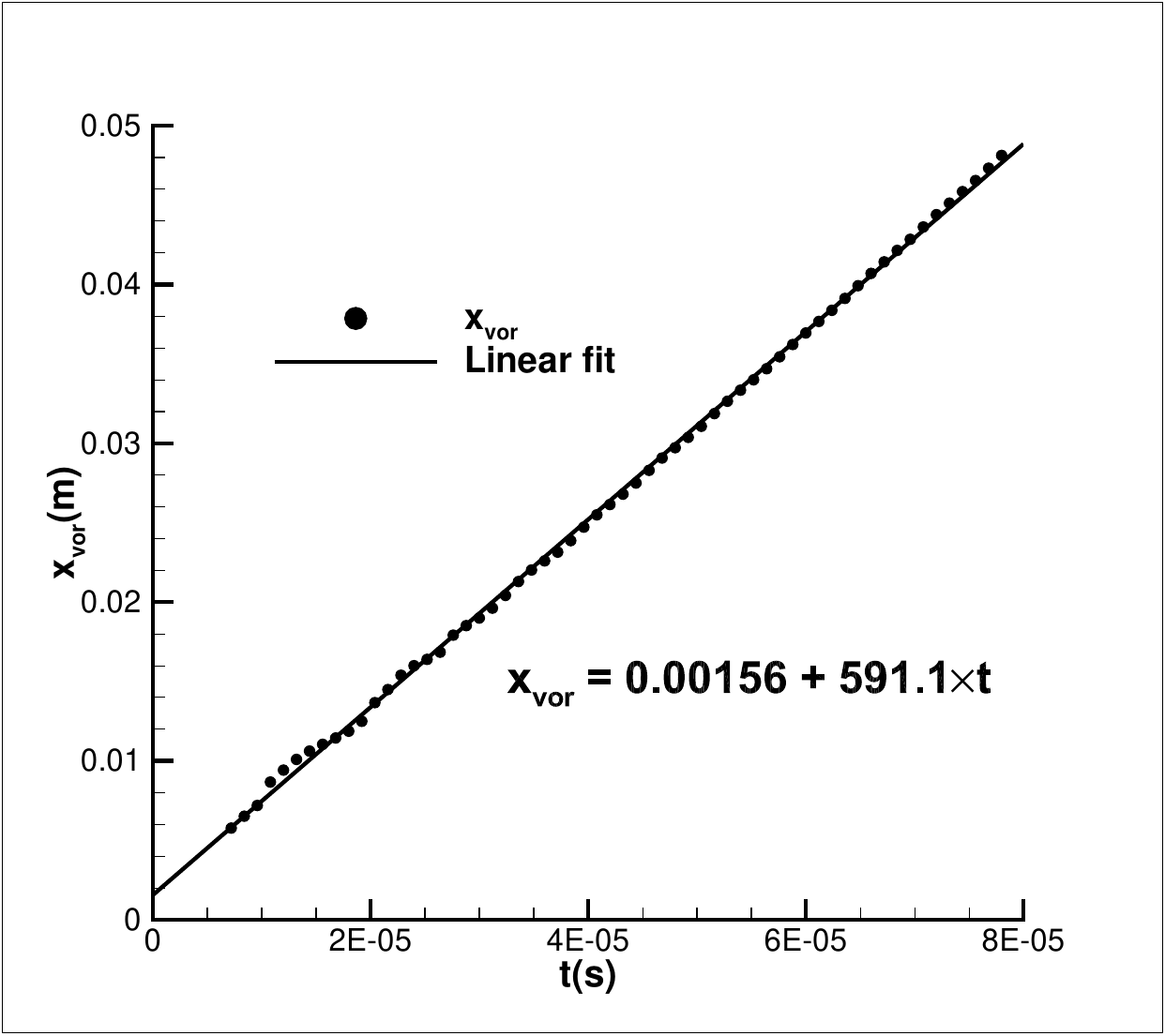}}
    \subfigure[]{
    \label{fig: Vthe-PS-line2}
    \includegraphics[clip=true,trim=10 10 15 15,width=.45\textwidth]{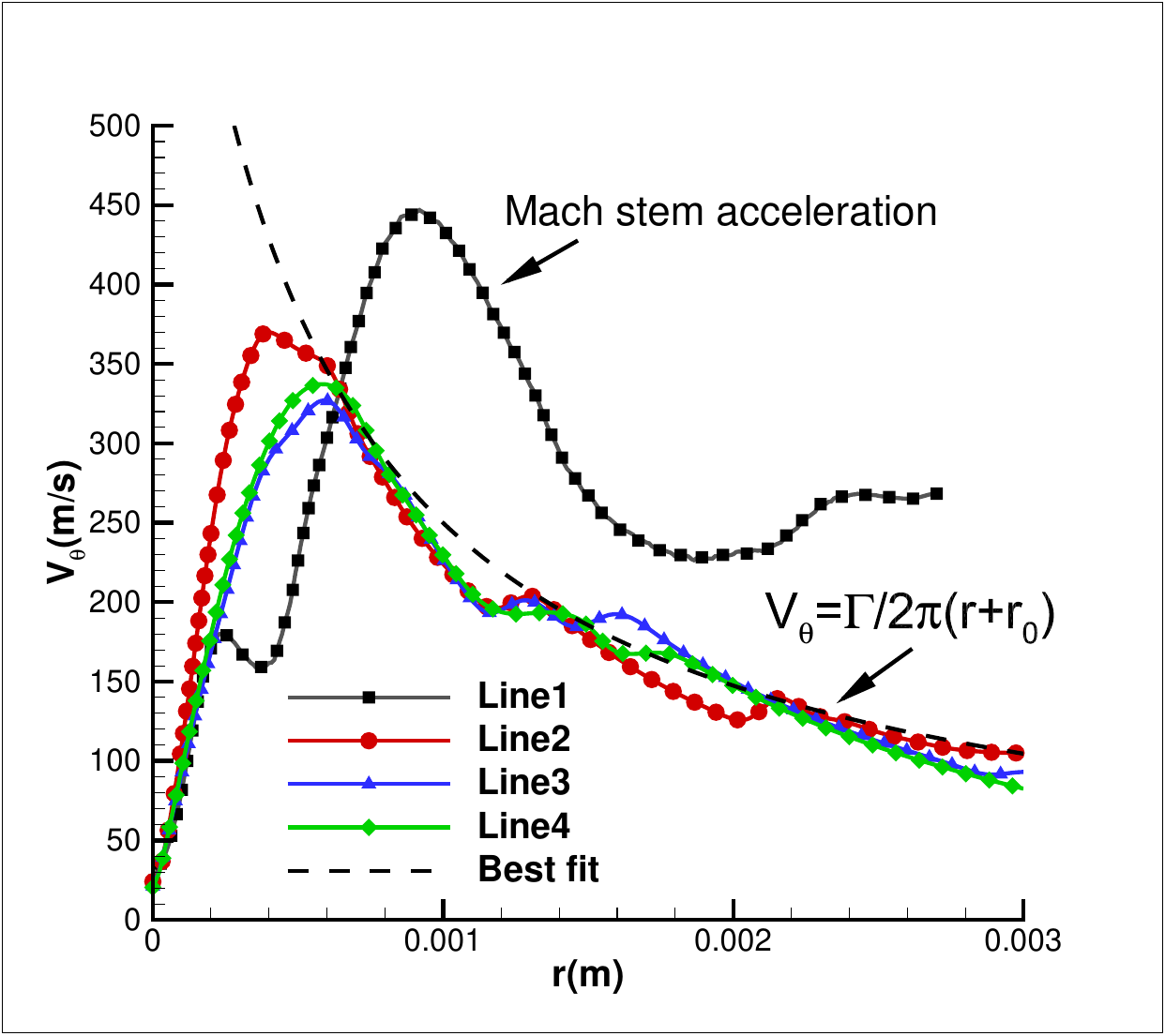}}
  \caption{PS SBI for $Ma$=2.4.
  (a) Azimuthal velocity contour in the frame of the vortex centre at $t=13.2$~$\umu$s (a1), $t=27.6$~$\umu$s (a2), $t=42$~$\umu$s (a3) and $t=69.6$~$\umu$s (a4). Isoline of scalar dissipation rate is plotted as blue line.
  (b) Motion of vortex centre with time.
  (c) Azimuthal velocity distribution along dashed lines in (a1)-(a4). Velocity distribution of a Lamb-Oseen type vortex (\ref{eq: vthe-theo}) is plotted as dashed line, obtained from the best curve fit of descending data in Line~2.  }\label{fig: Vthe-PS}
\end{figure}
Since passive scalar mixing does not influence vortical flow formation, the mixing is passively controlled by vortex stretching, as pointed out by \citet{marble1985growth} and \citet{meunier2003vortices}. So the flow structure is essential for mixing. After the shock impact, the passive scalar bubble obtains a translational velocity close to post-shock velocity $u'_1$. If one sets the origin on the moving vortex centre \citep{shariff1992vortex}, the azimuthal velocity around the vortex will be apparent:
\begin{equation}\label{eq: vthe}
  V_{\theta}=\sqrt{\left(u-V_v\right)^2+v^2},
\end{equation}
in which $u$ is the velocity of $x$-direction, and $v$ is the velocity of $y$-direction. A more rigorous way to stand on the vortex centre (detailed definition of vortex centre position can be found in appendix~\ref{App: vortex velo}) is to calculate the translational velocity of the vortex $V_v\approx591.1$~m/s plotted and linearly fitted in figure~\ref{fig: Vthe-PS-line1}.

Figure~\ref{fig: Vthe-PS-contour} shows the azimuthal velocity contour and vector defined by $u-V_v$ and $v$.
From the contour of azimuthal velocity $V_{\theta}$, the velocity vector is mainly around the vortex centre. To quantify the velocity distribution, figure~\ref{fig: Vthe-PS-line2} gives the $V_{\theta}$ along defined lines of different moments, as illustrated in figures~\ref{fig: Vthe-PS-contour}.
From figure~\ref{fig: Vthe-PS}(a1), it can be observed that Mach stem intrudes the vortex core, which is called ``embedded shock'' firstly found in compressible vortex ring~\citep{dora2014role,qin2020formation}. Therefore, the defined Line~1 emerges a high velocity region before encountering Mach stem.
Interestingly, after shock passage, the radial azimuthal velocity $V_{\theta}$ profiles of three later moments agree well with that of the spiral of a Lamb-Oseen vortex, which can be reduced to a point vortex model:
\begin{equation}\label{eq: vthe-theo}
  V_{\theta}\approx \frac{\Gamma}{2\upi(r+r_0)},
\end{equation}
where $\Gamma$ is the vortex circulation and $r_0=0.000439$ m obtained from the best curve fit of descending data Line~2.
From the isolines of scalar dissipation, the scalar is stretched by the vortex passively until the diffusion stage.
Thus, referring to~\citet{marble1985growth} and \citet{meunier2003vortices}, the PS SBI mixing time can be influenced by three main factors:
\begin{equation}\label{eq: tm-PS}
  t_\mathfrak{m}^{PS}=f(\Gamma, \mathscr{D}, r),
\end{equation}
where circulation $\Gamma$ is a measure of stirring, $\mathscr{D}$ denotes diffusivity, and distance $r$ between the passive scalar and vortex centre.

\begin{figure}
  \centering
  \includegraphics[clip=true,trim=0 30 0 0,width=.9\textwidth]{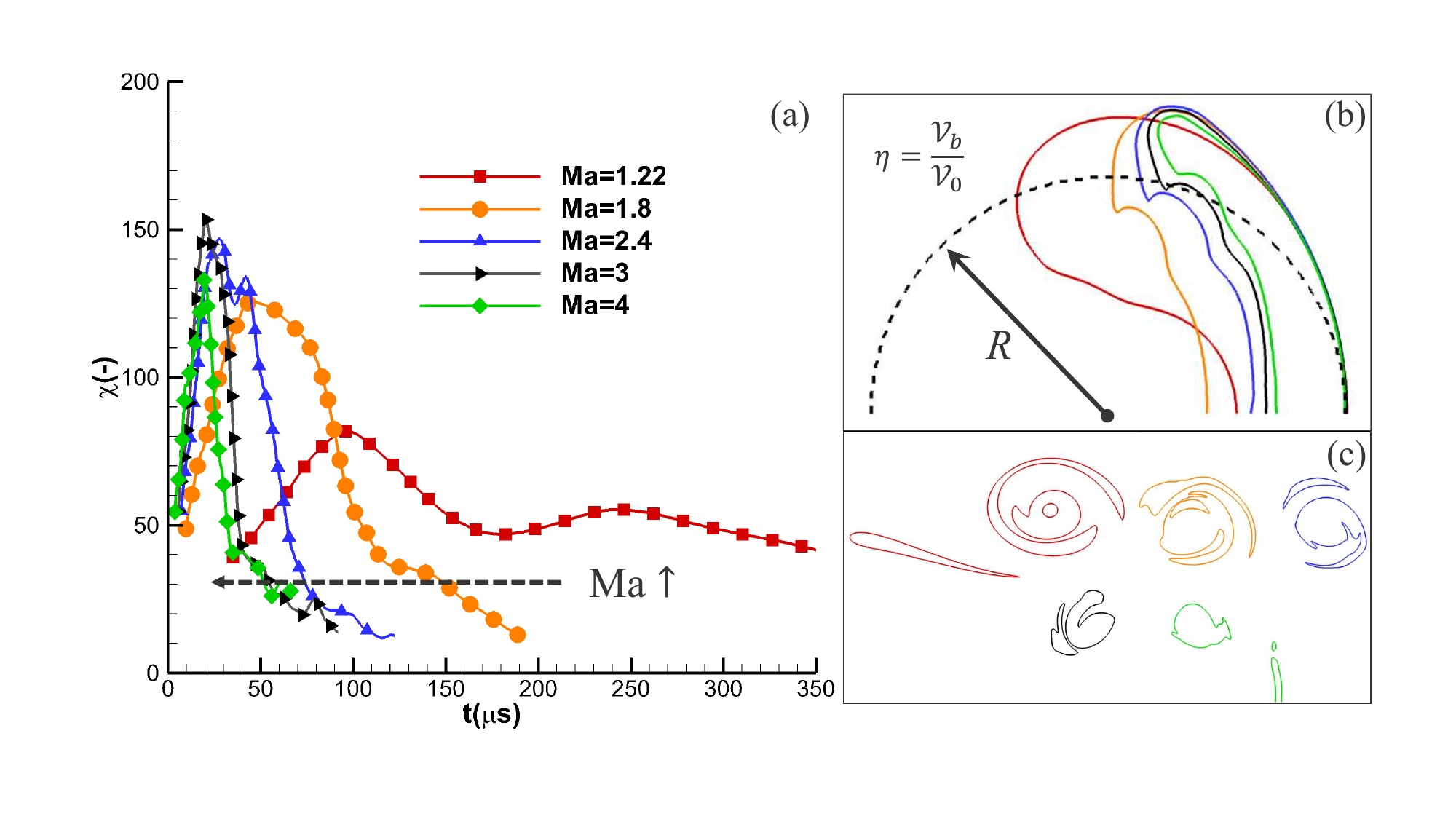}\\
  \caption{(a) Time history of scalar dissipation rate for PS SBI with different shock Mach number. (b) Isoline of cylindrical bubble boundary at initial time to show the compression effect on initial mixing region $\mathcal{V}_b=\eta \mathcal{V}_0\approx\eta\upi R^2$ and (c) final mixing region of PS SBI for different shock Mach number.
  Red: $Ma=1.22$, orange: $Ma=1.8$, blue: $Ma=2.4$, black: $Ma=3$, green: $Ma=4$.
  }\label{fig: Maeff-PS}
\end{figure}
Here, we further check the shock compression effect on PS SBI. As shown in figure~\ref{fig: Maeff-PS}(a), a higher Mach number leads to a shorter mixing time of scalar dissipation. It can be concluded that except for the circulation deposited by shock impact, the area that needs to be stirred is also essential for determining when mixing becomes stable.  As shown in figure \ref{fig: com-PS-VD}, the compression rate $\eta$ for both the PS and VD cases maintains a lower value at higher shock Mach number. In VD SBI, the compression rate is proposed and theoretically modelled by \citet{giordano2006richtmyer} and studied by \citet{niederhaus2008computational}.
Recently, a quantitative scaling of the final mixing extent proportional to the compression rate is built by \citet{yu2020scaling}.
Since the PS bubble's compression rate is nearly the same as the VD bubble's, the compression effect by shock impact is believed to be another factor for shorter mixing time at a higher shock Mach number.
A more intuitive explanation is the different helium areas for different shock Mach numbers in figures \ref{fig: Maeff-PS}(b) and \ref{fig: Maeff-PS}(c). The initial conditions for PS SBI show that the area for mixing shrinks to $\mathcal{V}_b=\eta \mathcal{V}_0$ (\ref{eq: com}), where $\mathcal{V}_0$ is the initial un-shocked cylindrical bubble volume. Higher Mach number leads to smaller compression rate and passive scalar bubble area, as indicated in figure~\ref{fig: Maeff-PS}(b).  Here, a general form of the PS SBI mixing time can be stated as:
\begin{equation}\label{eq: tm-PS-2}
  t_\mathfrak{m}^{PS}=f(\Gamma, \mathscr{D}, r, \eta).
\end{equation}

\subsection{VD mixing mechanism: Additional SBV enhanced stretching}
\label{subsec: VD mixing}
\begin{figure}
  \centering
    \subfigure[]{
    \label{fig: Vthe-VD-contour}
    \includegraphics[clip=true,trim=0 300 0 0,width=.99\textwidth]{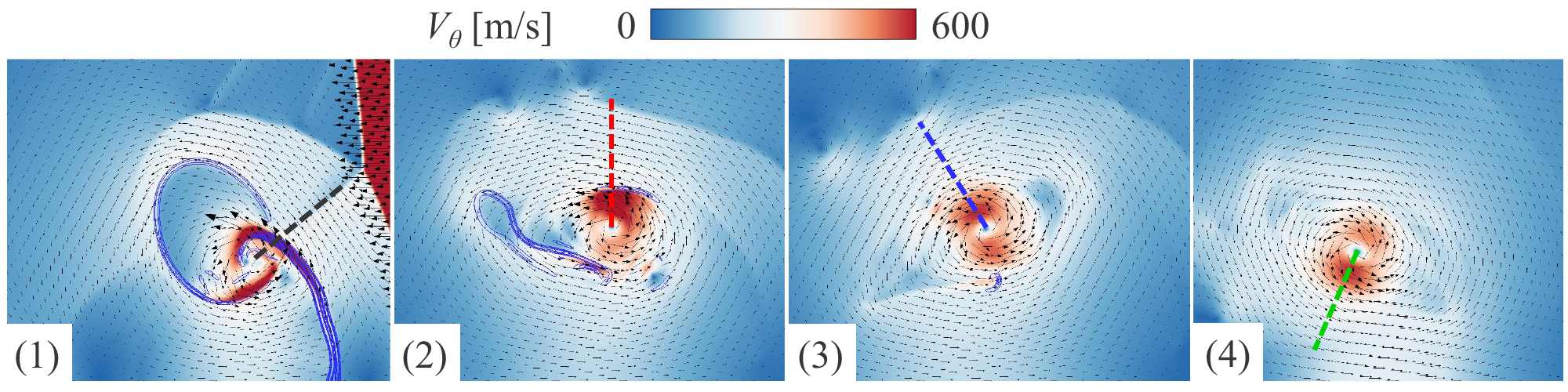}}\\
    \subfigure[]{
    \label{fig: Vthe-VD-line1}
    \includegraphics[clip=true,trim=10 10 15 15,width=.45\textwidth]{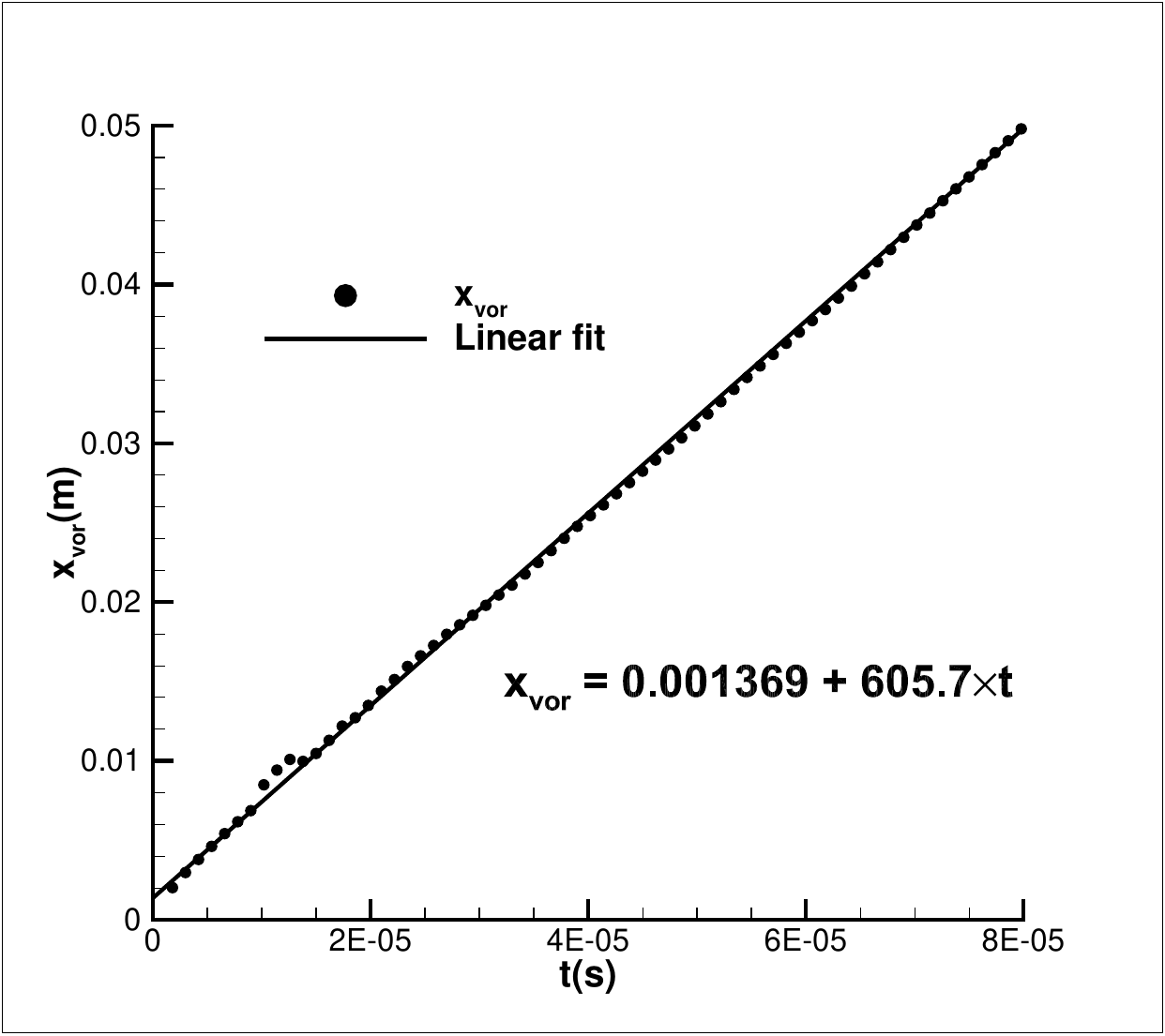}}
    \subfigure[]{
    \label{fig: Vthe-VD-line2}
    \includegraphics[clip=true,trim=10 10 15 15,width=.45\textwidth]{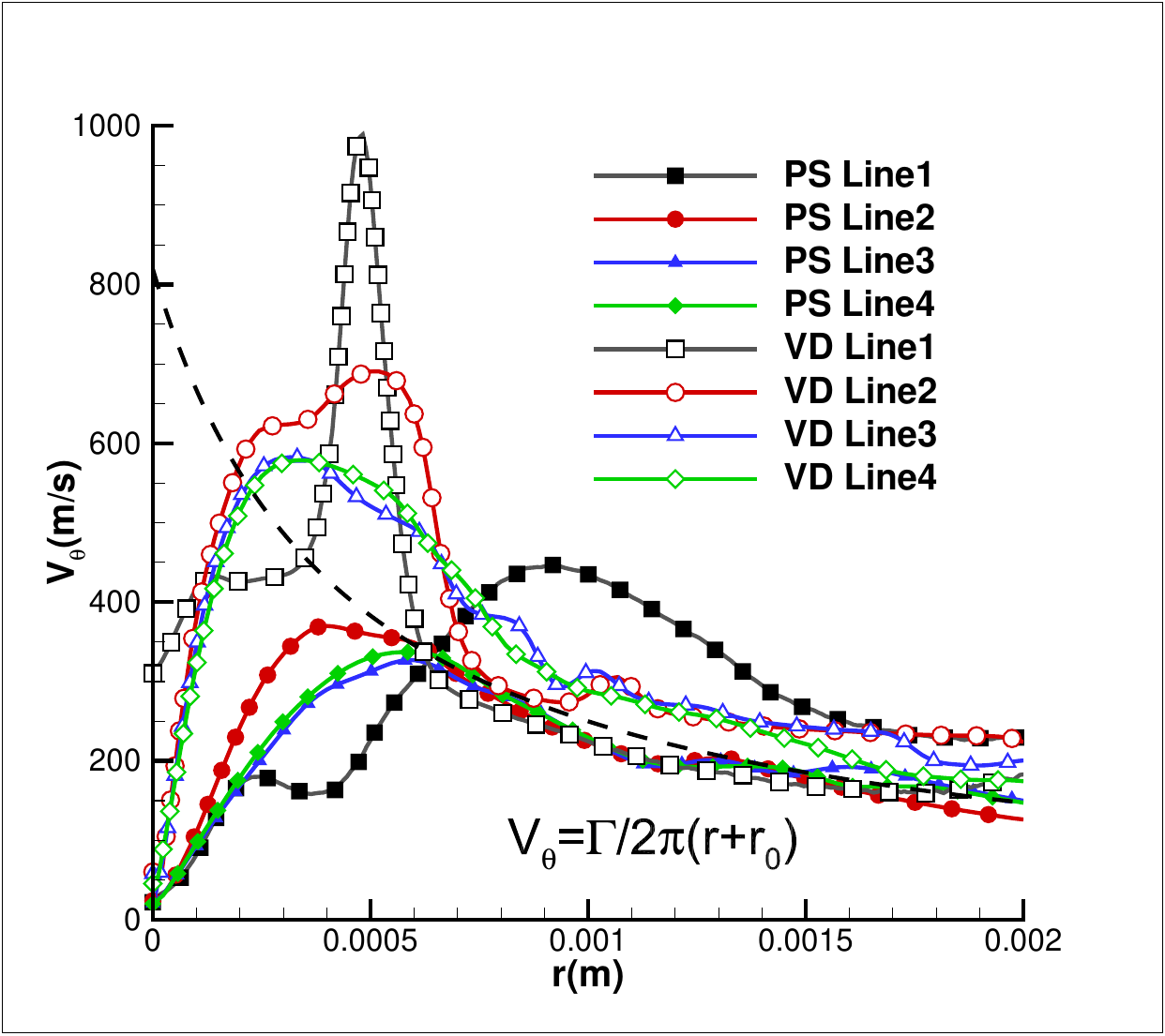}}
  \caption{VD SBI for $Ma$=2.4.
  (a) Azimuthal velocity contour in the frame of vortex centre at $t=13.2$~$\umu$s (a1), $t=27.6$~$\umu$s (a2), $t=42$~$\umu$s (a3) and $t=69.6$~$\umu$s (a4).
  (b) Motion of vortex centre with time.
  (c) Azimuthal velocity distribution along the defined lines captured from (a1) to (a4) compared with azimuthal velocity distribution at the same time obtained from figure~\ref{fig: Vthe-PS}. An increase of azimuthal velocity in VD case is obvious for each comparison. }\label{fig: Vthe-VD}
\end{figure}

%%% 这段分析中要加入已有对密度变化拉伸已有的研究。如密度确实被拉伸了，与应变张量来看。且要说明density influence的本质是inertial effect，轻密度介质受到力惯性更大，被加速
To render the mechanism causing faster mixing decay in VD SBI, we apply the same data treatment to the VD case, as shown in figure \ref{fig: Vthe-VD}. The vortex of the VD cases translates at nearly the same velocity as in the PS case at $V_v\approx605.7$ m/s, as shown in figure~\ref{fig: Vthe-VD-line1}. Under the vortex centre's coordinate, the azimuthal velocity contours of the VD case at four different moments are illustrated in figure~\ref{fig: Vthe-VD-contour}, where the legend is the same as the one in the PS case. The obvious difference is focused on the much greater azimuthal velocity $V_{\theta}$ at the bridge structure and the region around the vortex. In figure~\ref{fig: Vthe-VD-line2}, this velocity increase is clearly shown by the velocity profile along the defined lines of figure~\ref{fig: Vthe-VD-contour}.
The lines are characterized as one end originates from vortex centre pointing to the location of maximum azimuthal velocity.
For comparison, the PS SBI velocity profiles along the defined lines in figure~\ref{fig: Vthe-PS-line2} are also shown. It can be found that the increased velocity at the bridge structure (VD Line~1) is more than twice the velocity in PS SBI. At a distance far from the acceleration region where the bubble concentration is low, the azimuthal velocity returns to the Lamb-Oseen vortex model, as displayed in (\ref{eq: vthe-theo}). Thus, it can be inferred that this velocity increase is caused by the SBV, denoted by $\Delta V_{\theta}^b$:
\begin{equation}\label{eq: vthe-VD}
  \widetilde{V_{\theta}}\equiv V_{\theta}+\Delta V_{\theta}^{b},
\end{equation}
where $\widetilde{V_{\theta}}$ refers to azimuthal velocity distribution in VD SBI.
Furthermore, the increase in stretching velocity $\Delta V_{\theta}^b$ can be directly related to the local production of SBV $\omega_b$. One can easily derive the relation between local vorticity and its azimuthal velocity distribution:
\begin{equation}\label{eq: vthebaro-omega}
  \frac{\partial \widetilde{V_{\theta}}}{\partial r}=\frac{\partial V_\theta}{\partial r}+\frac{\partial \Delta V_{\theta}^{b}}{\partial r}\equiv\omega_{pv}+\omega_{b}= \omega,
\end{equation}
where $\omega_{pv}$ is vorticity attributed to a point vortex part, $\omega_b$ is the local SBV part and $\omega$ is the total vorticity magnitude along radius locally.
From the isolines of scalar dissipation in figure~\ref{fig: Vthe-VD-contour}, the azimuthal velocity increase stretches the bubble structure and enhances the mixing rate. That explains the faster mixing in VD SBI than in PS SBI and reveals the intrinsic variable density effect: the lighter gas responds faster than the heavy gas through local SBV production, implying the asymmetric mixing behaviour in VD flows.

\begin{figure}
  \centering
  \includegraphics[clip=true,trim=0 100 0 30,width=.99\textwidth]{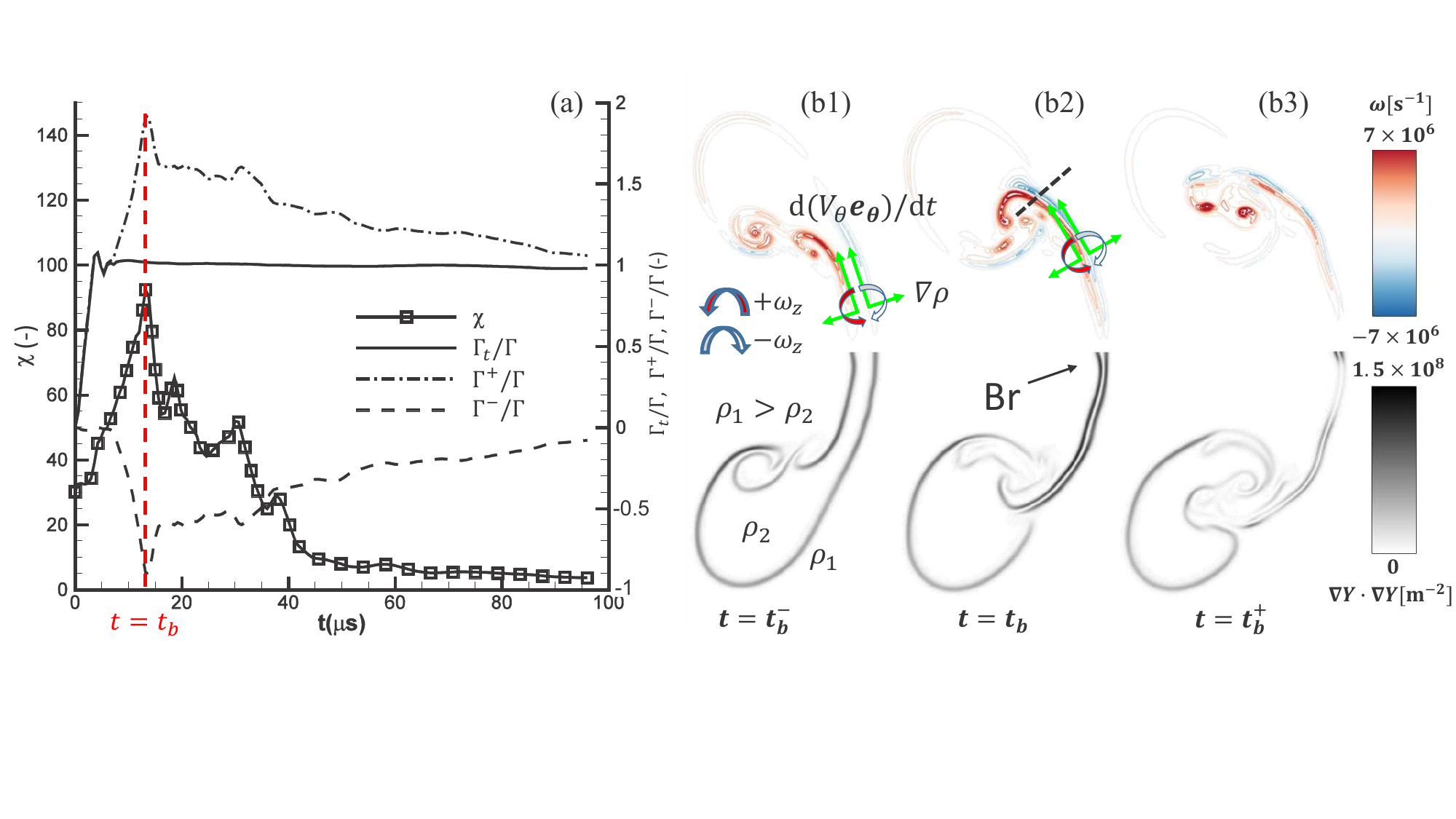}\\
  \caption{(a) Synchronous growth of opposite sign vorticty and scalar dissipation for $Ma$=2.4 VD SBI. (b) Isolines of vorticity (top) and contour of scalar dissipation (bottom) at instant near $t_{b}=13.2$~$\umu$s. (b1) $11.2$~$\umu$s;    (b2) $13.2$~$\umu$s; (b3) $15.2$~$\umu$s.}\label{fig: SBV-SDR}
\end{figure}

%%%%1. SBV产生机制；2. 共同生长：加速拉伸后，混合完成，SDR 下降，体现了环量与SDR的共同生长
In order to find the origin of the increase in velocity due to SBV, one should check the vorticity generation mechanism.
It has been proven that after the shock interaction, the main circulation is conserved inside the mixing region confined by the mass fraction contour, as shown in $\S$\ref{sec: num and IC}. The conservation of circulation indicates that the same quantitative positive vorticity is produced to balance the negative vorticity from SBV.
Here the total circulation, normalised by the final equilibrium circulation, is decomposed into the positive and negative ones:
\begin{equation}\label{eq: ciru-neg-pos}
  \Gamma_t/\Gamma=\Gamma^+/\Gamma+\Gamma^-/\Gamma,
\end{equation}
as validated in figure~\ref{fig: SBV-SDR}(a), showing the same up and down characteristic of opposite sign circulation.

The SBV production process in the VD SBI is illustrated in figure \ref{fig: SBV-SDR}(b).
The main vortex will come near the bubble's interface where the density gradient at the bridge structure exists. The roll-up process of the vortex will accelerate the interface leading to the acceleration $\mathrm{d}({V}_{\theta}\boldsymbol{e_\theta})/\mathrm{d}t$ in the azimuthal direction $\boldsymbol{e_\theta}$~\citep{peng2003vortex}. Assuming negligible gravitational force and incompressible flow, one obtains $\mathrm{d}({V}_{\theta}\boldsymbol{e_\theta})/\mathrm{d}t=-1/\rho\nabla{p}$~\citep{reinaud2000baroclinic}. Due to the existence of the density gradient $\nabla\rho$, the baroclinic vorticity production will follow:
\begin{equation}\label{eq: baro-mech}
  \frac{\mathrm{D}\boldsymbol{\omega_{b}}}{\mathrm{D}t}=\frac{1}{\rho^2}\nabla\rho\times\nabla p=\frac{1}{\rho} \frac{\mathrm{d}({V}_{\theta}\boldsymbol{e_\theta})}{\mathrm{d}t}\times\nabla\rho.
\end{equation}
From this equation, we can find that mixing will smear the density gradient $\nabla\rho$ and thus decrease the SBV production. SBV will give feedback to the mixing rate through the local stretching.

The integral effect of SBV on scalar dissipation is recorded in figure \ref{fig: SBV-SDR}(a).
The time history of the scalar dissipation rate is closely related to the SBV circulation profile, showing that the stretching from SBV controls the mixing rate.
A watershed of increase and decrease of scalar dissipation is at the moment $t=t_{b}$, when the SBV peak is attained, as shown in figure \ref{fig: SBV-SDR}(a).
From figure~\ref{fig: SBV-SDR}(b2), it can also be found that when SBV forms near the bridge structure, the local scalar dissipation becomes strong. In figure \ref{fig: SBV-SDR}(b3), after the production of SBV, the local baroclinic vorticity begins to decrease because mixing smears the density gradient $\nabla\rho$ and reduces the production of SBV, as analyzed.

%%%  introduce mixing and strain to explain others like Tomkins用别人的文献支撑说明其实也有类似现象，但没讲清楚，特别是关键性的周向速度没有体现出来，也无法进一步建模出混合的机制，从而反衬出文章价值大
It is noteworthy that the intense mixing rate at the bridge structure is also found in shock-heavy cylinder interactions~\citep{tomkins2008experimental}. The non-turbulent band of fluid is the most high-level mixing region. It is concluded that the vortex stretching effect at the bridge contributes 40\% of the mixing rate over all time. \citet{tomkins2008experimental} explained this high mixing rate as the strong strain rate.
In this study, the bridge structure's dissipation in the shock-light bubble interaction is further revealed as the additional stretching from the SBV. As soon as the SBV happens, the fast stretching of the bridge structure makes the density gradient across Br structure decrease, not like in the shock-heavy bubble interaction where the bridge structure can sustain mixing.

\begin{figure}
    \centering
    \subfigure[]{
    \label{fig: cirbaro-Ma}
    \includegraphics[clip=true,trim=15 15 35 50, width=.48\textwidth]{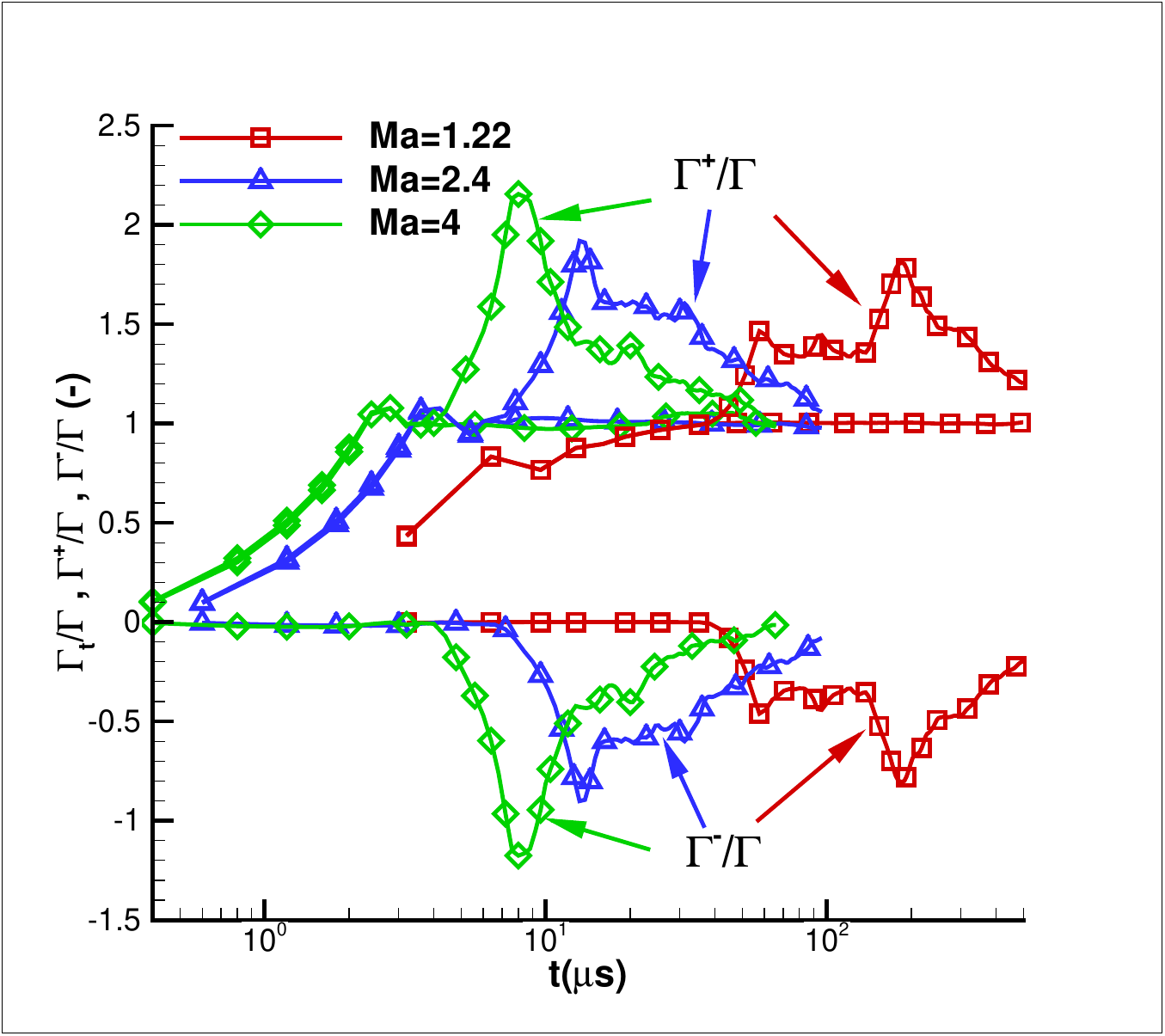}}
    \subfigure[]{
    \label{fig: cir-SDR-Ma}
    \includegraphics[clip=true,trim=15 15 12 45, width=.48\textwidth]{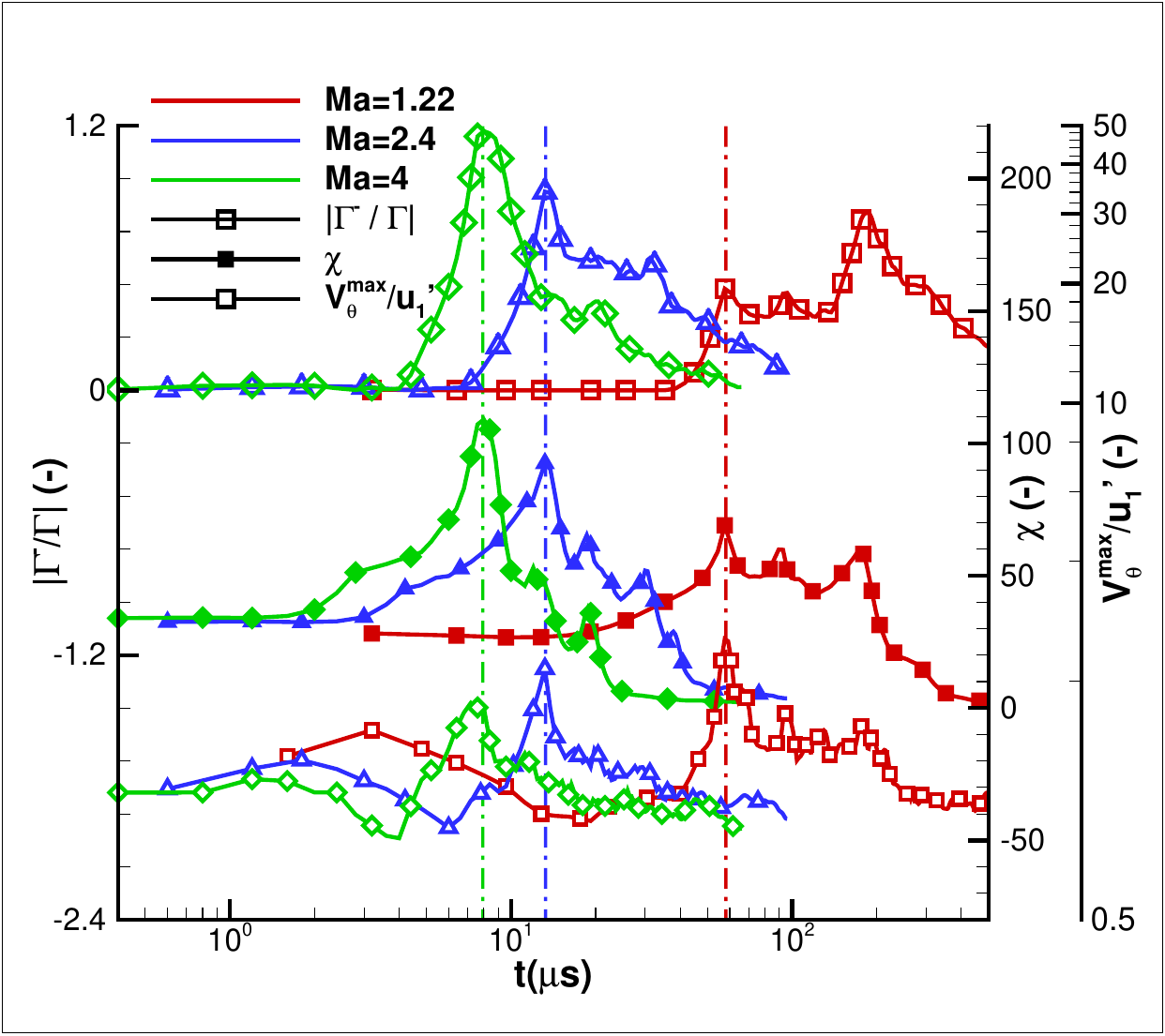}}
    \caption{(a) Dimensionless positive, negative and total circulation in different Mach number cases. (b) The synchronous growth of scalar dissipation and the secondary baroclinic circulation in different Mach number cases.  The maximum azimuthal velocity $V_\theta^{\textrm{max}}$ normalized by post-shock velocity $u_1'$ is also plotted. \label{fig: cir-baro} }
\end{figure}
Moreover, checking the normalised SBV circulation growth for different shock strength, one can found that the same magnitude of positive and negative vorticity occurs in all cases concerned, as shown in figure \ref{fig: cir-baro}(a).
Since positive circulation contains the initial shock induced vorticity, the absolute value of negative circulation,  $|\Gamma^-/\Gamma|$ solely attributed to negative SBV vorticity, can be denoted as SBV circulation.
Similar to the $Ma$=2.4 case, the synchronous growth of SBV circulation and scalar dissipation rate in other shock Mach number cases is also confirmed in figure~\ref{fig: cir-baro}(b). The SBV circulation peak value is higher as shock strength increases and happens at the time of bridge structure formation.
We further record the temporal evolution of maximum azimuthal velocity $V_\theta^{\textrm{max}}$ for each VD cases in figure~\ref{fig: cir-baro}(b). It can be found that when SBV circulation peak is reached, the peak value of azimuthal velocity occurs, in accordance to the azimuthal velocity profiles in figure~\ref{fig: Vthe-VD-line2}.
The relation between SBV peak and azimuthal velocity peak value will be discussed and modelled respectively in $\S$\ref{subsec: inertial velocity model}.

\subsection{Velocity difference model for azimuthal acceleration from SBV}
\label{subsec: inertial velocity model}
\subsubsection{SBV model: $\omega_b$}
\begin{figure}
  \centering
  \includegraphics[clip=true,trim=10 160 0 0,width=.9\textwidth]{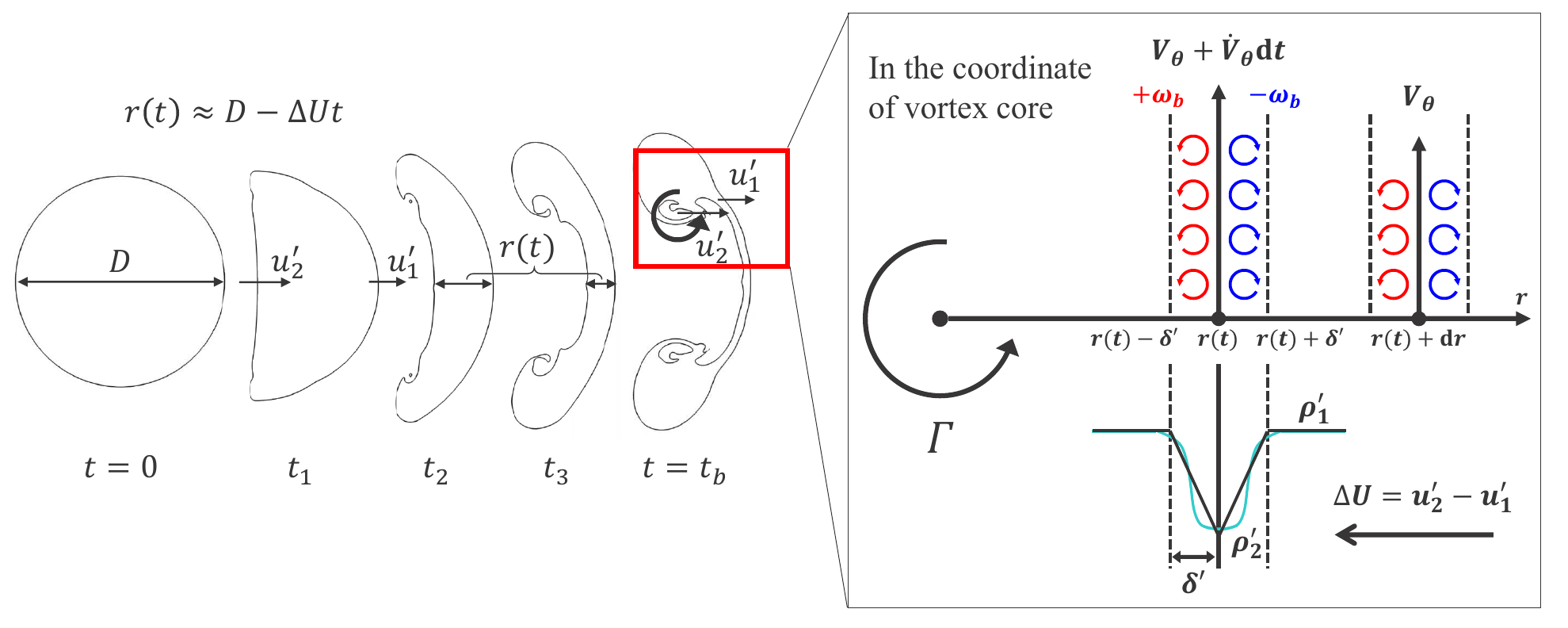}\\
  \caption{Illustration of the SBV production at bridge structure by a velocity difference model abstracted from the shocked light gas cylindrical bubble and shocked heavy ambient air.\label{fig: SBV_mech}}
\end{figure}
Figure~\ref{fig: SBV_mech} shows the illustration of the formation of the bridge structure and the SBV.
Since the cylindrical bubble is lighter than ambient air, the upstream edge's velocity $u'_2$ is much larger than the downstream edge's, the same as the shocked air $u'_1$~\citep{yu2020two}. A vortex is first formed in the upstream edge and moves at approximately $u'_2$ at the beginning. Then we can simplify this process as a model problem as illustrated in the magnified frame in figure \ref{fig: SBV_mech}. In the coordinate of the vortex core, the bridge structure containing a density interface moves towards the vortex with circulation $\Gamma$ at a speed of $\Delta U=u'_2-u'_1$.
The deposited SBV accumulates as the density interface moves.
As discussed in PS SBI, the velocity profile in the mixing region can be fitted by a point vortex before the SBV occurs.
Thus, it is appropriate to use the assumption that the $V_{\theta}$ is reciprocal to the distance to the main vortex core $r$:
\begin{equation}\label{eq: vavd-illu1}
  V_{\theta}=\frac{\Gamma}{2\upi r}.
\end{equation}
Thus, the acceleration of the bridge interface can be illustrated as:
\begin{equation}\label{eq: vavd-illu2}
  \frac{\mathrm{d}V_{\theta}}{\mathrm{d}t}=\frac{\mathrm{d}V_{\theta}}{\mathrm{d}r}
  \frac{\mathrm{d}r}{\mathrm{d}t}
  =\frac{\Gamma\Delta U}{2\upi r^2},
\end{equation}
since the density interface is approaching the main vortex core at the speed $\Delta U$, which is $\frac{\mathrm{d}r}{\mathrm{d}t}=-\Delta U$.
%The magnitude of density gradient along axis $r$, $|\nabla\rho|$ in (\ref{eq: baro-mech}), can be expressed as $\mathrm{d}\rho/\mathrm{d}r$ and the density gradient $\nabla\rho$ is presumed as perpendicular to the azimuthal acceleration $\mathrm{d}({V}_{\theta}\boldsymbol{e_\theta})/\mathrm{d}t$.
The density gradient $\nabla\rho$ in (\ref{eq: baro-mech}) is presumed as perpendicular to the azimuthal acceleration $\mathrm{d}({V}_{\theta}\boldsymbol{e_\theta})/\mathrm{d}t$ and directed along the $r$-axis (it has no component in any other direction).
Then magnitude of density gradient $|\nabla\rho|=\mathrm{d}\rho/\mathrm{d}r$, and
the local secondary baroclinic vorticity production rate can be expressed as:
\begin{equation}\label{eq: vavd-illu3}
  \dot{\omega}_{b}=\frac{\mathrm{D}\omega_{b}}{\mathrm{D}t}=\frac{1}{\rho}\frac{ \mathrm{d}V_\theta }{\mathrm{d}t} \frac{\mathrm{d}\rho}{\mathrm{d}r},
\end{equation}

From figure~\ref{fig: SBV_mech}, we can make several assumptions about $\mathrm{d}\rho/\mathrm{d}r$:
\begin{enumerate}
  \item inside the density interface, the density is invariant with time, which is light gas density $\rho_2'$ inside the density interface and heavy gas density $\rho_1'$ outside the density interface.
  \item the density interface is compact and symmetrical about $r(t)$.
  \item the thickness of density interface transition layer $\delta'$ at post-shock bridge structure is equal to $\sqrt{\eta}\delta$, where $\delta$ is the initial transition layer introduced in section~\ref{subsec: IC} (see appendix~\ref{App: drhodr} for details).
  \item the thickness $\delta'$ is much smaller than $r$:  $\delta'=\textit{O}(\mathrm{d}r)$.
\end{enumerate}
These four assumptions are made to describe the bridge structure formation. Therefore, $\mathrm{d}\rho/\mathrm{d}r$ can be expressed as:
\begin{equation}\label{eq:drhodr}
    \frac{\mathrm{d}\rho}{\mathrm{d}r}\approx\left\{ \begin{array}{ll}
    0 & r\in(0,r(t)-\delta'), \\
    \frac{\rho_1'-\rho_2'}{\delta^{'}} & r\in[r(t)-\delta',r(t)), \\
    0 & r=r(t), \\
    \frac{\rho_2'-\rho_1'}{\delta^{'}} & r\in(r(t),r(t)+\delta'], \\
    0 & r\in(r(t)+\delta',+\infty).
    \end{array}\right.
\end{equation}

For convenience, we only consider the value at point $r(t)$, which is related to the range $[r(t)-\delta',r(t))$ at the upwind side of the density interface transition layer.
Then the baroclinic vorticity production rate at this range can be further calculated as:
\begin{equation}\label{eq: vavd-illu4}
  \dot{\omega}_{b}=\frac{1}{\rho}\frac{\Gamma\Delta U}{2\upi r^2}\left(\frac{\mathrm{d}\rho}{\mathrm{d}r}\right)
  \approx\frac{\Gamma\Delta U|{At}^+|}{\upi\delta^{'} r^2},   \quad r\in[r(t)-\delta',r(t))
\end{equation}
where $\rho$ can be the average density $(\rho_1'+\rho_2')/2$ and
$r$ is the distance from the vortex core as shown in figure \ref{fig: SBV_mech}.
Then, the baroclinic vorticity production can be calculated:
\begin{equation}\label{eq: vor-baro}
  \omega_b(r,t+\mathrm{d}t)=\omega_b(r,t)+\dot{\omega}_{b}\mathrm{d}t\Rightarrow
  \omega_{b}=\int_0^{t}\dot{\omega}_{b}\mathrm{d}t',
\end{equation}
To complete the set of equations, the distance between the bridge and the main vortex $r$ in (\ref{eq: vavd-illu4}) should be modelled.
In figure \ref{fig: SBV_mech}, one can find that $r$ is the function of time:
\begin{equation}\label{eq: vavd-illu5}
 r(t)\approx D-\Delta{U}t.
\end{equation}
Hence, the baroclinic vorticity growth (\ref{eq: vor-baro}) can be expressed as:
\begin{eqnarray}\label{eq: vavd-illu6}
     \omega_{b}  & =  & \int_0^{t}\frac{\Gamma\Delta U|{At}^+|}{\upi\delta^{'} r^2} \mathrm{d}t'
       =\frac{\Gamma\Delta U|{At}^+|}{\upi\delta^{'}} \int_0^{t} \frac{1}{(D-\Delta Ut')^2}\mathrm{d}t' \nonumber\\
                          &  =  & \frac{\Gamma|{At}^+|}{\upi\delta^{'}}\frac{1}{D-\Delta Ut'} \bigg|_{0}^{t}
                          = \frac{\Gamma|{At}^+|}{\upi\delta^{'}}\left(\frac{1}{D-\Delta Ut}-\frac{1}{D}\right) \nonumber\\
                          &  =  & \frac{\Gamma|{At}^+|}{\upi\delta^{'}}\left(\frac{1}{r}-\frac{1}{D}\right), \quad r\in[r(t)-\delta',r(t)).
\end{eqnarray}
Using $D^2$ as the characteristic area, the normalized baroclinic vorticity at $t=t_b$, where $t_{b}$ is the time of the bridge formation and is also the first time of the synchronous growth of opposite sign circulation peak in figure~\ref{fig: cir-SDR-Ma}, can be obtained:
\begin{equation}\label{eq: vavd-illu8}
    \frac{\omega_{b}D^2}{\Gamma}=\frac{|{At}^+|D}{\upi\delta^{'}}\left(\frac{1}{1-\Delta Ut_b/D}-1\right).
\end{equation}

\begin{figure}
  \centering
    \subfigure[]{
    \label{subfig: SBV contour}
    \includegraphics[clip=true,trim=0 250 0 0,width=.99\textwidth]{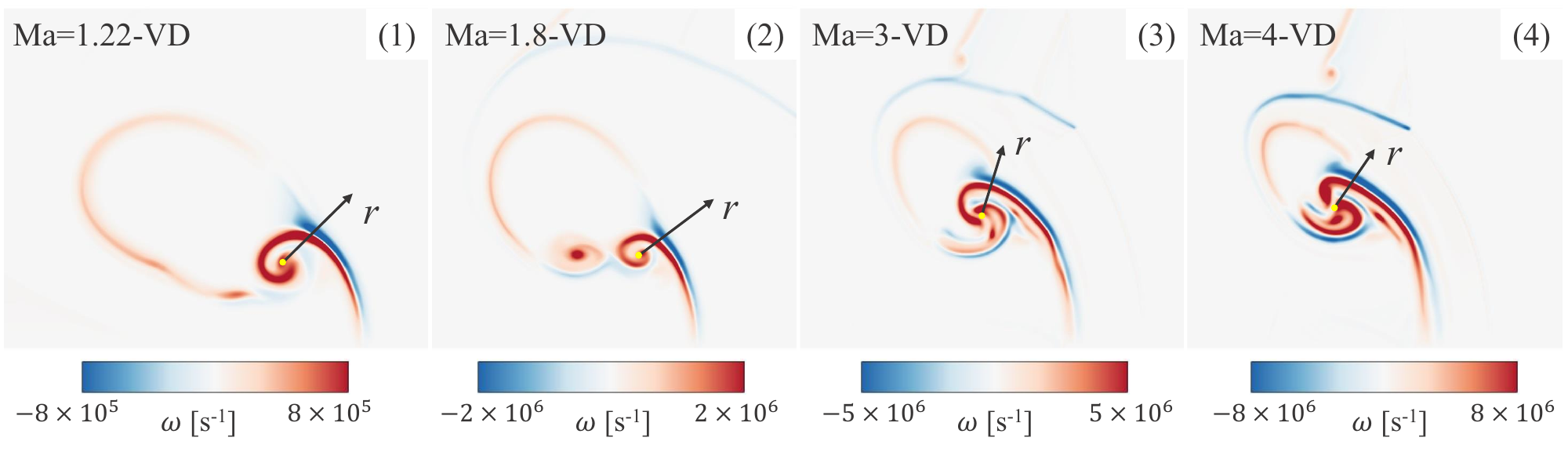}}\\
    \subfigure[]{
    \label{subfig: SBV profile}
    \includegraphics[clip=true,trim=15 15 15 15,width=.45\textwidth]{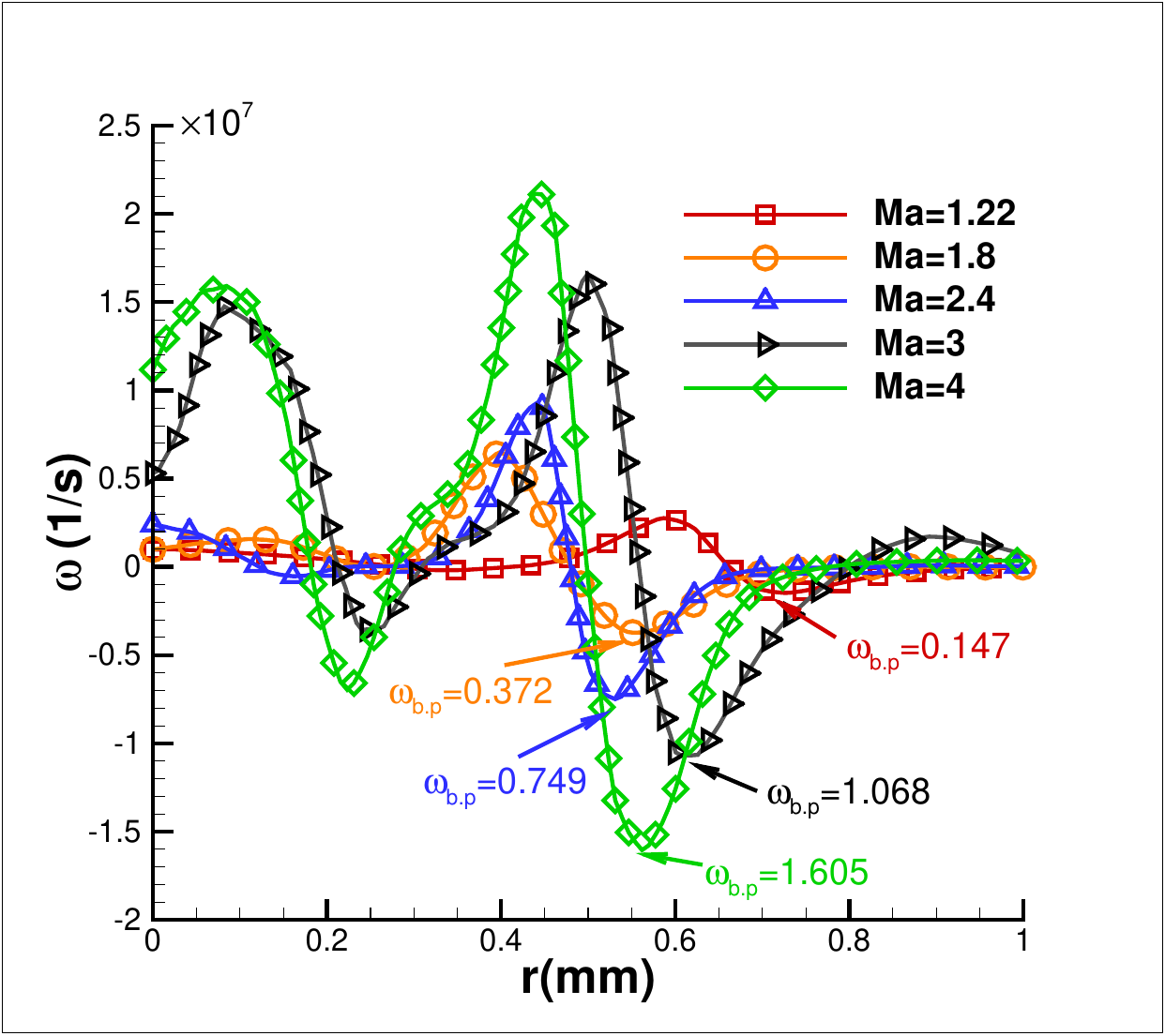}}
    \subfigure[]{
    \label{subfig: cir-dUt-D}
    \includegraphics[clip=true,trim=15 15 15 15,width=.45\textwidth]{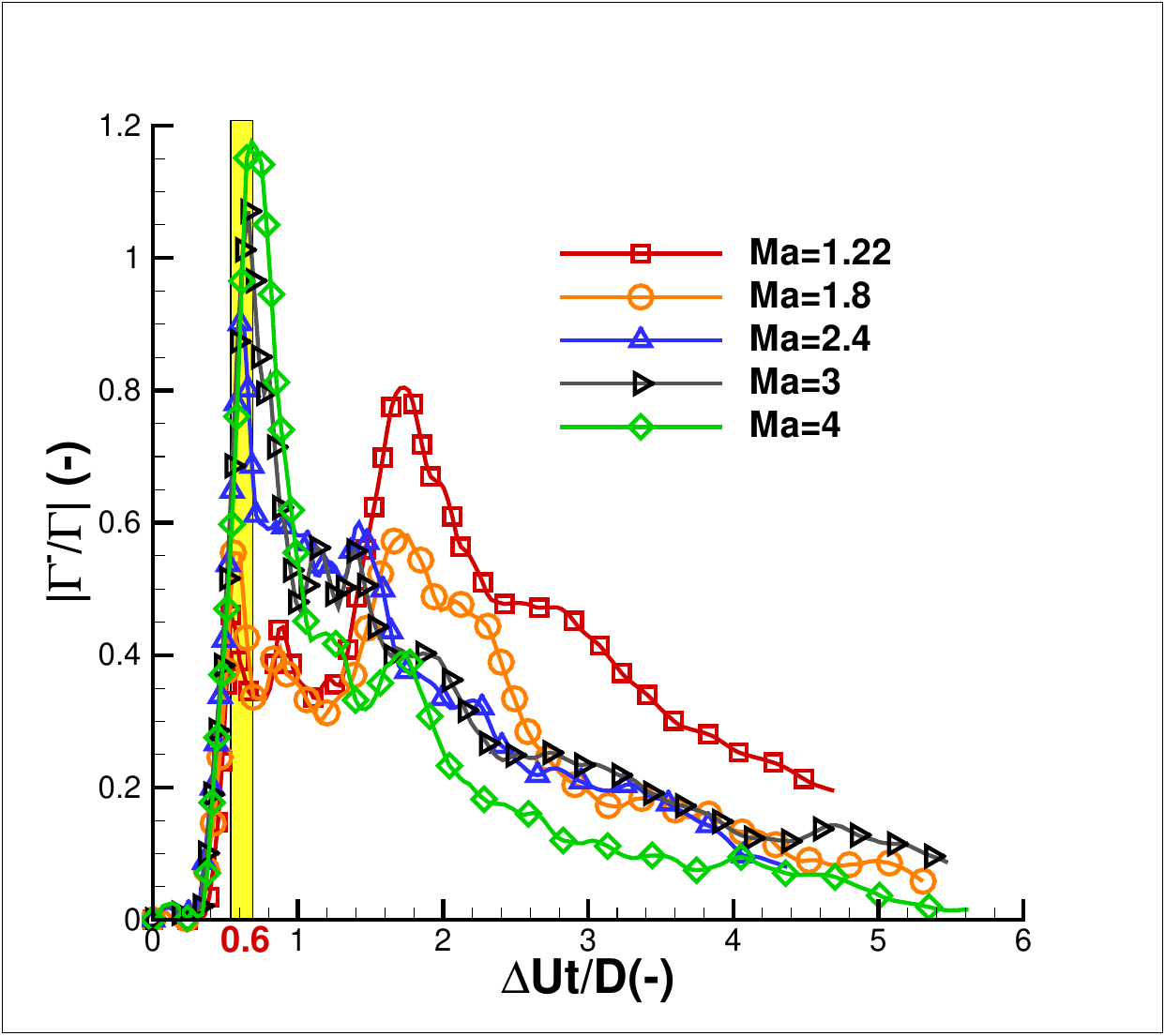}}
  \caption{
  (a) Vorticity contour for different Mach number VD SBI cases at $t=t_b$, namely, $Ma=1.22$, $t_b=58$~$\umu$s; $Ma=1.8$, $t_b=19.2$~$\umu$s; $Ma=3$, $t_b=11$~$\umu$s; $Ma=4$, $t_b=8$~$\umu$s. Coordinate $r$ with origin at vortex centre pointing to the location of maximum azimuthal velocity $V_\theta^{\textrm{max}}$ is built for each case.
  (b) Vorticity profile along axis $r$ built in (a) for each case. Profile for $Ma=2.4$ case is extracted as the data from VD Line~1 in figure~\ref{fig: Vthe-VD-line2} and vorticity contour in figure~\ref{fig: SBV-SDR}(b2).
  (c) SBV circulation $|\Gamma^-/\Gamma|$ for different shock Mach numbers vs. the scaled time $\Delta Ut/D$. }\label{fig: SBV profile}
\end{figure}
%%%%  验证公式4.17，对图13详细描述+找到最大涡量+介绍负环量+给出正比的证据，并给出表格证明！
The SBV model (\ref{eq: vavd-illu8}) predicts the normalized baroclinic vorticity at the bridge structure, which should be proportional to the peak value of SBV circulation $|\Gamma^-/\Gamma|$ (denoted as $\Gamma_{b}/{\Gamma}$ hereafter) in figure~\ref{fig: cir-SDR-Ma}.
To validate (\ref{eq: vavd-illu8}), we first provide evidence of the proportionality between $\Gamma_b$ and $\omega_b D^2$ from measured data and then compare with the SBV model.

The vorticity contours at $t=t_b$ for each Mach number are plotted in figure~\ref{subfig: SBV contour} ($Ma=2.4$ can be found in figure~\ref{fig: SBV-SDR}(b2)). A defined line is extracted from the vorticity contour to illustrate the vorticity distribution along the axis $r$ with origin at vortex core. Axis $r$ points to the location of maximum azimuthal velocity $V_\theta^{\textrm{max}}$. The distributions of vorticity along the axis $r$ for different cases are plotted in figure~\ref{subfig: SBV profile}. The crest and trough are observed, meaning the SBV layer. Since positive vorticity can not eliminate the shock induced initial vorticity deposition from SBV (same reason as SBV circulation), we measure the peak negative vorticity $\omega_{b,p}$ for each case as indicated in figure~\ref{subfig: SBV profile}.

As for the peak value of SBV circulation $|\Gamma^-/\Gamma|$, the baroclinic circulation under the scaling of $\Delta Ut_b/D$ is shown in figure~\ref{subfig: cir-dUt-D}.
From one-dimensional gas dynamics, we can theoretically obtain the velocity difference $\Delta{U}$ between the shocked bubble upstream interface $u'_2$ and the ambient air $u'_1$. The direct link between velocity difference $\Delta{U}$ and $t_{b}$ can be found:
\begin{equation}\label{eq: tU-D}
  \frac{\Delta{U}t_{b}}{D}=0.54\sim0.68 \quad \textrm{for }  Ma=1.22\sim 4.
\end{equation}
At approximately $\Delta Ut/D\approx 0.6$, the peak of baroclinic circulation is obtained and tabulated in table \ref{tab: t_baro}. From the last column of the table, the proportional relation between $\Gamma_b$ and $\omega_{b,p}D^2$ is evident.
\begin{table}
  \begin{center}
\def~{\hphantom{0}}
  \begin{tabular}{lccccccc}
      $Ma$  & $\Delta{U}$(m/s)   &   $t_{b}$($\umu$s) & $\Delta{U}t_{b}/D$(-)
                  &  $\omega_bD^2/\Gamma$(-)  &  $\Gamma_b/\Gamma$(-)  &  $k$(-) & $\omega_{b,p}D^2/\Gamma_b$(-) \\[3pt]
       1.22   & 48.60     & 58.0   &  0.54    &   4.51    &  0.46   &   9.8    &  $-110$   \\
       1.8     & 150.00   & 19.2   &  0.55    &   6.26    &  0.56   &   11.1  &    $-107$  \\
       2.4     & 236.88   & 13.2   &  0.60    &   8.88    &  0.90   &   9.9    &  $-99.6$  \\
       3        & 316.58   & 11.0   &  0.67    &   13.2    &  1.05   &   12.5   &  $-97.5$   \\
       4        & 442.92   &  8.0   &  0.68    &    15.0   &  1.20   &   12.5     &   $-98.0$  \\
  \end{tabular}
  \caption{$\Delta{U}=u'_2-u'_1$ obtained from one-dimensional gas dynamics. $t_{b}$ is the time at the first maximum of baroclinic circulation growth referring to figure \ref{fig: SBV_mech}. $k\equiv(\omega_bD^2)/\Gamma_b$ where $\omega_b$ is obtained from (\ref{eq: vavd-illu8}) at $t=t_b$ and $\Gamma_b/\Gamma$ is the peak value of $|\Gamma^-/\Gamma|$ from figure \ref{subfig: cir-dUt-D}. $\omega_{b,p}$ is the peak negative vorticity measured from figure \ref{subfig: SBV profile}.}
  \label{tab: t_baro}
  \end{center}
\end{table}
Moreover, through the value of $t_b$, SBV model (\ref{eq: vavd-illu8}) can be validated:
\begin{equation}\label{eq: vavd-illu9}
  \frac{\omega_bD^2}{\Gamma}\propto\frac{\Gamma_b}{\Gamma}\propto\frac{\omega_{b,p}D^2}{\Gamma}.
\end{equation}
The evidence of the proportionality between $\Gamma_b$ and $\omega_bD^2$ is (near) constancy of the parameter $k$ in table \ref{tab: t_baro}, showing that the simplified model, as illustrated in figure~\ref{fig: SBV_mech}, is appropriate.

\subsubsection{Azimuthal velocity increase model: $\Delta V_{\theta}^b$}
%% 0. 同时r=rm时涡量为0，此后切向速度开始下降；1. 说明rm及V_the的选取；2. V_the与omega关系验证(点到点涡涡量理论上为0)；3. 验证模型；
\begin{figure}
  \centering
  \includegraphics[clip=true,trim=0 90 0 0,width=.99\textwidth]{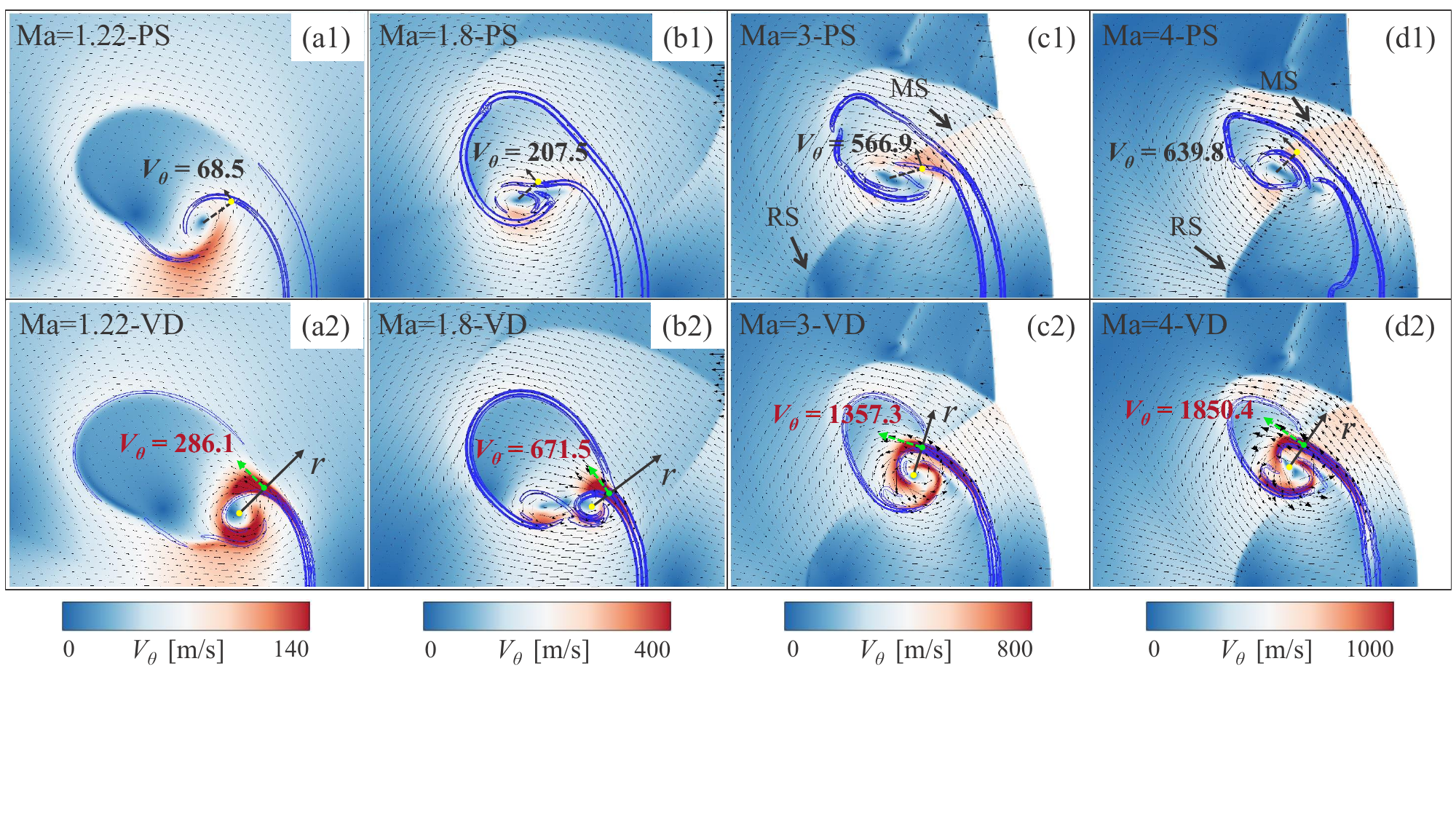}\\
  \caption{Azimuthal velocity distribution in the frame of vortex centre for different shock Mach numbers for VD SBI [shown as (a2) to (d2)] by comparing with results for PS SBI [shown as (a1) to (d1)]. The moments captured and coordinate $r$ are the same as the ones in figure~\ref{subfig: SBV contour}. }\label{fig: Vthe-Ma-VD}
\end{figure}
\begin{figure}
    \centering
    \subfigure[]{
    \label{subfig: 15a}
    \includegraphics[clip=true,trim=10 15 20 20, width=.45\textwidth]{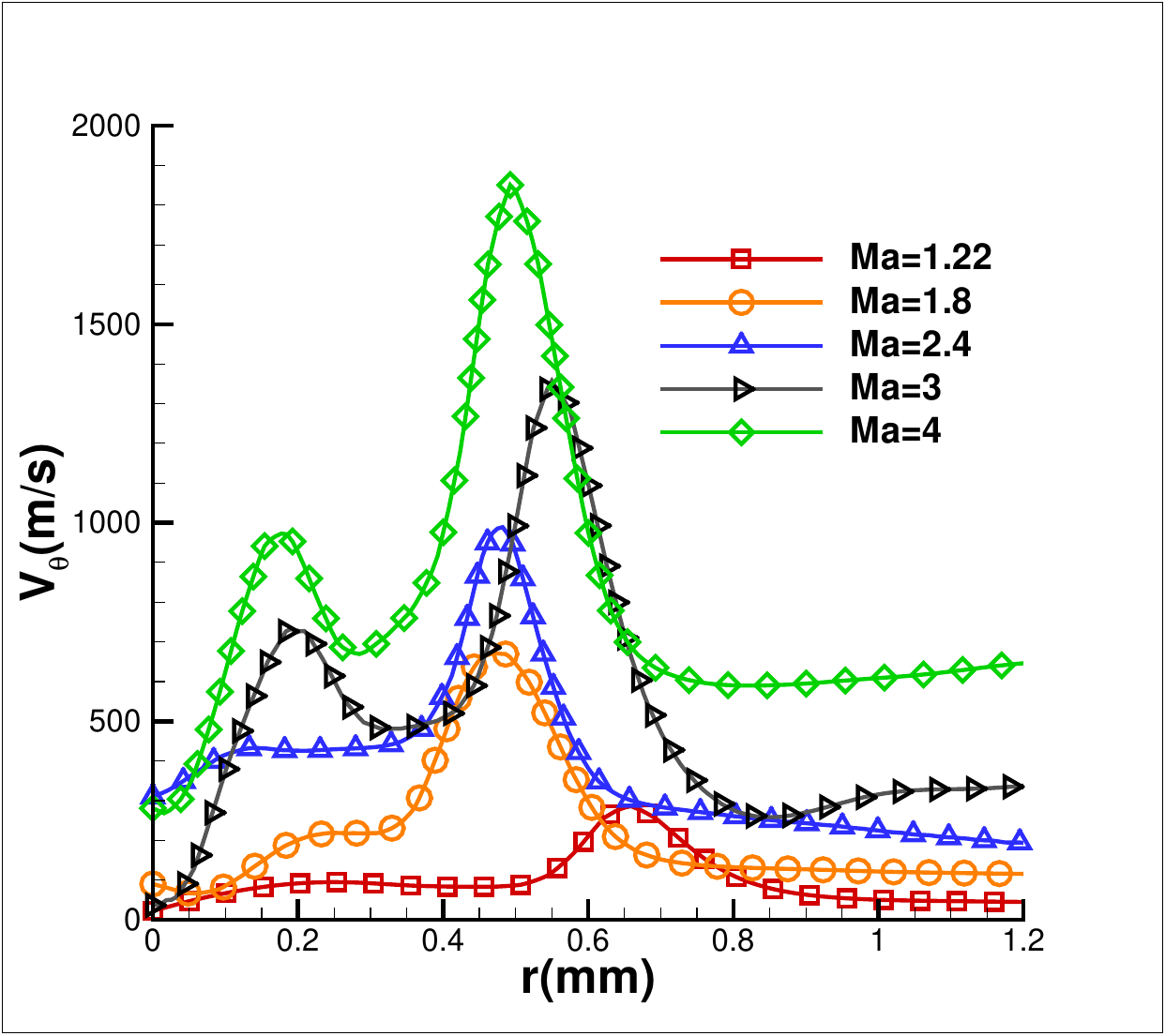}}
    \subfigure[]{
    \label{subfig: 15b}
    \includegraphics[clip=true,trim=10 15 20 20, width=.45\textwidth]{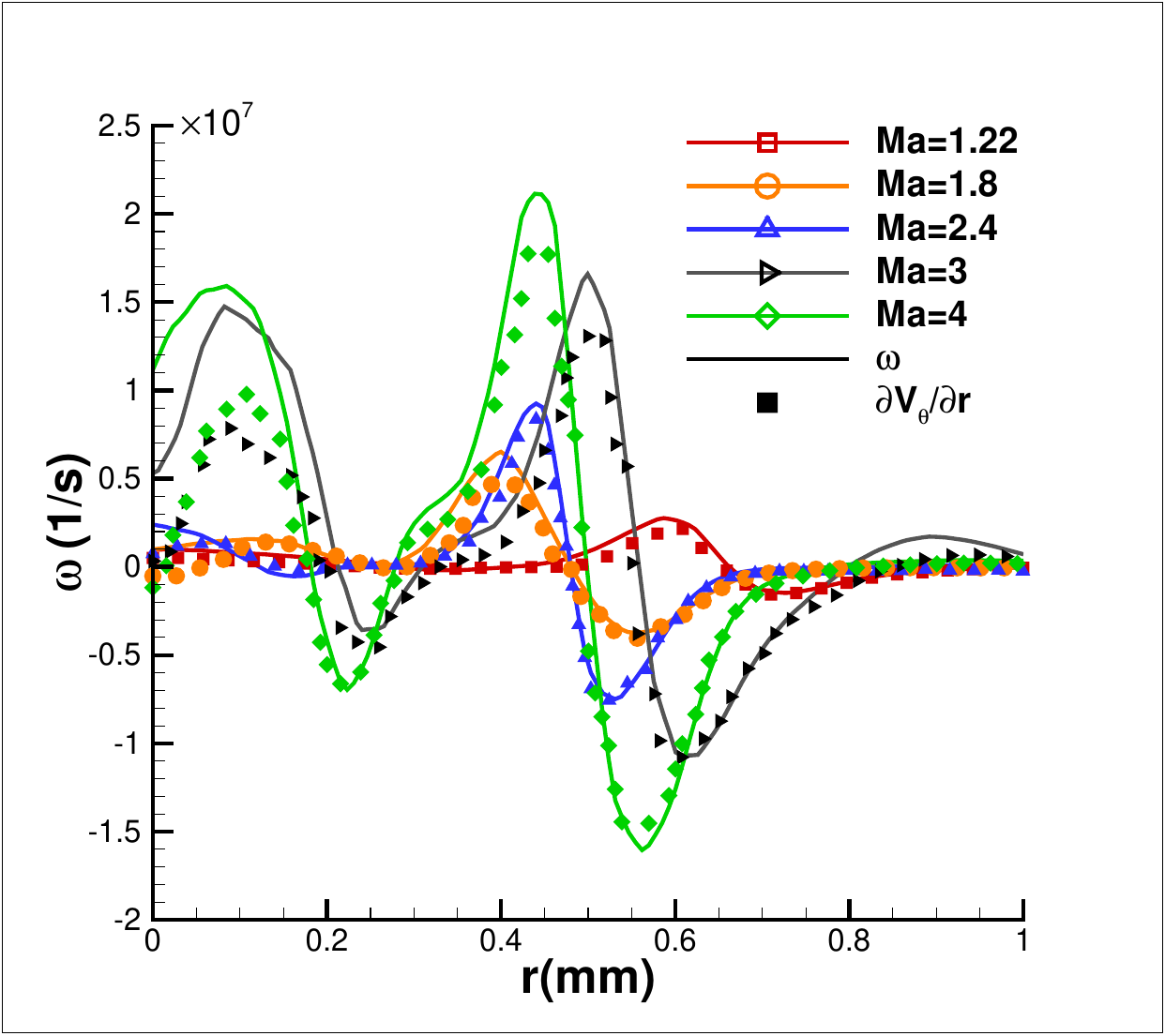}}
    \subfigure[]{
    \label{subfig: 15c}
    \includegraphics[clip=true,trim=10 15 20 20, width=.45\textwidth]{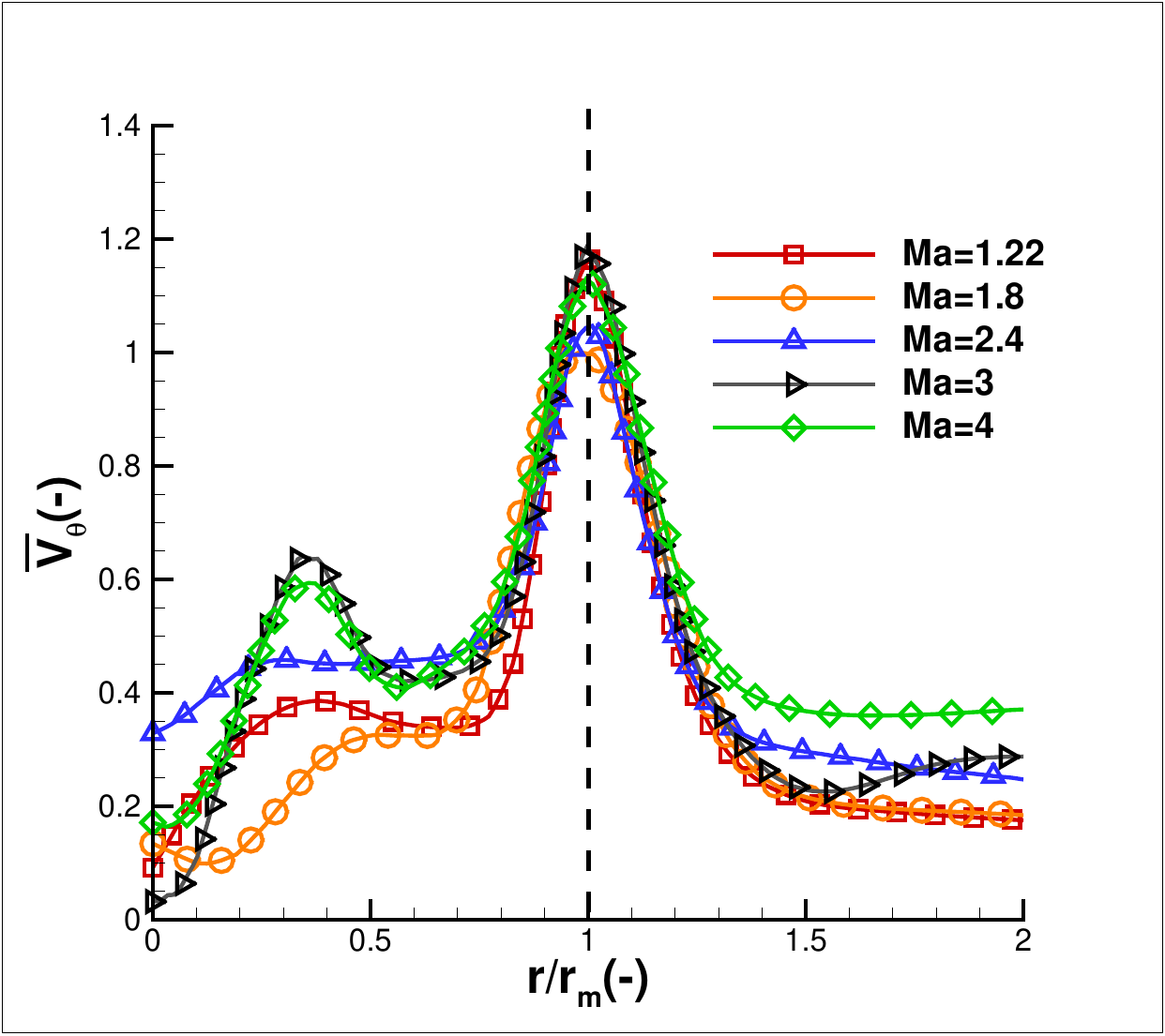}}
    \subfigure[]{
    \label{subfig: 15d}
    \includegraphics[clip=true,trim=10 15 20 20, width=.45\textwidth]{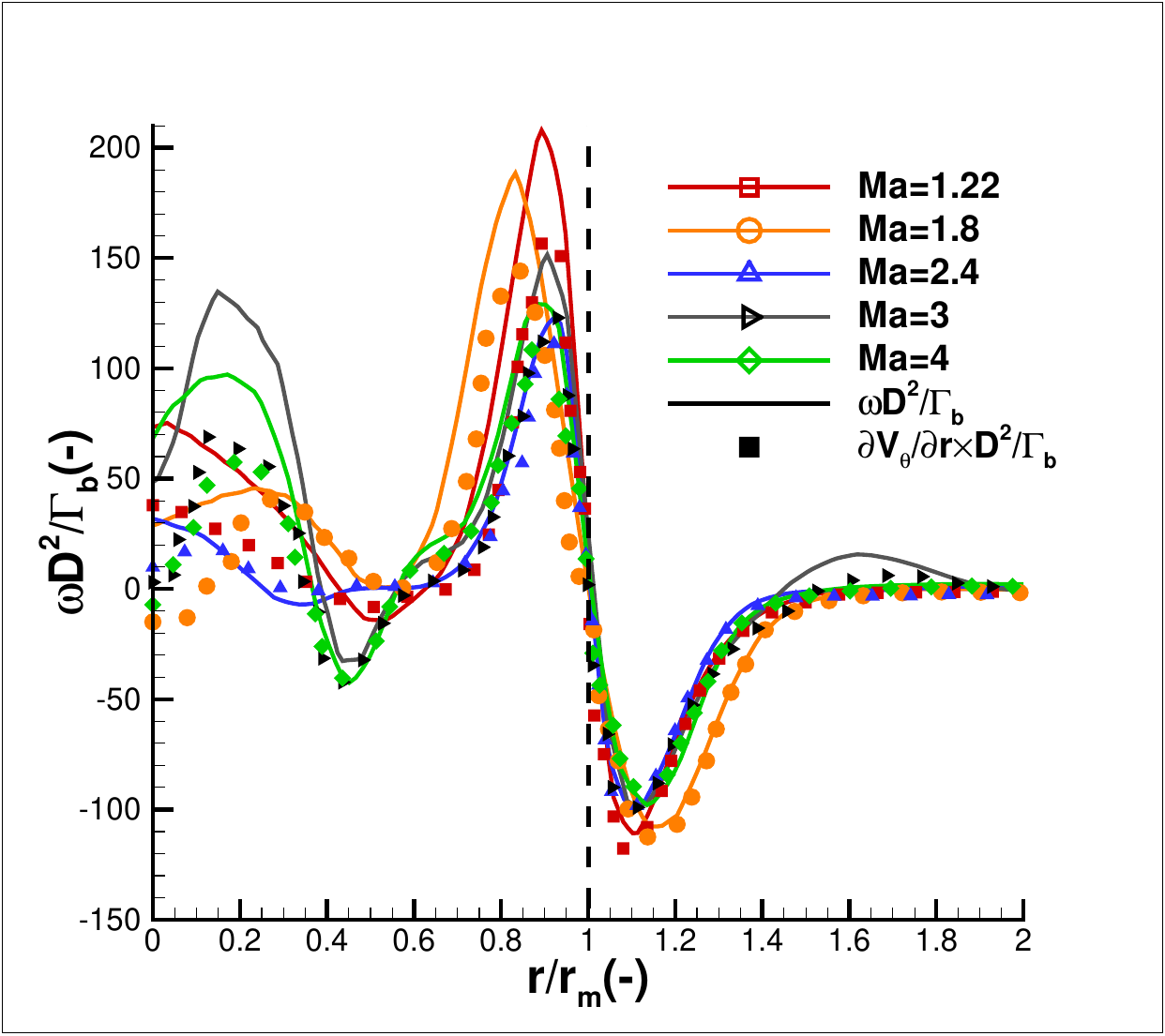}}
    \caption{
    (a) Azimuthal velocity distribution along the axis $r$ in figure~\ref{fig: Vthe-Ma-VD} for VD SBI with different shock Mach numbers.
    (b) The comparison between $\partial\widetilde{V}_{\theta}/\partial r$ and vorticity distribution along the axis $r$ in figure~\ref{subfig: SBV profile}.
    (c) Azimuthal velocity normalized by the modelled value (\ref{eq: vthe-2}) along the normalized axis $r/r_m$.
    (d) Normalized vorticity $\omega D^2/\Gamma_b$ along the normalized axis $r/r_m$.
    \label{fig: v_the-lines} }
\end{figure}
In this part, azimuthal velocity increased by SBV is modelled and validated. We have noted the derivative relation between baroclinic increased azimuthal velocity and SBV (\ref{eq: vthebaro-omega}):
\begin{equation}\label{eq: wb-vthe}
  \omega_b=\frac{\partial \Delta V_{\theta}^b}{\partial r}.
\end{equation}
Integrating SBV in (\ref{eq: vavd-illu6}) from origin to radius $r(t)$, we obtain:
\begin{equation}\label{eq: vthe-vd-theo}
    \Delta V_{\theta}^b  =  \int_0^{r(t)} \omega_b \mathrm{d}r
    =\frac{\Gamma|{At}^+|}{\upi\delta^{'}}\int_0^{r(t)} \left(\frac{1}{r}-\frac{1}{D}\right)\mathrm{d}r
 \end{equation}
Due to the fact that SBV at $(0,r(t)-\delta')$ is zero (density gradient is zero and see figure~\ref{fig: SBV_mech}), the SBV induced azimuthal velocity increase at $r(t_b)$ is:
\begin{eqnarray}\label{eq: vthe-vd-theo-2}
  \Delta V_{\theta}^b & = &  \frac{\Gamma|{At}^+|}{\upi\delta^{'}}\int_{r(t_b)-\delta^{'}}^{r(t_b)}\left(\frac{1}{r}-\frac{1}{D}\right)\mathrm{d}r \nonumber\\
        & = &
        \frac{\Gamma|{At}^+|}{\upi\delta^{'}}\left[\textrm{ln}\left(\frac{r(t_b)}{r(t_b)-\delta^{'}}\right)-\frac{\delta^{'}}{D}\right] \nonumber \\
        & = &
        \frac{\Gamma|{At}^+|}{\upi\delta^{'}}\left[\textrm{ln}\left(\frac{D-\Delta Ut_b}{D-\Delta Ut_b-\delta^{'}}\right)-\frac{\delta^{'}}{D}\right].
\end{eqnarray}
%It is noteworthy that if the integration (\ref{eq: vthe-vd-theo}) continues to $r(t)+\delta'$, the azimuthal velocity will gradually decrease to near zero along the axis $r$ due to the negative SBV during this range. This means that the maximum SBV induced azimuthal velocity occurs at $r=r(t_b)$ as indicated by (\ref{eq: vthe-vd-theo-2}).
Therefore, from (\ref{eq: vthe-VD}), total azimuthal velocity at $t=t_b$, when peak azimuthal velocity is reached, is composed of the point vortex induced part $V_{\theta}$ and SBV induced part $\Delta V_{\theta}^b$:
\begin{equation}\label{eq: vthe-2}
  \widetilde{V_{\theta}}=V_{\theta}+\Delta V_{\theta}^b=\frac{\Gamma}{2\upi r_m}+
  \frac{\Gamma|{At}^+|}{\upi\delta^{'}}\left[\textrm{ln}\left(\frac{D-\Delta Ut_b}{D-\Delta Ut_b-\delta^{'}}\right)-\frac{\delta^{'}}{D}\right],
\end{equation}
where $r_m$ is the distance from the vortex centre.

Figure \ref{fig: Vthe-Ma-VD} shows the azimuthal velocity contour for different shock Mach numbers for both PS SBI and VD SBI (the velocity of vortex $V_v$ is studied in appendix~\ref{App: vortex velo}). A higher azimuthal velocity for the VD cases is shown for all Mach numbers, which explains the faster mixing for the VD cases than PS cases.
To quantify the azimuthal velocity distribution of VD SBI,  the same coordinate $r$ as the one in figure~\ref{subfig: SBV contour} are examined, as shown in figure~\ref{subfig: 15a}. The location $r_m$ and the value of maximum azimuthal velocity can be obtained, which should agree with maximum azimuthal velocity recorded in figure~\ref{fig: cir-SDR-Ma}. Interestingly, the derivative of azimuthal velocity along axis $r$ is near the local vorticity distribution, as shown in figure~\ref{subfig: 15b}, validating the relation between SBV and its induced velocity increase (\ref{eq: vthebaro-omega}). Moreover, it can be inferred that the vorticity attributed to point vortex model is much smaller than SBV, which satisfies the zero vorticity of a standard point vortex model~\citep{wu2007vorticity}. The high vorticity near the vortex core for high Mach numbers is due to the entrainment of vorticity from the bottom of vortex centre.
The azimuthal velocity distribution normalized by the modelled value (\ref{eq: vthe-2}), $\overline{V_\theta}$, is presented in figure~\ref{subfig: 15c}. Good agreement can be observed when $r=r_m$, $\overline{V}_\theta\approx1$.
The general trend of the modelled value is consistent with the measured value, as tabulated in the last two columns of table~\ref{tab: vthe}.

%It is noteworthy that $V_\theta=\Gamma/2\upi r_m$ in table~\ref{tab: vthe} does not agree with the measured ones of PS SBI in figure~\ref{fig: Vthe-Ma-VD}(a1) to \ref{fig: Vthe-Ma-VD}(d1). The reason can be explained as the moment captured in the PS case is still early that a mature Lamb-Oseen type vortex has not been formed. In higher shock Mach number as $Ma$=4 PS case, Mach stem structure may also change the azimuthal velocity, as explained in $\S$\ref{subsec: PS mixing}, which is not considered in the point vortex model.
\begin{table}
  \begin{center}
  \begin{tabular}{lccccc}
      $Ma$  &  $\Gamma$(m$^2$/s)  & $r_m$(mm)  & $V_{\theta}$(m/s)    &   $\widetilde{V}_{\theta}$(m/s)(\ref{eq: vthe-2}) & $\widetilde{V}_{\theta}$(m/s)(meas.) \\[3pt]
       1.22 &  0.78  & 0.64  &  194.6    & 245.8  &  286.1         \\
       1.8   &  1.67  & 0.48  &  553.7    & 671.4   &  671.5     \\
       2.4   &  2.26  & 0.48  &  749.3    & 944.3   &  988.9         \\
       3      &  2.82  & 0.55  &  814.2    & 1142.9   &  1357.3       \\
       4      &  3.69  & 0.49  &  1186.4  & 1637.7   &  1850.4    \\
  \end{tabular}
  \caption{Validation of baroclinic velocity increase model (\ref{eq: vthe-2}). $V_\theta=\Gamma/2\upi r _m$, where
$r_m$ is the radius from vortex centre to the location of maximum azimuthal velocity $\widetilde{V}_{\theta}$, measured from figure~\ref{subfig: 15a}.  }
  \label{tab: vthe}
  \end{center}
\end{table}

In summary, $\S$\ref{subsec: VD mixing} and $\S$\ref{subsec: inertial velocity model} reveal the additional stretching of SBV in VD cases, explaining the shorter mixing time in VD SBI.  Since vortex stirring, compression effect and diffusion in PS SBI are shared in the VD cases, the mixing time for VD SBI can be implicitly expressed by adding the baroclinic azimuthal velocity increase $\Delta V_{\theta}^b$:
\begin{equation}\label{eq: cir-SDR}
  t_\mathfrak{m}^{VD}=f\left[\Gamma,\mathscr{D},r,\eta,\Delta V_{\theta}^b(\omega_b)\right].
\end{equation}
$\S$\ref{sec: theo} will discuss the mixing time for PS SBI and VD SBI based on the reduced-order model of baroclinic vorticity dynamics.

\section{A novel mixing time for VD SBI}
\label{sec: theo}
Based on the mixing mechanism for PS SBI revealed in (\ref{eq: tm-PS-2}) and for VD SBI revealed in (\ref{eq: cir-SDR}), stirring, diffusion, compression from shock, and density effect are systematically considered for modelling mixing behaviour within a vortical flow.
%The theoretical mixing time is derived from a point vortex mixing model combining the compression model in $\S$\ref{subsec: PS mixing} and the density inertial velocity model in $\S$\ref{subsec: inertial velocity model}.
%The characteristic mixing time for PS SBI considers the stirring of circulation, diffusion and compression effect, which coincides with mixing time measured in $\S$3. Then, by considering baroclinic azimuthal acceleration, mixing time for VD SBI adds the inertial effect revealed in last section. The general agreement is obtained from the theoretical value and measured one.

\subsection{Characteristic mixing time for PS SBI}
\label{subsec: PS model}
\begin{figure}
  \centering
  \includegraphics[clip=true,trim=0 0 180 0,width=.85\textwidth]{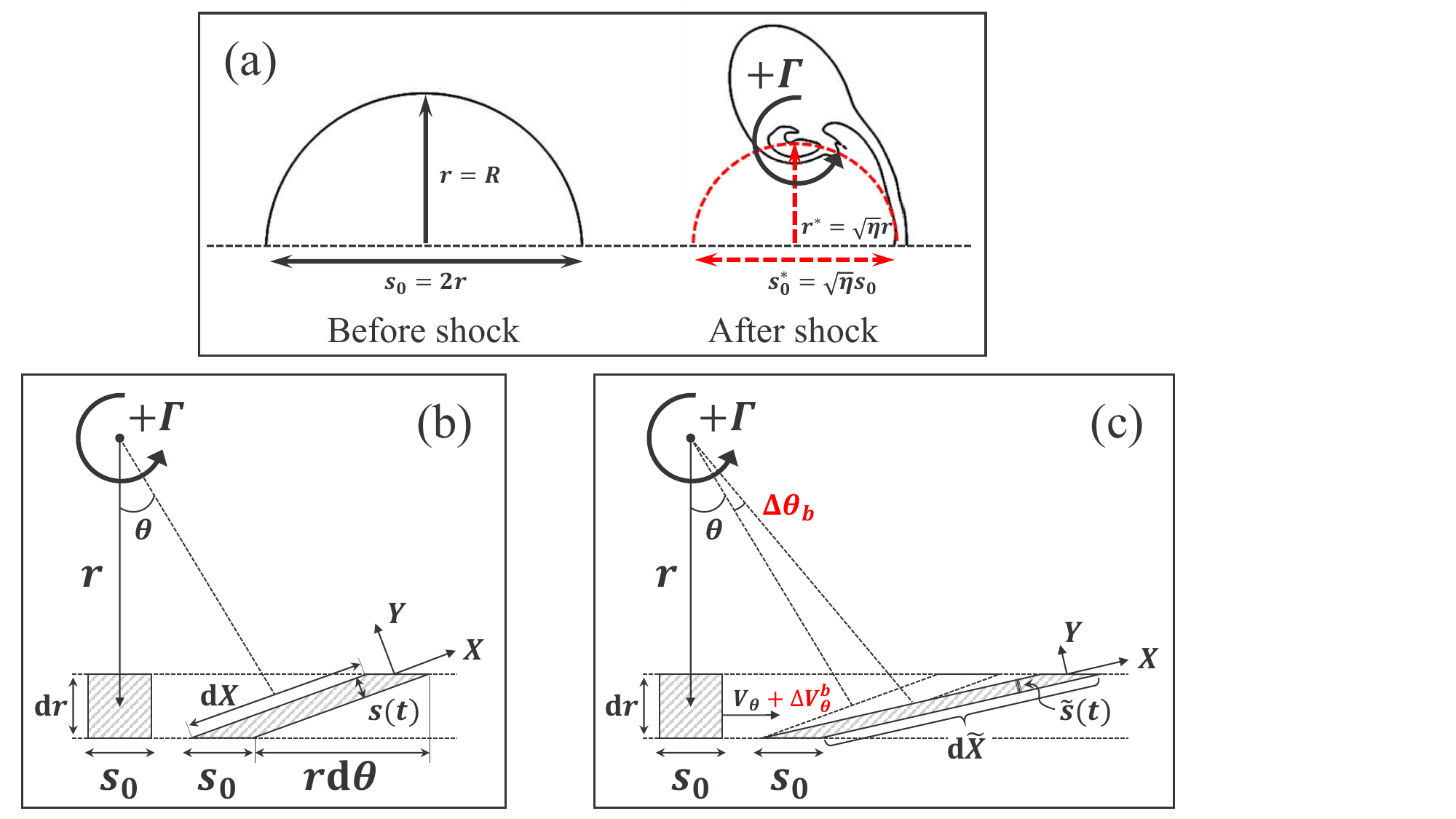}\\
  \caption{(a) Compression effect on initial mixing region; (b) stretching of passive scalar blob under the standard point vortex \citep{meunier2003vortices}; (c) SBV-enhanced stretching in variable density mixing.}\label{fig: mix-mech}
\end{figure}
From $\S$\ref{subsec: PS mixing}, it is observed that the mixing of PS SBI follows a typical vortical stretching enhanced mixing.
For the vortical flow stretching velocity $V_{\theta}$,  \citet{meunier2003vortices} theoretically proposed a characteristic mixing time for a passive scalar blob with length $s_0$, at distance $r$ away from vortex centre stretched by vortical flows, as illustrated in figure~\ref{fig: mix-mech}(b):
\begin{equation}\label{eq: t_s}
  t_\mathfrak{m}^{PS}=t_{m}=f(\Gamma, \mathscr{D}, r)=\frac{r^2}{\Gamma}\left(\frac{3\upi^2}{16}\right)^{1/3}\left(\frac{s_0}{r}\right)^{2/3}
  \left(\frac{\Gamma}{\mathscr{D}}\right)^{1/3}.
\end{equation}
When $t/t_m>1$, maximum concentration will be below the initial concentration, i.e. $\overline{Y}<1$, and mixing enters the diffusion-controlled stage.
Here, $t_m$ displays the P\'eclet number $Pe=\Gamma/\mathscr{D}$ dependence.
A similar $Pe^{1/3}$ dependency was also derived for a flame vortex interaction modelled by \citet{marble1985growth} and was used to evaluate the mixing time in supersonic streamwise vortex by \citet{waitz1997enhanced}.
To agree with the physical characteristic of (\ref{eq: t_s}), let $r=R$ be the cylindrical bubble radius and $s_0=2r$, as shown in figure \ref{fig: mix-mech}(a).
\begin{figure}
    \centering
    \subfigure[]{
    \label{fig: tm-ps-1}
    \includegraphics[clip=true,trim=15 15 50 50, width=.45\textwidth]{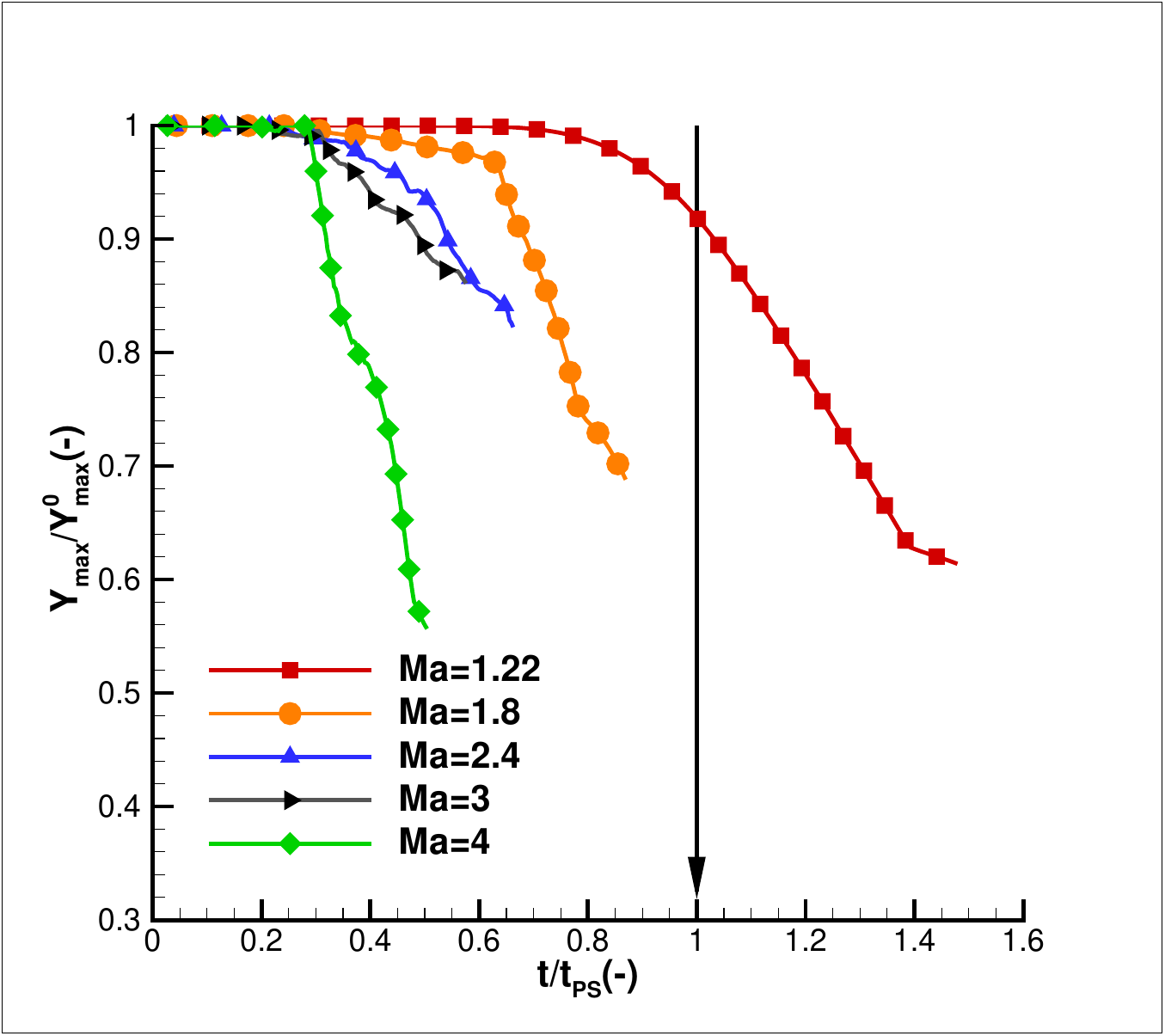}}
    \subfigure[]{
    \label{fig: tm-ps-2}
    \includegraphics[clip=true,trim=15 15 50 50, width=.45\textwidth]{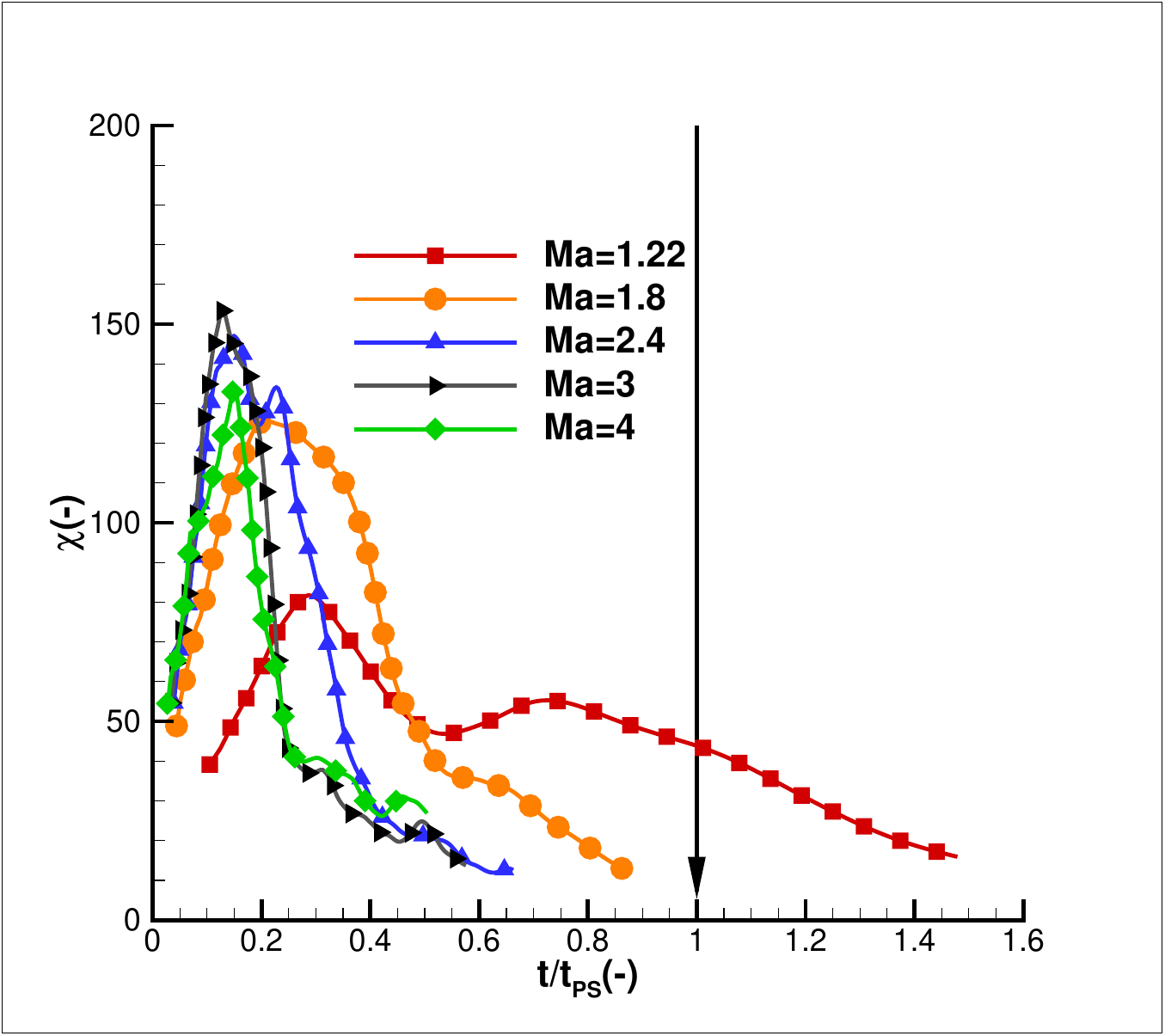}}
    \caption{(a) Maximum concentration decay and (b) scalar dissipation for PS SBI vs. scaled time $t/t_{m}$, where $t_m$ is expressed in (\ref{eq: t_s}).\label{fig: tm-ps} }
\end{figure}

Figure \ref{fig: tm-ps} shows the time history of normalised maximum concentration, $\overline{Y}$, and scalar dissipation, $\chi$, under the dimensionless scaling of $t/t_m$. For $Ma$=1.22, at $t/t_m\approx 1$, the maximum concentration decrease agrees well with the theoretical prediction. A similar trend is found in the scalar dissipation $\chi$ as most of the mixing rate is high before $t/t_m<1$ for PS SBI at $Ma$=1.22. This agreement shows that the scaling by (\ref{eq: t_s}) is appropriate for the general vortical mixing flows, such as PS SBI.
However, it is found that it can not scale the mixing of PS SBI for all Mach numbers, especially the cases of high Mach number.
The agreement for $Ma$=1.22 can be explained by the fact that shock compression effects on mixing are relatively weak in this scenario.
Thus, a definition of mixing time considering shock compression effects is needed.

\begin{figure}
    \centering
    \subfigure[]{
    \label{fig: tm*-ps-1}
    \includegraphics[clip=true,trim=15 15 50 50, width=.45\textwidth]{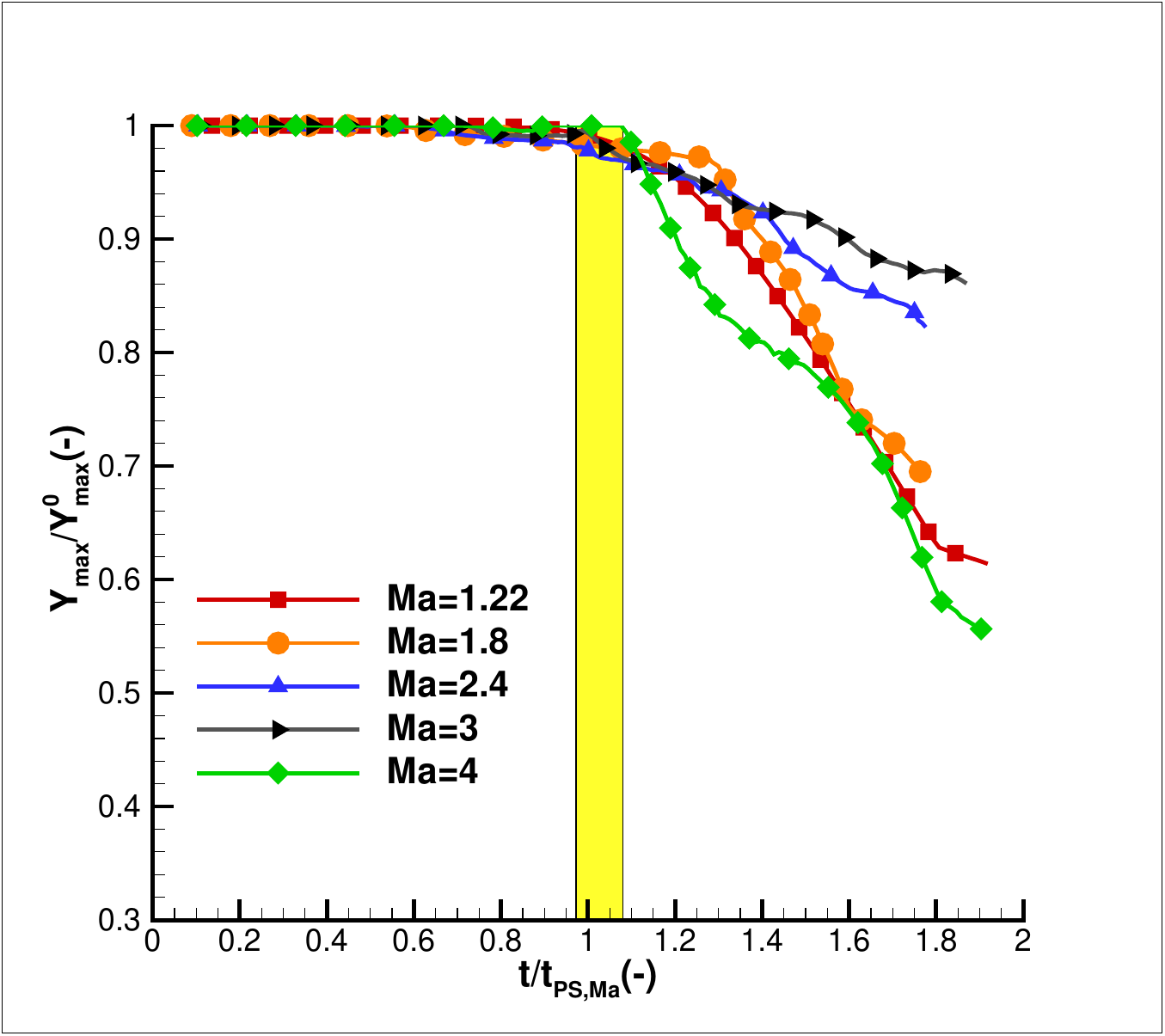}}
    \subfigure[]{
    \label{fig: tm*-ps-2}
    \includegraphics[clip=true,trim=15 15 50 50, width=.45\textwidth]{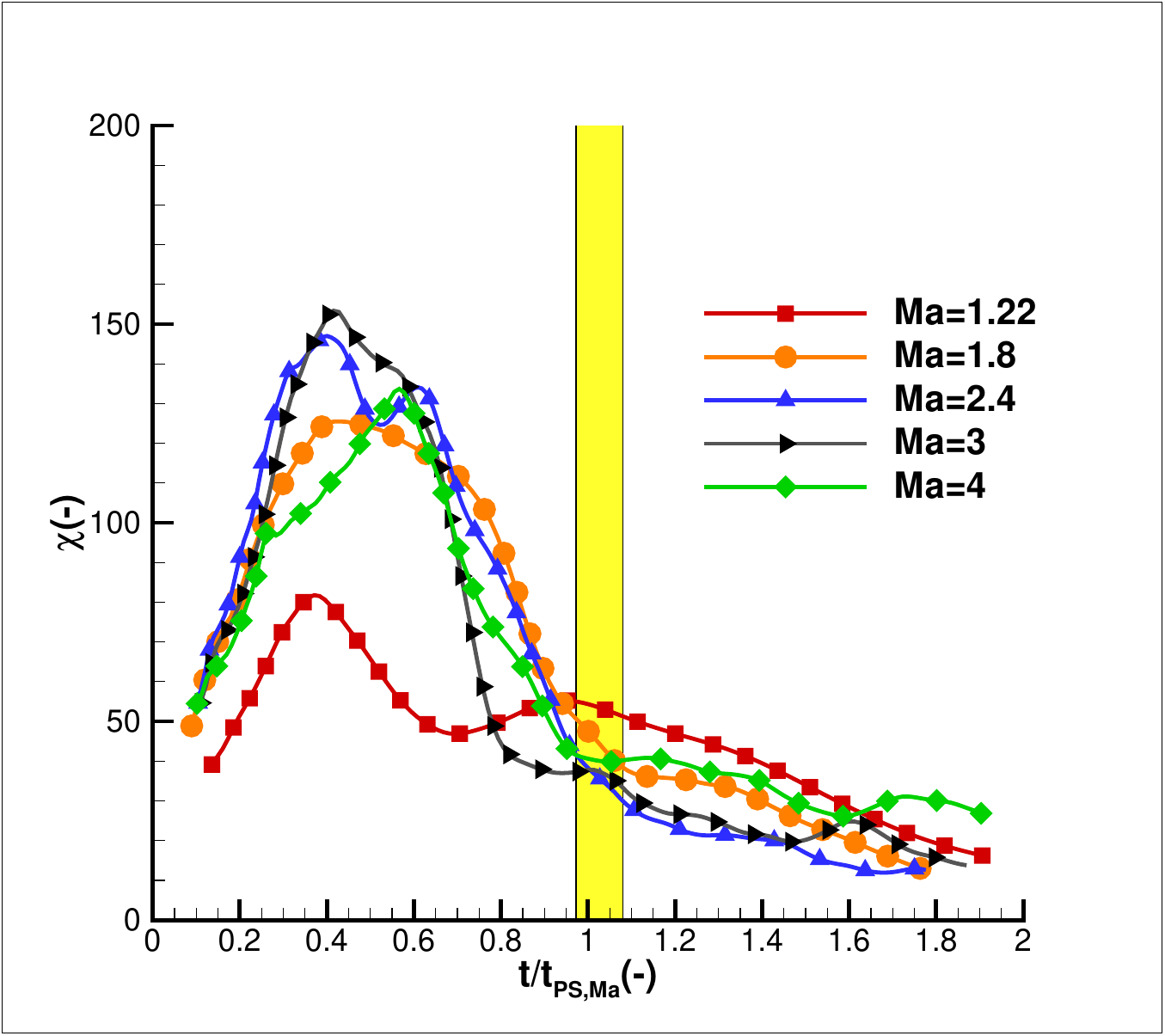}}
    \caption{(a) Maximum concentration decay and (b) scalar dissipation for PS SBI vs. scaled time $t/t^*_{m}$. By considering compression effects from the shock on the initial mixing region, mixing time can be correctly predicted at time scale of $t/t^*_{m}\approx 1$ where $t^*_m$ is expressed in (\ref{eq: t-s*}).  \label{fig: tm*-ps} }
\end{figure}

From $\S$\ref{subsec: PS mixing}, it is further found that after the initial shock, the cylindrical bubble is compressed and shrinks volumetrically, as depicted in figure~\ref{fig: Maeff-PS}(b). In other words, the amount of scalar left to be stirred by the vortex is reduced by a rate $\eta$ compared to initial bubble volume (\ref{eq: com}).
This finding can stimulate to revise the characteristic length $s_0$ and distance from vortex centre $r$:
\begin{equation}\label{eq: effect-radius}
  \mathcal{V}_b=\frac{\upi r^* s_0^*}{2}=\eta \frac{\upi r s_0}{2}=\eta \mathcal{V}_0 \Rightarrow \quad r^*=\sqrt{\eta}r  \text{ and }  s_0^*=\sqrt{\eta}s_0.
\end{equation}
As illustrated in the right part of figure~\ref{fig: mix-mech}(a), (\ref{eq: effect-radius}) denotes a compressed cylindrical bubble, with radius $r^*$ and centerline diameter $s_0^*$, stirred by a point vortex located at the bubble top.
This physical process can reflect the PS SBI mixing to some extent as discussed in appendix~\ref{App: PSmixing}.
After revising $s_0^*$ and $r^*$ in (\ref{eq: t_s}), the mixing time for passive scalar that considers compression can be expressed as:
\begin{equation}\label{eq: t-s*}
  t_\mathfrak{m}^{PS}=t_{m}^*=f(\Gamma, \mathscr{D}, r, \eta)=\frac{\eta r^2}{\Gamma}\left(\frac{3\upi^2}{16}\right)^{1/3}\left(\frac{\sqrt{\eta}s_0}{\sqrt{\eta}r}\right)^{2/3}
  \left(\frac{\Gamma}{\mathscr{D}}\right)^{1/3}.
\end{equation}
It reveals the proportional relationship between the scalar mixing time and the compression rate.
Figure \ref{fig: tm*-ps} shows the $\overline{Y}$ and $\chi$ versus the scaled time $t/t_m^*$, which scales well all shock Mach number cases.
At $t/t_m^*>1$, the maximum concentration begins to decrease, and the mixing rate tends to a low level of diffusion-controlled mixing, which validates that for vortical stretching, a compression effect by the initial shock impact is vital for the time that mixing can sustain.
\begin{table}
  \begin{center}
\def~{\hphantom{0}}
  \begin{tabular}{lcccccc}
      $Ma$  &  $\Gamma$(m$^2$/s)  & $\overline{\kappa}$(-)  & $\eta$(-)   &   $t_m$($\umu$s) & $t_m^*$($\umu$s) & $t_\mathfrak{m}^{PS}$($\umu$s) \\[3pt]
       1.22 &  0.78  & 1.4  &  0.771     & 335.3   &  258.6  &    292.2        \\
       1.8   &  1.67  & 1.1  &  0.489     & 219.0   &  107.1  &    128.3        \\
       2.4   &  2.26  & 1.0  &  0.373     & 184.7   &  69.0    &    74.1           \\
       3      &  2.82  & 1.0  &  0.308     & 159.4   &  49.1    &     53.1           \\
       4      &  3.69  & 1.0  &  0.265     & 133.2   &  35.3     &    39.5           \\
  \end{tabular}
  \caption{Comparison between theoretical mixing time from (\ref{eq: t_s}), (\ref{eq: t-s*}) and measured value referring to table \ref{tab: t_m}. Circulation $\Gamma$, diffusivity coefficient $\overline{\kappa}$ (see appendix \ref{App: diffusivity}) and compression rate $\eta$ are also listed. }
  \label{tab: tm-ps}
  \end{center}
\end{table}

Table \ref{tab: tm-ps} gives intuitive values that are required in calculating $t_m$ and $t_m^*$. For all Mach numbers, $t_m$ overestimates the mixing time for PS SBI, as discussed. Interestingly, although $t_m^*$ is much nearer to measured values, it always underestimates the mixing time, especially for low Mach number. The inferred reason is that the PS SBI cases set in this paper are not standard point vortex models. It takes some time before PS SBI forms into a quasi-standard point vortex model; that explains the extra time in the measured values compared to the theoretically modelled ones. A comparison between scalar mixing under a standard Lamb-Oseen vortex and $Ma=1.22$ PS SBI confirm the deduction, as discussed in appendix~\ref{App: PSmixing}.

\subsection{Characteristic mixing time for VD SBI}
\label{subsec: VD model}
\begin{figure}
    \centering
    \subfigure[]{
    \label{fig: tm-vd-1}
    \includegraphics[clip=true,trim=15 15 50 50, width=.45\textwidth]{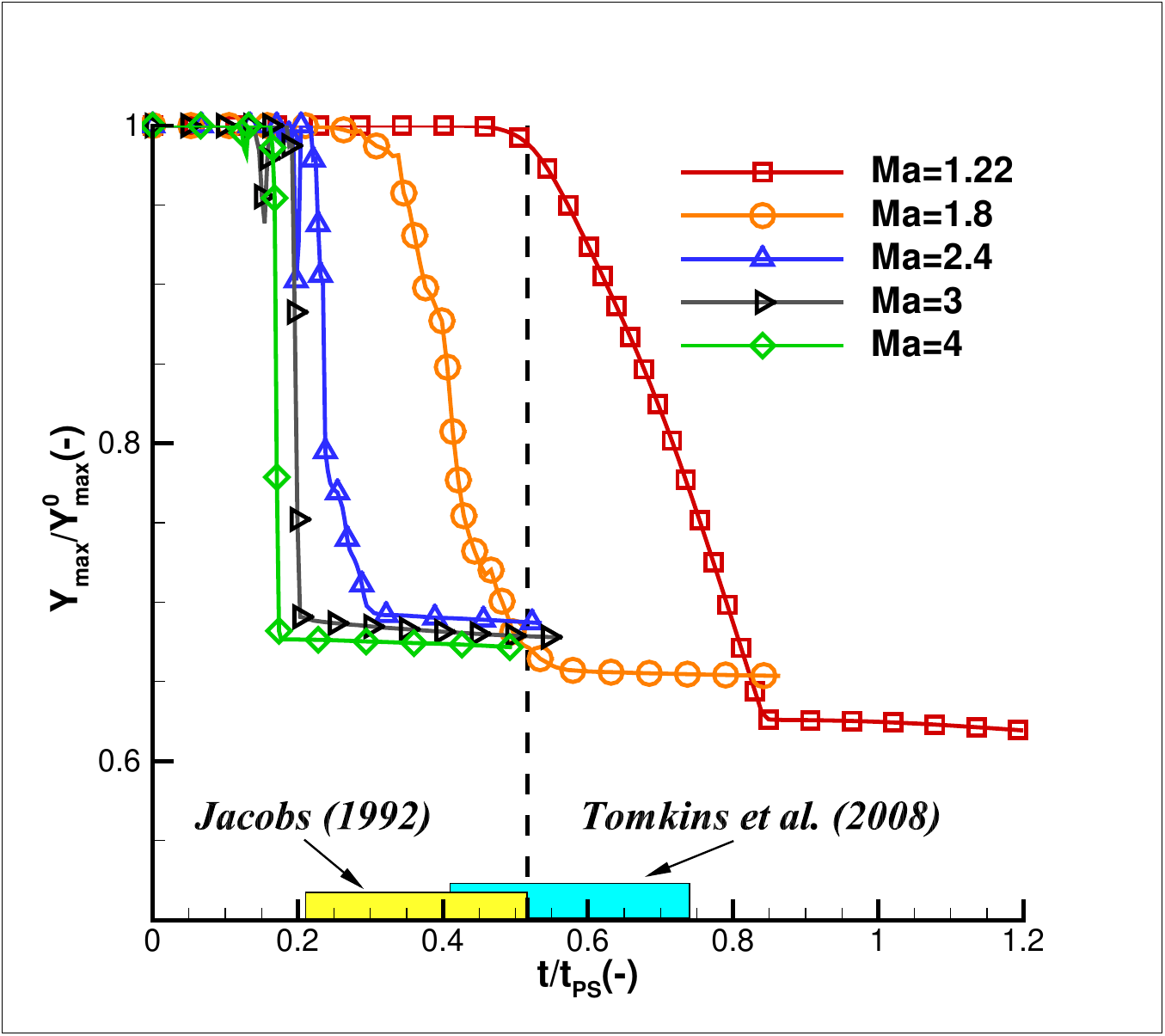}}
    \subfigure[]{
    \label{fig: tm-vd-2}
    \includegraphics[clip=true,trim=15 15 50 50, width=.45\textwidth]{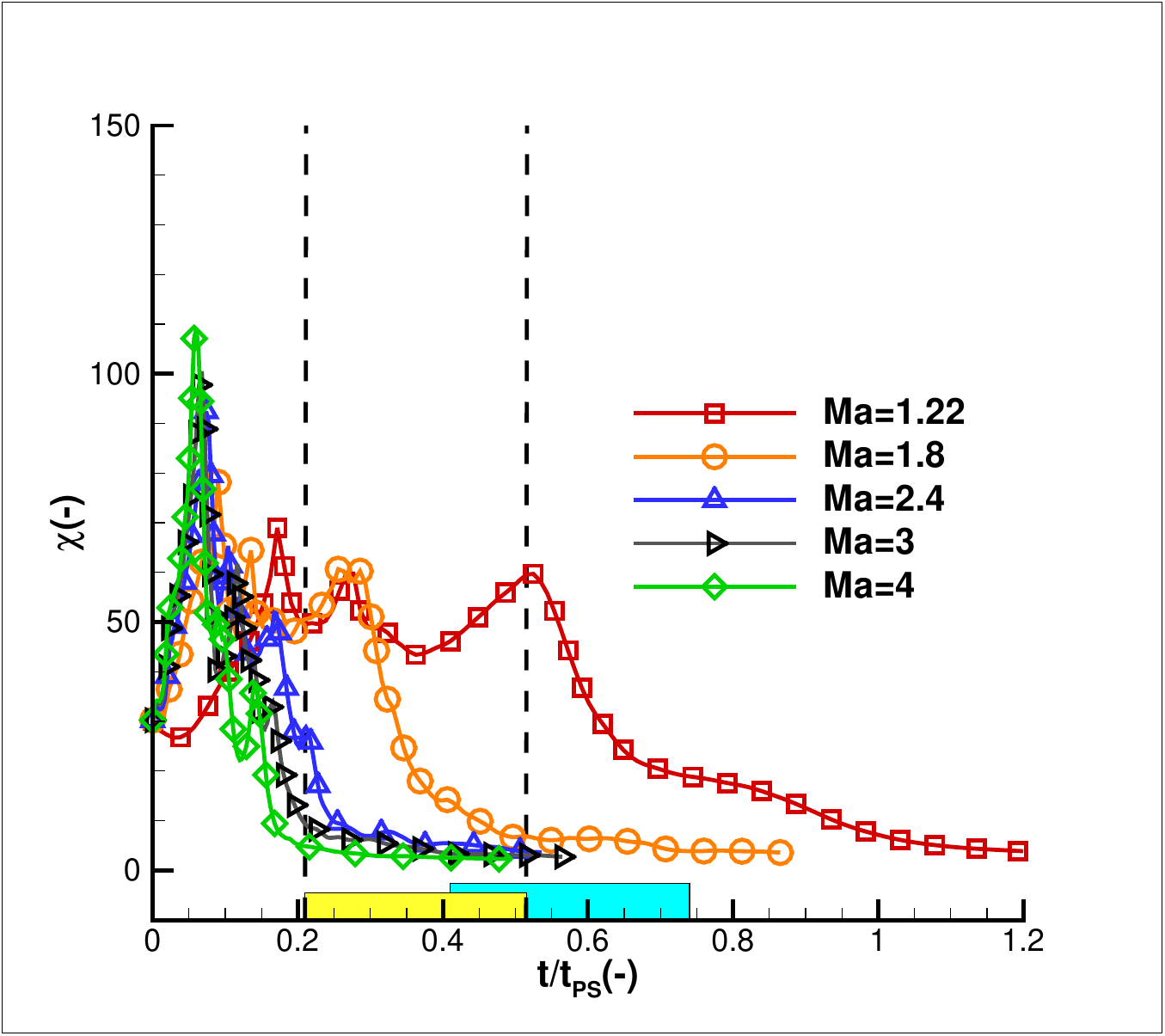}}
    \caption{(a) Maximum concentration decay and (b) scalar dissipation for VD SBI vs. scaled time $t/t_{m}$. Under the scaling of mixing time $t_{m}$ (\ref{eq: t_s}), the VD SBI cases show large discrepancy for different shock Mach number cases.  \label{fig: tm-vd} }
\end{figure}

%%  1. 原有理论不能归一；2. Jacobs的case和本文很接近，将其认为的混合时间轴转换过来，差了一个系数，并将区域标出；3. 结果发现a. 混合结束时刻就是最大浓度下降，b. 混合发生区域与本文一致；4. Tomkins也做了相关的工作，虽然有部分重叠，但归一性比较差。5. 总的来说，即使是低马赫数，这个预测依然高估了VD SBI的混合特征时间，需要构建新的理论来统一变密度引起的加速拉伸对混合的影响。
Figure~\ref{fig: tm-vd} shows the time history of two mixing indicators versus the dimensionless timescale by the canonical mixing time (\ref{eq: t_s}).
In the previous research of SBI, \citet{jacobs1992shock} first introduced this mixing time in shock-light cylindrical bubble interaction, which is close to the case studied in the VD SBI with $Ma$=1.22. It was concluded that mixing happens between $0.10<t/t_m<0.25$ by setting $r=D$ and $t_m=r^2/(\Gamma^{2/3}\mathscr{D}^{1/3})$, only with a different coefficient $2^{-4/3}(3\upi^2/16)^{1/3} \approx0.487$ compared to (\ref{eq: t_s}). Thus, we convert the mixing sustaining time from \citet{jacobs1992shock} into $0.205<t/t_m<0.51$. Interestingly, the mixing time revealed by \citet{jacobs1992shock} coincides with the case in this work, as presented in figure~\ref{fig: tm-vd}(a). When $t/t_m>0.51$, the normalised maximum concentration $\overline{Y}$ begins to decrease. Moreover, a high mixing rate is found during the sustaining mixing period, as shown in figure~\ref{fig: tm-vd}(b). Further study on shock-heavy cylindrical bubble by \citet{tomkins2008experimental} showed similar results with mixing happening in  $0.41<t/t_m<0.74$, although the overlap region is relatively weak.
In short summary, (\ref{eq: t_s}) still largely overestimates the mixing time in VD SBI in general.

%Figure \ref{fig: tm*-vd} shows the $\overline{Y}$ and $\chi$ behaviour under the scaling of $t/t_m^*$. After shock compression effect is considered, $t_m^*$ becomes close to the mixing time of VD SBI. However, $t_m^*$ of all Mach number overestimates the mixing time of VD SBI due to the faster mixing time phenomena as illustrated in previous sections. Thus, one needs to develop the mixing time approximation based on the acceleration of azimuthal velocity due to secondary baroclinic vorticity production, which reflects the important effect of variable density on mixing.
%\begin{figure}
%    \centering
%    \subfigure[]{
%    \label{fig: tm*-vd-1}
%    \includegraphics[clip=true,trim=15 15 50 50, width=.48\textwidth]{Fig23-VD-NMF-t_sMa}}
%    \subfigure[]{
%    \label{fig: tm*-vd-2}
%    \includegraphics[clip=true,trim=15 15 50 50, width=.48\textwidth]{Fig23-VD-t_sMa}}
%    \caption{(a) Maximum concentration decay of VD SBI at time scale of $t/t^*_{m}$ and (b) Time history of scalar dissipation under time scale of $t/t^*_{m}$. $t^*_{m}$ (\ref{eq: t-s*}) overestimated the mixing time in VD SBI.  \label{fig: tm*-vd} }
%\end{figure}

%% 1. 先从4.2开始回顾说对比PS，VD有SBV导致加速拉伸特征；2. 而这种变密度效应带来的加速拉伸特征是（5.1）所没有考虑到的；3. 4.3节我们建立了加速拉伸的很好模型，但没有体现其对混合的影响。我们这里基于Villermaux的单涡拉伸模型进一步提出加速拉伸模型。
It has been observed that SBV-induced additional stretching occurs in VD SBI, compared to PS SBI. The additional stretching is not considered in a single point vortex velocity distribution, as modelled in (\ref{eq: t_s}). In $\S$\ref{subsec: inertial velocity model}, we have built the SBV increased azimuthal velocity model $\widetilde{V}_{\theta}$ for VD SBI. Here, we combine this model with the advection-diffusion equation to quantify the effect of additional stretching on the mixing time.
Starting from azimuthal velocity $\widetilde{V}_{\theta}$ of VD SBI, referring back to figure~\ref{fig: SBV_mech}, the shocked light bubble moves faster than the shocked ambient air with a velocity increase $\Delta U$. This SBV-accelerated stretching leads to an intuitive model, as shown in figure~\ref{fig: mix-mech}(c). Thus from (\ref{eq: vthe-VD}), the azimuthal velocity can be expressed as:
\begin{equation}\label{eq: vthe-mix}
  \widetilde{V}_{\theta}=\frac{\Gamma}{2\upi r}+\Delta{V}^b_{\theta}.
\end{equation}
Noting that the azimuthal velocity from point vortex is steady with time, the acceleration is mainly attributed to SBV enhanced stretching.
Here, we can obtain an abbreviation of SBV enhanced stretching $\Delta V_{\theta}^b$ from (\ref{eq: vthe-vd-theo-2}):
\begin{eqnarray}\label{eq: dot-vthe}
  \Delta{V}^b_{\theta} & = & \frac{\Gamma|{At}^+|}{\upi\delta^{'}}\left[\textrm{ln}\left(1+\frac{\delta^{'}}{r-\delta^{'}}\right)-\frac{\delta^{'}}{D}\right] \nonumber\\
  & \approx & \frac{\Gamma|{At}^+|}{\upi}\left(\frac{1}{r}-\frac{1}{D}\right),
\end{eqnarray}
by noting that $\delta'=\textit{O}(\mathrm{d}r)$ and $r=\textit{O}(R)$, meaning $\delta'\ll r$, $\textrm{ln}(1+\delta'/(r-\delta'))\approx \delta'/r$.
%Interestingly, the azimuthal velocity acceleration is not related to the density gradient thickness distribution $\delta'$.
Thus, the azimuthal velocity can be further expressed as:
\begin{equation}\label{eq: dot-vthe-3}
  \widetilde{V}_{\theta}=\frac{\Gamma}{2\upi r}\left(1+2|{At}^+|\right)-\frac{\Gamma|{At}^+|}{\upi D}.
\end{equation}
%The most obvious difference between PS SBI and VD SBI is that azimuthal velocity is not only the function of $r$ but also the acceleration from SBV by time $t$ as the second term in (\ref{eq: vthe-vd}).
%It is noteworthy that since the distance $r$ is not a variable here, the characteristic time $t$ is the only unknown.
As shown in figure \ref{fig: mix-mech}(c), a blob is stretched under this velocity $\widetilde{V}_{\theta}$. The turning angle $\widetilde{\theta}$ of this bubble at distance $r$ from the vortex centre is \citep{meunier2003vortices,marble1985growth}:
\begin{equation}\label{eq: theta}
  \widetilde{\theta}(r,t)=\int_0^t\frac{\widetilde{V}_{\theta}}{r}\mathrm{d}t=\theta(r,t)+\Delta\theta_b(r,t)=
  \frac{\Gamma t}{2\upi r^2}\left(1+2|{At}^+|\right)-\frac{\Gamma|{At}^+|t}{\upi Dr}.
\end{equation}
Comparing with PS SBI, $\Delta\theta_b(r,t)$, is added due to the SBV enhanced stretching.
Then the derivative of $\widetilde{\theta}(r,t)$ with respect to distance $r$ is:
\begin{equation}\label{eq: dthedt}
  \frac{\mathrm{d}\widetilde{\theta}(r,t)}{\mathrm{d}r}=-\frac{\Gamma t}{\upi r^3}\left(1+2|{At}^+|\right)+\frac{\Gamma|{At}^+|t}{\upi Dr^2}.
\end{equation}
The cylindrical bubble at distance $r$ has been stretched to the length, as shown in figure~\ref{fig: mix-mech}(c):
\begin{equation}\label{eq: DX}
  \mathrm{d}\widetilde{X}=\sqrt{\mathrm{d}r^2+(r\mathrm{d}\widetilde{\theta})^2}
  =\mathrm{d}r\sqrt{1+r^2\left(\frac{\mathrm{d}\widetilde{\theta}}{\mathrm{d}r}\right)^2}.
\end{equation}
Since the scalar surface $s(t)\mathrm{d}X=s_0\mathrm{d}r$ remains constant in the absence of diffusion, \citet{meunier2003vortices} introduce the transverse or striation thickness evolution $s(t)$ of the strip under the deformation field of a standard point vortex, as shown in figure~\ref{fig: mix-mech}(b):
\begin{equation}\label{eq: st-viller}
  {s}(t)=\frac{s_0\mathrm{d}r}{\mathrm{d}{X}}=\frac{s_0}
  {\sqrt{1+\Gamma^2 t^2/(\upi^2 r^4)}}.
\end{equation}
Here, under the SBV enhanced stretching, the scalar surface should still maintain the conserved characteristic as $\widetilde{s}(t){\mathrm{d}\widetilde{X}}=s_0\mathrm{d}r$, which leads to:
\begin{equation}\label{eq: st}
  \widetilde{s}(t)=\frac{s_0\mathrm{d}r}{\mathrm{d}\widetilde{X}}=\frac{s_0}
  {\sqrt{1+\left[\frac{\Gamma t}{\upi r^2}\left(1+2|{At}^+|\right)-\frac{\Gamma|{At}^+|t}{\upi Dr}   \right]^2 }}.
\end{equation}
It can be found that the existence of SBV enhanced stretching velocity in VD flows results in a thinner striation $\widetilde{s}(t)$ than ${s}(t)$, as compared in figure~\ref{fig: mix-mech}(c).
In order to model the mixing of this cylindrical bubble blob, it is convenient to set the advection-diffusion equation for concentration $Y$ in the coordinate frame $(O,x,y)$, as shown in figure \ref{fig: mix-mech}(c):
\begin{equation}\label{eq: onedim-ADE}
  \frac{\partial Y}{\partial t}+U\frac{\partial Y}{\partial x}+V\frac{\partial Y}{\partial y}=\mathscr{D}\left(\frac{\partial^2Y}{\partial x^2}+\frac{\partial^2Y}{\partial y^2}\right).
\end{equation}
Here, the origin of the system $(O,x,y)$ is a Lagrangian frame set on a moving scalar. Its direction is changing temporally with the motion of scalar stretching.
Then the local velocity can be described as \citep{villermaux2019mixing}:
\begin{equation}\label{eq: U axis}
  U=\frac{\mathrm{d}x}{\mathrm{d}t}=\frac{\mathrm{d}x}{\mathrm{d}\widetilde{s}}\frac{\mathrm{d}\widetilde{s}}{\mathrm{d}t}
  =-\frac{x}{\widetilde{s}}\frac{\mathrm{d}\widetilde{s}}{\mathrm{d}t} \text{ and  } V=\frac{\mathrm{d}y}{\mathrm{d}t}=\frac{\mathrm{d}y}{\mathrm{d}\widetilde{s}}\frac{\mathrm{d}\widetilde{s}}{\mathrm{d}t}
  =\frac{y}{\widetilde{s}}\frac{\mathrm{d}\widetilde{s}}{\mathrm{d}t},
\end{equation}
and considering that stretching along the $x$-direction is much larger than in the $y$-direction, then (\ref{eq: onedim-ADE}) turns to:
\begin{equation}\label{eq: onedim-ADE-abbrev}
  \frac{\partial Y}{\partial t}+\frac{y}{\widetilde{s}}\frac{\mathrm{d}\widetilde{s}}{\mathrm{d}t}
  \frac{\partial Y}{\partial y}=\mathscr{D}\frac{\partial^2Y}{\partial y^2}.
\end{equation}
Using the canonical transformation developed by \citet{ranz1979applications}:
\begin{equation}\label{eq: changeframe}
  \xi=\frac{y}{\widetilde{s}(t)} \text{   and   }  \tau(r)=\int_0^t\frac{\mathscr{D}\mathrm{d}t'}{\widetilde{s}(t')^2},
\end{equation}
then, (\ref{eq: onedim-ADE-abbrev}) transforms to a simple diffusion equation:
\begin{equation}\label{eq: ADE-abbrev}
  \frac{\partial Y}{\partial\tau}=\frac{\partial^2Y}{\partial \xi^2}.
\end{equation}
The initial conditions at $\tau=0$ are:
\begin{equation}\label{eq: initial condi}
    \left\{ \begin{array}{ll}
    Y=Y_{\max} & \textrm{for } |\xi|<1/2, \\
    Y=0             & \textrm{for } |\xi|<1/2.
    \end{array}\right.
\end{equation}
Then the concentration at the radial position in the frame of the vortex will diffuse and smear as the following solution \citep{socolofsky2005special}:
\begin{equation}\label{eq: concen}
  Y(\xi,\tau)=\frac{Y_{\max}}{2}\left[\textrm{erf}\left(\frac{\xi+1/2}{2\sqrt{\tau}}\right)
                                                           -\textrm{erf}\left(\frac{\xi-1/2}{2\sqrt{\tau}}\right)\right].
\end{equation}
The maximum concentration that is mixed can be regarded as the concentration at centerline:
\begin{equation}\label{eq: Y_m}
  Y(r=R,\tau)=Y_{\max}\textrm{erf}\left(\frac{1}{4\sqrt{\tau}}\right).
\end{equation}
The mixing time for $Y(r=R,\tau)$ for VD SBI is reached when the argument of error function is of order unity (i.e. $4\sqrt{\tau}>1$).
It is noteworthy that from Ranz transformation (\ref{eq: changeframe}), $\tau$ should satisfy:
\begin{equation}\label{eq: time}
   \tau(r)=\int_0^t\frac{\mathscr{D}\mathrm{d}t'}{\widetilde{s}(t')^2}=\frac{\mathscr{D}t}{s_0^2}+
    \frac{\mathscr{D}t^3}{3s_0^2}\left[\frac{\Gamma}{\upi r^2}\left(1+2|{At}^+|\right)-\frac{\Gamma|{At}^+|}{\upi Dr}   \right]^2>\frac{1}{16}.
\end{equation}
Considering that the first part from pure diffusion contribution can be ignored and $D=2r$, we can obtain the mixing time for VD SBI:
\begin{equation}\label{eq: t_b}
  t_\mathfrak{m}^{VD}=\widetilde{t}_m=\frac{r^2}{\Gamma}\left(\frac{3\upi^2}{16}\right)^{1/3}\left(\frac{s_0}{r}\right)^{2/3}
  \left(1+\frac{3}{2}|{At}^+|\right)^{-2/3}\left(\frac{\Gamma}{\mathscr{D}}\right)^{1/3}.
\end{equation}
Here, one can find that mixing time model (\ref{eq: t_s}) is extended to variable density scenario by considering post-shock Atwood number ${At}^+$.
The addition expression ($1+3/2|{At}^+|$) illustrates that the SBV-enhanced stretching of light gas is proportional to the stretching from a basic vortical flow by a ratio $\frac{3}{2}|{At}^+|$, which shows a $-2/3$ scaling on mixing time.
\begin{figure}
    \centering
    \subfigure[]{
    \label{fig: tw-vd-1}
    \includegraphics[clip=true,trim=15 15 50 50, width=.45\textwidth]{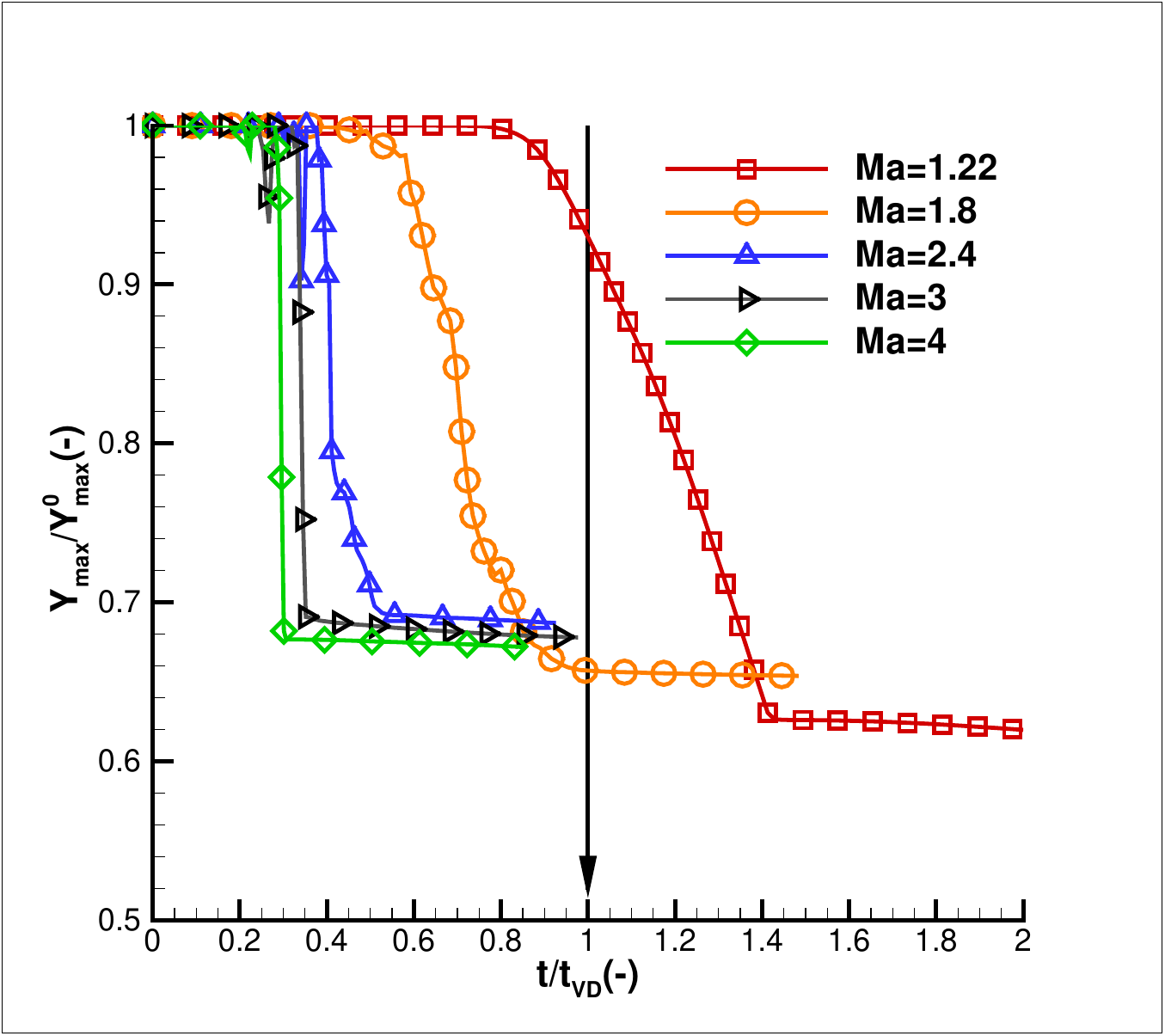}}
    \subfigure[]{
    \label{fig: tw-vd-2}
    \includegraphics[clip=true,trim=15 15 50 50, width=.45\textwidth]{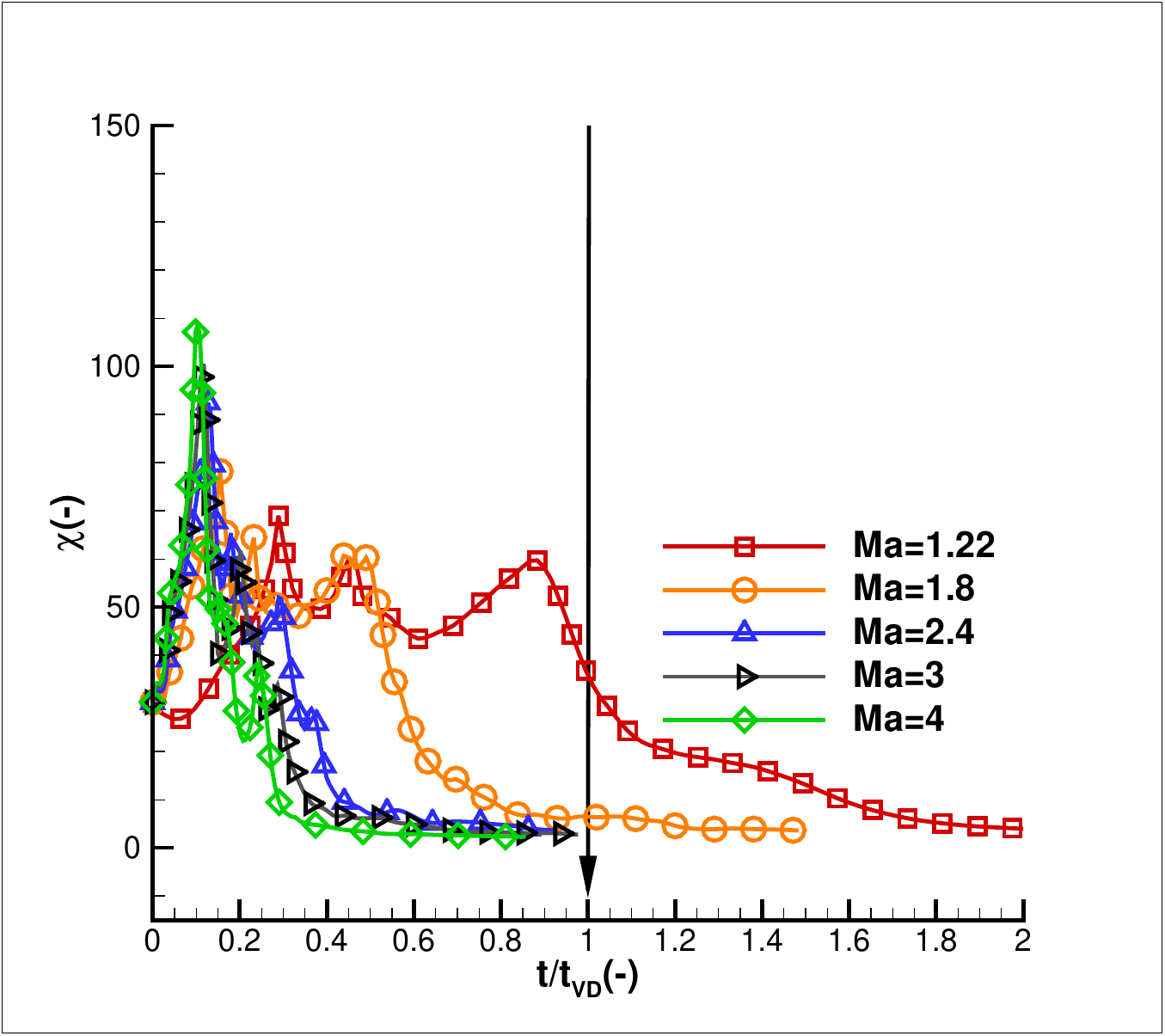}}
    \caption{(a) Maximum concentration decay and (b) scalar dissipation for VD SBI vs. scaled time $t/\widetilde{t}_{m}$. Mixing time $\widetilde{t}_{m}$ (\ref{eq: t_b}) considering only baroclinic enhanced stretching, fails to predict the correct scaling.   \label{fig: tw-vd} }
\end{figure}

Figure \ref{fig: tw-vd} shows the maximum mass fraction $\overline{Y}$ and scalar dissipation $\chi$ versus the scaled time $t/\widetilde{t}_m$.
For low shock Mach number, $Ma$=1.22, (\ref{eq: t_b}) slightly over-predicts the mixing time. However, after considering the SBV-enhanced stretching term, it performs better than the prediction from (\ref{eq: t_s}).
Like PS SBI, $\widetilde{t}_m$ still overestimates the mixing time, especially for higher shock Mach number cases, as tabulated in table \ref{tab: tm-vd}.
Thus, the compression effect needs to be taken into account in (\ref{eq: t_b}).

Considering compression effects in the same way as in the PS SBI, by using (\ref{eq: effect-radius}) to revise the cylindrical bubble length and distance from the vortex centre, equation (\ref{eq: t_b}) is further derived as:
\begin{equation}\label{eq: t_b*}
  t_\mathfrak{m}^{VD}=\widetilde{t}_m^*=\frac{\eta r^2}{\Gamma}\left(\frac{3\upi^2}{16}\right)^{1/3}\left(\frac{s_0}{r}\right)^{2/3}
  \left(1+\frac{3}{2}|{At}^+|\right)^{-2/3}\left(\frac{\Gamma}{\mathscr{D}}\right)^{1/3},
\end{equation}
where $r=R$ and $s_0=2r$.
It can be found that (\ref{eq: t_b*}) is the generalized formula of (\ref{eq: t_s}) by taking the compression and SBV enhance stretching effect. When $\eta=1$ and ${At}^+=0$,  (\ref{eq: t_b*}) degenerates to (\ref{eq: t_s}).

\begin{figure}
    \centering
    \subfigure[]{
    \label{fig: tw*-vd-1}
    \includegraphics[clip=true,trim=15 15 50 50, width=.45\textwidth]{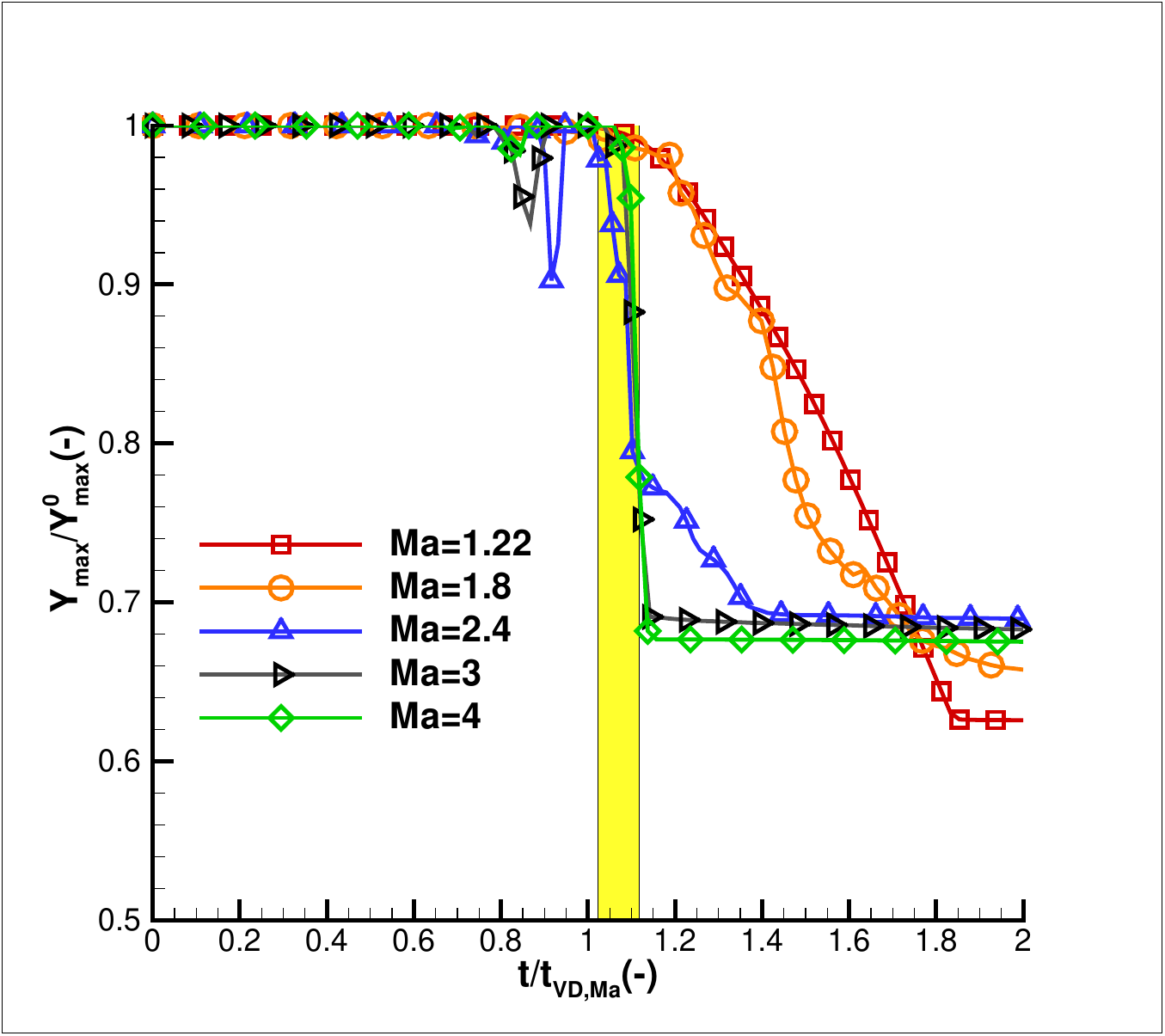}}
    \subfigure[]{
    \label{fig: tw*-vd-2}
    \includegraphics[clip=true,trim=15 15 50 50, width=.45\textwidth]{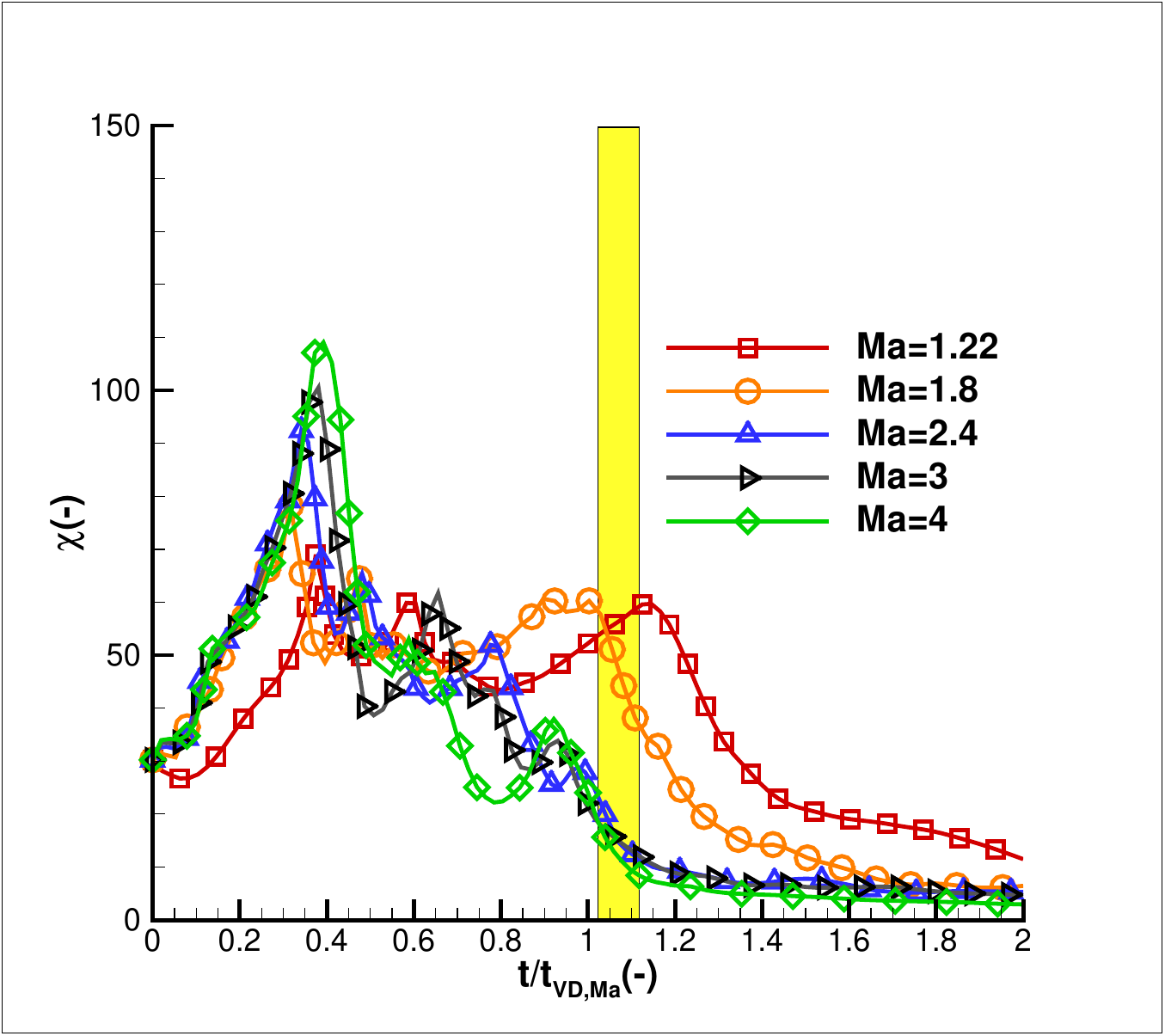}}
    \caption{(a) Maximum concentration decay and (b) scalar dissipation for VD SBI vs. scaled time $t/\widetilde{t}^*_{m}$. By combining the baroclinic enhanced stretching model and compression effect, satisfactory prediction of mixing time $\widetilde{t}^*_{m}$ (\ref{eq: t_b*}) can be found for VD SBI in general. When $t/\widetilde{t}_m^*>1$, maximum mass fraction begins to decrease rapidly and scalar dissipation enters the diffusion stage with low-level mixing.  \label{fig: tw*-vd} }
\end{figure}

With $t/\widetilde{t}_m^*$ scaling, figure \ref{fig: tw*-vd} shows that $\widetilde{t}_m^*$ predicts well the mixing time for VD SBI for all Mach numbers in general. When $t/\widetilde{t}_m^*>1$, the maximum mass fraction begins to decrease, and scalar dissipation enters the diffusion stage.
Table \ref{tab: tm-vd} illustrates the values from the theoretical model and as measured from figure \ref{fig: mixing time}.
It can be found that the model still slightly underestimate the VD mixing time due to the unsatisfactory of the standard point vortex in accordance to the model behaviour in PS SBI.
However, the fast mixing decay for VD mixing can be observed by comparing values of $t_m$ and $\widetilde{t}_m$, and also by values of $t_m^*$ and $\widetilde{t}_m^*$. That variable density cases have generally shorter mixing time than passive scalar cases can be predicted by the novel mixing time model.
\begin{table}
  \begin{center}
\def~{\hphantom{0}}
  \begin{tabular}{lcccccc}
      $Ma$ & $\overline{\kappa}$(-) & $t_m$($\umu$s) & $t_m^*$($\umu$s)
      & $\widetilde{t}_m$($\umu$s) & $\widetilde{t}_m^*$($\umu$s)
      & $t_\mathfrak{m}^{VD}$($\umu$s) \\[3pt]
       1.22 &  1.4  & 335.3  &  258.6     & 199.3   &  153.6   &   184.1        \\
       1.8   &  1.2  & 212.7  &  104.0     & 124.0   &  60.6     &    72.1        \\
       2.4   &  1.1  & 179.0  &  66.8       & 103.6   &  38.6     &    40.3           \\
       3      &  1.0  & 159.4  &  49.1       & 92.1     &  28.4      &     30.7           \\
       4      &  1.0  & 133.2  &  35.3       & 77.0     &  20.4       &    22.2           \\
  \end{tabular}
  \caption{Comparison between theoretical mixing time of $t_m$ (\ref{eq: t_s}), $t_m^*$ (\ref{eq: t-s*}), $\widetilde{t}_m$ (\ref{eq: t_b}), $\widetilde{t}_m^*$ (\ref{eq: t_b*}) and measured value $t_\mathfrak{m}^{VD}$ referring to table \ref{tab: t_m}. Diffusivity coefficient $\overline{\kappa}$ (see appendix \ref{App: diffusivity}) is also listed. }
  \label{tab: tm-vd}
  \end{center}
\end{table}

\subsection{Scaling analysis on mixing time}
\label{subsec: model sum}
%%%% 总结这段内容，对比PS和VD几次方关系，引出后面要研究的几个对象
\begin{figure}
    \centering
    \subfigure[]{
    \label{fig: scale-1}
    \includegraphics[clip=true,trim=5 5 50 50, width=.32\textwidth]{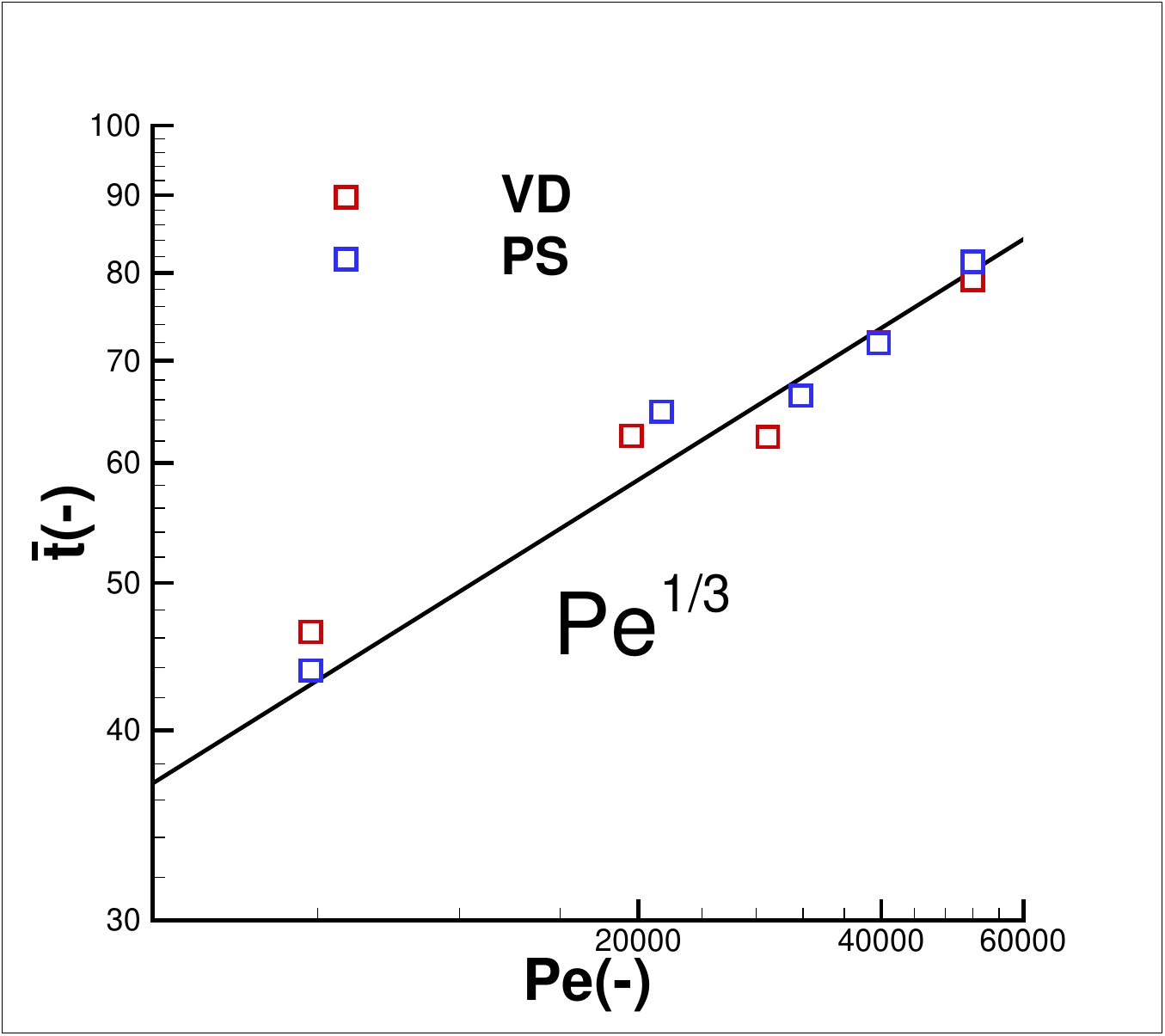}}
    \subfigure[]{
    \label{fig: scale-2}
    \includegraphics[clip=true,trim=5 5 50 50, width=.32\textwidth]{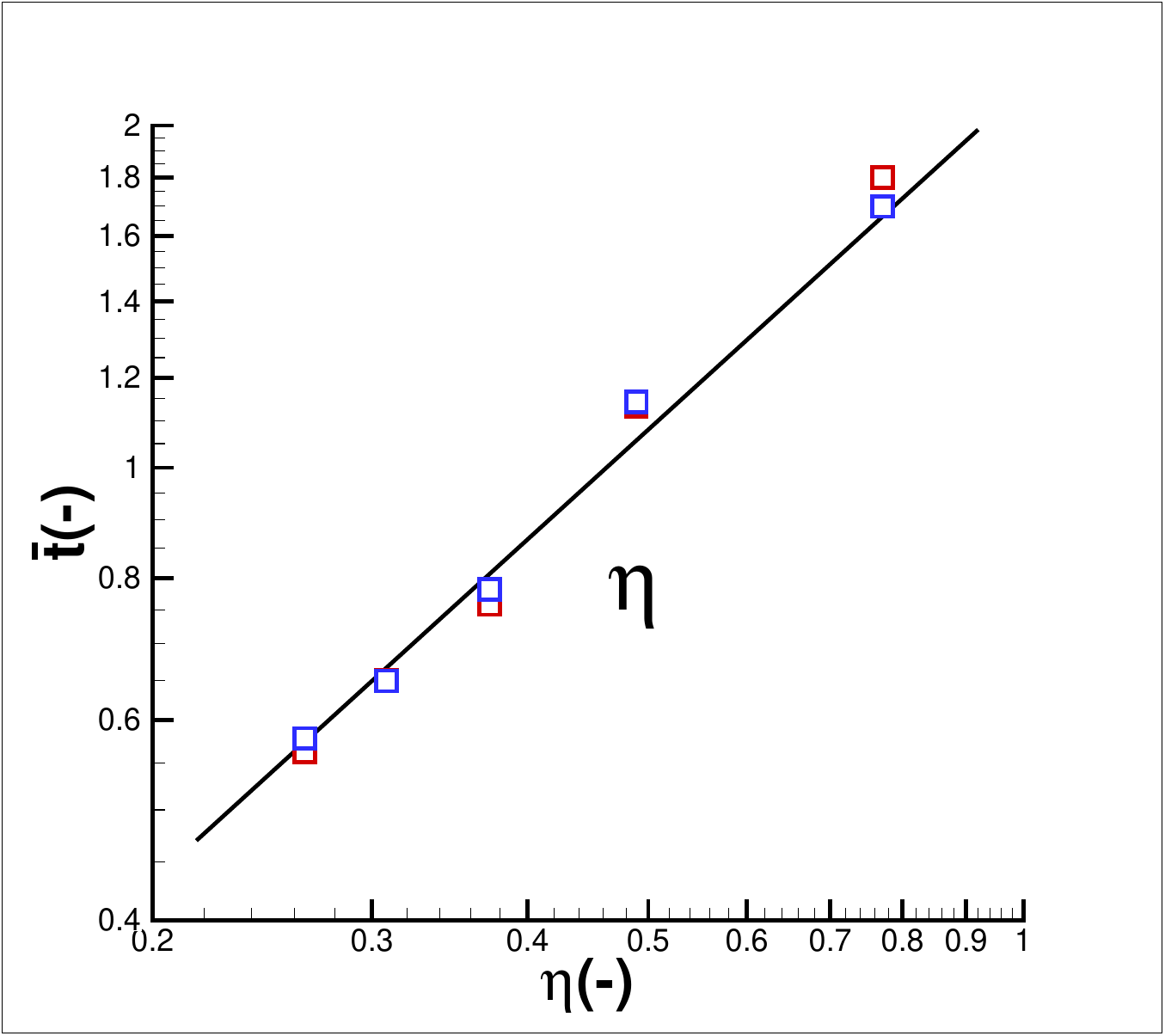}}
    \subfigure[]{
    \label{fig: scale-3}
    \includegraphics[clip=true,trim=5 5 50 50, width=.32\textwidth]{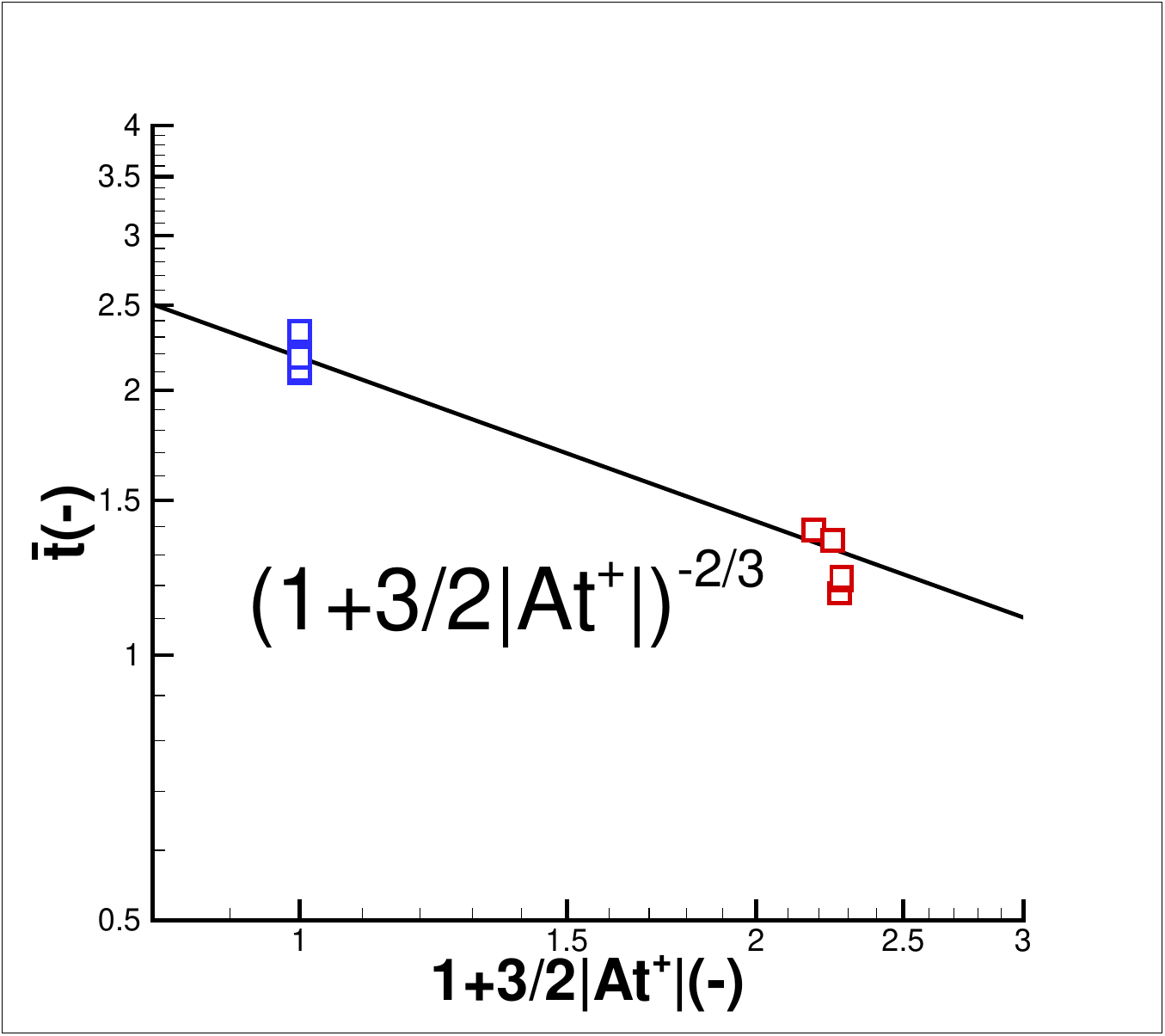}}
    \caption{Scaling of mixing time for PS SBI and VD SBI under three dimensionless controlling parameters. (a) Dependence of $\overline{t}$ (\ref{eq: tbar1}) on $\Pen$ number. (b) Dependence of $\overline{t}$ (\ref{eq: tbar2}) on compression rate $\eta$. (c) Dependence of $\overline{t}$ (\ref{eq: tbar3}) on $1+3/2|{At}^+|$.  \label{fig: scale} }
\end{figure}
By considering the compression rate $\eta$, the canonical mixing time theory of P\'eclet number $\Pen$ can be applied for the PS SBI. Further, taking SBV enhanced stretching into account, the VD inertial effect represented by ${At}^+$ is modelled. To compare the scaling of passive scalar and variable density, figure~\ref{fig: scale} shows the relationship of three dimensionless numbers with mixing time $\overline{t}$.
For the $Pe\sim\overline{t}$ relation, $\overline{t}$ is defined as:
\begin{equation}\label{eq: tbar1}
    \overline{t}\equiv\frac{\Gamma t_\mathfrak{m}}{\eta R^2}\left(1+\frac{3}{2}|{At}^+|\right)^{\frac{2}{3}}.
\end{equation}
where $|{At}^+|=0$ for PS SBI cases and $t_\mathfrak{m}$ is the measured mixing time for PS or VD SBI referring to table~\ref{tab: t_m}.
$Pe^{1/3}$ scaling is observed for both passive scalar cylindrical bubble and variable density ones in figure \ref{fig: scale-1}, which satisfies the scaling law proposed by \citet{meunier2003vortices} and \cite{marble1985growth}.
%However, in variable density cases, $Pe^{1/4}$ is dominating the overall range of mixing time, which means $\Pen$ is less critical in the VD case than in the PS case.
Next, compression rate scaling is examined. For the $\eta\sim\overline{t}$ relation, $\overline{t}$ is defined as:
\begin{equation}\label{eq: tbar2}
    \overline{t}\equiv\frac{\Gamma t_\mathfrak{m}}{R^2}\left(\frac{\Gamma}{\mathscr{D}}\right)^{-\frac{1}{3}}
    \left(1+\frac{3}{2}|{At}^+|\right)^{\frac{2}{3}}
\end{equation}
The compression rate for PS SBI and VD SBI shows a nearly linear relation with mixing time in figure \ref{fig: scale-2}.
Finally, the dimensionless number $1+3/2|{At}^+|$ is derived to reflect the density effect from SBV enhanced stretching added to a vortical flow.
For the $(1+3/2|{At}^+|)\sim\overline{t}$ relation, $\overline{t}$ is defined as:
\begin{equation}\label{eq: tbar3}
  \overline{t}\equiv\frac{\Gamma t_\mathfrak{m}}{\eta R^2}\left(\frac{\Gamma}{\mathscr{D}}\right)^{-\frac{1}{3}}.
\end{equation}
From figure \ref{fig: scale-3}, $-2/3$ scaling of $1+3/2|{At}^+|$ on mixing time is shown, which means that the density effect decreases the mixing time through SBV enhanced stretching in variable density flows.
%When the bubble density increases, i.e. small $|{At}^+|$, the mixing time becomes longer as predicted by (\ref{eq: t_b*}).

The relationship between the mixing time models under a vortical flow with circulation $\Gamma$ is summarized in (\ref{eq: theo-sum}).
Three independent dimensionless parameters indicate three mixing mechanism: $Pe$ number representing the ratio of stirring over diffusion, $\eta$ presenting shock compression effect and ${At}^+$ representing SBV enhanced stretching.
For a shock-free, constant-density mixing, $t_m$ only considers the effect of $Pe$ number, which is the original model proposed by~\citet{marble1985growth} and~\citet{meunier2003vortices}.
For a shock-compressed, constant-density mixing, $t_m^*$ considers the combined effect of $Pe$ and shock compression.
For a shock-free, variable-density mixing, $\widetilde{t}_m$ considers the combined effect of $Pe$ and SBV enhanced stretching from variable density effect.
For a shock-compressed, variable-density mixing, $\widetilde{t}_m^*$ considers all the possible effects which influence mixing time.
\begin{eqnarray}\label{eq: theo-sum}
  t_m=f(Pe)\frac{r^2}{\Gamma} & \xrightarrow{\eta} & t_m^*=f(Pe,\eta)\frac{r^2}{\Gamma} \nonumber\\
  {{At}^+}\downarrow & \qquad & \downarrow{{At}^+} \\
  \widetilde{t}_m=f(Pe,{At}^+)\frac{r^2}{\Gamma} & \xrightarrow[\eta] & \widetilde{t}^*_m=f(Pe,\eta,{At}^+)\frac{r^2}{\Gamma} \nonumber
\end{eqnarray}
When the density difference decreases (${At}^+\rightarrow0$) and shock compression is absent ($\eta\rightarrow0$), equation (\ref{eq: t_b*}) shows that the scaling of variable density mixing can degenerate to pure passive scalar mixing.
In other words, under the same conditions of the circulation, diffusivity, and compression, the mixing time for VD SBI comes near that for PS SBI, and the density effect has a limited influence on mixing.

\section{A theory of mixing enhancement number based on mixing time}
\label{sec: modulation}
For a low drag injection system based on a shock impingement mixing enhancement strategy, \citet{marble1990shock} proposed a characteristic scaling time as:
\begin{equation}\label{eq: marble scaling}
  t_m=\frac{R^2}{\Gamma}\sim\left(\frac{R}{c_0}\right)\frac{1}{Ma^2-1},
\end{equation}
which shows a strong dependence on the shock Mach number. By increasing the shock Mach number from 1.10 to 1.30, the total pressure losses are still small, and the mixing time can be reduced by three, which is promising in supersonic combustion problems. However, this Mach number scaling only considers the passive scalar stirring effect from the vortex, but not the compression and SBV-enhanced stretching contribution as analysed in $\S$\ref{subsec: VD model}. Here, we further decompose (\ref{eq: t_b*}) to a Mach number dependence expression to illustrate the mixing enhancement contribution from shock strength and variable density effect.
%\begin{figure}
%    \centering
%    \subfigure[]{
%    \label{subfig: Ma_scale}
%    \includegraphics[clip=true,trim=10 15 15 50, width=.485\textwidth]{Fig22a-Ma_scale}}
%    \subfigure[]{
%    \label{subfig: mixing-enhance}
%    \includegraphics[clip=true,trim=10 15 15 50, width=.485\textwidth]{Fig22b-mixing-enhance}}
%    \caption{(a) Mach number scaling for mixing time of VD and PS SBI obtained from table~\ref{tab: t_m}. (b) Variation of mixing enhancement number $\mathcal{M}_{sbv}$ with density inertial coefficient $\mathscr{K}$. Insert: dependency between Atwood number and density inertial coefficient. \label{fig: ma scale} }
%\end{figure}
\begin{figure}
    \centering
    \includegraphics[clip=true,trim=5 15 15 40, width=.65\textwidth]{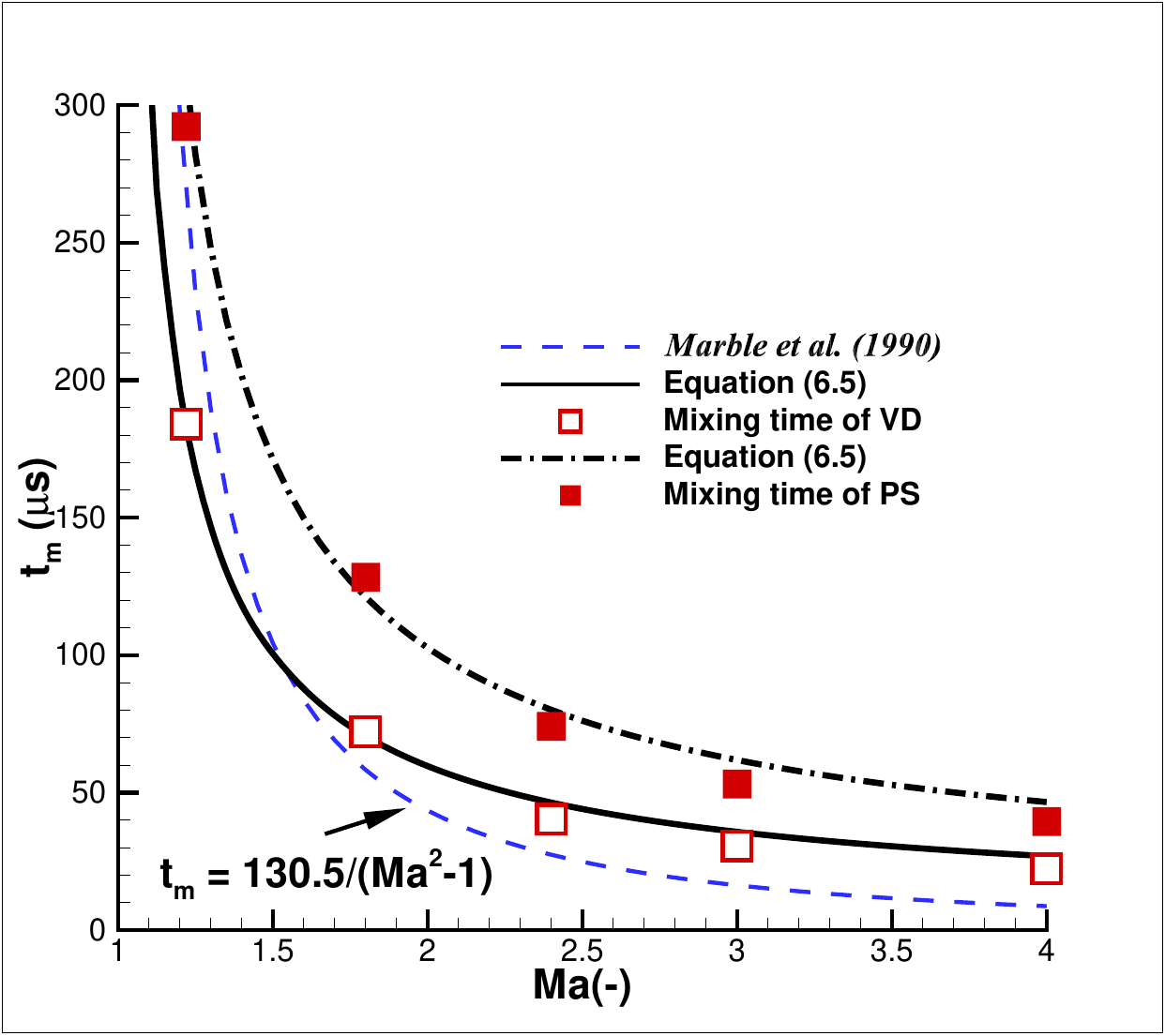}
    \caption{Mach number scaling for mixing time for VD and PS SBI obtained from table~\ref{tab: t_m}. \label{fig: ma scale} }
\end{figure}

%%图7.1分析：Marble的Ma number scaling低马赫数的确反映出ma数增加后混合特征时间的快速下降，但高马赫数并不在遵循Ma scaling。
%%同时指出由于二次斜压导致的加速拉伸特征的存在，低马赫数可以实现比Marble90预测更短的混合特征时间，这说明二次斜压特征引起的密度效应在低马赫数更加有效缩短混合特征时间，可实现低阻快速增强混合的目的。举个例子：Ma数从1.1增加到1.3和密度比增加相同的量，带来的混合时间的缩短差别不大。
We can transform (\ref{eq: t_b*}) into:
\begin{equation}\label{eq: tm-Ma-1}
  \widetilde{t}_m^*=\frac{\eta r^{2}}
  {\Gamma^{2/3}\mathscr{D}^{1/3}(1+3/2|{At}^+|)^{2/3}}
  \left(\frac{3\upi^2}{16}\right)^{\frac{1}{3}}\left(\frac{s_0}{r}\right)^{\frac{2}{3}}.
\end{equation}
Then the compression rate $\eta$ can be expressed in the form of Mach number \citep{yu2020scaling}:
\begin{equation}\label{eq: tm-Ma-2}
  \eta\approx\frac{\rho_2}{\rho'_2}\sim\frac{\rho_1}{\rho'_1}=\frac{(\gamma-1)Ma^2+2}{(\gamma+1)Ma^2}.
\end{equation}
For the circulation, \citet{yang1994model} proposed a theoretical model to predict the circulation magnitude:
\begin{equation}\label{eq: tm-Ma-3}
  \Gamma\approx\Gamma_{YKZ}=\frac{4R}{W_i}\frac{p'_1-p_1}{\rho'_1}|{At}|=\frac{8\left[(\gamma-1)Ma^2+2\right](Ma^2-1)}{(\gamma+1)^2Ma^3}|{At}|
 c_0R,
\end{equation}
where ${At}=(\rho_2-\rho_1)/(\rho_2+\rho_1)$ is the Atwood number.
%The SBV induced inertial velocity difference $\Delta U$ is:
%\begin{equation}\label{eq: tm-Ma-4}
%  \Delta U=u'_2-u'_1=\mathscr{K}u'_1=\mathscr{K}\left(1-\frac{\rho_1}{\rho'_1}\right)Ma\cdot c_0
%           =2\mathscr{K}\cdot Ma\cdot c_0\frac{Ma^2-1}{(\gamma+1)Ma^2},
%\end{equation}
%where $\mathscr{K}\equiv u'_2/u'_1-1$ can be defined as the density inertial coefficient. \citet{yu2020two} found that for a fixed density ratio between the bubble and ambient air (or the same Atwood number), $u'_2/u'_1$ is constant and is approximately 1.42 for helium bubble in the air under the different shock strength scenarios (also see inserted figure~\ref{fig: ma scale}(b)).
By substituting the compression rate $\eta$ (\ref{eq: tm-Ma-2}) and circulation $\Gamma$ (\ref{eq: tm-Ma-3}) into (\ref{eq: tm-Ma-1}), we obtain:
\begin{equation}\label{eq: tm-Ma-5}
  \widetilde{t}_m^*=\frac{\big\{\left[(\gamma-1)Ma^2+2\right](\gamma+1)\big\}^{1/3}}{(Ma^2-1)^{2/3}}\bigg\{
       \frac{r^{4/3}(3\upi^2)^{1/3} (s_0/r)^{2/3}}{2^{10/3}\mathscr{D}^{1/3}c_0^{2/3}[|{At}|(1+3/2|{At}^+|)]^{2/3}}\bigg\},
\end{equation}
This shows not only a complicated dependence on Mach number but also on diffusivity and Atwood number.
Moreover, the theoretical mixing time for passive scalar SBI (\ref{eq: t-s*}) equals to $t_m^*=\widetilde{t}_m^*({At}^+=0)$.

The scaling proposed by \citet{marble1990shock} is plotted in figure~\ref{fig: ma scale} against the mixing time evaluated for VD and PS SBI in table~\ref{tab: t_m}. The prediction in (\ref{eq: marble scaling}) that a small increase in shock strength leads to a large decrease in mixing time is reasonable for lower Mach numbers. However, deviation occurs in a high Mach number, showing the underestimation of the $1/(Ma^2-1)$ scaling.
The theoretical model (\ref{eq: tm-Ma-5}) is also compared. Good quantitative agreement is found between the simulation and the model.
As for shock Mach number effect, the shock's influence on the mixing time is weakening since the mixing time can hardly decrease further for higher shock strength.
In the low shock Mach number region, the existence of SBV-enhanced stretching causes a shorter characteristic mixing time than the one (\ref{eq: marble scaling}) predicted. In other words, weaker shock impingement with lower pressure loss can condense the mixing time by amplifying the density difference between the mixture and ambient air.

%%定义混合增强数；说明表/cue前文；如果想在低马赫数更加短的混合特征时间，就采用加热喷注的方式，可以实现密度进一步下降，增快混合特征时间。我们这里给出了可能的设计方法（Marble90中的一句话！）。
\begin{table}
  \begin{center}
\def~{\hphantom{0}}
  \begin{tabular}{lccccc}
              & $Ma$=1.22  & $Ma$=1.8  & $Ma$=2.4  & $Ma$=3  & $Ma$=4\\[3pt]
       $\mathcal{M}_{sbv}$(meas.)     & 0.63      & 0.56       &  0.54       &   0.57    &  0.56       \\
       $\mathcal{M}_{sbv}$(\ref{eq: ratio-PS-VD})     & 0.59      & 0.58       &  0.57       &   0.57    &  0.57       \\
       $\varepsilon$  & $-5.6\%$      & $3.7\%$       &  $5.5\%$       &   $-0.1\%$    &  $2.8\%$       \\
  \end{tabular}
  \caption{Mixing enhancement number of measured value obtained from table \ref{tab: t_m} and from (\ref{eq: ratio-PS-VD}). The relative errors $\varepsilon$ between modelled value and measured ones for different cases are offered at the last row. }
  \label{tab: mixing enhance}
  \end{center}
\end{table}
Here, we may define the ratio between the VD mixing time~(\ref{eq: t_b*}) and PS mixing time~(\ref{eq: t-s*}) as a mixing enhancement number:
\begin{equation}\label{eq: ratio-PS-VD}
  \mathcal{M}_{sbv}\equiv\frac{\widetilde{t}_m^*}{t_m^*}=\left(1+\frac{3}{2}|{At}^+|\right)^{-\frac{2}{3}}.
\end{equation}
As we have mentioned in section~\ref{subsec: faster decay}, the mixing time for VD SBI is shorter than the one for PS SBI as a near constant ratio for all shock Mach number.
The defined mixing enhancement number~$\mathcal{M}_{sbv}$ (\ref{eq: ratio-PS-VD}) predicts a near constant ratio between PS and VD SBI, which agrees well with measured value as shown in table~\ref{tab: mixing enhance}.
The scaling of mixing enhancement number indicates that
compared with normally used hydrocarbon fuel, the Atwood number of hydrogen will increase to $|{At}|=0.87$ compared to $|{At}|=0.285$ of methane. In this case, the hydrogen bubble's mixing time will be reduced by SBV enhanced stretching near 40\% in a low shock Mach number if all other conditions are keep the same. Therefore, the mixing enhancement number $\mathcal{M}_{sbv}$ based on the mixing time offers the possibility of further enhancing mixing through controlling the intrinsic SBV-enhanced stretching mechanism, rather than increasing the shock strength at the cost of higher wave drag.

\section{Conclusions}
\label{sec: conclusions}
Spurring from the demands of mixing time estimation for variable density flows in scramjet combustors, SBI is chosen as the typical problem to study the influence of an unsteady streamwise vortex on compressible variable-density mixing enhancement.
Through high-resolution simulations, a wide range of shock Mach numbers from 1.22 to 4 is set to interact with a cylindrical helium bubble, and compared to a passive scalar counterpart.

It is interesting to find that the maximum concentration of mass fraction decays much faster in VD SBI than in PS SBI. The mixing rate, represented by scalar dissipation, also enters into a diffusion-controlled steady state more quickly in the VD scenarios.
The phenomenon, i.e. a shorter mixing time for the VD cases than PS cases, occurs at all shock Mach numbers. By investigating the azimuthal velocity, which is the stretching source of concentration decay, we observe that PS SBI demonstrates a quasi-standard Lamb-Oseen type velocity distribution, while an apparent local acceleration occurs in VD SBI.
The local velocity increase originates from the SBV production, which also explains the local mixing rate increase.
Through analyzing the formation process of SBV, we propose a velocity difference model, assuming that the shocked cylindrical bubble moves at a higher velocity $\Delta U$ than the ambient air. It is the combination of velocity difference $\Delta U$ and post-shock Atwood number ${At}^+$ that yields the local SBV-enhanced additional stretching.

Based on the observation of SBV-enhanced stretching, we further build a mixing time estimation model for both PS and VD SBI.
As for PS SBI, the model proposed by \citet{meunier2003vortices} is modified to consider the inherent shock compression effect. It shows a relatively good prediction after revision from compression that when $t=t^*_m$, maximum concentration begins to fade away, and scalar dissipation transits into a low-level mixing.
%The mixing time $t^*_m$ shows the scaling dependence on $\Pen$ number and compression rate as $t^*_m\sim\eta Pe^{1/3}R^2/\Gamma$.
As for VD SBI, we have proposed a theoretical estimation for VD SBI mixing time $\widetilde{t}_m^*$ by considering the SBV-enhanced stretching and the compression effect on the initial mixing region.
The mixing time model extends the model in PS mixing by considering the variable density difference.
Thus, a generalized form of mixing time is expressed as $\widetilde{t}^*_m\sim\eta (R^2/\Gamma)Pe^{1/3}(1+3/2|{At}^+|)^{-2/3}$, which reveals the dependence of mixing time on $\Pen$ number, compression rate $\eta$, and post-shock Atwood number ${At}^+$ suggesting the underlying SBV induced variable density effect.

Since the SBV-enhanced stretching mechanism is ubiquitous in VD flows and its influence on mixing time is revealed,
a Mach number analysis shows the importance of SBV-enhanced stretching effect on mixing enhancement in shock-accelerated inhomogeneity flows.
A mixing enhancement number defined by the ratio of VD and PS mixing time further illustrates the scaling of the SBV-enhanced stretching, which may offer a new way to increase mixing behaviour in a supersonic streamwise vortex of a scramjet through controlling the variable-density effect in general.
\\

\section*{Acknowledgments}
The authors thank Professor H. Xu, Post-doc S.Y. Qin, and L.Y. Li for discussing and checking the structure and the contents.
Professor G. Wang and doctor H.Y. Liu are appreciated for the support of simulation method.
The authors also thank the anonymous referees for their valuable comments.

\section*{Funding}
This work was supported by the NSFC Project (91941301, 91441205) and the National Science Foundation for Young Scientists of China (Grant No.51606120).
Besides, the Center for High-Performance Computing of SJTU has provided the supercomputer $\upi$~2.0 that has contributed to the present research.

\section*{Declaration of interests}
The authors report no conflict of interest.

\section*{Author ORCIDs}
\noindent
Hong Liu https://orcid.org/0000-0001-9011-8309;\\
Bin Yu https://orcid.org/0000-0001-6632-3468;\\
Bin Zhang https://orcid.org/0000-0002-1307-2404;\\
Yang Xiang https://orcid.org/0000-0002-1820-8622.

\appendix
\section{Mesh independence study}\label{App: mesh}
\begin{figure}
    \centering
    \subfigure[]{
    \label{fig: grid depend}
    \includegraphics[clip=true,width=.99\textwidth]{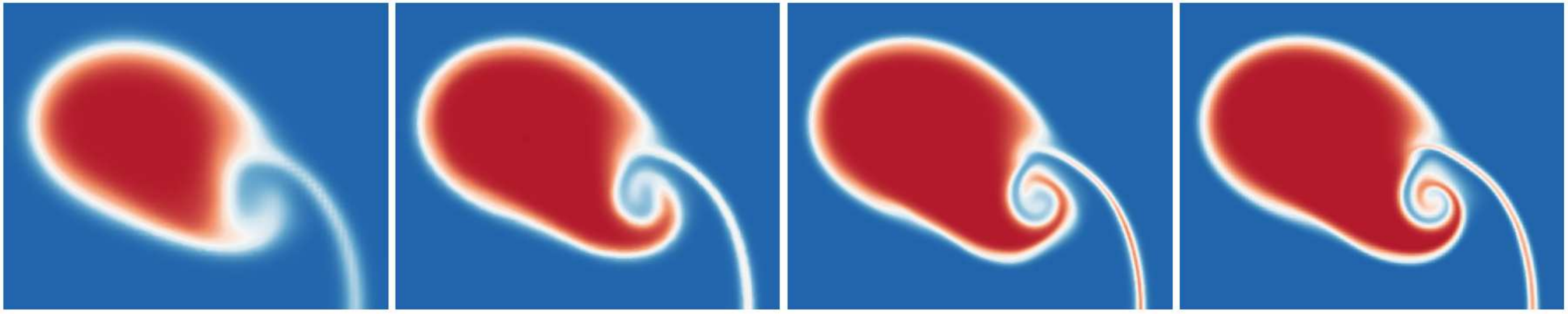}}
    \subfigure[]{
    \label{fig: grid depend2-1}
    \includegraphics[clip=true,trim=10 10 5 25, width=.45\textwidth]{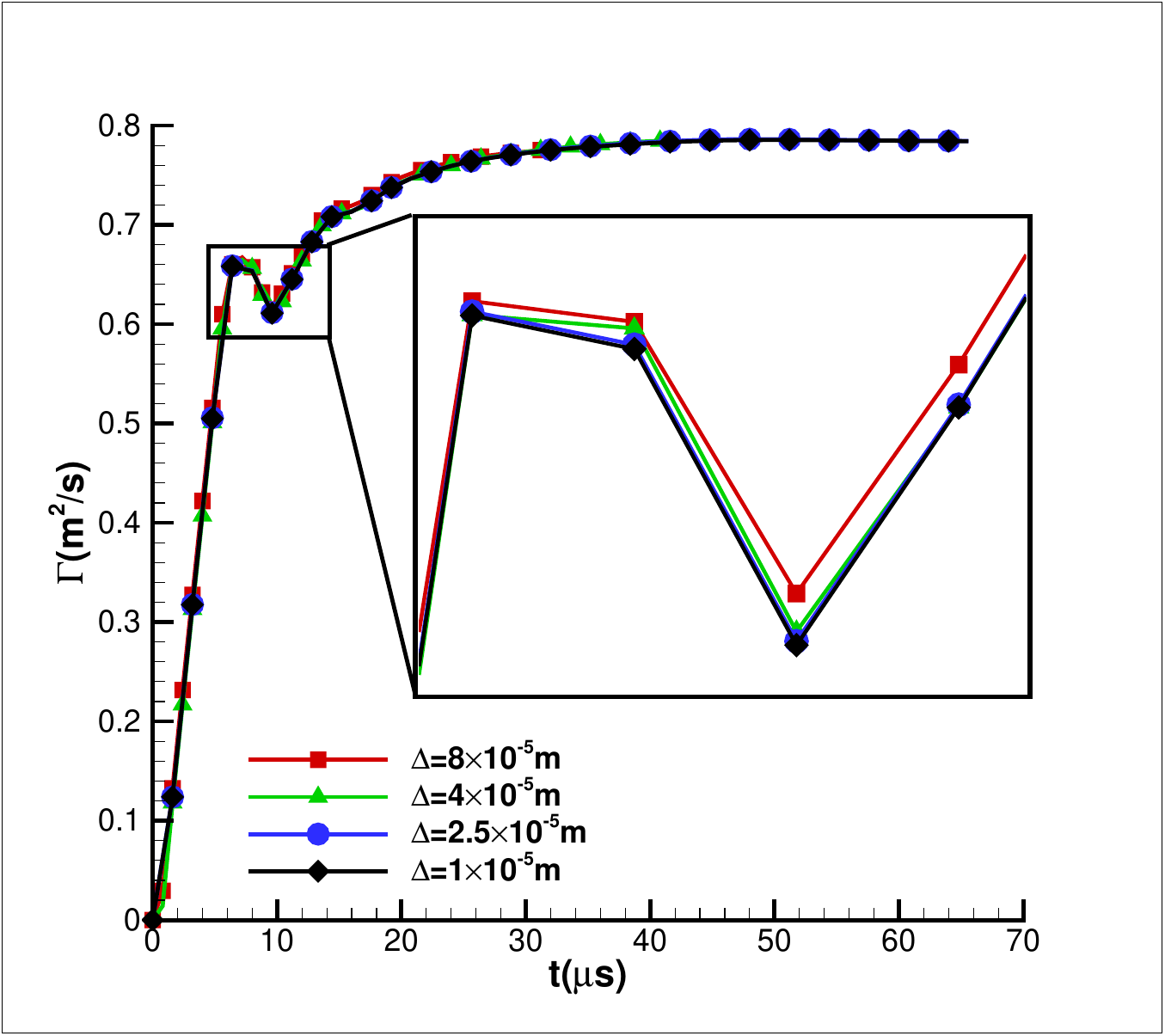}}
    \subfigure[]{
    \label{fig: grid depend2-2}
    \includegraphics[clip=true,trim=10 10 5 25, width=.45\textwidth]{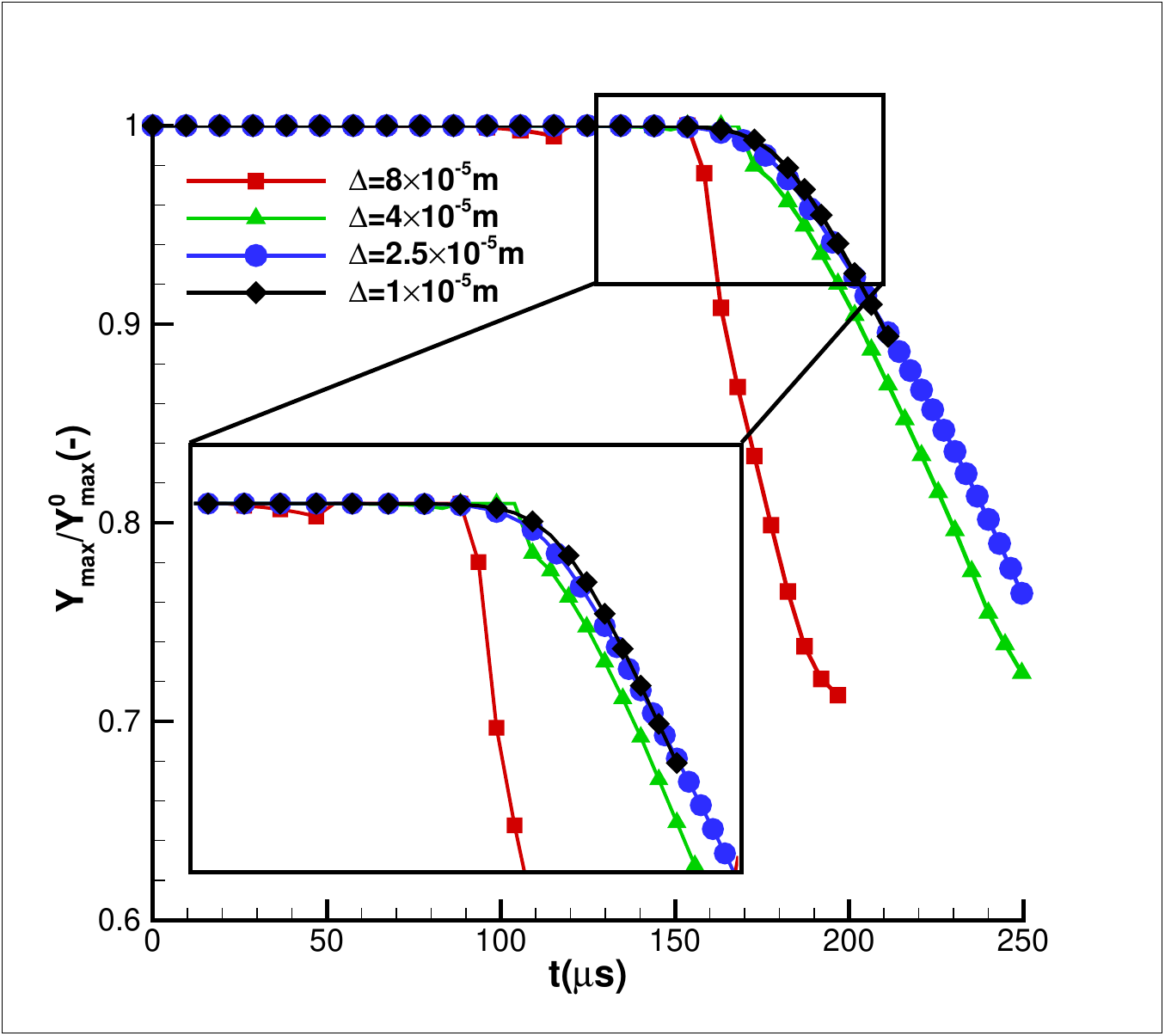}}
    \caption{
    (a) Grid dependence study on the helium mass fraction contour at $t=65.6$~$\umu$s.
    From left to right, mesh-1: $\Delta=8\times10^{-5}$~m;  mesh-2: $\Delta=4\times10^{-5}$~m; mesh-3: $\Delta=2.5\times10^{-5}$~m; mesh-4: $\Delta=1\times10^{-5}$~m.
Circulation (b) and maximum concentration (c) of different mesh resolutions.  \label{fig: grid depend2} }
\end{figure}

% mesh dependence
Four kinds of mesh resolutions are compared, which are mesh-1, $\Delta=7.5\times10^{-5}$~m; mesh-2, $\Delta=4\times10^{-5}$~m; mesh-3, $\Delta=2.5\times10^{-5}$~m, and mesh-4, $\Delta=1\times10^{-5}$~m. The case of $Ma=1.22$ shock interacting with the pure helium cylindrical bubble is examined.
The contours of the helium mass fraction from different mesh resolutions are presented in figure~\ref{fig: grid depend}. With the increase of the mesh resolution, the spiral of the vortex becomes more evident. It can be found that a good consistency of flow structures exists between mesh-3 and mesh-4.

The quantitative parameters from different resolutions are further compared in figure~\ref{fig: grid depend2}.  As for the circulation $\Gamma$, defined in (\ref{eq: circu}), figure~\ref{fig: grid depend2-1} shows that the circulation trend of mesh-3 is similar to that for the finest mesh-4. As for the mixing characteristic, the normalised maximum concentration from different resolutions is compared in figure~\ref{fig: grid depend2-2}. The maximum concentration will decay faster in the coarse mesh resolution, such as mesh-1, due to the larger numerical viscosity. On the contrary, the results from mesh-3 can be regarded as nearly the same as those from mesh-4. In general, considering the balance between the computational burden and accuracy, we choose the resolution of mesh-3 in this work, which is sufficient to reflect the flow structures on the mixing behaviour.

\section{Numerical uncertainty from different schemes}\label{App: WENO}
The choice of different orders numerical schemes is vital for capturing physical features of RMI flow fields~\citep{mosedale2007assessment,drikakis2009large}. This appendix discusses the numerical uncertainty from different high-order WENO schemes.
To simplify the problem, it is feasible to solve
the dimensionless Navier-Stokes equations~\citep{zhang2003numerical}\footnote{Only in this appendix, a symbol without a superscript denotes dimensionless variable and with a superscript denotes dimensional variable. }:
\begin{eqnarray}\label{eq: dim-NS}
\frac{\partial\rho}{\partial t}+\frac{\partial\rho u}{\partial x}
+\frac{\partial\rho v}{\partial y}
&=&0 \label{eq: NS1}\\
\frac{\partial\rho u}{\partial t}+\frac{\partial\left(\rho u^2+p\right)}{\partial x}
+\frac{\partial\rho uv}{\partial y}
&=&\frac{\mu}{\text{Re}}\left(\frac{4}{3}u_{xx}+u_{yy}+\frac{1}{3}v_{xy}\right) \label{eq: NS2}\\
\frac{\partial\rho v}{\partial t}+\frac{\partial\rho uv}{\partial x}
+\frac{\partial\left(\rho v^2+p\right)}{\partial y}
&=&\frac{\mu}{\text{Re}}\left(v_{xx}+\frac{4}{3}v_{yy}+\frac{1}{3}u_{xy}\right)\label{eq: NS3}\\
\frac{\partial e_0}{\partial t}+\frac{\partial u\left(e_0+p\right)}{\partial x}
+\frac{\partial v\left(e_0+p\right)}{\partial y}
&=&\frac{\mu}{\text{Re}}\left\{\frac{\gamma}{\text{Pr}}\left(e_{xx}+e_{yy}\right)\right. \nonumber \\
& &+\frac{2}{3}\left[\left(u^2\right)_{xx}+\left(v^2\right)_{yy}-\left(uv_y\right)_x-\left(vu_x\right)_y\right] \nonumber \\
& &\left.+\frac{1}{2}\left[\left(v^2\right)_{xx}+\left(u^2\right)_{yy}+\left(vu_y\right)_x+\left(uv_x\right)_y\right]\right\},  \label{eq: NS4}
\end{eqnarray}
where $\rho,p,e_0,e$ represent the dimensionless gas density, pressure, total energy and internal energy respectively, $u,v$ are the dimensionless speed of gas. The specific heat ratio $\gamma=1.4$, which is generally estimated as the characteristic of ideal air gas. $Re$ and $Pr$ are Reynolds number and Prandtl number respectively.
The dimensionless viscosity $\mu=1$, meaning that the viscosity is uniform in the computational domain.
Uniform viscosity is generally accepted in simplified RMI simulation~\citep{zhang2003numerical}. We will show later that the results obtained from simplified dimensionless equations (\ref{eq: NS1}) to (\ref{eq: NS4}) can represents the ones obtained from complete form of Navier-Stokes equations (\ref{eq: NS}).

The third-order Total-Variation-Diminishing (TVD) Runge-Kutta (RK) scheme are applied to solve the temporal iteration.
Different orders of the WENO scheme~\citep{liu1994weighted,jiang1996efficient} are compared. In general, we denote the numerical results from the third-order, fifth-order, seventh-order, and ninth-order WENO schemes on  dimensionless equations (\ref{eq: NS1}) to (\ref{eq: NS4}) as WENO~3, WENO~5, WENO~7 and WENO~9 separately.
\begin{figure}
    \centering
    \subfigure[]{
    \label{fig: sod-rho}
    \includegraphics[clip=true,trim=5 5 15 20, width=.48\textwidth]{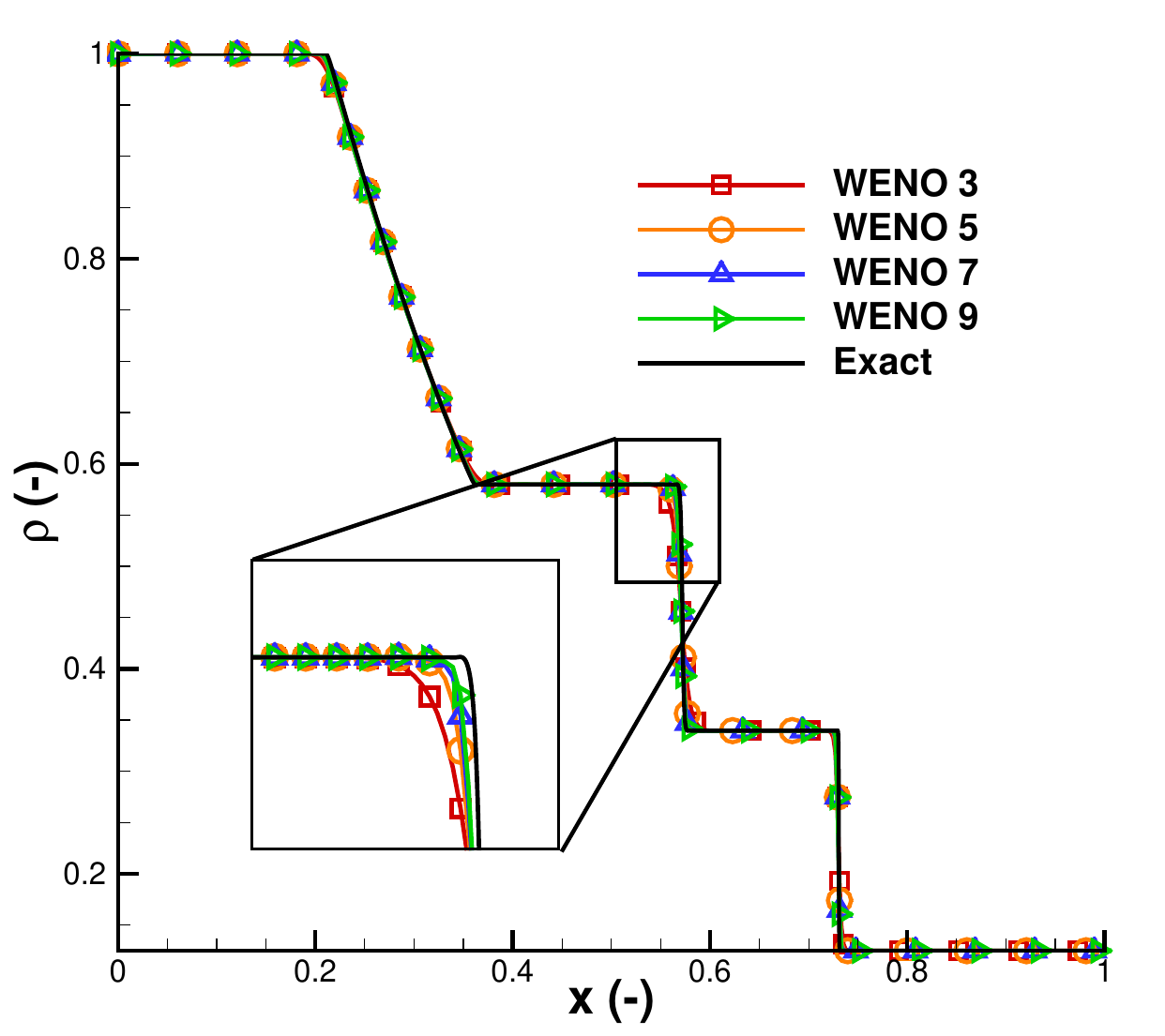}}
    \subfigure[]{
    \label{fig: 1dsbi-rho}
    \includegraphics[clip=true,trim=5 5 15 20, width=.48\textwidth]{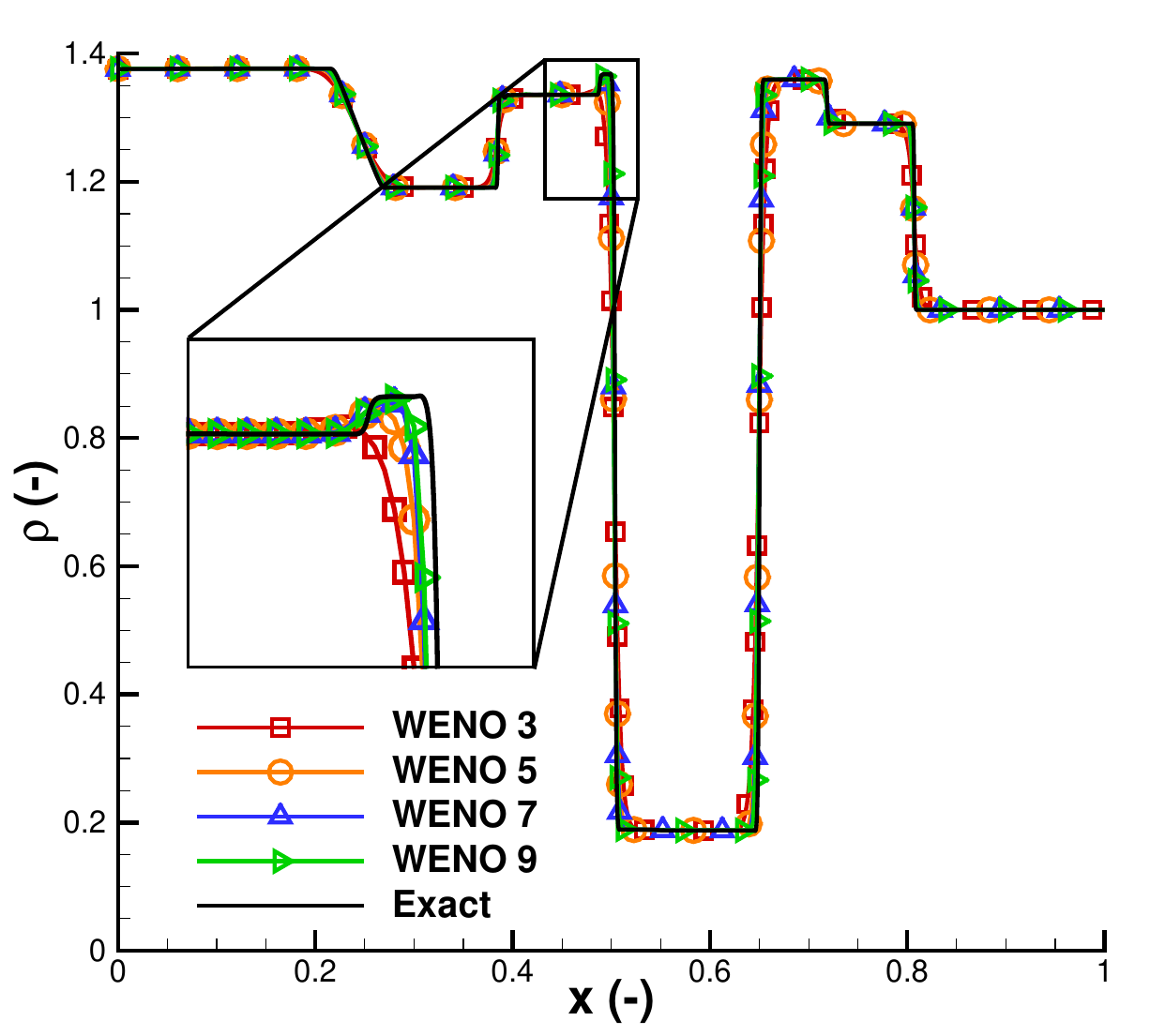}}
    \caption{The results of density profile for (a) Sod's shock tube at $t=0.2$ and (b) one-dimensional shock-slab interaction at $t=0.3$ with different order WENO schemes.
    Different schemes on a 480-point gird compared to the reference solutions obtained with the WENO 3 scheme  on a high-resolution 4800-point grid.  \label{fig: sod-1dsbi} }
\end{figure}

\subsection{Numerical uncertainty in one-dimensional problem}
The widely used Sod's shock-tube problem~\citep{tritschler2013numerical} is chosen as the first validation case. The initial conditions are:
\begin{equation}\label{eq: sod-IC}
  (\rho, u, p)=\begin{cases}(1.0,0.75,1.0) & \text {if} \quad 0.0<x<0.3, \\
(0.125,0.0,0.1) & \text {if} \quad 0.3<x<1.0.\end{cases}
\end{equation}
The second validation case is the shock bubble interaction problem in one dimension, which is proposed by~\citet{quirk1996dynamics}. It consists of a $Ma=1.22$ moving shock, initially at $x=0.25$, interacting with a helium gas slab set at $0.4< x < 0.6$.
The detailed initial conditions are:
\begin{equation}\label{eq: 1D-SBI}
  (\rho, u, p)=
  \begin{cases}(1.3765,0.3948,1.57) & \text {if} \quad 0.0<x<0.25 \\
(1.0,0.0,1.0) & \text {if} \quad 0.25<x<0.4, \\
& \text {or} \quad 0.6<x<1.0, \\
(0.138,0.0,1.0) & \text {if} \quad 0.4<x<0.6.\end{cases}
\end{equation}
Reynolds number and Prandtl number are set as $Re=10^5$ and $Pr=0.7$.

The results of density distribution for two cases are shown in figure~\ref{fig: sod-1dsbi}.
The simulation from different schemes are sampled on a 480-point gird, compared to the reference solutions obtained with the WENO~3 scheme on a high-resolution 4800-point grid.
It can be found that the high-order WENO schemes (WENO 5, WENO 7 and WENO 9) fit the exact results better than the ones from the WENO 3 scheme.
Generally, the performance of WENO~5 scheme agrees well with the one of WENO~9.

\subsection{Shock-cylindrical bubble interaction}
%\subsubsection{Initial conditions for $Ma=1.22$ VD SBI}
To compare the two-dimensional SBI simulation results from dimensionless NS equations (\ref{eq: NS1}) to (\ref{eq: NS4}) and from NS equations (\ref{eq: NS}) with multi-components, one should transform the initial conditions with physical value in \emph{ParNS} code to dimensionless variables.
The variables in (\ref{eq: NS1}) to (\ref{eq: NS4}) are dimensionless as follows:
\begin{equation}
x=\widetilde{x}/L^*,u=\widetilde{u}/u^*, v=\widetilde{v}/u^*,p=\widetilde{p}/p^*, t=\widetilde{t}/t^*,\rho=\widetilde{\rho}/{\rho}^*,
\end{equation}
in which the variables with tilde above represent the primary values with physical units, and the variables with star represent the reference values.
Here, the reference length $L^*=0.0026$ m, equal to the radius of bubble. The reference pressure $p^*=101325$~Pa and  the reference density $\rho^*=1.189$~kg/$\text{m}^3$, equal to the pressure and density of the ambient air. The reference velocity $u^*=\sqrt{p^*/\rho^*}=291.9$~m/s, and the reference time $t^*=L^*/u^*=8.564$~$\umu$s. Furthermore, dimensionless mesh resolution and calculation time step are set to maintain the same as the ones in \emph{ParNS} code.

If we choose $Ma=1.22$ VD SBI calculated from \emph{ParNS} code for comparison, it can be found that the corresponding initial values of the dimensionless pressure, density and velocity are the same as ones in 1D shock-bubble interaction (\ref{eq: 1D-SBI}). Moreover, the specific heat ratio $\gamma$ is unified to 1.4. Prandtl number is set as $Pr=0.72$, which is the same as the one set in \emph{ParNS} code. The reasonable value of $Re$ number in dimensionless equations is vital, since it determines the comparability between the two sets of equations.
As for $Re=\rho^* u^* L^*/\mu^*=u^* L^*/\nu^*$, the global kinetic viscosity
$\nu^*$ for $Ma=1.22$ VD SBI in \emph{ParNS} results can be estimated from the Schmidt number as $Sc=\nu^*/\mathscr{D}^*=0.5$. We find that the global diffusivity in $Ma=1.22$ VD SBI is $\overline{\kappa}\mathscr{D}_m\approx99.4\times10^{-6}$~m$^2$/s (see appendix~\ref{App: diffusivity}). Therefore, a comparable $Re$ number should set as $Re=14596$ from $\nu^*\approx50\times10^{-6}$~$\text{m}^2$/s.

It is noteworthy that a diffuse interfacial transition layer at the initial bubble boundary (\ref{eq:Gaussian}) is set in the calculation of \emph{ParNS} code, as shown in figure~\ref{fig:initial-conditions}.
By using the canonical correlation between the density and mass fraction (\ref{eq: den-MF}),
we can obtain the corresponding initial density distribution profile, $\rho(r)$. Therefore, the density distribution across the diffuse interfacial transition layer is derived as follows:
\begin{equation}\label{eq:Gaussian-rho}
    \rho(r)=\left\{ \begin{array}{ll}
    \rho_2 & {r\leq{R}}, \\
    \frac{\rho_2 \rho_1}{\rho_2+\left(\rho_1 - \rho_2\right)\mathrm{e}^{-\alpha\left[(r-R)/\delta\right]^2}} & {R<r\leq{R+\delta}}, \\
    \rho_1 & r>R+\delta,
    \end{array}\right.
\end{equation}
where $\rho_1$ and $\rho_2$ are the density of ambient air and helium.

%\subsubsection{Qualitative and quantitative comparisons}
\begin{figure}
\centering
\includegraphics[width=0.99\textwidth]{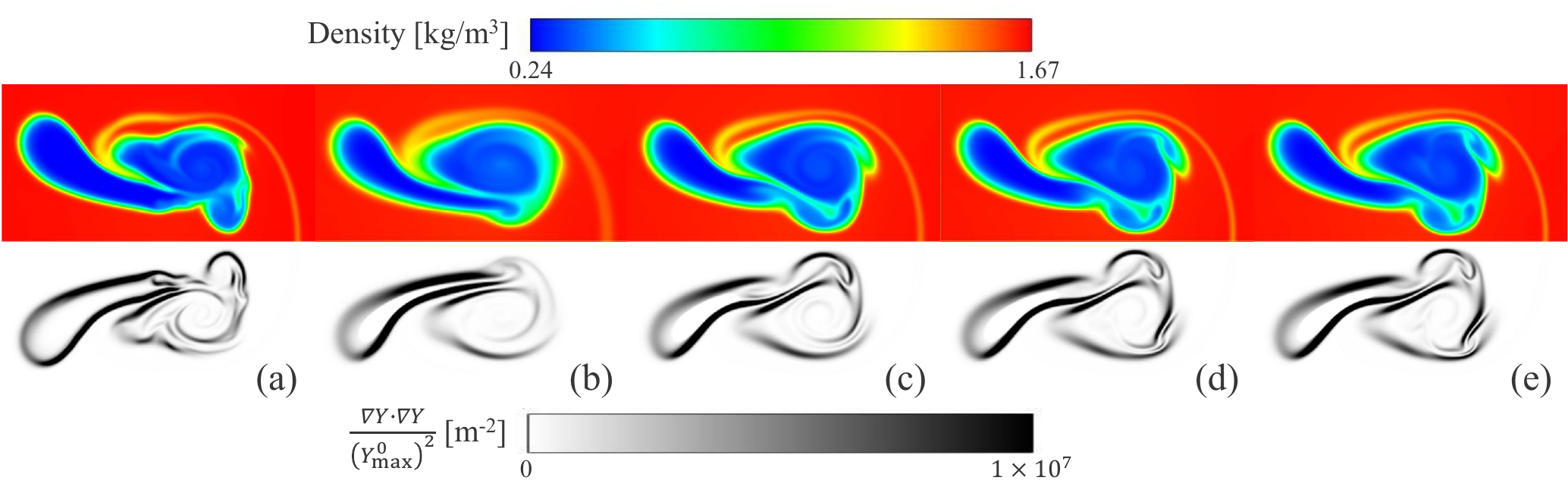}
\caption{Comparison of density (top) and scalar dissipation (bottom) at $t=16.8$ (144~$\umu$s)  in the $Ma=1.22$ VD SBI simulated by (a) \emph{ParNS}, (b) the third-order WENO, (c) the fifth-order WENO, (d) the seventh-order WENO, and (e) the ninth-order WENO. }\label{fig: den-SDR-WENO}
\end{figure}
Qualitative comparisons of flow field with density and scalar dissipation at one moment (144~$\umu$s) between different schemes are plotted in figure~\ref{fig: den-SDR-WENO}. Here, the mass fraction in scalar dissipation calculation (\ref{eq: scalar energy}) is transformed from local density by (\ref{eq: den-MF}).
Two observations can be found: first, the flow structures of \emph{ParNS} are similar to the ones simulated from dimensionless equations, despite that the component computation is absent in dimensionless NS equations. The similarity indicates the $Re$ number is set reasonably to compare with the multi-components NS equation (\ref{eq: NS}).
Second, general similarity is obtained between the results from WENO~5, WENO~7 and WENO~9, which behave less numerical uncertainty than WENO~3 does.

\begin{figure}
    \centering
    \subfigure[]{
    \label{fig: cir-com-WENO}
    \includegraphics[clip=true,trim=5 5 0 20, width=.48\textwidth]{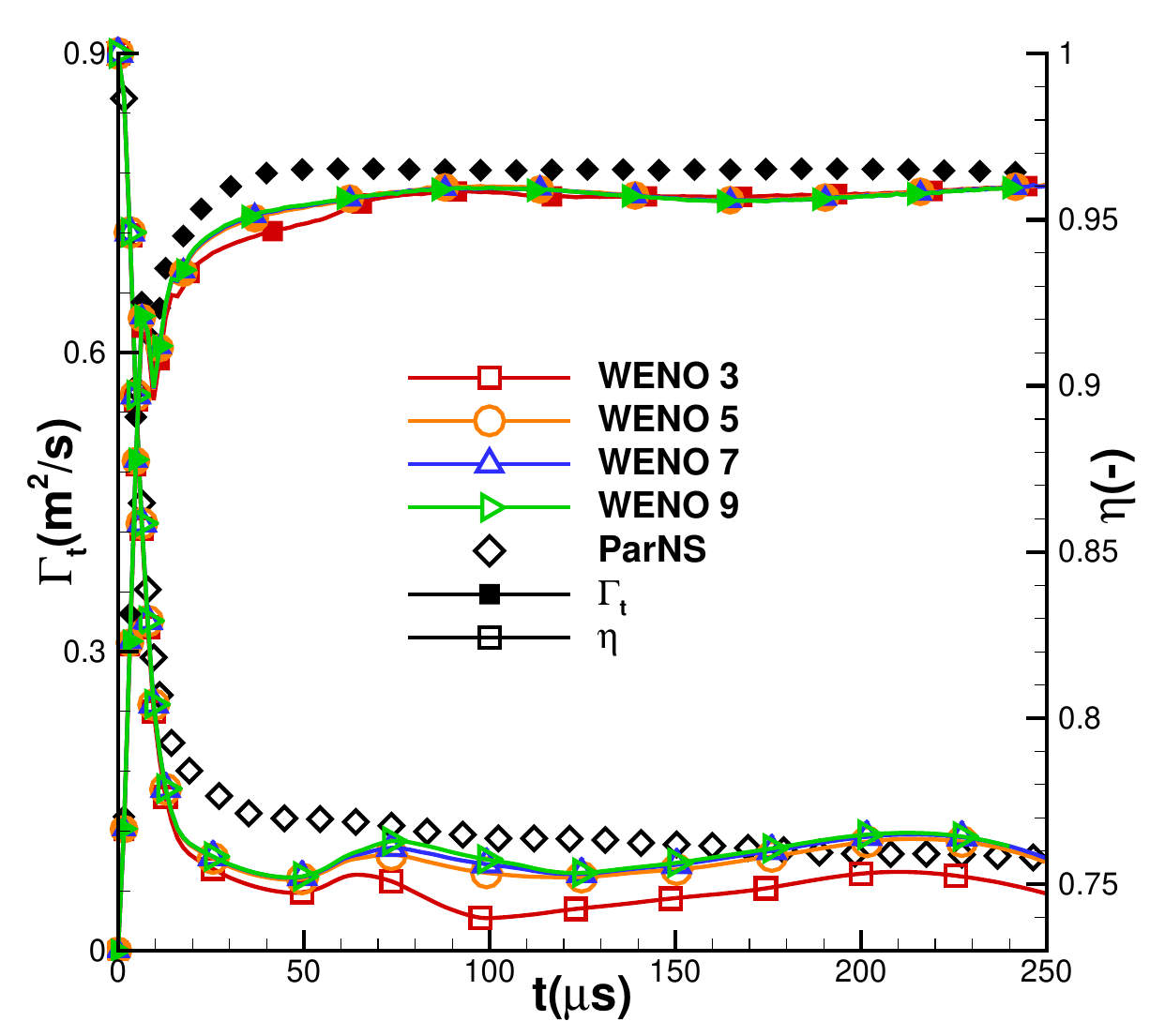}}
    \subfigure[]{
    \label{fig: SDR-MaxCon}
    \includegraphics[clip=true,trim=5 5 0 20, width=.48\textwidth]{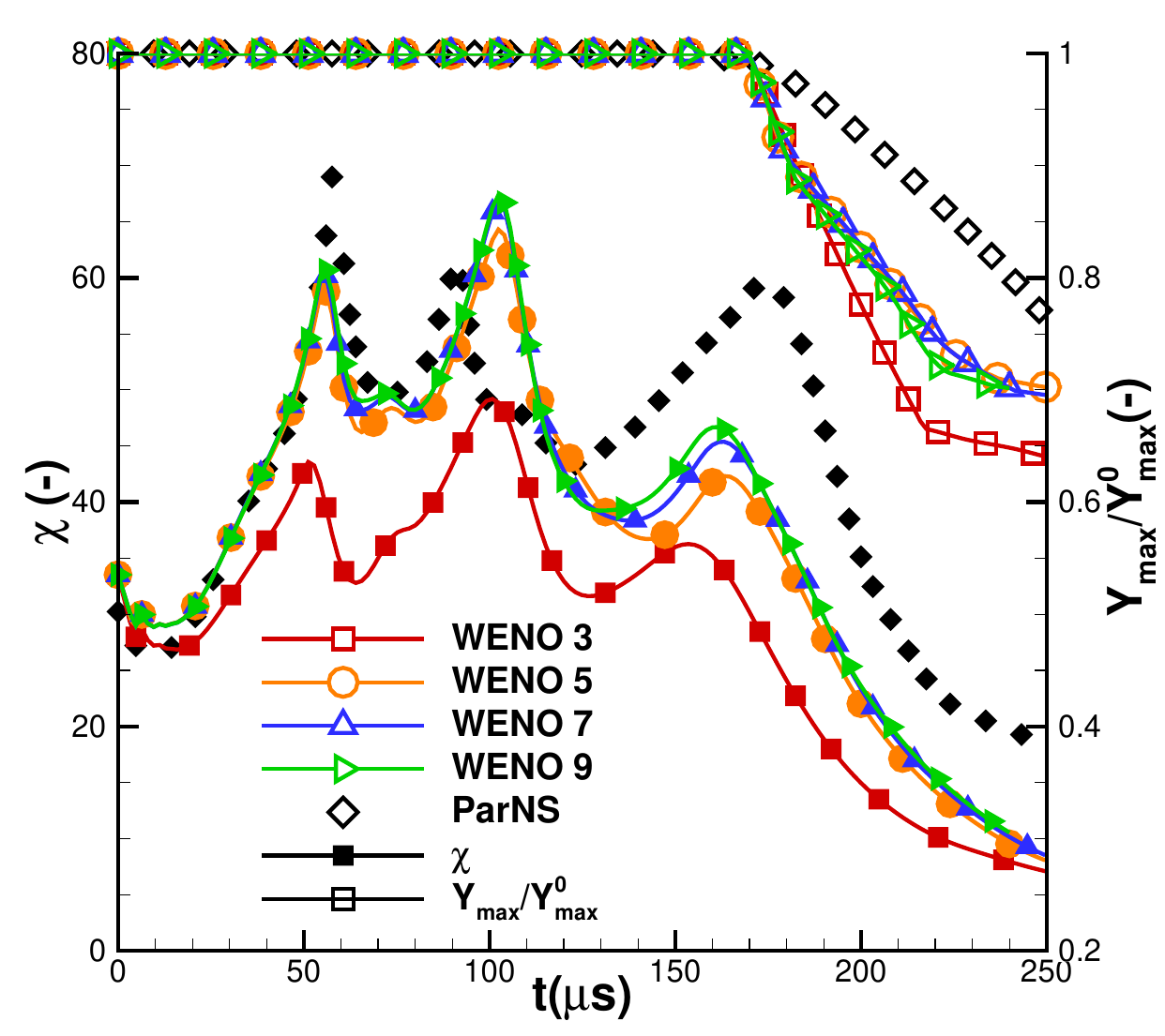}}
    \caption{(a) Time history of total circulation and compression rate between different schemes.
    (b) Time history of scalar dissipation and maximum mass fraction between different schemes.
    \label{fig: WENO-SBI-line} }
\end{figure}
Given that total circulation and compression rate are important systematic parameters,
figure~\ref{fig: cir-com} offers the comparisons between the results from \emph{ParNS} and from different WENO schemes.
Particularly, the similar circulation value indicates that another form of $Re$ number in vortical flows as $Re=\Gamma/\nu$~\citep{glezer1988formation} also maintains the same in \emph{ParNS} simulation and different WENO schemes simulations.
As for temporal evolution of total circulation, compression rate, scalar dissipation and maximum mass fraction shown in figure~\ref{fig: WENO-SBI-line}, it is obvious that the high-order WENO schemes perform better than the third-order WENO scheme. However, the results related to mass fraction with the high-order WENO schemes have differences with that with the \emph{ParNS} code, which may explained as some important physical process including gaseous diffusion elimination in dimensionless NS equations.

%%% discussions
From the qualitative and quantitative comparisons between different schemes, we can found that the numerical uncertainty from the widely accepted WENO~5 behaves nearly the same as that from higher-order schemes.
Therefore, plenty of research in SBI or RMI are numerically studied though WENO schemes (see~\citet{johnsen2006implementation,shankar2011two,lombardini2012transition,
hejazialhosseini2013vortex,tritschler2014richtmyer,Ding2017On,li2019circulation}).
In present study, considering the balance between increasing the resolution accuracy~\citep{thornber2010influence} and reducing the computational burden~\citep{mosedale2007assessment}, we select the fifth-order WENO scheme for this type of flows, which is suitable for capturing flow structures and mixing pattern in general from the comparisons above.

\section{Diffusivity approximation for PS and VD SBI}\label{App: diffusivity}
%% （对PS扩散系数的修正（直接改组分会降低扩散系数，无法实现扩散系数与VD接近，因此通过提高粘性系数\mu_l为原来的五倍，保证Re数及Pe数及Sc数一致，以及Pr也一致）。计算方法，解释一下为什么高Ma 扩散系数变低；最后说一下扩散系数随时间有剧烈变化对结果的影响会在之后的研究中进一步研究，及与常扩散系数结果对比差异有多大
\begin{figure}
    \centering
    \includegraphics[clip=true,trim=0 180 0 0,width=.99\textwidth]{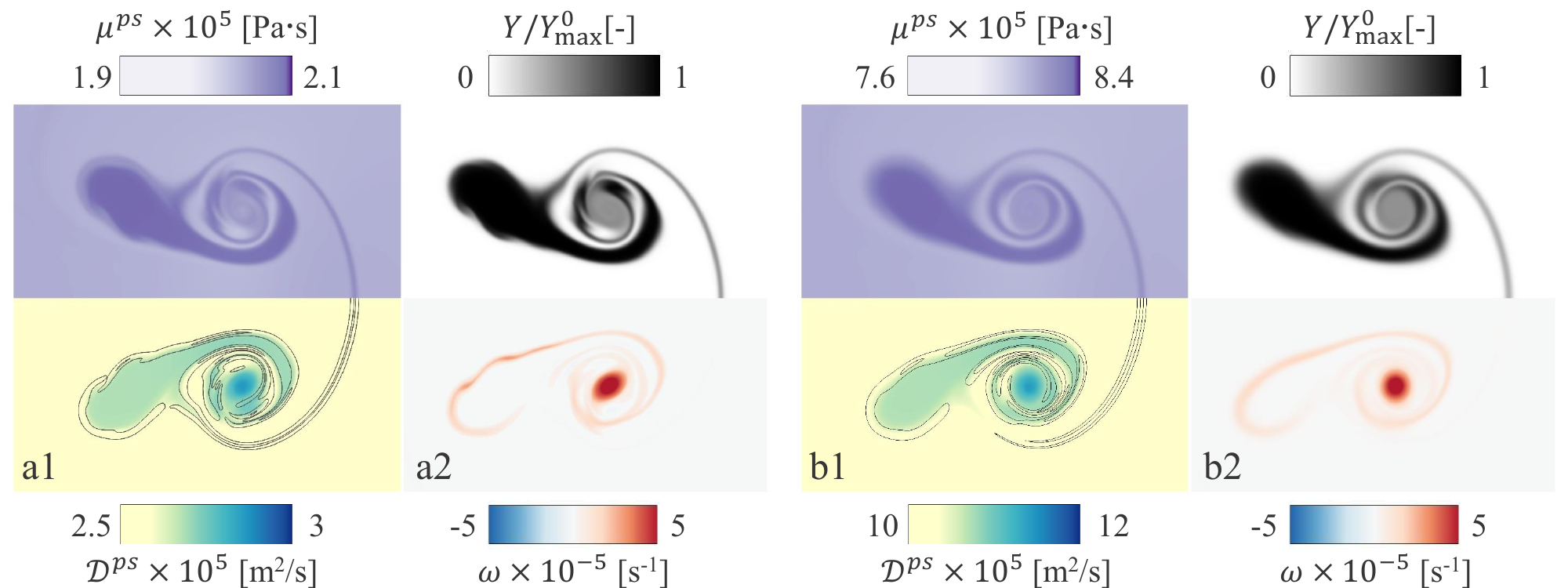}
    \caption{Dynamic viscosity (1 top), diffusivity with scalar dissipation rate isolines (1 bottom), normalized mass fraction (2 top) and vorticity (2 bottom) for original PS SBI with $\mu^{ps}_{\textrm{origin}}$ (a) and viscosity elevated PS SBI with $\mu^{ps}$ (b) at $t=153.6$~$\umu$s and $Ma=1.22$.  }\label{fig: diffusivity contour}
\end{figure}
%%%原因；改变Miu是通用的方式+不影响NS方程，本质改变Re数；云图，增加与不增加云图差别不大，由于粘性变小，原来的Re数变高出现二次不稳定性；由于全场扩散系数不一致，过渡到需要定义一个扩散系数，混合发生区域的扩散系数。
The initial conditions introduced in $\S$\ref{subsec: IC for PS} for PS SBI meet the requirement that the bubble density $\rho^{ps}$ should be nearly the same as the shocked ambient air $\rho_{air}$, which can eliminate the variable density effect. However, because the diffusivity of the PS case is calculated as $\mathscr{D}^{ps}=\nu/Sc=\mu^{ps}/(\rho^{ps} Sc)$, if $\rho^{ps}$ is elevated, the diffusivity will become smaller. Thus, P\'eclet number, $\Pen=\Gamma/\mathscr{D}$ and Reynolds number $\Rey=\Gamma/\nu$ will become larger in PS cases than in the VD case.
After calculation, we find that $\rho^{ps}\approx 4\rho'_{\textrm{He}}$ in all shock Mach number cases, while the viscosity $\mu$ is hardly changed after the components alteration. Thus, it is appropriate to set $\mu^{ps}$ four times larger than the original value $\mu^{ps}_{\textrm{origin}}$. In that case, equal diffusivity in PS SBI and VD SBI, $\mathscr{D}^{ps}\approx\mathscr{D}^{vd}$, will be satisfied. Moreover, this makes $Pe$, $\Rey$, $Sc$, and $\Pran$ near equal in both PS and VD cases except that density difference is negligible in PS SBI. As we have found out, it is essential to control these dimensionless numbers due to their effect on mixing behaviour.

Controlling $Re$ number through changing viscosity is a common way to study the influence of systematic parameters on flow fields. Here, the way of elevating viscosity $\mu$ is similar to changing $Re$ number in dimensionless NS equations to exhibit viscosity effect on Rayleigh-Taylor instability~\citep{hu2019effect}, Richtmyer-Meshkov instability~\citep{walchli2017reynolds, groom2021reynolds} and Kelvin$-$Helmholtz instability~\citep{rahmani2014effect}.
To elucidate the difference between original viscosity $\mu^{ps}_{\textrm{origin}}$ and elevated viscosity $\mu^{ps}$ in PS SBI, we further examine the viscosity and diffusivity contour, as depicted in figure~\ref{fig: diffusivity contour}.
Several observations can be outlined.
Viscosity and diffusivity are near constant for PS case after the components alteration from VD SBI.
The general flow structures are similar between two cases from the mass fraction and vorticity contour. Since the viscosity is smaller (higher $Re$ number) in original one, second instability occurs at the boundary of trailing lobe structure.
After elevating viscosity, the diffusivity increases by the same multiple without large deviation of flow structures from original PS case, which means that it is physically justifiable to set a higher viscosity for the problem concerned.

\begin{figure}
    \centering
    \subfigure[]{
    \label{fig: effdiff-ps}
    \includegraphics[clip=true,trim=10 10 5 25, width=.45\textwidth]{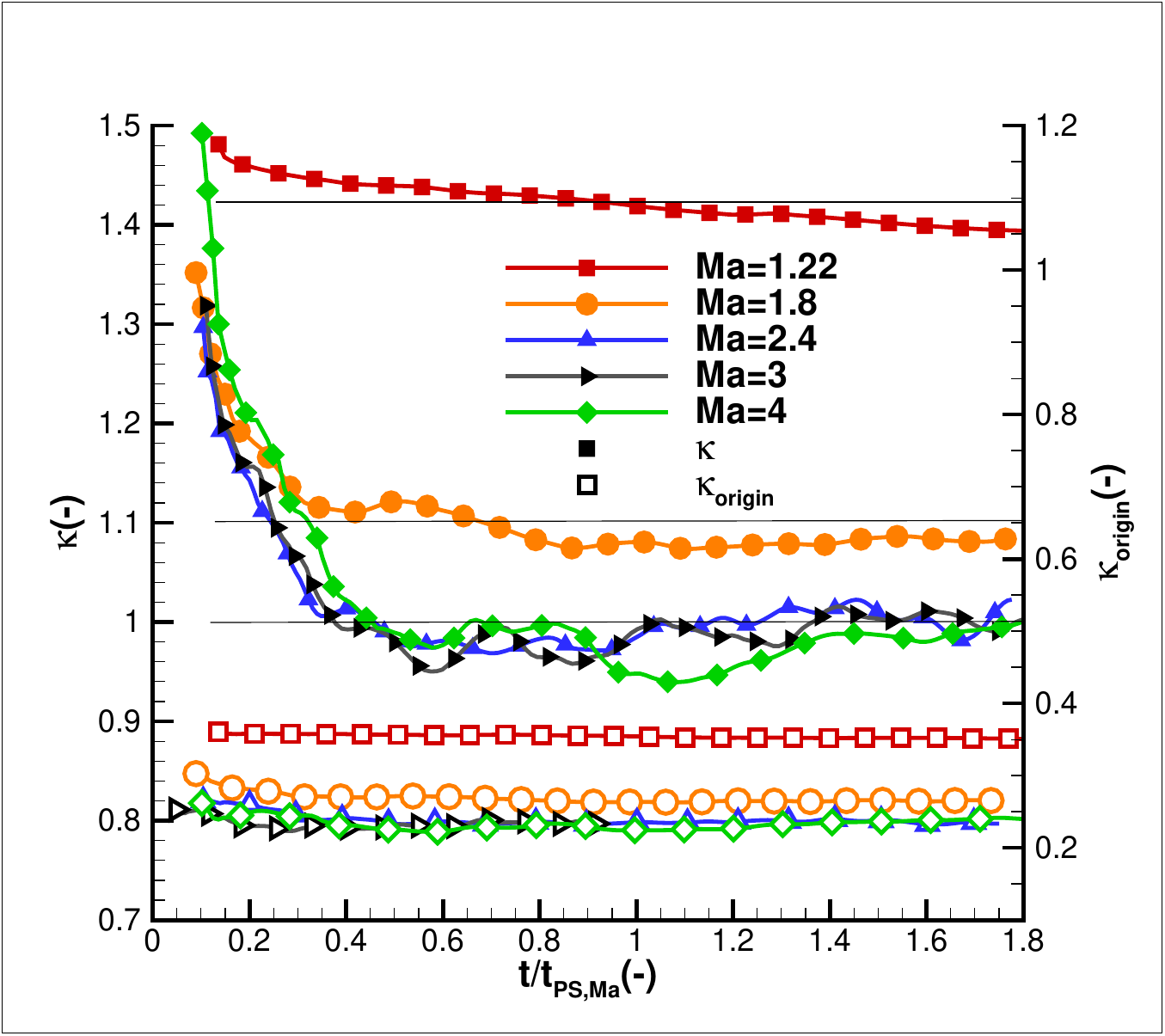}}
    \subfigure[]{
    \label{fig: effdiff-vd}
    \includegraphics[clip=true,trim=10 10 5 25, width=.45\textwidth]{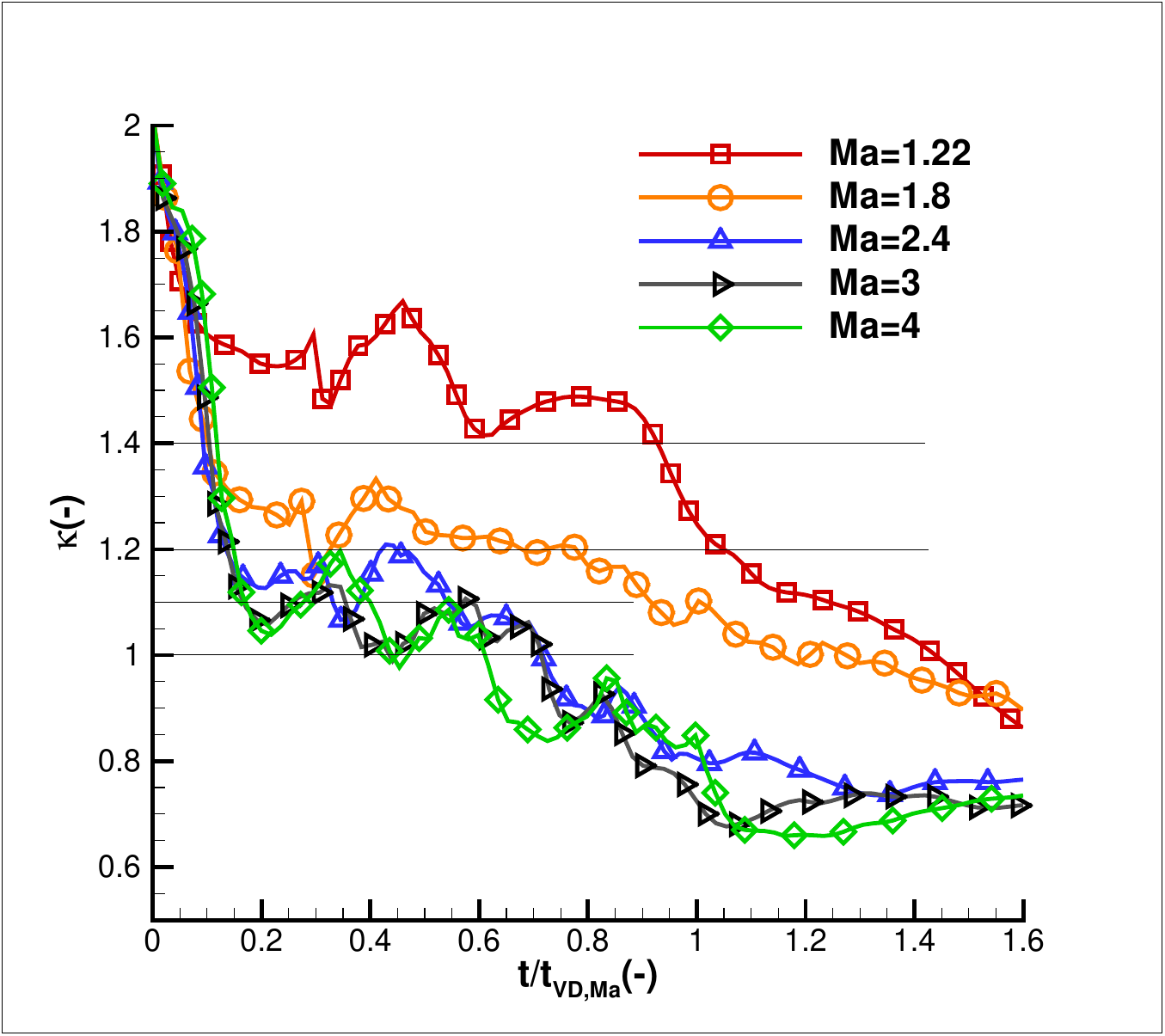}}
    \caption{Effective diffusivity coefficient $\kappa(t)$ in PS cases (a) and VD cases (b) for different shock Mach numbers.  Note that the effective diffusivity coefficient of the original viscosity without elevation, $\kappa_{\textrm{origin}}$ for PS SBI are also compared in (a). \label{fig: effdiff} }
\end{figure}
Thus, we further introduce effective diffusivity $\mathscr{D}_e$, which is crucial for estimating diffusivity in VD SBI with large diffusivity difference between helium and air.
Because the diffusion is happening on the edge of the cylindrical bubble, where the scalar dissipation is large as shown in figure~\ref{fig: diffusivity contour}, it is reasonable to evaluate diffusivity via the mixing indicator, scalar dissipation $\chi$.
Firstly, we introduce the diffusivity coefficient $\kappa(t)$ as:
\begin{equation}\label{eq: Kappa}
  \kappa(t)=\frac{\int\mathscr{D}\nabla{Y}\cdot\nabla{Y}d{V}}{ \mathscr{D}_{m} \int\nabla{Y}\cdot\nabla{Y}d{V}},
\end{equation}
where $\mathscr{D}_m=71\times10^{-6}$ m$^2$/s is the standard diffusivity of helium in air at standard atmospheric conditions~\citep{wasik1969measurements}.
Figure~\ref{fig: effdiff} shows the time history of $\kappa(t)$ of PS SBI and VD SBI. It can be found that before the influence of shock impact becomes small, diffusivity keeps a high value due to the pre-shock diffusivity being larger in the lower density environment. After the shock passes, the diffusivity decreases to a steady value for different shock Mach numbers.
The original PS SBI with lower viscosity $\mu^{ps}_\textrm{origin}$ for different Mach numbers are also compared in figure~\ref{fig: effdiff-ps}. Without elevated viscosity, diffusivity is much smaller than the one in VD cases.
The viscosity elevated PS SBI obtains a similar magnitude of diffusivity with that of VD SBI, which satisfies the same controlling numbers between VD and PS cases.

To determine a diffusivity value during the whole mixing process of SBI, which is necessary for the modelling of the mixing time in (\ref{eq: t_s}), a time averaging $\kappa$ is defined as:
\begin{equation}\label{eq: kappabar}
  \overline{\kappa}=\frac{1}{t_\mathfrak{m}-t_{\textrm{sh}}}\int_{t_{\textrm{sh}}}^{t_\mathfrak{m}}\kappa(t)\mathrm{d}t\Rightarrow \mathscr{D}_{e}=\overline{\kappa}\mathscr{D}_m,
\end{equation}
where $t_{\textrm{sh}}$ is the time when shock compression is finished, and the compression rate is steady in figure \ref{fig: com-PS-VD}.
Effective diffusivity $\mathscr{D}_e$ is then obtained for each case in PS and VD SBI. Different values of $\overline{\kappa}$ in PS and VD SBI are tabulated in tables \ref{tab: tm-ps} and \ref{tab: tm-vd}, which validates the nearly equal diffusivity in the same Mach number cases for PS and VD scenarios.

\begin{figure}
    \centering
    \includegraphics[clip=true,trim=0 65 0 0,width=.99\textwidth]{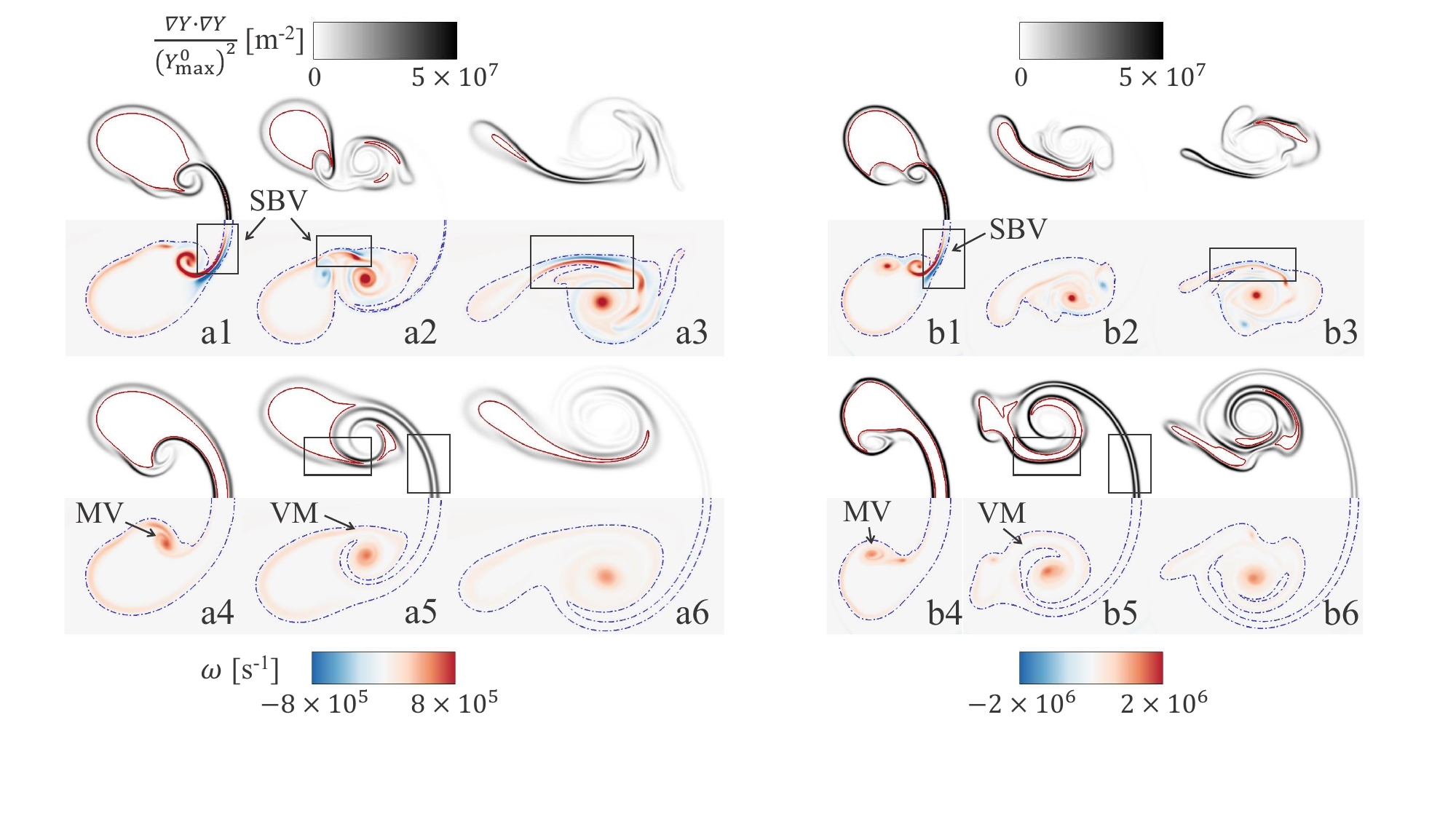}\\
    \includegraphics[clip=true,trim=0 65 0 0,width=.99\textwidth]{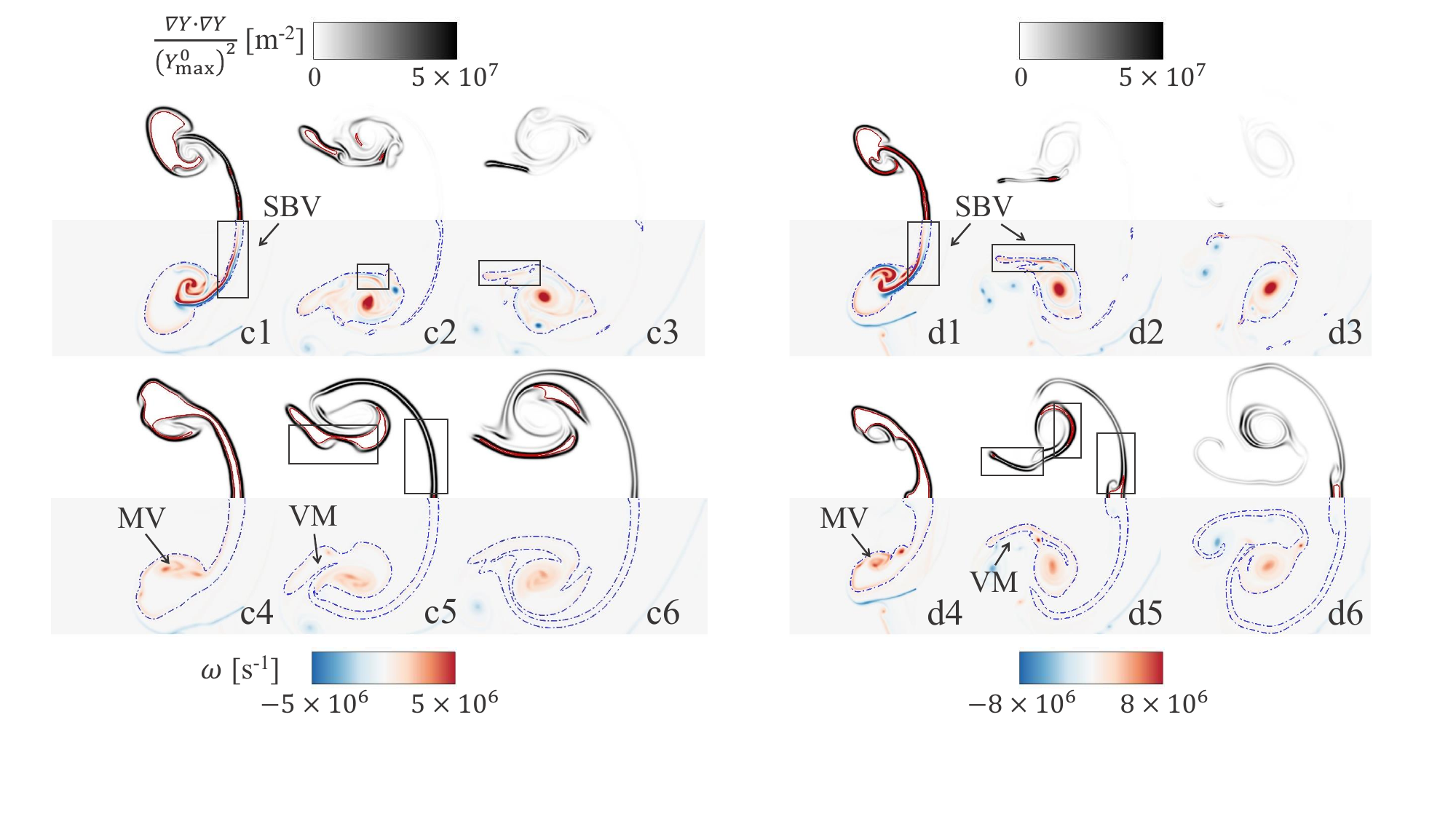}
    \caption{Scalar dissipation rate with red isoline of $Y_{\textrm{He}}=95\%Y_{\max}^0$ (top) and vorticity with blue isoline of $Y_{\textrm{He}}=5\%Y_{\max}^0$ (bottom) for (a) $Ma=1.22$; (b) $Ma=1.8$; (c) $Ma=3$; (d) $Ma=4$.
    For each Mach number, $1-3$ indicate VD SBI and $4-6$ indicate PS SBI at three same moments marked as three blue circles in figure~\ref{fig: mixing time}.
    Specifically, $t=58$~$\umu$s, 96~$\umu$s, 176~$\umu$s for $Ma=1.22$;
    $t=19.2$~$\umu$s, 43.2~$\umu$s, 59.2~$\umu$s for $Ma=1.8$;
    $t=11$~$\umu$s, 19.8~$\umu$s, 26.4~$\umu$s for $Ma=3$;
    $t=8$~$\umu$s, 20~$\umu$s, 27.2~$\umu$s for $Ma=4$.
       }\label{fig: SDR evolu diff Ma}
\end{figure}
\section{Discussions on scalar dissipation behaviour}\label{App: SDR evolu}
%% 对照云图讲VD SDR变化与SBV相关，PSSDR变化与涡量合并相关；进一步深挖机理；进一步线图验证，并总结何时出现SDR peak；可能存在其他物理机制，作为未来研究重点，review几篇相关文章
Several peaks occur during temporal evolution of scalar dissipation in figure~\ref{fig: mixing time}. Here, we attempt to explain the scalar dissipation behaviour difference between VD and PS SBI, namely the source of scalar dissipation growth.
%In section~\ref{subsec: faster decay}, PS and VD SBI for $Ma=2.4$ have been detailed examined. As for VD SBI, it can be found that scalar dissipation peaks are closely related to SBV appearance. The first and second peak of $\chi$ is in accordance to the occurrence of SBV at bridge structure and trailing lobe.
The moments when $\chi$ peaks of VD SBI occur are denoted as the blue circles for each Mach number in figure~\ref{fig: mixing time}. The scalar dissipation and vorticity contours at these moments are depicted in figure~\ref{fig: SDR evolu diff Ma}.
As for VD SBI, scalar dissipation peaks are closely related to SBV appearance, which can be found through high scalar dissipation with high SBV locally.  The high SBV indicates high level stretching, which mainly happens at bridge structure and trailing lobe.
Actually, the synchronous growth of baroclinic circulation and scalar dissipation presented in figure~\ref{fig: cir-SDR-Ma} is the quantitative evidence for the close relation between SBV and scalar dissipation in VD SBI.

As for PS SBI, the mixing process is less fierce than VD SBI due to the absence of SBV, as shown in figure~\ref{fig: SDR evolu diff Ma}. The source of stretching comes from the merging of vorticity and the azimuthal velocity from main vortex. It should be noted that in PS mixing, flow field is not altered by mixing process, meaning that high scalar dissipation should satisfy the co-existence of local stretching and high concentration scalar. Since the initial conditions for PS SBI introduced in $\S$\ref{subsec: IC for PS} share the similarity with VD SBI, the high scalar dissipation occurs still at bridge structure and trailing lobe where some vorticity remains from the initial conditions. With the merging of these vorticity into main vortex, the local scalar is stretched and dissipated.

Further, we invoke the evolutionary source of scalar dissipation rate from the advection-diffusion equation~\citep{buch1996experimental}:
\begin{equation}\label{eq: SDR-evo}
  \left(\frac{\partial}{\partial t}+\boldsymbol{u}\cdot\nabla-\mathscr{D}\nabla^2\right)\frac{1}{2}\nabla Y\cdot\nabla Y=-\nabla Y\cdot\epsilon\cdot\nabla Y-\mathscr{D}\nabla(\nabla Y):\nabla (\nabla Y),
\end{equation}
where $\epsilon\equiv\frac{1}{2}\left(\nabla u+\nabla u^{\textrm{T}}\right)$ is the strain rate tensor, the symmetric part of velocity gradient tensor and $\nabla:\nabla$ is the Frobenius inner product of gradient tensor.
The time derivative of scalar dissipation $\chi$ (\ref{eq:TMR}) can be expressed as:
\begin{eqnarray}\label{eq: SDR-evo-2}
  \frac{\mathrm{d}\chi}{\mathrm{d}t} & = &
  \frac{\mathrm{d}}{\mathrm{d}t}\left[\frac{1}{{\left(Y^0_{\textrm{max}}\right)}^2}\int\nabla{Y}\cdot\nabla{Y} \mathrm{d}V\right] \nonumber\\
  & = & \frac{1}{{\left(Y^0_{\textrm{max}}\right)}^2}\left(
\underbrace{-\int2\nabla Y\cdot\epsilon\cdot\nabla Y\mathrm{d}V}_{S_\textrm{strain}}
\underbrace{-\int2\mathscr{D}\nabla(\nabla Y):\nabla (\nabla Y)\mathrm{d}V}_{S_\textrm{diffusion}}
\right),
\end{eqnarray}
by noting that $\int\nabla^2(\nabla Y\cdot\nabla Y)\mathrm{d}V=0$~\citep{yu2020scaling}.
Thus, it can be found that the source of scalar dissipation is composed by strain term and strictly negative diffusion term.
\begin{figure}
    \centering
    \subfigure[]{
    \label{subfig: strain-SDR-PS}
    \includegraphics[clip=true,trim=10 10 5 25, width=.45\textwidth]{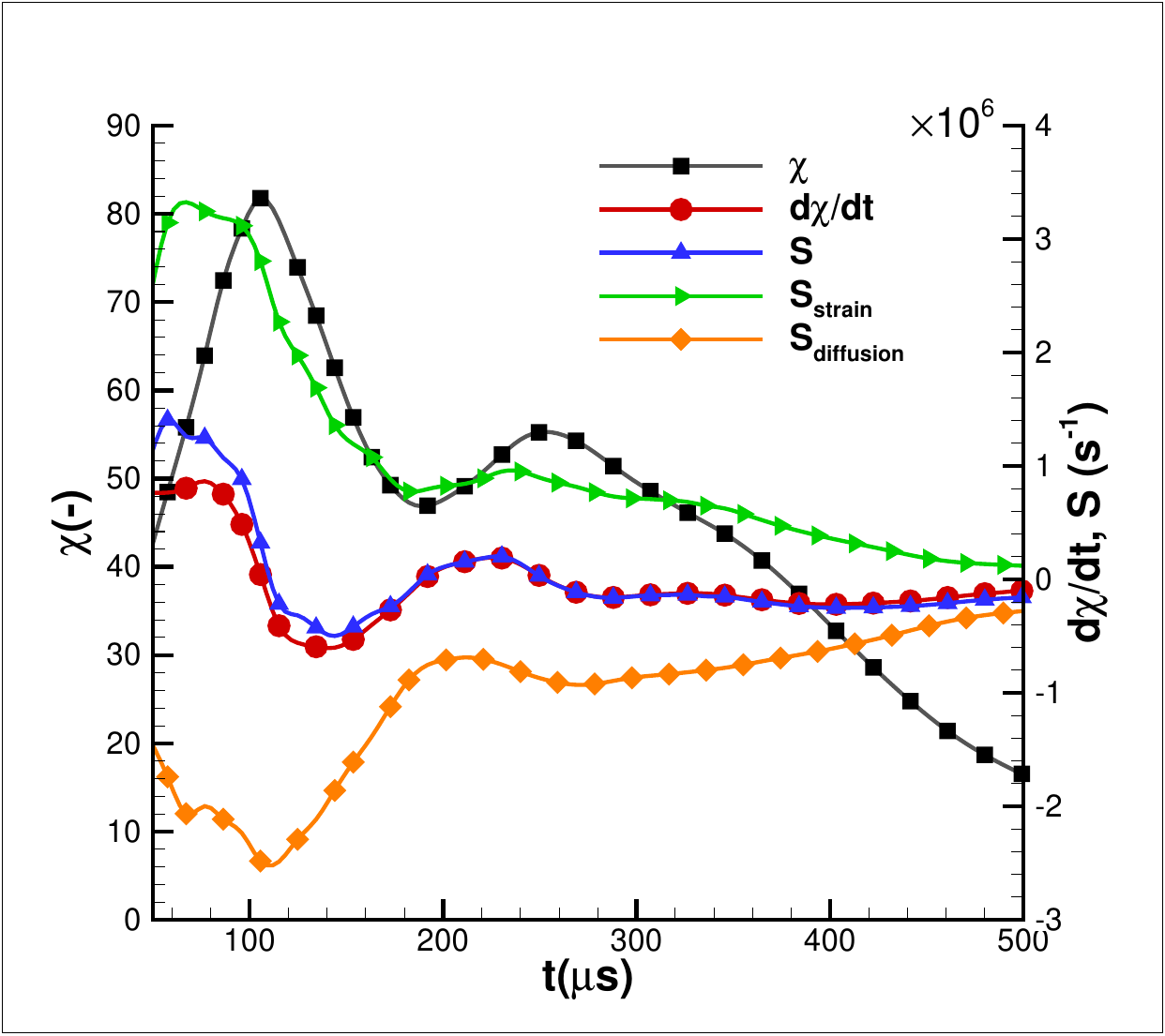}}
    \subfigure[]{
    \label{subfig: strain-SDR-VD}
    \includegraphics[clip=true,trim=10 10 5 25, width=.45\textwidth]{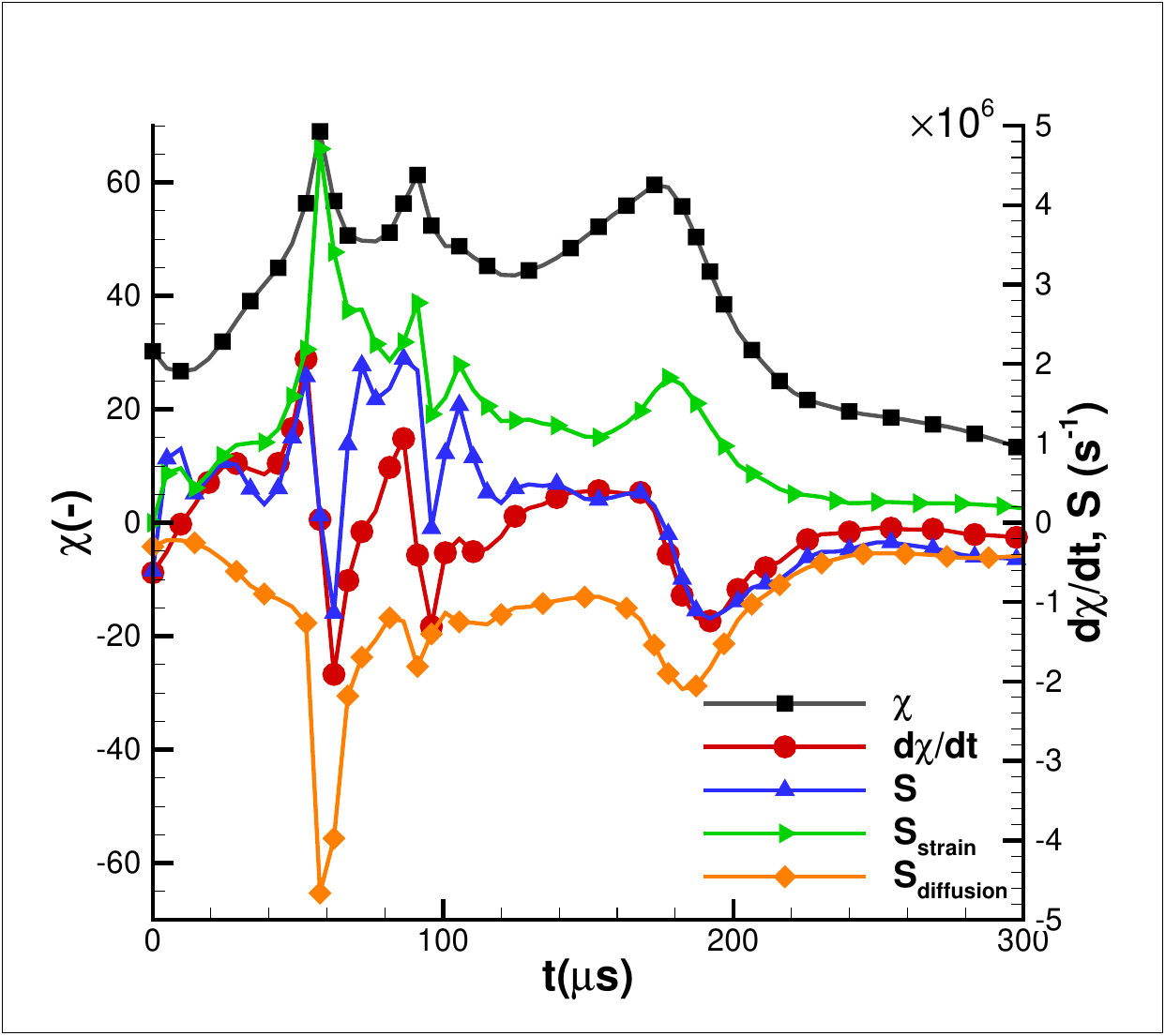}}
    \caption{Temporal evolution of scalar dissipation and its time derivative comparing with the source term in (\ref{eq: SDR-evo-2}) for $Ma=1.22$ PS SBI  (a) and VD SBI (b).  \label{fig: strain-SDR} }
\end{figure}
Figures~\ref{subfig: strain-SDR-PS} and \ref{subfig: strain-SDR-VD} show the strain source term and diffusion source term for $Ma=1.22$ PS and VD SBI respectively. The sum of $S_\textrm{strain}$ and $S_\textrm{diffusion}$, denoted as $S$, shows the same trend as the time derivative of scalar dissipation, validating decomposition (\ref{eq: SDR-evo-2}).
%Slight deviation appears in VD SBI, which can be explained by the hyperbolic conservation violation of advection diffusion equation in variable density flows~\citep{yu2020scaling}.
The detailed evolution mechanism of scalar dissipation and its source term in both PS and VD flows for different shock Mach numbers are worthy for future study.

\section{Approximation of vortex propagation velocity}\label{App: vortex velo}
\begin{figure}
    \centering
    \subfigure[]{
    \label{fig: vv-1.22}
    \includegraphics[clip=true,trim=10 10 5 25, width=.23\textwidth]{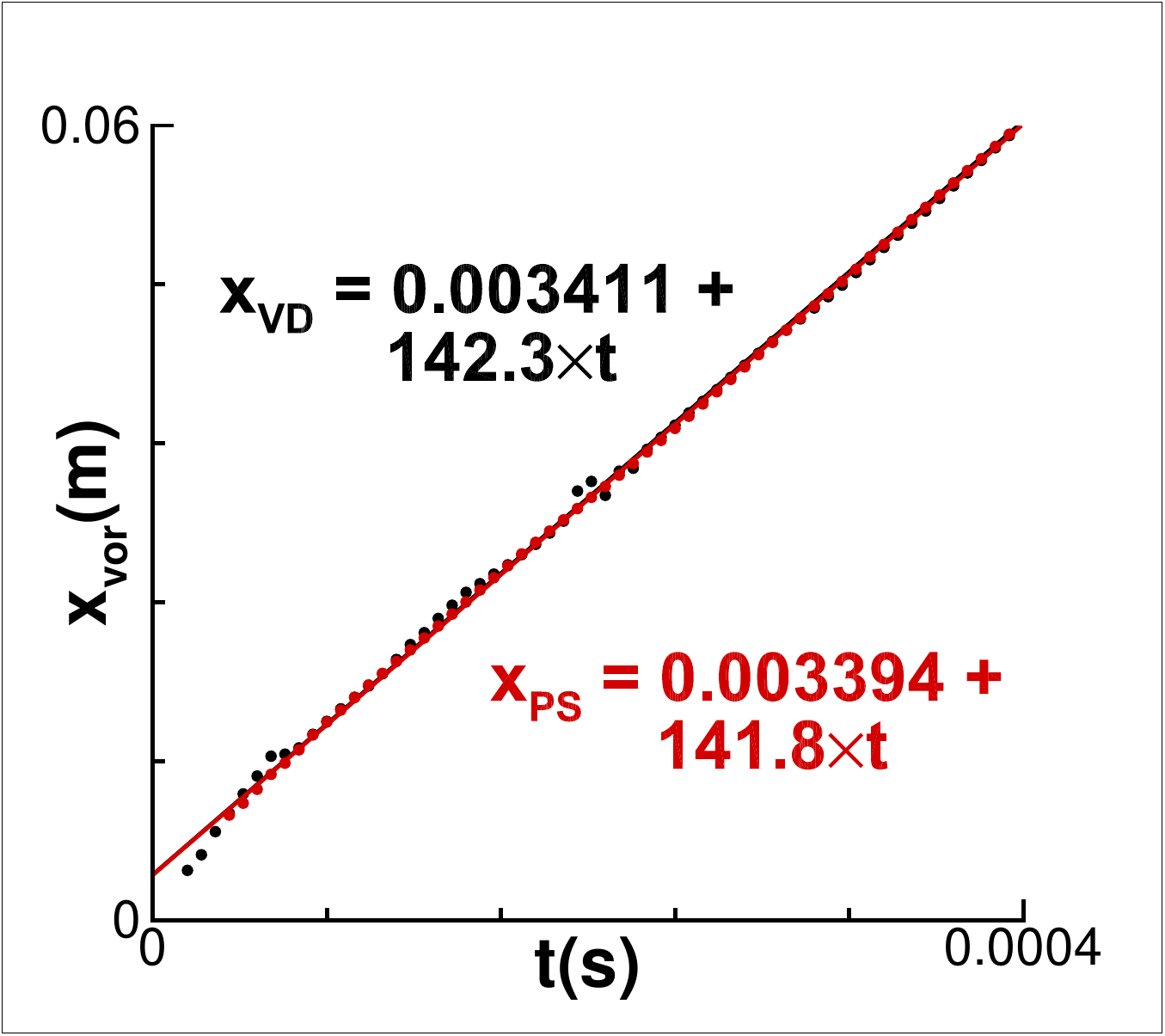}}
    \subfigure[]{
    \label{fig: vv-1.8}
    \includegraphics[clip=true,trim=10 10 5 25, width=.23\textwidth]{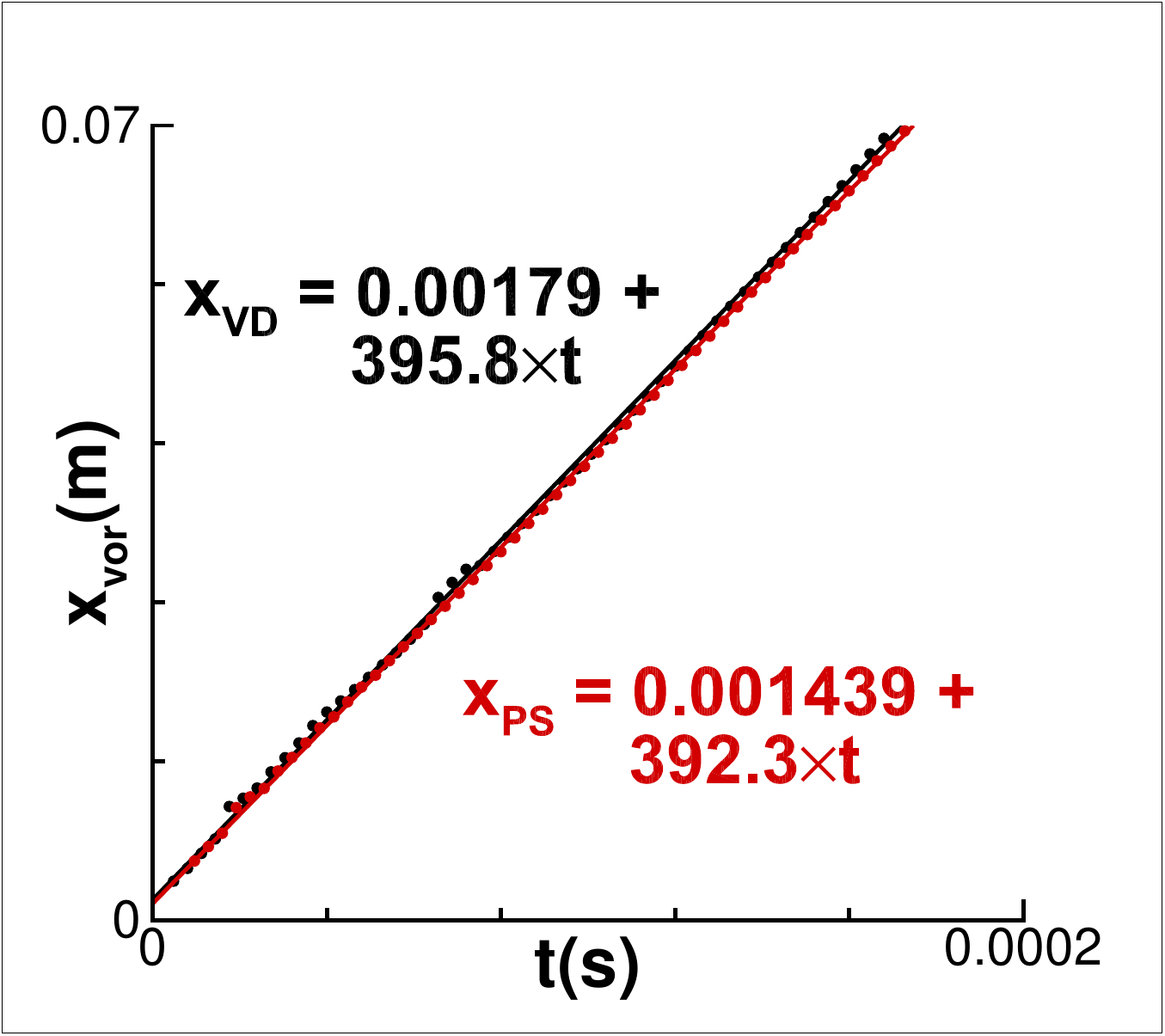}}
    \subfigure[]{
    \label{fig: vv-3}
    \includegraphics[clip=true,trim=10 10 5 25, width=.23\textwidth]{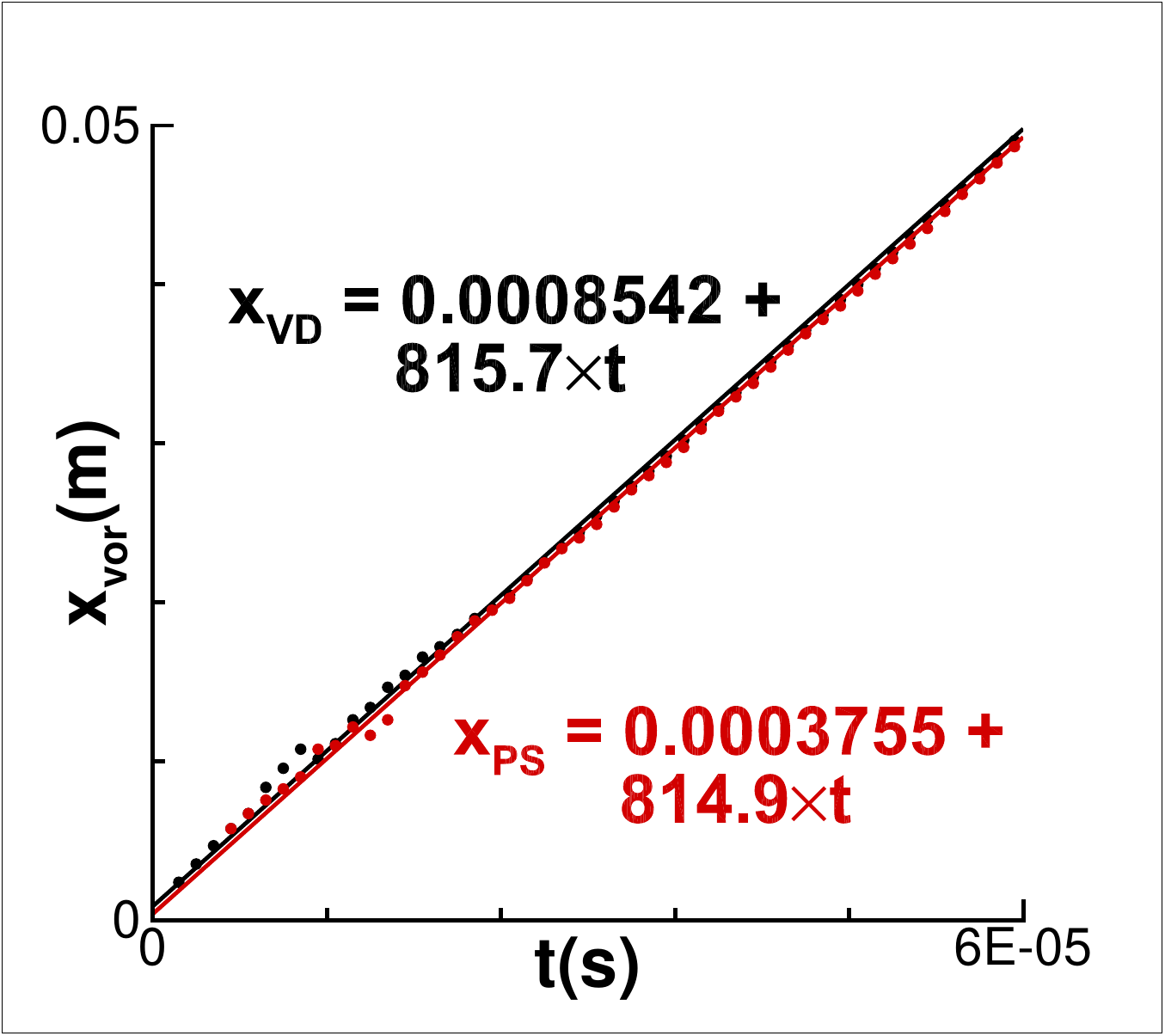}}
    \subfigure[]{
    \label{fig: vv-4}
    \includegraphics[clip=true,trim=10 10 5 25, width=.23\textwidth]{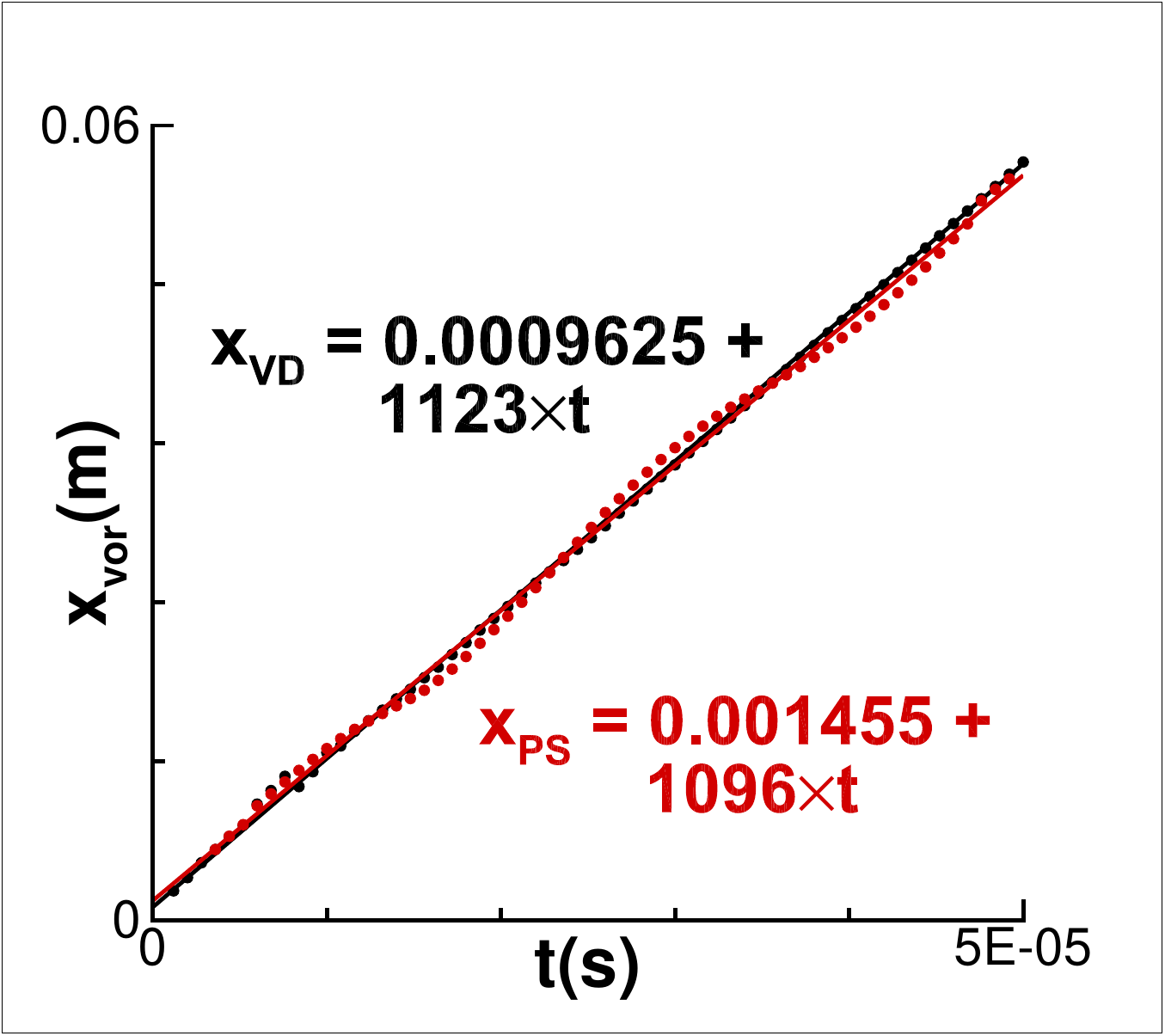}}
    \caption{Time history of vortex centre position to obtain vortical translational velocity for (a) $Ma=1.22$, (b) $Ma=1.8$, (c) $Ma=3$ and (d) $Ma=4$. Linear fit are plotted as solid lines for both VD (black) and PS cases (red).  \label{fig: vv-diffma} }
\end{figure}
In $\S$\ref{subsec: inertial velocity model}, the velocity of the vortex $V_v$ for both the passive scalar and variable density cases is required to set the local coordinate system on the vortex centre. The motion of the vortex centre is recorded as the position of peak vorticity:
\begin{equation}\label{eq: x_v}
  x_v\equiv x\mid_{\omega=\omega_{peak}}.
\end{equation}
Figure \ref{fig: vv-diffma} shows the position of $x_v$ for different shock Mach number cases. A linear fit is applied in the estimation of the vortex velocity. Higher shock Mach number leads to faster motion of the vortex as expected. For all cases, the velocities of the passive scalar vortex and variable density vortex are similar at each Mach number. \\

\section{Linear approximation model of $\mathrm{d}\rho/\mathrm{d}r$}\label{App: drhodr}
\begin{figure}
    \centering
    \subfigure[]{
    \label{fig: drhodr-1}
    \includegraphics[clip=true,trim=0 0 300 0, width=.32\textwidth]{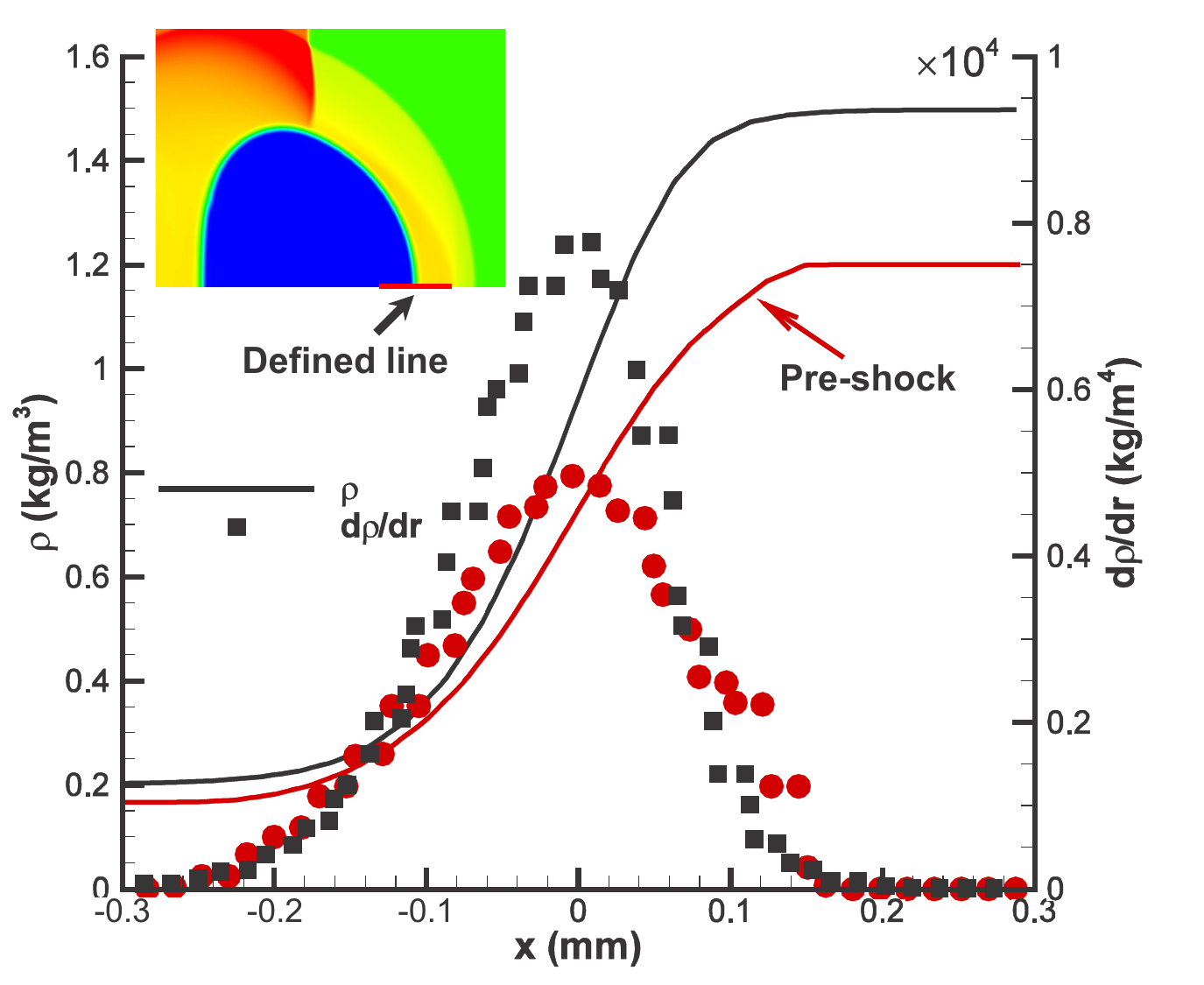}}
    \subfigure[]{
    \label{fig: drhodr-2}
    \includegraphics[clip=true,trim=0 0 300 0, width=.32\textwidth]{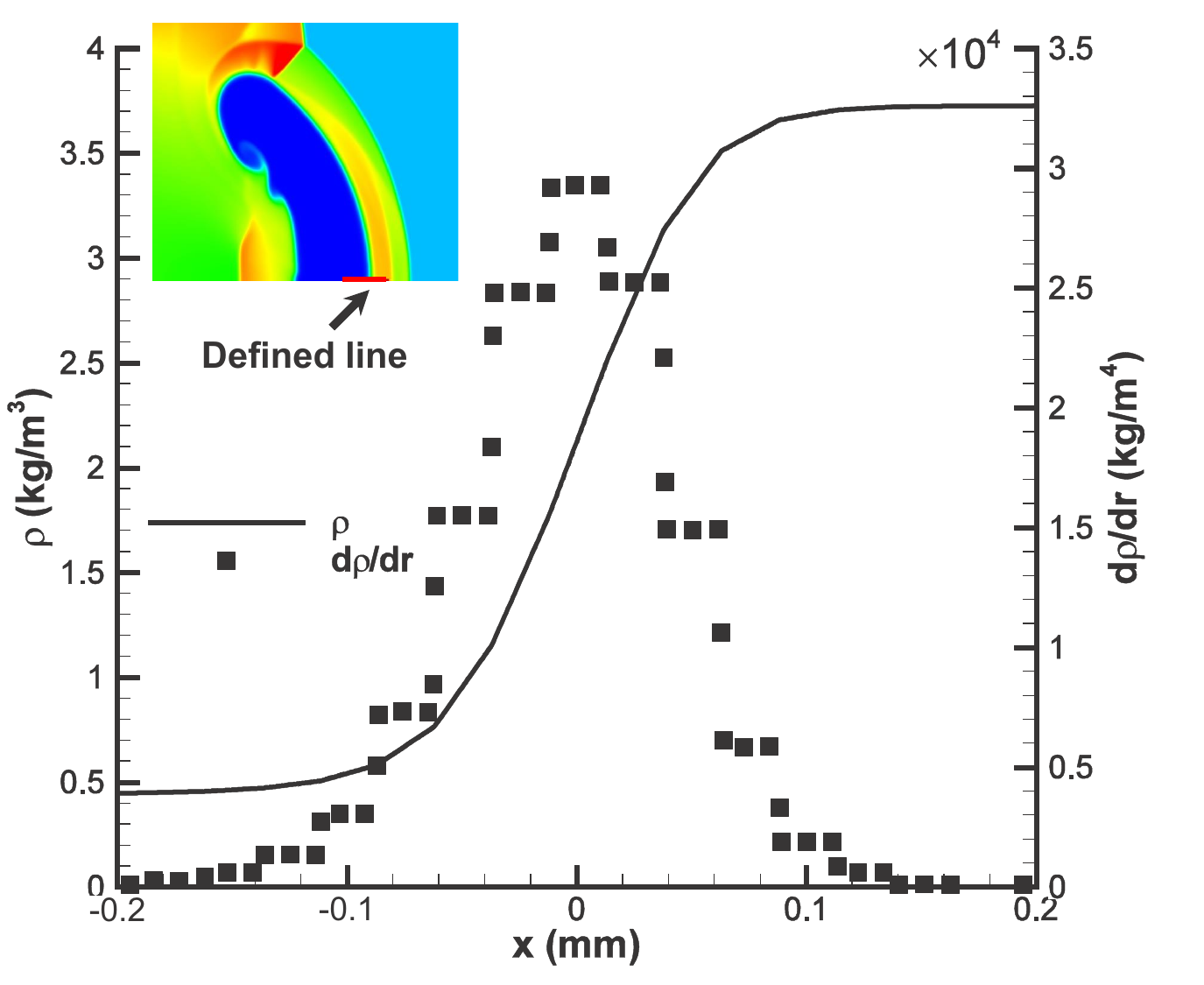}}
    \subfigure[]{
    \label{fig: drhodr-3}
    \includegraphics[clip=true,trim=0 0 300 0, width=.32\textwidth]{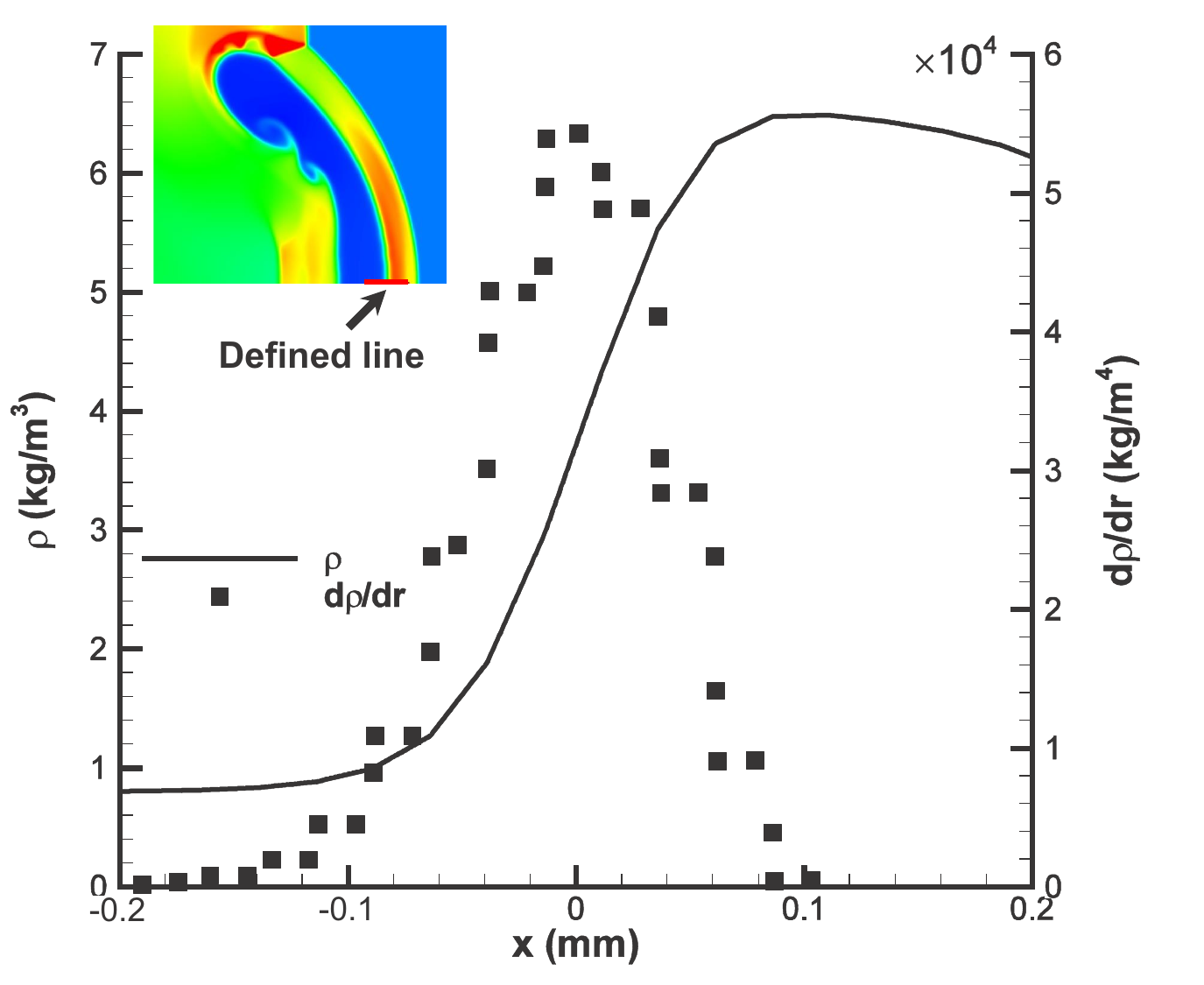}}
    \caption{Post-shock density distribution and its spatial derivative at bridge structure for (a) $Ma=1.22$; (b) $Ma=2.4$ and (c) $Ma=4$.  The extracted lines for each case are indicated through the inserted density contour.
    Specifically, the pre-shock density distribution and its spatial derivative are plotted in (a) for comparison.
    \label{fig: drhodr} }
\end{figure}
%%目的；图1 preshock情况；真实结果分析；模型假设；表介绍
In SBV model (\ref{eq: vavd-illu8}) and azimuthal acceleration model (\ref{eq: vthe-2}), the post-shock density interface transition layer $\delta'$ is used. This appendix validates the assumption (\ref{eq:drhodr}) made in $\S$\ref{subsec: inertial velocity model}. Since the density interface thickness forms from the initial transition layer, it is justifiable to regard the spatial derivative of density $\mathrm{d}\rho/\mathrm{d}r$ as the post-shock initial transition layer.

Figure~\ref{fig: drhodr-1} shows the density profile of pre-shock interfacial layer. As the initial conditions (\ref{eq:Gaussian}), the thickness of transition layer $\delta=0.15R=0.39$~mm, which is quite near to the region with high $\mathrm{d}\rho/\mathrm{d}r$.
Density profiles of post-shock transition layer for $Ma=1.22$, 2.4 and 4 are illustrated in figures~\ref{fig: drhodr-1} to \ref{fig: drhodr-3}. The data are extracted from the defined line along symmetric axis of downstream cylindrical bubble edge, as plotted in the inserted figure.
With the increase of shock Mach number, the peak value of gradient of density $\mathrm{d}\rho/\mathrm{d}r$ raises as well. This raise of gradient comes from both compression of transition layer $\delta'$ and the increase of post-shock air/helium density difference $\Delta\rho'$. Here, we use the half of $\mathrm{d}\rho/\mathrm{d}r$ peak value in figure~\ref{fig: drhodr} as an average of density gradient distribution, $(\mathrm{d}\rho/\mathrm{d}r)_m$, as listed in table~\ref{tab: drhodr}.

Since a narrow-band quasi-gaussian density distribution can be found for all cases, it is reasonable to approximate $(\mathrm{d}\rho/\mathrm{d}r)_m$ as $\Delta\rho'/\delta'=(\rho_1'-\rho_2')/\delta'$. The density of bubble $\rho_2'$ and ambient air $\rho_1'$ increase after shock, obtained from one-dimensional shock dynamics.
After shock impact, the cylindrical bubble is compressed by ratio $\eta$ in volume. Following (\ref{eq: effect-radius}), the length scale of cylindrical bubble is compressed by ratio $\sqrt{\eta}$, leading to the thickness of transition layer $\delta'=\sqrt{\eta}\delta$.
From comparing with profile in figure~\ref{fig: drhodr}, post-shock helium density $\rho_2'$, post-shock air density $\rho_1'$ and transition layer thickness $\delta'$ can approximate density distribution along the density interface.
%Moreover, it can be found that the modelled value are near the half of the measured ones.
Therefore, the general trend of $\mathrm{d}\rho/\mathrm{d}r$ can be reasonably expressed by the simple linear density distribution model, as validated in table~\ref{tab: drhodr}.
\begin{table}
  \begin{center}
\def~{\hphantom{0}}
  \begin{tabular}{lcccccc}
      $Ma$  &  $\rho_1'$(kg/m$^3$)  & $\rho_2'$(kg/m$^3$)  & $\delta'$(mm)   &   $\Delta\rho'/\delta'$(kg/m$^4$) & $(\mathrm{d}\rho/\mathrm{d}r)_m$(kg/m$^4$) & Ratio \\[3pt]
       1.22      & 1.65  & 0.20  & 0.342  & 4249        & 3890  & 0.92  \\
       1.8       & 2.83  & 0.26  & 0.273  & 9420        & 9300 & 0.98  \\
       2.4       & 3.86  & 0.32  & 0.238  & 14858       & 14700 & 0.99  \\
       3         & 4.64  & 0.37  & 0.216  & 19703       & 19600 & 1.00  \\
       4         & 5.50  & 0.44  & 0.201  & 25184       & 27100 & 1.08  \\
Pre-shock & 1.20  & 0.17  & 0.390  & 2650        & 2500  & 0.94
  \end{tabular}
  \caption{Validation of (\ref{eq:drhodr}) by comparing with the measured value $(\mathrm{d}\rho/\mathrm{d}r)_m$ from figure \ref{fig: drhodr}. Here, $\delta'=\sqrt{\eta}\delta$ is the thickness of post-shock density transition layer near bridge structure. $\Delta\rho'=\rho_1'-\rho_2'$ is the density difference between post-shock air $\rho_1'$ and post-shock helium $\rho_2'$. Ratios between measured value and model value are listed in last column.
  For comparison, last row shows the pre-shock data with compression rate $\eta=1$. }
  \label{tab: drhodr}
  \end{center}
\end{table}

\section{Scalar mixing under a standard Lamb-Oseen vortex}\label{App: PSmixing}
In this appendix, we compare the scalar mixing under a standard Lamb-Oseen vortex with that of $Ma=1.22$ PS SBI to confirm the assumption of a compressed cylindrical bubble made in (\ref{eq: effect-radius}), as shown in figure~\ref{fig: mix-mech}(a).
\subsection{Validation of mixing time theory by~\citet{meunier2003vortices}}
To faithfully track the mixing behaviour of a passive scalar blob, the canonical advection-diffusion equation (\ref{eq: ADE}) is numerically solved,
where $Y$ is the local scalar concentration, $\mathscr{D}$ is the scalar diffusivity, $\boldsymbol{u}=(u,v)$ are the velocity in the $x$- and $y$-direction respectively.
We use the standard fifth-order WENO scheme~\citep{liu1994weighted,jiang1996efficient} to discretisize the equation for obtaining the simulation results with high precision. Solution time is marched by
the third-order TVD Runge-Kutta method~\citep{gottlieb1998total}.

\begin{figure}
    \centering
    \subfigure[]{
    \label{fig: ADE-viller}
    \includegraphics[clip=true,trim=0 0 555 0, width=0.445\textwidth]{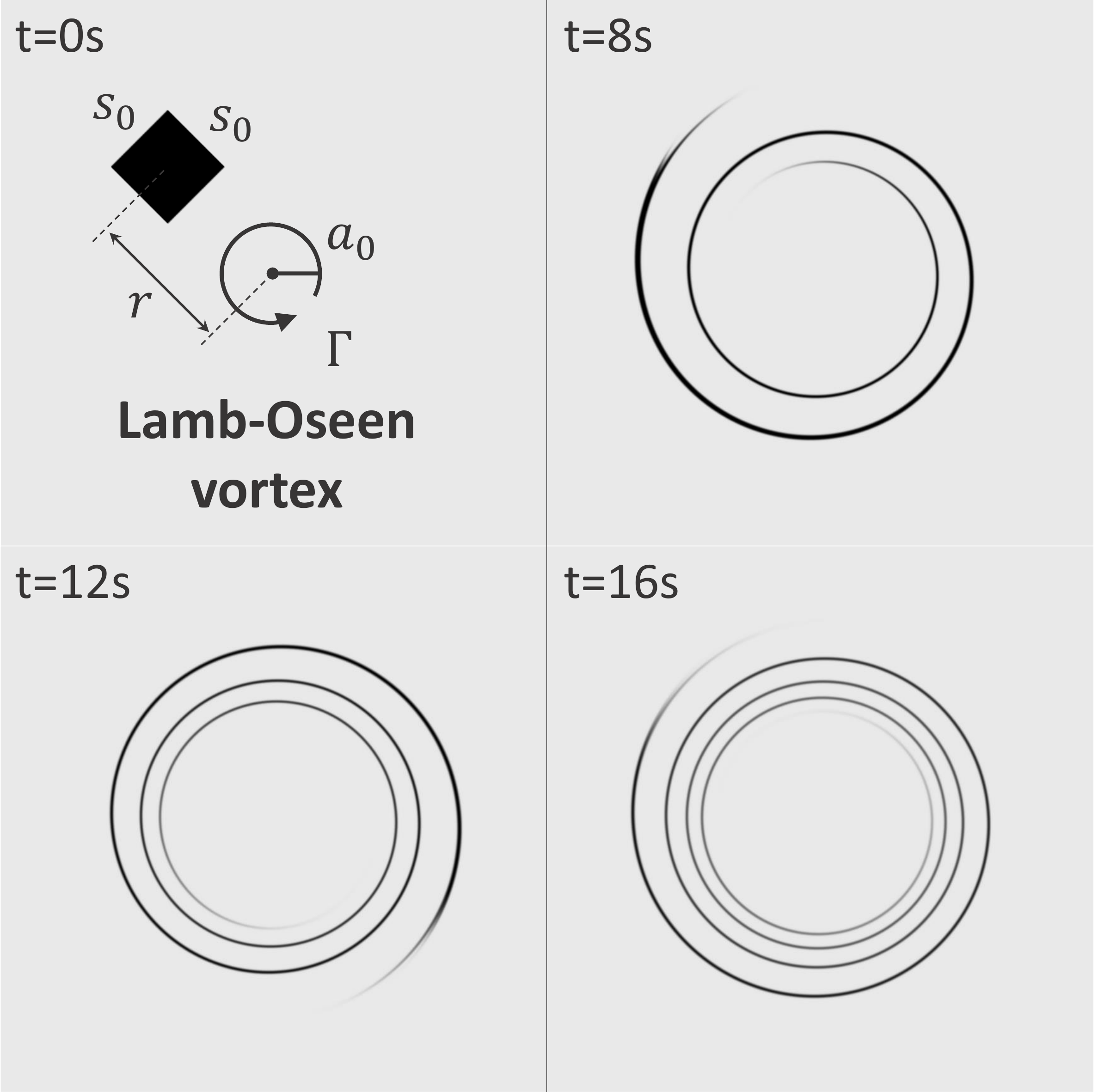}}
    \subfigure[]{
    \label{fig: ADE-viller-line}
    \includegraphics[clip=true,trim=5 5 20 20, width=0.5\textwidth]{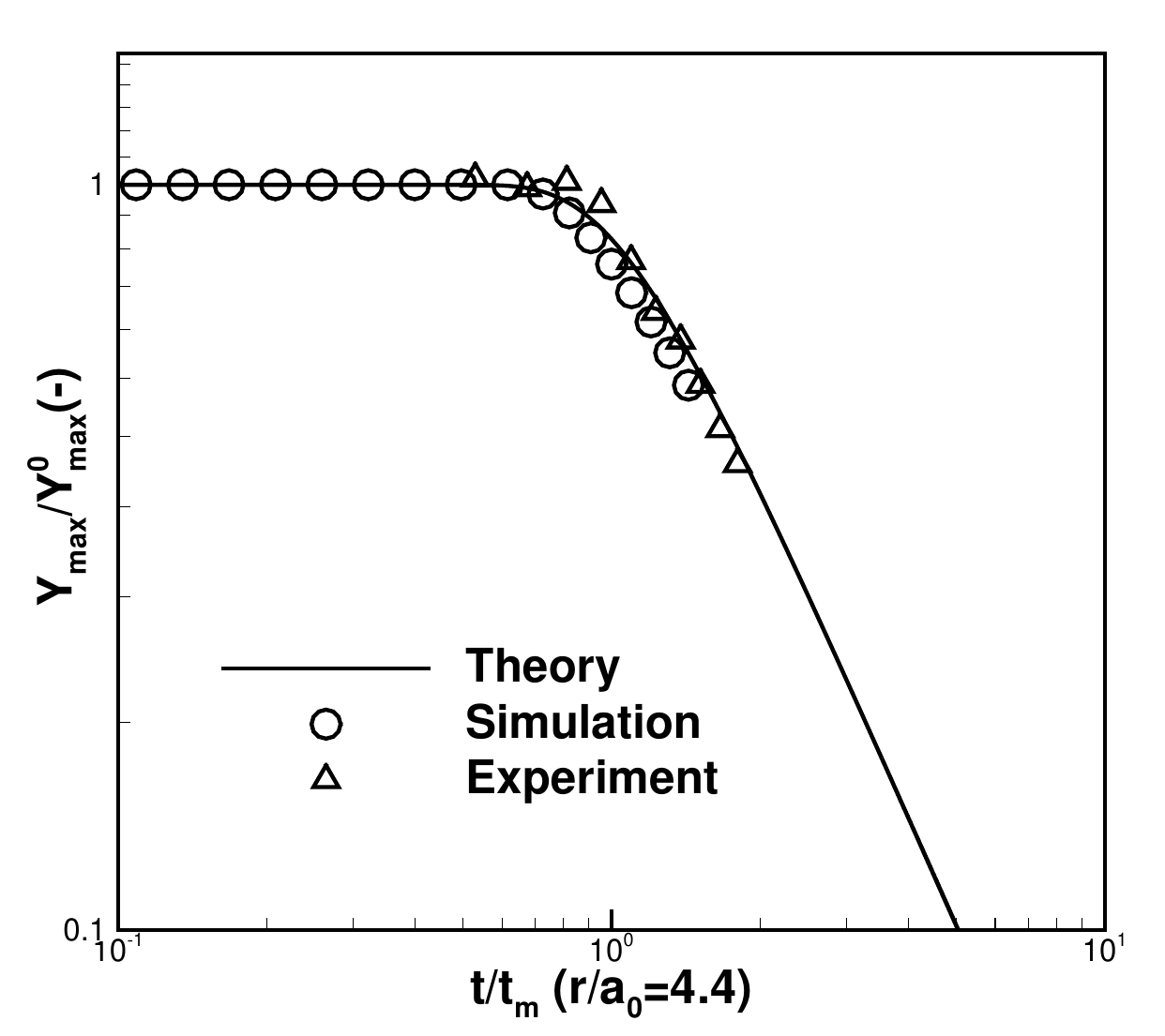}}
    \caption{
    (a) Spiral-type passive scalar mixing of an initial square-shape dye under a deformation field of a Lamb-Oseen vortex.
    (b) Decay of maximum dye concentration from present simulation, experiment~\citep{meunier2003vortices} and theoretical mixing time model~(\ref{eq: theo}).
    \label{fig: ADE comp} }
\end{figure}

Here, a dye mixing in a standard Lamb-Oseen type vortex generated in a water tank experiment by~\citet{meunier2003vortices} is numerically studied.
In the experiment, the measured temporal varying azimuthal velocity profile $v_{\theta}$ agrees well with that of a Lamb-Oseen vortex:
\begin{equation}\label{eq: lamb-oseen vor}
v_{\theta}=\frac{\Gamma}{2\upi r}\left(1-\mathrm{e}^{-r^2/a^2}\right),
\end{equation}
where $\Gamma=1.42\times10^{-3}$~m$^2$/s is the vortex circulation. $a$ is the vortex core radius:
\begin{equation}\label{eq: vorcoreR}
a^2=a^2_0+4\nu t,
\end{equation}
where $a_0=3$~mm is the initial vortex core radius and $\nu=10^{-6}$~m$^2$/s is the water kinetic viscosity at temperature 20$^{\circ}$C.
To validate the numerical scheme for (\ref{eq: ADE}) and compare with the experiment, we set a square-shape passive scalar blob with length $s_0=2.2$~mm under the deformation flow field (\ref{eq: lamb-oseen vor}). The scalar diffusivity is measured as $\mathscr{D}=5\times10^{-10} \text{ m}^2$/s. The distance between the center of the scalar blob and the vortex center $r=4.4a_0=13.2$~mm, as shown in figure~\ref{fig: ADE-viller}.

The grid size should be smaller than the Batchelor scale~\citep{batchelor1959small}, $\lambda_b=\sqrt{\mathscr{D}/\gamma_s}$, to resolve the intermolecular diffusion of species being mixed, if the scalar is deformed at a stretching rate $\gamma_s$~\citep{villermaux2019mixing}.
In a point vortex assumption, the deformation rate can be obtained from (\ref{eq: st-viller}) as:
\begin{equation}
\gamma_s=-\frac{\dot{s}}{s}=\frac{\Gamma^2 t}{\upi^2 r^4+\Gamma^2 t^2}.
\end{equation}
Since the expression shows a temporal varying deformation rate, the minimal Batchelor scale can be estimated with the maximal $\gamma_{s,\max}$ at $t=\frac{\upi r^2}{\Gamma}$:
\begin{equation}
\gamma_{s,\max}=\frac{\Gamma}{2\upi r^2}\Rightarrow\lambda_{b,\min}=\sqrt{\frac{\mathscr{D}}{\gamma_{s,\max}}} =\sqrt{\frac{2\mathscr{D} \upi r^2}{\Gamma}}
\end{equation}
Therefore, the minimal Batchelor scale is estimated as $\lambda_{b,\min}\approx2.8\times10^{-6}$ m in the present case, and the grid size is then determined to be $\Delta=2.4\times10^{-6}$ m, smaller than $\lambda_{b,\min}$. The corresponding time step is set as $10^{-5}$~s in the simulation.

The temporal evolution of square-shape dye concentration is illustrated in figure~\ref{fig: ADE-viller}. The maximum concentration decays inside the stretched spiral arms of deformed scalar due to diffusion. Time history of maximum concentration decay is plotted in figure~\ref{fig: ADE-viller-line}. Generally, the results from present simulation agrees well that from water tank experiment, validating the numerical method. For comparison, the theoretical prediction for the maximal concentration decline is also plotted~\citep{meunier2003vortices}:
\begin{equation}\label{eq: theo}
Y_M(r,t)=Y_0 \text{erf} \left[\frac{1}{4\sqrt{\mathscr{D} t / s_0^2 + \mathscr{D} \Gamma^2 t^3 / \left(3 \upi^2 r^4 s_0^2 \right)}}\right].
\end{equation}
It can be found that the maximum concentration begins to decay if the dimensionless time $t/t_m\approx1$ (\ref{eq: t_s}) is reached.

\subsection{Comparisons of mixing in a Lamb-Oseen vortex and in PS SBI}
In $\S$\ref{subsec: PS mixing}, the velocity field of PS SBI is observed similar to the one of a point vortex model. Since we have validated the numerical scheme for advection diffusion equation (\ref{eq: ADE}), it is feasible to compare the mixing behaviour of a scalar in a Lamb-Oseen vortex and in PS SBI. It is noteworthy that the controlling parameters in Lamb-Oseen vortex is vital and should be comparable to the one in PS SBI. Here, we choose $Ma=1.22$ PS SBI as an example for comparison.

The key parameters required in a velocity profile of Lamb-Oseen vortex (\ref{eq: lamb-oseen vor}) are extracted from the numerical results of PS SBI.
For $Ma=1.22$ PS SBI, the circulation $\Gamma=0.78 \text{ m}^2$/s, the diffusivity $\mathscr{D}=\overline{\kappa}\mathscr{D}_m\approx9.94\times10^{-5} \text{ m}^2$/s and the kinetic viscosity can therefore be obtained by a constant Schmidt number $Sc=\nu/\mathscr{D}=0.5$ as $\nu\approx5\times10^{-5}$~m$^2$/s.
It is essential to set an initial vortex core radius $a_0$, since azimuthal velocity will be spuriously supersonic near vortex centre under the circulation magnitude if $a_0$ is zero.
To estimate the initial vortex core radius $a_0$, a nominal vortex core radius $a$ in (\ref{eq: vorcoreR}) is obtained from the maximum azimuthal velocity (\ref{eq: vthe}) from $Ma=1.22$ PS SBI.

\begin{figure}
    \centering
    \subfigure[]{
    \label{fig: velo-vec}
    \includegraphics[clip=true,trim=15 15 25 25, width=.435\textwidth]{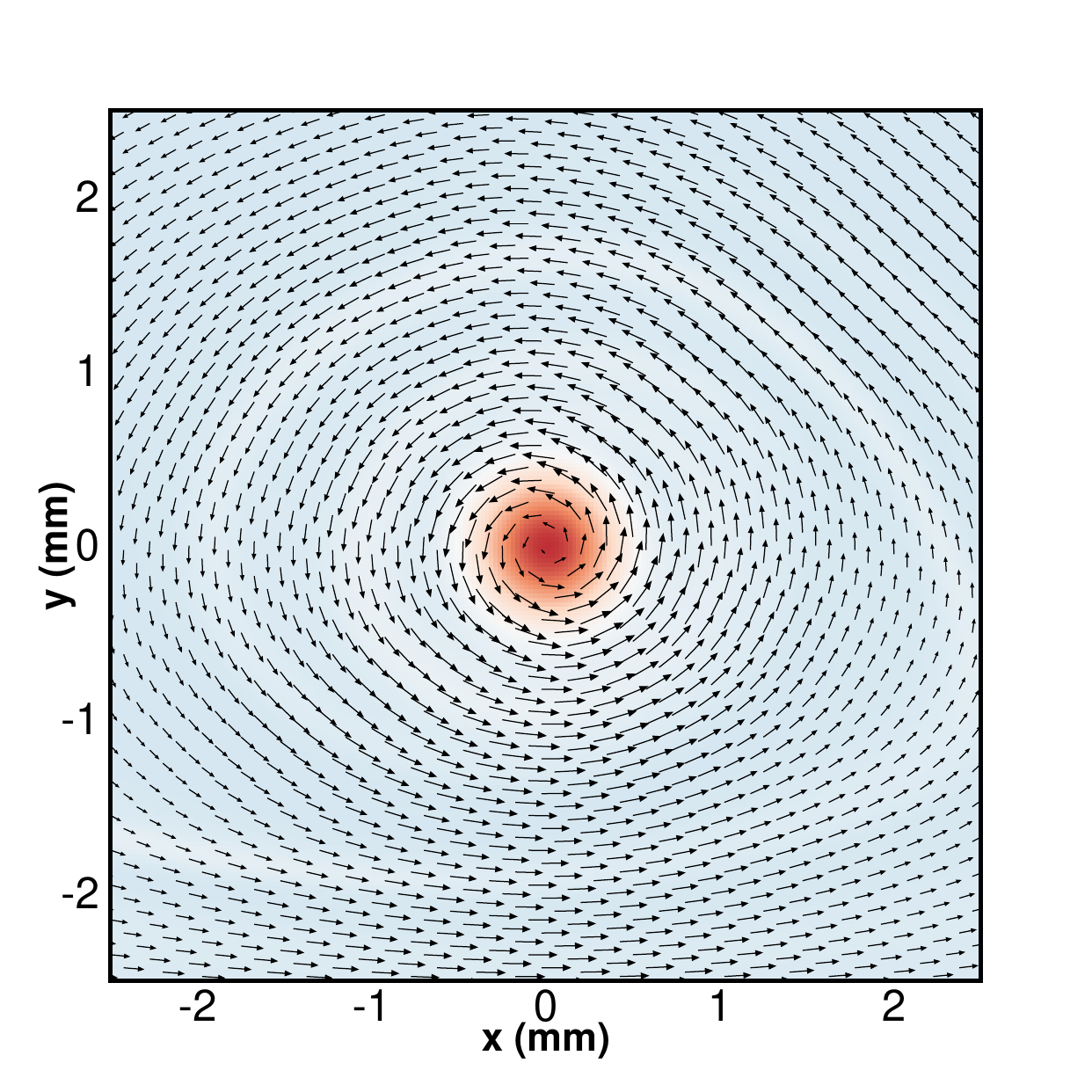}}
    \subfigure[]{
    \label{fig: a0}
    \includegraphics[clip=true,trim=25 20 40 40, width=.48\textwidth]{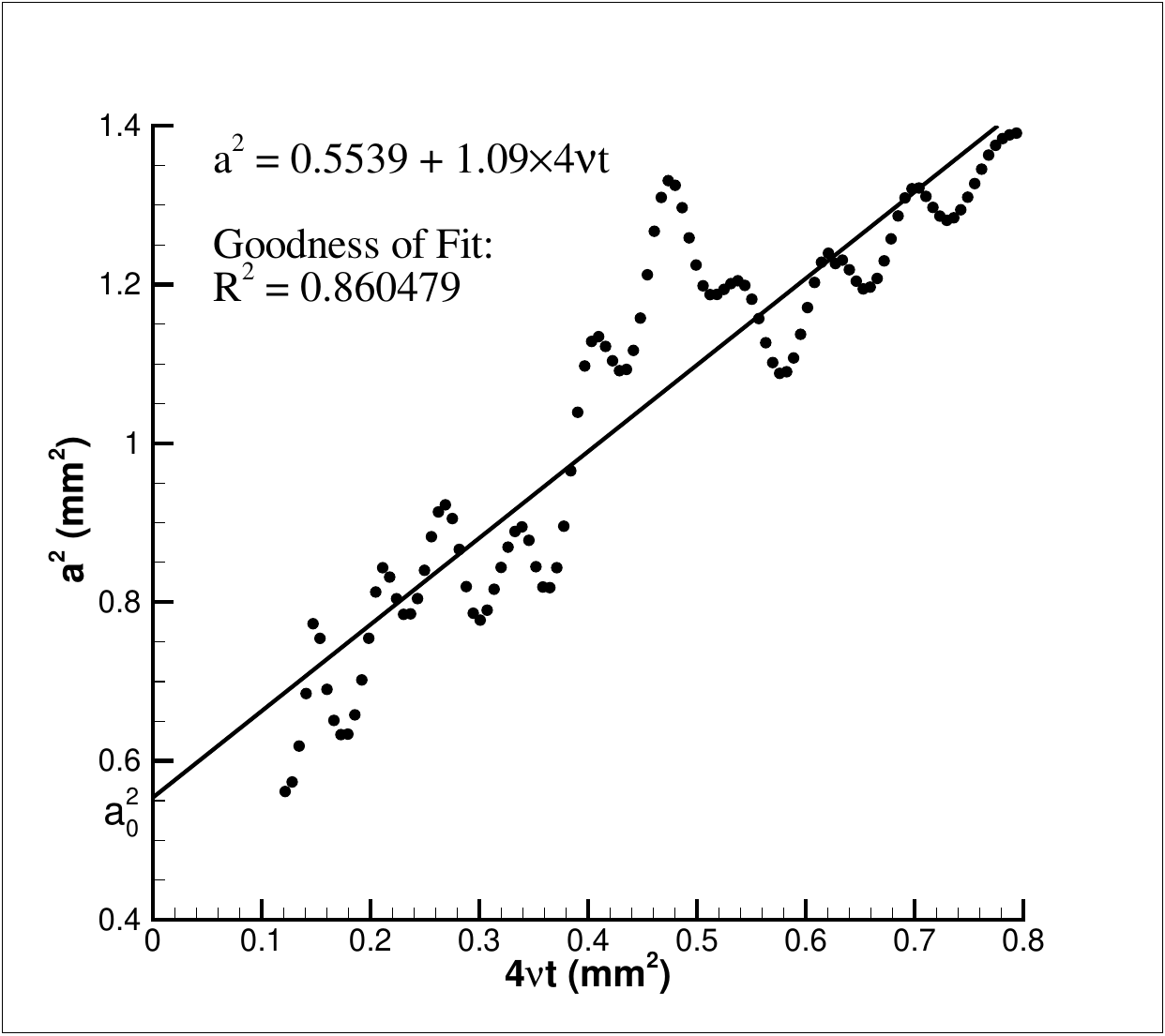}}
    \caption{
    (a) The azimuthal velocity vector (\ref{eq: vthe}) and the vorticity contour of $Ma=1.22$ PS SBI at $t=297.6$~$\umu$s.
    (b) Nominal core size of the vortex pattern in $Ma=1.22$ PS SBI measured by a least-square fit of the transformed maximum azimuthal velocity field and compared to (\ref{eq: a0}) (solid line). \label{fig: fit-a0} }
\end{figure}

From (\ref{eq: lamb-oseen vor}), the azimuthal velocity at one moment, such as in figure~\ref{fig: velo-vec}, will achieve its maximum along $r$ where the local derivative is zero:
\begin{equation}\label{eq: x-r}
\frac{\partial v_{\theta}}{\partial r}=-\frac{\Gamma}{2\upi r^2} \left(1 - \mathrm{e}^{-r^2/a^2} - \frac{2r^2}{a^2} \mathrm{e}^{-r^2/a^2} \right)=0.
\end{equation}
By defining $x_0 = \frac{r_0^2}{a^2}$, we can obtain the solution of transcendental equation (\ref{eq: x-r}) at $x_0\approx 1.25643$.
Therefore, the maximal azimuthal velocity is reached at $r=r_0$ in (\ref{eq: lamb-oseen vor}):
\begin{equation}
v_{\theta,\max}=\frac{\Gamma}{2\upi a \sqrt{x_0}} \left(1 - \mathrm{e}^{-x_0}\right)=\frac{K}{\sqrt{a_0^2+4\nu t}},
\end{equation}
where $K=\frac{\Gamma}{2\upi \sqrt{x_0}} \left(1 - \mathrm{e}^{-x_0}\right)\approx0.07922$.
It can be found that the maximum azimuthal velocity gradually decays with time due to the viscosity.
Therefore, the relationship of the nominal vortex core measured from the maximal azimuthal velocity and the initial vortex core radius $a_0$ is:
\begin{equation}\label{eq: a0}
a_0^2+4\nu t=\frac{K^2}{v_{\theta,\max}^2}.
\end{equation}
The temporal varying maximum azimuthal velocity of $Ma=1.22$ PS SBI is recorded and transformed into nominal vortex core radius $a$, as shown in figure~\ref{fig: a0}. From the least-square linear fit of measured data, we can obtain the initial vortex core radius $a_0\approx0.74$~mm.

\begin{figure}
\centering
\includegraphics[clip=true,trim=0 25 0 0, width=0.99\textwidth]{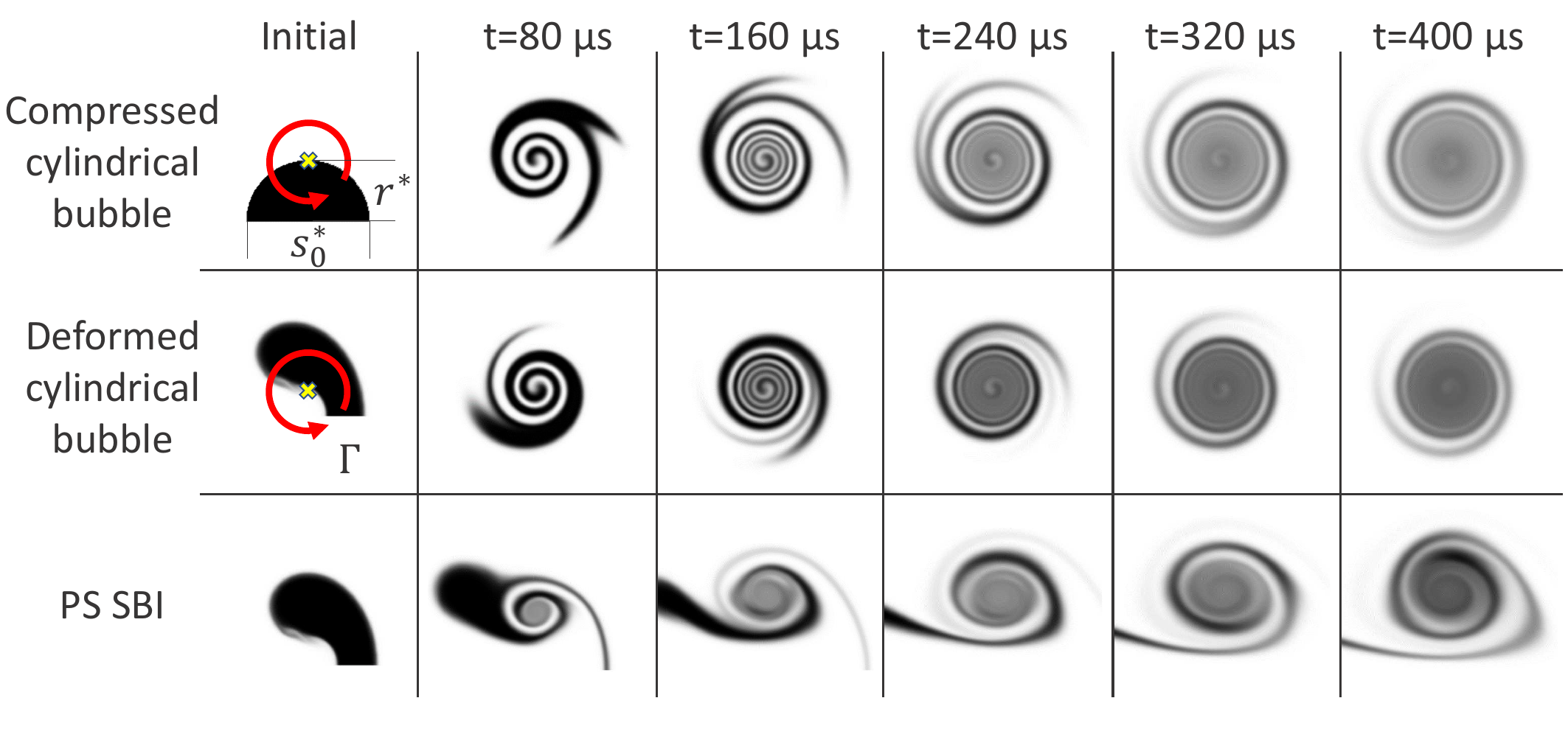}
\caption{Comparisons of three kinds passive scalar mixing temporal evolution. The first and second row is passive scalar mixing under a standard Lamb-Oseen vortex respectively, compared to mixing of $Ma=1.22$ PS SBI.
Note that the first kind is in accordance to the assumption in figure~\ref{fig: mix-mech}(a).
}\label{fig: compress-deform}
\end{figure}

The initial conditions for passive scalar are set as a compressed cylindrical bubble with radius $r^*=\sqrt{\eta}r=2.28$~mm and width $s_0^*=\sqrt{\eta}s_0=4.57$~mm, where $r=2.6$~mm is the pre-shocked cylindrical bubble radius and $\eta=0.771$ is compression rate for $Ma=1.22$ PS SBI. Figure~\ref{fig: compress-deform} plots the initial conditions for a compressed scalar bubble under an ideal Lamb-Oseen vortex.
Based on the estimation of Batchelor scale in this case $\lambda_b\approx9.12\times10^{-5}$~m, the grid size is chosen as $4\times10^{-5}$~m, which is sufficient to capture the intermolecular mixing. The corresponding time step is selected as $8\times10^{-8}$~s.

Figure~\ref{fig: compress-deform} compares three kinds of passive scalar mixing, namely the compressed cylindrical bubble, deformed cylindrical bubble induced by shock and PS SBI. The first and second kind is under the same ideal Lamb-Oseen vortex with conditions comparable to a $Ma=1.22$ PS SBI. As for the compressed bubble case, the vortex center is set on the top of the bubble, as the basic assumption for mixing time model in figure~\ref{fig: mix-mech}(a).
To compare with PS SBI, we set another deformed bubble case, which presumes that the pattern of scalar is initially the same the one in PS SBI, except that a comparable mature vortex is formed immediately.
The vortex center is set on the location of the maximum vorticity. As for PS SBI, the vorticity is initially deposited along the deformed bubble edge and evolves into the main vortex by itself at some later time.

From the qualitative comparisons of scalar mixing characteristics between three cases in figure~\ref{fig: compress-deform}, two observations can be found.
First, by comparing compressed bubble case with deformed bubble case, the temporal patterns of scalar mixing are similar in general. The scalar is stirred by vortex and forms into a solenoid shape. The concentration decay at the edge of bubble away from vortex centre is slowest, which validates that using $s_0^*$ and $r^*$ in mixing time model (\ref{eq: t-s*}) is suitable.
Second, by comparing deformed bubble case and PS SBI case, a remarkable similarity can be observed particularly at the region around the vortex of PS SBI. This means that using a point vortex model to estimate the mixing behaviour in PS SBI is reasonable to some extent.  Since the main vortex is formed through vorticity merging at later time in PS SBI, the mixing process is slower than the one in deformed bubble case.
\begin{figure}
\centering
\includegraphics[clip=true,trim=5 5 18 20, width=0.55\textwidth]{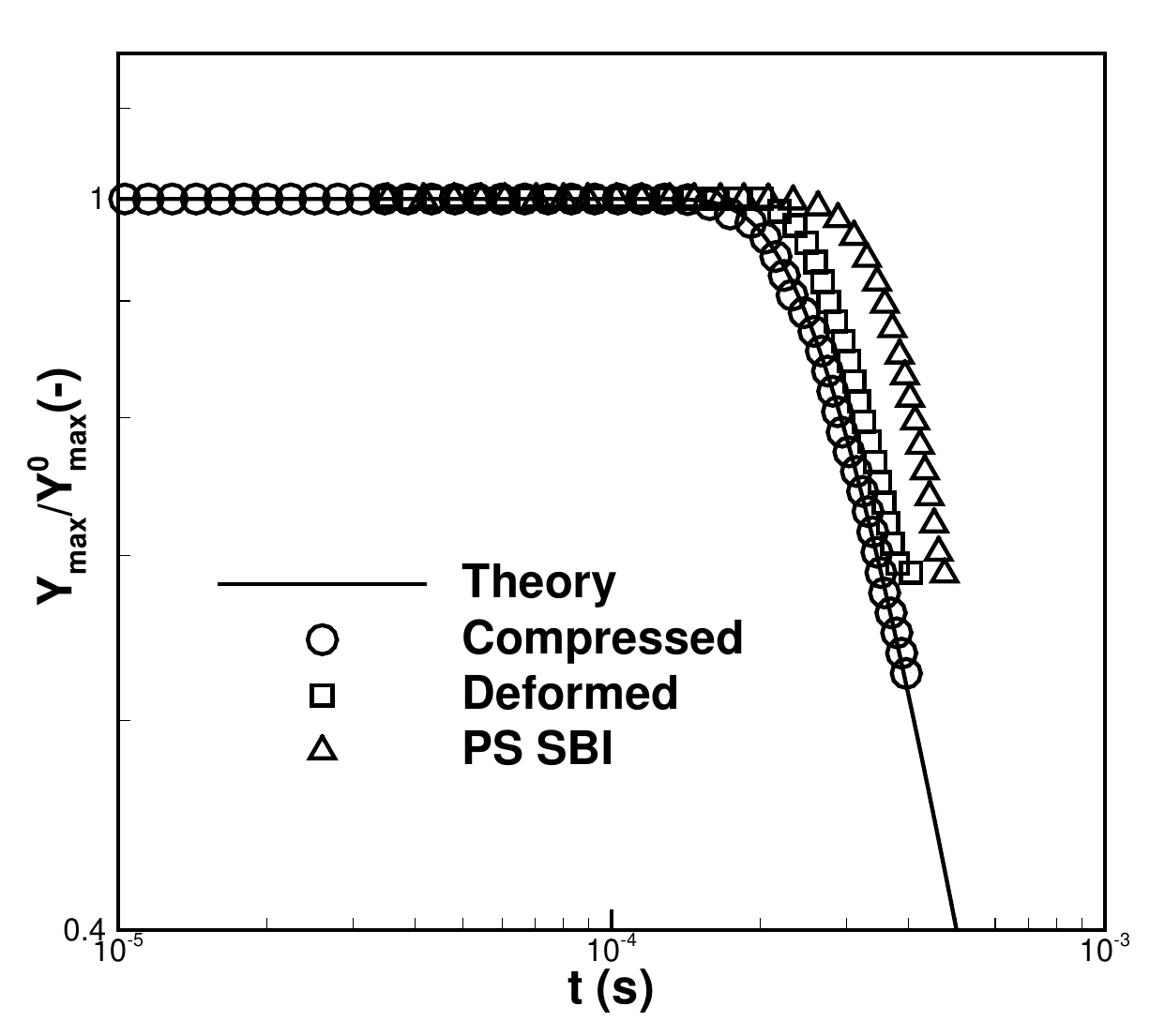}
\caption{Decay of maximum mass fraction from three kinds of passive scalar mixing in figure~\ref{fig: compress-deform}. Theoretical mixing model~(\ref{eq: theo}) with $r^*$ and $s_0^*$ is plotted for comparison. }
\label{fig: com-deform-line}
\end{figure}

Quantitative comparison of maximal concentration decay between three cases is depicted in figure~\ref{fig: com-deform-line}. The theoretical prediction (\ref{eq: theo}) with $r^*$ and $s_0^*$ (corresponding to $t_m^*$ (\ref{eq: t-s*})) is also plotted.
The decrease pattern of concentration in compressed bubble agrees well with that of theory, which indicates that mixing time model behaves well even in scalar with irregular shape (note that a cylindrical bubble in this case and a square-shape scalar in the experiment of~\citet{meunier2003vortices}). The maximum concentration decay in deformed bubble case is slightly slower than the model prediction, while it becomes closer to the result of PS SBI. The minor differences between the three cases illustrate that if a mature vortex forms earlier, the theoretical predicted mixing time is more accurate. In general, the mixing behaviour in PS SBI shares the similarity with that in an ideal Lamb-Oseen vortex, which supports the basic point vortex model assumption for present study.

\bibliographystyle{jfm}
\bibliography{jfm-instructions}

\end{document}